\documentclass[final,3p,times]{elsarticle}
\usepackage{graphics}
\usepackage{amssymb}
\usepackage[nodots,nocompress]{numcompress}
\biboptions{super}
\journal{Physics Reports}
\begin{document}
\begin{frontmatter}
\title{Superconducting Metamaterials}
\author[a,b]{N. Lazarides}
\author[a,b,c]{G. P. Tsironis}
\address[a]{National University of Science and Technology "MISiS", Leninsky 
            Prospekt 4, Moscow, 119049, Russia}
\address[b]{Department of Physics, University of Crete, P. O. Box 2208, 71003 
            Heraklion, Greece}
\address[c]{School of Engineering and Applied Sciences, Harvard University, 
            Cambridge, Massachusetts 02138, USA}
\begin{abstract}
Metamaterials, i.e. artificial, man-made media designed to achieve properties 
not available in natural materials, have been the focus of intense research 
during the last two decades. Many properties have been discovered and multiple 
designs have been devised that lead to multiple conceptual and practical 
applications. Superconducting metamaterials made of superconducting metals have 
the advantage of ultra low losses, a highly desirable feature. The additional 
use of the celebrated Josephson effect and SQUID (superconducting quantum 
interference device) configurations enrich the domain of superconducting 
metamaterials and produce further specificity and functionality. SQUID-based 
metamaterials are both theoretically investigated but also  fabricated and 
analyzed experimentally in many laboratories and exciting new phenomena have 
been found both in the classical and quantum realms. The enticing feature of a 
SQUID is that it is a unique nonlinear oscillator that can be actually 
manipulated through multiple external means. This domain flexibility is 
inherited to SQUID-based metamaterials and metasurfaces, i.e. extended units 
that contain a large arrangement of SQUIDs in various interaction 
configurations. Such a unit can be viewed theoretically as an assembly of weakly 
coupled nonlinear oscillators and as such presents a {\it nonlinear dynamics 
laboratory} where numerous, classical as well as quantum complex, 
spatio-temporal phenomena may be explored. In this review we focus primarily on 
SQUID-based  superconducting metamaterials and present basic properties related 
to their individual and collective responses to external drives; the work 
summarized here is primarily theoretical and computational with nevertheless 
explicit presentation of recent experimental works. We start by showing how a 
SQUID-based system acts as a genuine metamaterial with right as well as left 
handed properties, demonstrate that the intrinsic Josephson nonlinearity leads 
to wide-band tunability, intrinsic nonlinear as well as flat band localization. 
We explore further exciting properties such as multistability and 
self-organization and the emergence of counter-intuitive chimera states of 
selective, partial organization. We then dwell into the truly quantum regime and 
explore the interaction of electromagnetic pulses with superconducting qubit 
units where the coupling between the two yields phenomena such as self-induced 
transparency and superradiance. We thus attempt to present the rich behavior of 
coupled superconducting units and point to their basic properties and practical 
utility.
\end{abstract}
\end{frontmatter}
\newpage

\section*{Contents}
\label{Contents-S0}
\begin{table}[h!]
  \centering  \label{table1}
        \begin{tabular}{lrlr}  \hline  
  &   & {\em Abstract}      & 1 \\
  &   & {\em Contents}      & 2 \\
1.&   & {\em Introduction}  & 3 \\
  &1.1& \hspace{1cm} Metamaterials $\&$ Synthetic Media: Concepts and Perspectives.   & 3 \\
  &1.2& \hspace{1cm} Nonlinear, Superconducting, and Active Metamaterials.            & 4 \\
  &1.3& \hspace{1cm} Superconducting Metamaterials from Zero to Terahertz Frequencies.& 5 \\
  &1.4& \hspace{1cm} Summary of Prior Work in Superconducting Metamaterials.          & 6 \\
  &1.5& \hspace{1cm} SQUID Metamaterials.                                             & 9 \\

2.&   & {\em SQUID-Based Metamaterials I: Models and Collective Properties}           &11 \\
  &2.1& \hspace{1cm} The rf-SQUID as an artificial magnetic "atom".                   &11 \\
  &2.2& \hspace{1cm} SQUID Metamaterial Models and Flux Wave Dispersion.              &18 \\
  &2.3& \hspace{1cm} Wide-Band SQUID Metamaterial Tunability with dc Flux.            &23 \\
  &2.4& \hspace{1cm} Energy Transmission in SQUID Metamaterials.                      &26 \\
  &2.5& \hspace{1cm} Multistability and Self-Organization in Disordered SQUID Metamaterials.&30 \\

3.&   & {\em SQUID-Based Metamaterials II: Localization and Novel Dynamic States}     &36 \\
  &3.1& \hspace{1cm} Intrinsic Localization in Hamiltonian and Dissipative Systems.   &36 \\
  &3.2& \hspace{1cm} Dissipative Breathers in SQUID Metamaterials.                    &36 \\
  &3.3& \hspace{1cm} Collective Counter-Intuitive Dynamic States.                     &39 \\
  &3.4& \hspace{1cm} Chimera States in SQUID Metamaterials.                           &40 \\
  &   & 3.4.1 \hspace{1cm} SQUID Metamaterials with Non-Local Coupling.               &40 \\
  &   & 3.4.2 \hspace{1cm} SQUID Metamaterials with Local Coupling.                   &46 \\

4.&   & {\em SQUID Metamaterials on Lieb Lattices.}                                   &51 \\
  &4.1& \hspace{1cm} Nearest-Neighbor Model and Frequency Spectrum.                   &51 \\
  &4.2& \hspace{1cm} From flat-Band to Nonlinear Localization.                        &52 \\

5.&   & {\em Quantum Superconducting Metamaterials.}                                  &56 \\
  &5.1& \hspace{1cm} Introduction.                                                    &56 \\
  &5.2& \hspace{1cm} Superconducting Qubits.                                          &57 \\
  &5.3& \hspace{1cm} Self-Induced Transparency, Superradiance, and Induced Quantum Coherence. & 59 \\
  &   & 5.3.1 \hspace{1cm}Description of the Model System.                            &59 \\
  &   & 5.3.2 \hspace{1cm}Second Quantization and Reduction to Maxwell-Bloch Equations.&60 \\ 
  &   & 5.3.3 \hspace{1cm}Approximations and Analytical Solutions.                    &62 \\
  &   & 5.3.4 \hspace{1cm}Numerical Simulations.                                      &64 \\

6.&   & {\em Summary}                                                                 &69 \\
  &   & {\em Acknowledgements}                                                        &70 \\
  &   & {\em Appendix}: Derivation of the Maxwell-Bloch-sine-Gordon equations.        &71 \\
  &   & {\em References}                                                              &76 \\
\hline
\end{tabular} 
\end{table}
\newpage

\section{Introduction}
\label{Introduction-S1}
\subsection{Metamaterials $\&$ Synthetic Media: Concepts and Perspectives}
\label{Introduction-S1.1}
\emph{Metamaterials} represent a new class of materials generated by the 
arrangement of artificial structural elements, designed to achieve advantageous 
and/or unusual properties that do not occur in natural materials. In particular, 
naturally occurring materials show a limited range of electrical and magnetic 
properties, thus restricting our ability to manipulate light and other forms of 
electromagnetic waves. The functionality of metamaterials, on the other hand, 
relies on the fact that their constitutive elements can be engineered so that 
they may achieve access to a widely expanded range of electromagnetic properties.
Although metamaterials are often associated with negative refraction, this is 
only one manifestation of their possible fascinating behaviors; they also 
demonstrate negative permittivity or permeability, cloaking capabilities 
\cite{Schurig2006}, perfect lensing \cite{Pendry2000}, high frequency magnetism 
\cite{Linden2006}, classical electromagnetically induced transparency
\cite{Papasimakis2008,Kurter2011,Jin2013,CaihongZhang2017}, as well as dynamic
modulation of Terahertz (THz) radiation \cite{ChunLi2017}, among other 
properties. High-frequency magnetism, in particular, exhibited by {\em magnetic 
metamaterials}, is considered one of the "forbidden fruits" in the Tree of 
Knowledge that has been brought forth by metamaterial research 
\cite{Zheludev2010}. Their unique properties allow them to form a material base 
for other functional devices with 
tuning and switching capabilities \cite{Zheludev2010,Zheludev2011,Tong2018}. 
The scientific activity on metamaterials which has exploded since their first 
experimental demonstration \cite{Smith2000,Shelby2001}, has led to the emergence 
of a new, rapidly growing interdisciplinary field of science. This field has 
currently progressed to the point where physicist, material scientists and 
engineers are now pursuing applications, in a frequency region that spans several 
orders of magnitude, from zero \cite{Wood2007,Magnus2008,Navau2009,Mawatari2012,
Mach-Batlle2017} to THz \cite{Yen2004,Linden2004,Withayachumnankul2009,Gu2010,
Jin2010,Chen2010,Zhang2013a} and optical \cite{Linden2006,Shalaev2007,
Soukoulis2007,Litchinitser2008}. Historically, the metamaterial concept goes 
back to 1967 \cite{Veselago1967}, when V. Veselago investigated hypothetical 
materials with simultaneously negative permeability and permittivity with 
respect to their electromagnetic properties. He showed that simultaneously 
negative permeability and permittivity result in a negative refractive index 
for such a medium, which would bend the light the "wrong" way. The realization 
of materials with simultaneously negative permeability and permittivity, 
required for negative refractive index, had however to wait until the turn of 
the century, when D. Smith and his collaborators demonstrated for the first 
time a structure exhibiting negative refraction in the microwaves 
\cite{Smith2000}. The first metamaterial was fabricated by two interpenetrating 
subsystems, one them providing negative 
permittivity while the other negative permeability within the same narrow 
frequency band. Specifically, an array of thin metallic wires and an array of 
metallic rings with a slit (split-ring resonators), which were fabricated 
following the "recipies" in the seminal works of J. B. Pendry, provided the 
negative permeability \cite{Pendry1999} and the negative permittivity 
\cite{Pendry1996}, respectively. The wires and the split-rings act as 
electrically small resonant "particles", undertaking the role of atoms in 
natural materials; however, they are themselves made of conventional materials 
(highly conducting metals). Accordingly, a metamaterial represents a higher 
level of structural organization of matter, which moreover is man-made. 

The key element for the construction of metamaterials has customarily been the 
split-ring resonator (SRR), which is a subwavelength "particle"; in its simplest 
version it is just a highly conducting metallic ring with a slit. The SRR and 
all its subsequent versions, i.e., U particles, H particles, $\Omega$ or 
$\Omega-$like particles, double and/or multislit SRR molecules, are resonant 
particles which effectively act as artificial "magnetic atoms" 
\cite{Caputo2012}. The SRRs can be regarded as inductive-resistive-capacitive 
($RLC$) oscillators, featuring a self-inductance $L$, a capacitance $C$, and a 
resistance $R$, in an electromagnetic field whose wavelength much larger than 
their characteristic dimension. As long as a metamaterial comprising SRRs is 
concerned, the wavelength of the electromagnetic field has to be much larger 
than its unit cell size; then the field really "sees" the structure as a 
homogeneous medium at a macroscopic scale and the macroscopic concepts of 
permittivity and permeability become meaningful. The (effective) homogeneity is 
fundamental to the metamaterial concept, as it is the ability to structure a 
material on a scale less than the wavelength of the electromagnetic field of 
interest. Although in microwaves this is not a problem, downsizing the scale of 
metamaterial elements to access the optical frequency range may be a non-trivial 
issue. The advent of metamaterials has led to structures with many different 
designs of elemental units and geometries, that may extend to one 
\cite{Shamonina2004,Butz2013a}, two \cite{Shelby2001,Mawatari2012}, or three 
dimensions \cite{Zagoskin2012}. One of the most investigated metamaterial 
designs which does not contain SRRs is the fishnet structure and its versions in 
two \cite{Kafesaki2007}, quasi-two \cite{Wuestner2010}, and three dimensions 
\cite{NaLiu2008,Valentine2008}. However, all these metamaterials have in common 
that they owe their extraordinary electromagnetic properties more to their 
carefully designed and constructed internal structure rather than, e.g., 
chemical composition of their elements. Metamaterials comprising of split-rings 
or some other variant of resonant elements, are inherently discrete; 
discreteness effects do not however manifest themselves as long as the 
metamaterial responds linearly (low-field intensities) and the homogeneous 
medium approximation holds. The coupling effects, however, in relatively dense 
SRR metamaterials are of paramount importance for a thorough understanding of 
certain aspects of their behavior, since they introduce spectral splitting 
and/or resonant frequency shifts \cite{Gay-Balmaz2002,Hesmer2007,Penciu2008,
NaLiu2009,Sersic2009,NaLiu2010,Feth2010}. The SRRs are coupled to each other 
through non-local magnetic and/or electric dipole-dipole interaction, with 
relative strength depending on the relative orientation of the SRRs in an array. 
However, due to the nature of the interaction, the coupling energy between 
neighboring SRRs is already much less than the characteristic energy of the 
metamaterial; thus in most cases next-nearest and more distant neighbor 
interactions can be safely neglected. SRR-based metamaterials support a new kind 
of propagating waves, referred to as magnetoinductive waves, for metamaterials 
where the magnetic interaction between its units is dominant. They exhibit 
phonon-like dispersion curves and they can transfer energy \cite{Shamonina2004,
Syms2010}, and they have been experimentally investigated both in linear and 
nonlinear SRR-based metamaterials \cite{Wiltshire2003,Dolling2006,Shadrivov2007}. 
It is thus possible to fabricate contact-free data and power transfer devices 
which make use of the unique properties of the metamaterial structure, and may 
function as a frequency-selective communication channel for devices via their 
magneto-inductive wave modes \cite{Stevens2010}.

Unfortunately, metamaterials structures comprising of resonant metallic elements 
revealed unfavorable characteristics that render them unsuitable for most 
practical applications. The SRRs, in particular, suffer from high Ohmic losses 
at frequencies close to their resonance, where metamaterials acquire their 
extraordinary properties. Moreover, those properties may only appear within a 
very narrow band, that is related to the weak coupling between elements. High 
losses thus hamper any substantial progress towards the practical use of these 
metamaterials in novel devices. Many applications are also hampered by the lack 
of tuning capabilities and relatively bulky size. However, another breakthrough 
came with the discovery of non-resonant, transmission line negative refractive 
index metamaterials \cite{Eleftheriades2002,Caloz2002}, which very quickly led 
to several applications, at least in the microwaves \cite{Caloz2009}. 
Transmission line metamaterials rely on the appropriate combination of 
inductive-capacitive ($L\,C$) lumped elements into large networks. The tremendous 
amount of activity in the field of metamaterials since $\sim 2000$ has been 
summarized in various reviews \cite{Smith2004,Caloz2005,Linden2006,Padilla2006,
Shalaev2007,Litchinitser2008,Anlage2011,Soukoulis2011,YLiu2011,Simovski2012} 
and books \cite{Eleftheriades2005,Caloz2006,Engheta2006,Pendry2007,Marques2007,
Krowne2007,Ramakrishna2009,TJCui2010,Cai2010,Solymar2009,Noginov2012,Tong2018}. 

\subsection{Nonlinear, Superconducting, and Active Metamaterials}
\label{Introduction-S1.2}
Dynamic tunability is a property that is required for applications \cite{Fan2015}; 
in principle, one should be able to vary the effective (macroscopic) parameters 
of a metamaterial in real time, simply by varying an applied field. Tunability 
provides the means for fabricating meta-devices with switching capabilities 
\cite{Zheludev2010,Zheludev2011}, among others, and it can be achieved by the 
introduction of nonlinearity. Nonlinearity adds new degrees of freedom for 
metamaterial design that allows for both tunability and multistability - another 
desired property, that may offer altogether new functionalities and 
electromagnetic characteristics \cite{Lapine2017}, as well as wide-band 
permeability \cite{Lapine2012}. It was very soon after the first demonstration 
of metamaterials, named at that time as {\em negative refractive index materials},
when it became clear that the SRR structure has considerable potential to 
enhance nonlinear effects due to the intense electric fields which can be 
generated in their slits \cite{OBrien2004}. Following these ideas, several 
research groups have demonstrated nonlinear metamaterial units, by filling the 
SRR slits with appropriate materials, e.g., with a strongly nonlinear dielectric 
\cite{Hand2008}, or with a photo-sensitive semiconductor. Other approaches have 
made use of semiconducting materials, e.g., as substrates, on which the actual 
metamaterial is fabricated, that enables modulation of THz transmission by 
$50\%$ \cite{HTChen2006}. However, the most convenient method for introducing 
nonlinearity in SRR-based metamaterials was proved to be the insertion of 
nonlinear electronic components into the SRR slits, e.g., a variable capacitance 
diode (varactor) \cite{Powell2007,Wang2008}. The dynamic tunability of a 
two-dimensional metamaterial comprising varactor-loaded SRRs by the power of an 
applied field has been demonstrated experimentally \cite{Shadrivov2008}. Both 
ways of introducing nonlinearity affect the capacitance $C$ of the SRRs which 
becomes field-dependent; in the equivalent electrical circuit picture, in which 
the SRRs can be regarded as lumped element electrical oscillators, the 
capacitance $C$ acquires a voltage dependence and in turn a field-dependent 
magnetic permeability. Nonlinear transmission line metamaterials are reviewed 
in Ref. \cite{Kozyrev2008}.

Nonlinearity does not however help in the reduction of losses; in nonlinear 
metamaterials the losses continue to be a serious problem. The quest for 
{\em loss compensation} in metamaterials is currently following two different 
pathways: a "passive" one, where the metallic elements are replaced by 
superconducting ones \cite{Anlage2011,Jung2014}, and an "active" one, where 
appropriate constituents are added to metallic metamaterials that provide gain 
through external energy sources. In order to fabricate both nonlinear and 
{\em active metamaterials}, gain-providing electronic components such as tunnel 
(Esaki) diodes \cite{Esaki1958} or particular combinations of other 
gain-providing devices have to be utilized. The Esaki diode, in particular, 
features a negative resistance part in its current-voltage characteristics, and 
therefore can provide both gain and nonlinearity in a conventional (i.e., 
metallic) metamaterial. Tunnel diodes which are biased so that they operate at 
the negative resistance region of their characteristics may also be employed for 
the construction of ${\cal PT}-$symmetric metamaterials, that rely on balanced 
gain and loss \cite{Lazarides2013a}. ${\cal PT}-$symmetric systems correspond 
to a new paradigm in the realm of artificial or "synthetic" materials that do 
not obey separately the parity ($\cal P$) and time ($\cal T$) symmetries; 
instead, they do exhibit a combined ${\cal PT}$ symmetry \cite{Ruter2010,
Regensburger2012}. The notions of ${\cal PT}-$symmetric systems originate for 
non-Hermitian quantum mechanics \cite{Bender1998,Bender2002}, but they have been 
recently extended to optical lattices \cite{ElGanainy2007,Makris2008}. The use 
of active components which are incorporated in metamaterial unit elements has 
been actually proposed several years ago \cite{Boardman2007}, and it is currently 
recognized as a very promising technique of compensating losses 
\cite{Boardman2010}. Low-loss and active negative index metamaterials by 
incorporating gain material in areas with high local field have been demonstrated 
in the optical \cite{Xiao2010}. Recently, transmission lines with periodically 
loaded tunnel diodes which have the negative differential resistance property 
have been realized and tested as low-loss metamaterials, in which intrinsic 
losses are compensated by gain \cite{Jiang2011}. Moreover, a combination of 
transistors and a split-ring has been shown to act as a loss-compensated 
metamaterial element \cite{Xu2012}. In the latter experiment, the quality factor 
for the combined system exhibits huge enhancement compared with that measured 
for the split-ring alone.      

The "passive" approach to loss reduction employes superconducting materials, 
i.e, materials exhibiting absence of dc resistance below a particular 
temperature, known as the critical temperature, $T_c$. A rough classification of 
the superconducting materials is made on the basis of their critical temperature; 
according to that, there are low-$T_c$ and high-$T_c$ superconducting materials. 
The former include primarily elemental and binary compounds, like Niobium (Nb), 
Niobium di-Selenide (NbSe$_2$) and more recently Niobium Nitride (NbN), while 
the most known representatives of the latter are the superconducting perovskites 
such as Yttrium-Barium-Copper-Oxide (YBCO). The latter is the most commonly used 
perovskite superconductor which typically has a critical temperature 
$T_c \sim 90 K$, well above the boiling point of liquid Nitrogen. The last few 
years, there has been an increasing interest in {\em superconducting 
metamaterials} that exploit the zero resistance property of superconductivity, 
targeting at severe reduction of losses and the emergence of intrinsic 
nonlinearities due to the extreme sensitivity of the superconducting state to 
external stimuli \cite{Zheludev2011,Anlage2011}. The direct approach towards 
fabrication of superconducting metamaterials relies on the replacement of the 
metallic split-rings of the conventional SRR-based metamaterials by 
superconducting ones. More shopisticated realizations of superconducting 
metamaterials result from the replacement of the metallic SRRs by rf SQUIDs 
(Superconducting QUantum Interference Devices) \cite{Likharev1986}; those SQUID 
metamaterials are discussed below.

Superconducting metamaterials are not however limited to the above mentioned 
realizations, but they also include other types of artificial metamaterials;
thin superconducting plates have been used in a particular geometrical 
arrangement to "beat the static" \cite{Narimanov2008} and make possible a zero 
frequency metamaterial (dc metamaterial) \cite{Wood2007,Magnus2008,Navau2009,
Gomory2012,Mawatari2012}. Other types of superconducting metamaterials in the 
form of heterostructures, where superconducting layers alternate with 
ferromagnetic or non-magnetic metallic layers have been shown to exhibit 
electromagnetically induced transparency \cite{Kurter2011,Wu2011a,Jin2013}, 
switching capabilities \cite{Kurter2012}, magnetic cloaking, and concentration 
\cite{Prat-Camps2013}. Recently, tunable electromagnetically induced transparency
has been demonstrated in a Niobium Nitride (NbN) terahertz (THz) superconducting 
metamaterial. An intense THz pulse is used to induce nonlinearities in the NbN 
thin film and thereby tune the electromagnetically induced transparency-like 
behavior \cite{CaihongZhang2017}. Furthermore, the dynamic process of parity-time
($\cal PT$) symmetry breaking was experimentally demonstrated in a hybridized 
metasurface which consists of two coupled resonators made from metal and NbN
\cite{Wang2017}. Negative refraction index metamaterials in the visible 
spectrum, based on MgB$_2$/SiC composites, have been also realized 
\cite{Limberopoulos2009}, following prior theoretical investigations 
\cite{Kussow2007}. Moreover, there is substantial evidence for negative 
refraction index in layered superconductors above the plasma frequency of the 
Josephson plasma waves \cite{Golick2010}, that was theoretically investigated 
by several authors \cite{Pimenov2005,Rakhmanov2010}. Other types of 
superconducting metamaterials include those made of magnetically active planar 
spirals \cite{Kurter2011b}, as well as those with rather special ("woodcut") 
geometries \cite{Savinov2012c}, two-dimensional arrays of Josephson junctions 
\cite{Adams2013}, as well as superconducting "left-handed" transmission lines 
\cite{Salehi2005,Wang2006}. Recently, in a novel one-dimensional Josephson 
metamaterial composed of a chain of asymmetric SQUIDs, strong Kerr nonlinearity
was demonstrated \cite{WenyuanZhang2017}. Moreover, the Kerr constant was 
tunable over a wide range, from positive to negative values, by a magnetic flux
threading the SQUIDs.

\subsection{Superconducting Metamaterials from Zero to Terahertz Frequencies}
\label{Introduction-S1.3}
There are several demonstrations of superconducting metamaterial elements which 
exhibit tunability of their properties by varying the temperature or the applied 
magnetic field \cite{Ricci2005,Ricci2007,Gu2010,Fedotov2010,Chen2010,Jung2013,
Trang2013}. We should also mention the theoretical investigations (nonlinear 
circuit modeling) on a multi-resonant superconducting split-ring resonator
\cite{Mazdouri2017}, and on a "meta-atom" composed of a direct current (dc)
SQUID and a superconducting rod attached to it, which exhibits both electric
and magnetic resonant response \cite{Shramkova2017}. Superconducting split-rings 
combined into two-dimensional planar arrays form superconducting metamaterials 
exhibiting tunability and switching capabilities at microwave and THz
frequencies \cite{Gu2010,Ricci2007,Kurter2010,Jin2010,Chen2010,Zhang2011,Wu2011b,
Zhang2012,Zhang2013b,Zhang2013a,Grady2013}. Up to the time of writing, 
metamaterials comprising superconducting SRRs employ one of the following 
geometries:

(i)   square SRRs with rectangular cross-section in the double, narrow-side 
      coupled SRR geometry \cite{Ricci2005,Ricci2006,Ricci2007};

(ii)  circular, asymmetrically split-rings 
      \cite{Fedotov2010,Savinov2012,Savinov2013};

(iii) square SRRs with rectangular cross-section in the single SRR geometry 
      \cite{Gu2010};

(iv)  electric inductive-capacitive SRRs of two different types 
      \cite{Singh2012a}. 

Also, novel metamaterial designs including a "woodcut" type superconducting 
metamaterial, and niobium-connected asymmetrically split-ring metamaterials were 
demonstrated \cite{Savinov2013}. All these metamaterials were fabricated in the 
planar geometry, using either conventional, low$-T_c$ superconductors such as 
niobium (Nb) and niobium nitride films, or the most widely used member of the 
high$-T_c$ superconductor family, i.e., the yttrium-barium-copper-oxide ($YBCO$). 
The experiments were performed in microwaves and in the (sub-)THz range 
($\sim 0.1 -2~THz$).

All these superconducting metamaterials share a common feature: they all 
comprise resonant sub-wavelength superconducting elements, that exhibit a strong 
response at one particular frequency, i.e., the resonance frequency, $f_0$. That 
resonance frequency is tunable under external fields, such as temperature, 
constant (dc) and time-periodic (ac) magnetic fields, and applied current, due 
to the extreme sensitivity of the superconducting state to external stimuli. 
(Note however that for some geometries there can be more than one strong 
resonances.) The experimental investigation of the resonances and their ability 
for being shifted either to higher or lower frequencies relies on measurements 
of the complex transmission spectrums, with dips signifying the existence of 
resonances. However, not only the frequency of a resonance but also its quality 
is of great interest in prospective applications. That quality is indicated by 
the depth of the dip of the complex transmission magnitude in the corresponding 
transmission spectrum, as well as its width, and quantified by the corresponding 
quality factor $Q$. In general, the quality factor increases considerably as the 
temperature decreases below the critical one at $T_c$. Other factors, related to 
the geometry and material issues of the superconducting SRRs that comprise the 
metamaterial, also affect the resonance frequency $f_0$. Thus, the resonance 
properties of a metamaterial can be engineered to achieve the desired values, 
just like in conventional metamaterials. However, for superconducting 
metamaterials, the thickness of the superconducting film seems to be an 
important parameter, because of the peculiar magnetic properties of 
superconductors. Using proper design, it is possible to switch on and off the 
resonance in superconducting metamaterials in very short time-scales, providing 
thus the means of manufacturing devices with fast switching capabilities.    

\subsection{Summary of earlier work in superconducting metamaterials}
\label{Introduction-S1.4}
In this Subsection, a brief account is given on the progress in the development 
and applications of superconducting (both classical and quantum) metamaterials, 
i.e., metamaterials utilizing either superconducting materials or devices, is 
given. A more detailed and extended account is given in two review articles on
the subject \cite{Anlage2011,Jung2014}, as well as in Chapter $5.5$ of a recently
published book \cite{Tong2018}. The status of the current research on SQUID 
metamaterials is discussed separately in the next Subsection (Subsection $1.5$).
In the older of the two review articles \cite{Anlage2011}, the properties of 
superconductors which are relevant to superconducting metamaterials, and the
advantages of superconducting metamaterials over their normal metal counterparts
are discussed. The author reviews the status of superconductor-ferromagnet 
composites, dc superconducting metamaterials, radio frequency (rf) superconducting 
metamaterials, and superconducting photonic crystals (although the latter fall 
outside the domain of what are usually called metamaterials). There is also a 
brief discussion on SQUID metamaterials, with reference to the theoretical works 
in which it was proposed to use an array of rf SQUIDs as a metamaterial 
\cite{Du2006,Lazarides2007}. In the second review article \cite{Jung2014}, a more 
detailed account on the advantages of superconducting metamaterials over their 
normal metal counterparts was given, along with an update on the status of 
superconducting metamaterials. Moreover, analogue electromagnetically-induced 
transparency superconducting metamaterials and superconducting SRR-based 
metamaterials are also reviewed. In this review article, there is also a 
discussion on the first experiments on SQUID metamaterials 
\cite{Butz2013a,Trepanier2013,Jung2014b} which have confirmed earlier theoretical
predictions. However, a lot of experimental and theoretical work on SQUID 
metamaterials has been performed after the time of writing of the second review
article in this field. The present review article aims to give an up-to-date and 
extended account of the theoretical and experimental work on SQUID metamaterials 
and reveal their extraordinary nonlinear dynamic properties. SQUID metamaterials 
provide a unique testbed for exploring complex spatio-temporal dynamics. In the 
quantum regime, a prototype model for a "basic" SCQMM which has been investigated 
by several authors is reviewed, which exhibits novel physical properties. Some of 
these properties are discussed in detail.

Superconductivity is a macroscopic quantum state of matter which arises from the 
interaction between electrons and lattice vibrations; as a results, the electrons 
form pairs (Cooper pairs) which condense into a single macroscopic ground state. 
The latter is the equilibrium thermodynamic state below a transition (critical) 
temperature $T_c$. The ground state is separated by a temperature-dependent 
{\em energy gap} $\Delta$ from the excited states with quasi-particles 
(quasi-electrons). As mentioned earlier, the superconductors are roughly 
classified into high and low critical temperature ones (high$-T_c$ and low$-T_c$, 
respectively). In some circumstances, the Cooper pairs can be described in terms 
of a single macroscopic quantum wavefunction $\Psi =\sqrt{n_s}\, \exp[i \theta_s]$, 
whose squared magnitude is interpreted as the local density of superconducting
electrons ($n_s$), and whose phase $\theta_s$ is coherent over macroscopic 
dimensions. Superconductivity exhibits several extraordinary properties, such as 
zero dc resistance and the Meissner effect. Importantly, it also exhibits 
macroscopic quantum phenomena including fluxoid quantization and the Josephson 
effects at tunnel (insulating) barriers and weak links. When two superconductors
$S_L$ and $S_R$ are brought close together and separated by a thin insulating 
barrier, there can be tunneling of Cooper pairs from one superconductor to the 
other. This tunneling produces a supercurrent (Josephson current) between $S_L$ 
and $S_R$, $I_J = I_c \, \sin( \phi_J )$, where $I_c$ is the critical current of 
the Josephson junction and $\phi_J =\phi_L (t) -\phi_R (t) 
-\frac{2\pi}{\Phi_0} \int_{S_L}^{S_R} {\bf A} ({\bf r},t) d{\bf l}$ is the 
gauge-invariant Josephson phase, with $\phi_L$ and $\phi_R$ being the phases of 
the macroscopic quantum wavefunctions of $S_L$ and $S_R$, respectively, 
${\bf A} ({\bf r},t)$ the electromagnetic vector potential in the region between 
$S_L$ and $S_R$, and $\Phi_0 =\frac{h}{2 e} \simeq 2.07\times 10^{-15} ~Wb$ the 
flux quantum ($h$ is Planck's constant and $e$ the electron's charge). Depending 
on whether $\phi_J$ is time-dependent or not, the appearance of the supercurrent 
$I_J$ is referred to as the {\em ac} or the {\em dc Josephson effect}.

Superconductors bring three unique advantages to the development of 
metamaterials in the microwave and sub-THz frequencies which have been 
analyzed in Refs. \cite{Anlage2011,Jung2014}. Namely, (i) low losses (one 
of the key limitations of conventional metamaterials), (ii) the possibility for 
higher compactification of superconducting metamaterials compared to other 
realizations (superconducting SRRs can be substantially miniaturized while still 
maintaining their low-loss properties), and (iii) strong nonlinearities inherent 
to the superconducting state, which allow for tunability and provide switching 
capabilities. The limitations of superconducting metamaterials arise from the 
need to create and maintain a cryogenic environment, and to bring signals to and 
from the surrounding room temperature environment. Quite fortunately, 
closed-cycle cryocoolers have become remarkably small, efficient and inexpensive 
since the discovery of high$-T_c$ superconductors, so that they are now able to 
operate for several years unattended, and moreover they can accommodate the heat 
load associated with microwave input and output transmission lines to room 
temperature. Superconductors can also be very sensitive to variations in 
temperature, stray magnetic field, or strong rf power which can alter their 
properties and change the behavior of the metamaterial. Thus, careful 
temperature control and high quality magnetic shielding are often required for 
reliable performance of superconducting metamaterials.

Superconducting metamaterials exhibit intrinsic nonlinearity because they are 
typically made up of very compact elementary units, resulting in strong currents 
and fields within them. Nonlinearity provides tunability through the variation
of external fields. For example, a connected array of asymmetrically-split Nb 
resonators shows transmission tunable by current at sub-THz frequencies due to 
localized heating and the entrance of magnetic vortices \cite{Savinov2012}. 
The change in superfluid density by a change in temperature was demonstrated for 
a superconducting thin film SRR \cite{Ricci2006}. Later, it was demonstrated that 
the resonant frequency of a Nb SRR changes significantly with the entry of 
magnetic vortices \cite{Ricci2007}. Similar results showing complex tuning with 
magnetic field were later demonstrated at microwaves using high$-T_c$ 
superconducting SRRs \cite{Trang2013} and at sub-THz frequencies with similarly
designed Nb SRRs \cite{Savinov2013}. The nonlinearity associated with the 
resistive transition of the superconductor was exploited to demonstrate a 
bolometric detector at sub-THz frequencies using the collective properties of an
asymmetrically split-ring array made of Nb \cite{Savinov2013}. Thermal tuning 
has been accomplished at THz frequencies by varying the temperature in 
high$-T_c$ (YBCO) metamaterial \cite{Chen2010} and NbN electric
inductive-capacitive thin film resonators \cite{Wu2011b}. Enchancement of 
thermal tunability was accomplished by decreasing the thickness of the 
high$-T_c$ superconducting films which make up square SRRs \cite{Singh2013}. 
Fast nonlinear response can be obtained in superconducting films in THz 
time-domain experiments. In such an experiment it was found that intense THz 
pulses on a NbN metamaterial could break significant number of Cooper pairs to
produce a large quasi-particle density which increases the effective surface 
resistance of the film and modulates the depth of the SRR resonance
\cite{Zhang2013a,Zhang2013b}. The tuning of high$-T_c$ (YBCO) SRR metamaterial
with variable THz beam intensity has demonstrated that the resonance strength
decreases and the resonance frequency shifts as the intensity is increased
\cite{Grady2013}.

A natural opportunity to create a negative real part of the effective magnetic 
permeability $\mu_{eff}$ is offered by a gyromagnetic material for frequencies 
above the ferromagnetic resonance \cite{Chui2002}. However, the imaginary part 
of $\mu_{eff}$ is quite large near the resonance and limits the utility of such
a metamaterial. A hybrid metamaterial, resulting from the combination of the
gyromagnetic material mentioned earlier with a superconductor can help to 
reduce losses \cite{Nurgaliev2008}. A superlattice consisting of high$-T_c$
superconducting and manganite ferromagnetic layers (YBCO/LSMO) was created and 
it was shown to produce a negative index band in the vicinity of the 
ferromagnetic resonance ($\sim 90~GHz$) at magnetic fields between $2.9$ and 
$3.1~T$ \cite{Pimenov2005}. More recently, a metamaterial composed of permanent 
magnetic ferrite rods and metallic wires was fabricated. This metamaterial 
exhibits not only negative refraction but also near-zero refraction, without 
external magnetic field \cite{KeBi2014}.

The concept of a dc metamaterial operating at very low frequencies that could 
make up a dc magnetic cloak has been proposed and investigated in Ref.
\cite{Wood2007}. The first realization of such a metamaterial, which is based 
on non-resonant elements, consists of an array of superconducting plates 
\cite{Magnus2008}. The superconducting elements exclude a static magnetic 
field, and provide the foundation for the diamagnetic effect, when that field is 
applied normal to the plates. The strength of the response depends on the ratio 
between the dimension of the plates and the lattice spacing. An experimental
demonstration of a dc metamaterial cloak was made using an arrangement of Pb 
thin film plates \cite{Magnus2008}. Subsequent theoretical work has refined the 
dc magnetic cloak design and suggested that it can be implemented with 
high$-T_c$ superconducting thin films \cite{Navau2009}. It was later 
demonstrated experimentally that a specially designed cylindrical 
superconductor-ferromagnet bilayer can exactly cloak uniform static magnetic 
fields \cite{Gomory2012}.

Superconducting rf metamaterials consisting of two-dimensional Nb spirals 
developed on quartz substrates show strong tunability as the transition 
temperature is approached \cite{Kurter2010,Kurter2011b}. Rf metamaterials have 
great potential in applications such as magnetic resonance imaging devices for 
non-invasive and high resolution medical imaging \cite{Wiltshire2001}. The
superconducting rf metamaterials have many advantages over their normal metal
counterparts, such as reducing considerably the Ohmic losses, compact structure, 
and sensitive tuning of resonances with temperature or rf magnetic field, which 
makes them promising for rf applications. Similar, high$-T_c$ rf metamaterials
(in which the spirals are made by YBCO) were also fabricated, which enable
higher operating temperatures and greater tunability \cite{Ghamsari2013}.

The elementary units (i.e., the meta-atoms) in a metamaterial can be combined
into meta-molecules so that the interactions between the meta-atoms can give 
rise to qualitatively new effects, such as the classical analogue of the 
electromagnetically induced transparency (EIT). This effect has been observed 
in asymmetrically-split ring metamaterials in which Fano resonances have been
measured as peaks in the transmission spectrum, corresponding to metamaterial
induced transparency \cite{Fedotov2010}. Metamaterials consisting of normal 
metal - superconductor hybrid meta-molecules can create strong classical 
EIT effects \cite{Kurter2011}. The meta-molecule consists of a gold (Au) strip 
with end caps and two superconducting (Nb) SRRs (the "bright" and the "dark" 
element, respectively). A tunable transparency window which could even be 
switched off completely by increasing the intensity of the signal propagating 
through the meta-molecule was demonstrated \cite{Kurter2011,Kurter2012}. 
EIT effects were also observed in the THz domain utilizing NbN bright and
dark resonators to create a transparency window \cite{Wu2011a}. Further 
experiments on all-superconducting (NbN) metamaterials utilizing strongly
coupled SRR-superconducting ring elements showed enhanced slow-light features
\cite{Jin2013}.

A {\em quantum metamaterial} is meant to be an artificial optical medium that 
(a) comprise quantum coherent unit elements whose parameters can be tailored,
(b) the quantum states of (at least some of) these elements can be directly 
controlled, and (c) maintain global coherence for sufficiently long time. These 
properties make a quantum metamaterial a qualitatively different system 
\cite{Zagoskin2011,Zagoskin2016}. Superconducting quantum metamaterials offer 
nowadays a wide range of prospects from detecting single microwave photons to 
quantum birefringence and superradiant phase transitions \cite{Jung2014}. They 
may also play a role in quantum computing and quantum memories. The last few 
years, novel superconducting devices, which can be coupled strongly to external 
electromagnetic field, can serve as the quantum coherent unit elements of 
superconducting quantum metamaterials (SCQMMs). For example, at ultra-low 
temperatures, superconducting loops containing Josephson junctions exhibit a 
discrete energy level spectrum and thus behave in many aspects as quantum 
meta-atoms. It is very common to approximate such devices as two-level quantum 
systems, referred to as {\em superconducting qubits}, whose energy level 
splitting corresponds to a frequency of the order of a few GHz. The interaction 
between light and a SCQMM is described by photons coupling to the artificial 
two-level systems, i.e., the superconducting qubits. The condition of keeping 
the energy of thermal fluctuations $k_B\, T$, where $k_B$ is Boltzmann's 
constant and $T$ the temperature, below the energy level splitting $h\, f$ of 
the qubit, where $h$ is Planck's constant and $f$ the transition frequency, 
requires temperatures well below $1 ~K$. In the past few years, research on 
superconducting qubits has made enormous progress that paves the way towards 
superconducting qubit-based quantum metamaterials.

There are several theoretical investigations on the physics of one-dimensional 
arrays of superconducting qubits coupled to transmission-line resonators
\cite{Rakhmanov2008,Zagoskin2009,Ian2012,Viehmann2013,Shvetsov2013,Volkov2014,
Asai2015,Ivic2016,Asai2018}. Moreover, two-dimensional \cite{Zueco2012} and 
three-dimensional \cite{Zagoskin2012} SCQMMs based on Josephson junction 
networks were proposed. A more extended discussion of the theoretical works 
on SCQMMs is given in Subsection $5.1$. Still there is little progress in the 
experimental realization of such systems. The first SCQMM which was implemented 
in 2014 \cite{Macha2014}, and comprises $20$ flux qubits arranged in a double 
chain geometry. In that prototype system, the dispersive shift of the resonator  
frequency imposed by the SCQMM was observed. Moreover, the collective resonant 
coupling of groups of qubits with the quantized mode of a photon field was
identified, despite of the relatively large spread of the qubit parameters.
Recently, an experiment on an SCQMM comprising an array of $15$ {\em twin flux 
qubits}, was demonstrated \cite{Shulga2018}. The qubit array is embedded directly 
into the central electrode of an Al coplanar waveguide; each qubit contains $5$ 
Josephson junctions, and it is strongly coupled to the electromagnetic waves 
propagating through the system. It was observed that in a broad frequency range,
the transmission coefficient through that SCQMM depends periodically on the 
external magnetic field. Moreover, the excitation of the qubits in the array 
leads to a large resonant enhancement of the transmission. We undoubtedly expect 
to see more experiments with arrays of superconducting qubits placed in 
transmission lines or waveguides in the near future.

\subsection{SQUID Metamaterials}
\label{Introduction-S1.5}
The rf SQUIDs, mentioned above, are highly nonlinear superconducting devices 
which are long known in the Josephson community and encompass the Josephson 
effect \cite{Josephson1962}. The simplest version of a SQUID is made by a 
superconducting ring which is interrupted by a Josephson junction (JJ); the 
latter is typically formed by two superconductors separated by a thin insulating 
(dielectric) layer. The current through the insulating layer and the voltage 
across the junction are then determined by the celebrated Josephson relations 
and crucially affect the electromagnetic behavior of the rf SQUID. SQUIDs have 
found numerous technological applications in modern science \cite{Kleiner2004,
Clarke2004a,Clarke2004b,Fagaly2006}; they are most commonly used as magnetic 
field sensors, since they can detect even tiny magnetic fields and measure their 
intensity with unprecedented precision. SQUID metamaterials constitute the direct 
superconducting analogue of conventional (metallic) nonlinear (i.e., varactor 
loaded) SRR-based magnetic metamaterials, which result from the replacement of 
the nonlinear SRRs by rf SQUIDs. The latter possess inherent nonlinearity due to 
the Josephson element. Similarly to the conventional (metallic), SRR-based 
magnetic metamaterials, the SQUIDs are coupled magnetically to each other through 
magnetic dipole-dipole interactions. Several years ago, theoretical 
investigations have suggested that rf SQUID arrays in one and two dimensions can 
operate as magnetic metamaterials both in the classical \cite{Lazarides2007} and 
in the quantum regime \cite{Du2006}, and they may exhibit negative and/or 
oscillating effective magnetic permeability in a particular frequency band which 
encloses the resonance frequency of individual SQUIDs. Recent experiments on 
single rf SQUIDs in a waveguide demonstrated directly the feasibility of 
constructing SQUID-based thin-film metasurfaces \cite{Jung2013}. Subsequent 
experiments on one-dimensional, quasi-two-dimensional, and truly two-dimensional 
SQUID metamaterials have revealed a number of several extraordinary properties 
such as negative diamagnetic permeability \cite{Jung2013,Butz2013a}, broad-band 
tunability \cite{Butz2013a,Trepanier2013}, self-induced broad-band transparency 
\cite{Zhang2015}, dynamic multistability and switching \cite{Jung2014b}, as well 
as coherent oscillations \cite{Trepanier2017}. Moreover, nonlinear localization 
\cite{Lazarides2008a} and nonlinear band-opening (nonlinear transmission) 
\cite{Tsironis2014b}, as well as the emergence of dynamic states referred to as 
{\em chimera states} in current literature \cite{Lazarides2015b,Hizanidis2016a}, 
have been demonstrated numerically in SQUID metamaterial models. Those 
counter-intuitive dynamic states, which have been discovered numerically in rings 
of identical phase oscillators \cite{Kuramoto2002}, are reviewed in Refs.
\cite{Panaggio2015,Yao2016}. Moreover, numerical investigations on SQUID 
metamaterials on Lieb lattices which posses a flat band in their frequency 
spectrum, reveal the existence of flat-band localized states in the linear 
regime and the more well-known nonlinearly localized states in the nonlinear 
regime \cite{Lazarides2017}. The interaction of an electromagnetic wave with a 
diluted concentration of a chain of SQUIDs in a thin film suggests a mechanism 
for the excitation of magnetization waves along the chain by a normally incident 
field \cite{Maimistov2010}. In the linear limit, a two-dimensional array of rf
SQUIDs acts as a {\em metasurface} that controls the polarization of an 
electromagnetic wave \cite{Caputo2015}. 

SQUID arrays have been also integrated in larger devices in order to take 
advantage of their extraordinary properties; notably, amplification and 
squeezing of quantum noise has been recently achieved with a tunable 
SQUID-based metamaterial \cite{Castellanos2008}. Other important developments 
demonstrate clearly that SQUID-based metamaterials enable feedback  control of 
superconducting cubits \cite{Riste2012}, observation of Casimir effects 
\cite{Lahteenmaki2013}, measurements of nanomechanical motion below the standard
quantum limit \cite{Teufel2009}, and three-wave mixing \cite{Roch2012}. At 
sufficiently low (sub-Kelvin) temperatures, SQUID metamaterials provide access 
to the quantum regime, where rf SQUIDs can be manipulated as flux and phase 
qubits \cite{Poletto2009,Castellano2010}. Recently, the technological advances 
that led to nano-SQUIDs make possible the fabrication of SQUID metamaterials at 
the nanoscale \cite{Wernsdorfer2009}.

From the above discussion it should be clear that the field of superconducting 
metamaterials, in which superconductivity plays a substantial role in 
determining their properties, has expanded substantially. In this review, we 
focus on the SQUID metamaterials, that represent an area of the field of 
superconducting metamaterials, which however has already reached a level of 
maturity. We also focus on SCQMMs, and in particular on a prototype model for 
a chain of charge qubits in a transmission-line resonator \cite{Rakhmanov2008}.
The SCQMMs are related to the (classical, i.e., not truly quantum) SQUID 
metamaterials in that they also encompass the Josephson effect. In Section 2, 
we describe the SQUID metamaterial models used for simulating real systems in 
current research, we provide the corresponding dispersion of flux waves which 
can propagate in SQUID metamaterials, and we present numerical results (along 
with selected experimental ones), which reveal novel properties such as wide-band 
tunability, energy transmission, and multistability. In Section 3, we present 
and discuss results on nonlinear localization in SQUID metamaterials, which leads
to the generation of states referred to as discrete breathers. In that Section, 
we also emphasize the possibility for the emergence of chimera states in SQUID 
metamaterial models with either nonlocal or local (nearest-neighbor) couplng 
between their elements (i.e., the SQUIDs). In Section 4, the dynamical model for 
SQUID metamaterials on a Lieb lattice is presented, along with its full frequency 
spectrum. The latter contains a flat band, which allows for the formation of 
flat-band localized states in the linear regime. The case of nonlinearly 
localized states, which can be formed in the nonlinear regime, as well as the 
transition between the two regimes, is investigated. In Section 5, we describe a 
model SCQMMs (a chain of charge qubits in a superconducting transmission-line
resonator) and discuss the possibility for having propagating self-induced 
transparent or superradiant pulses in that medium. Most importantly, those 
pulses induce quantum coherence effects in the medium itself, by exciting 
population inversion pulses in the qubit subsystem. Moreover, the speed of the 
propagating pulses can be controlled by proper engineering of the parameters of 
the qubits. The most important points made in this review are summarized in 
Section 6. 
\newpage

\section{SQUID-Based Metamaterials I: Models and Collective Properties}
\subsection{The rf-SQUID as an artificial magnetic "atom"}
\begin{figure}[t!]
\includegraphics[angle=0, width=0.9\linewidth]{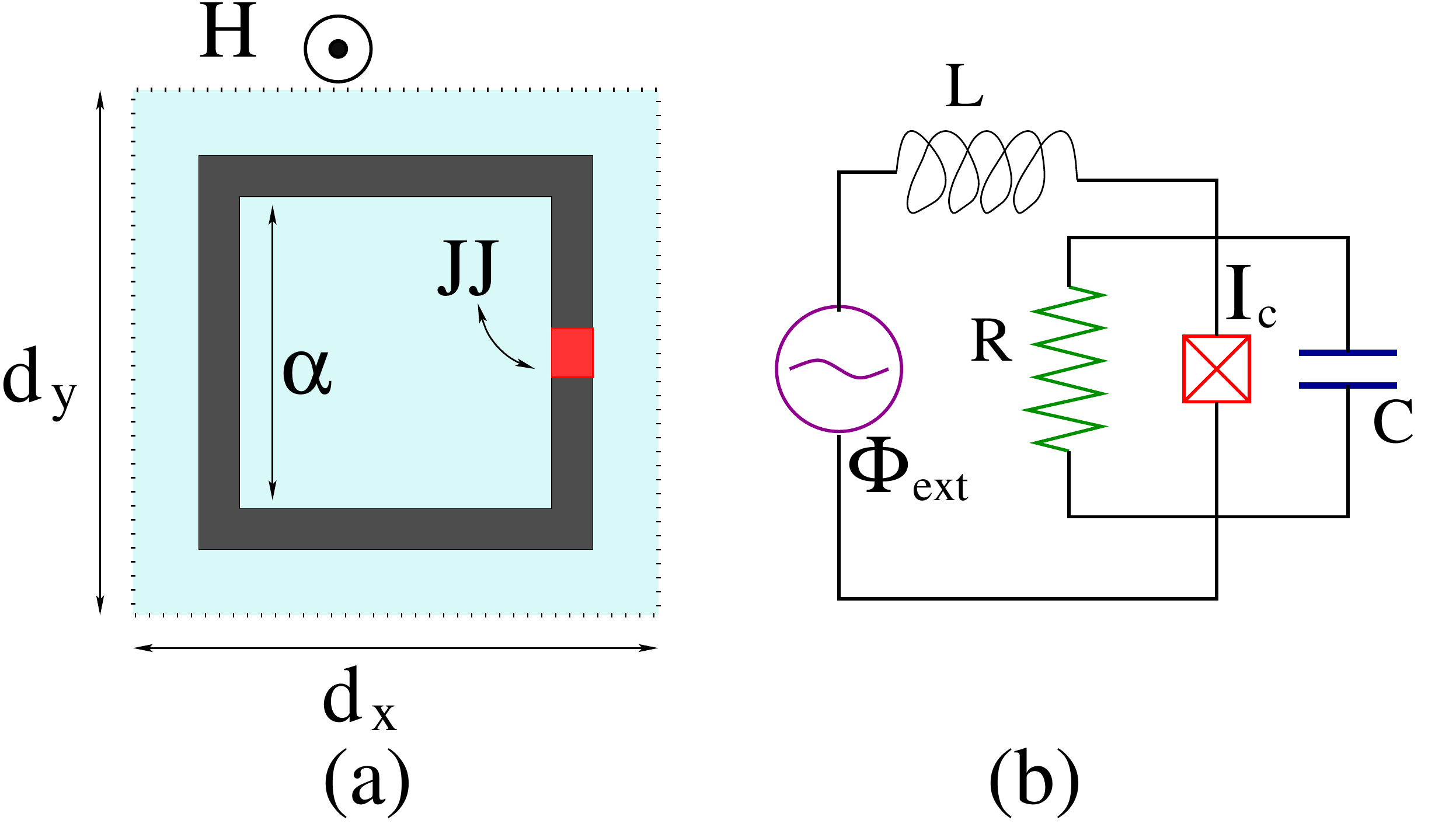}
\caption{
(a) Schematic drawing of an rf SQUID in a time-dependent magnetic field 
    ${\bf H}(t)$ perpendicular to its loop.
(b) Equivalent electrical circuit for an rf SQUID driven by a time-dependent
    flux source $\Phi_{ext} (t)$.
\label{fig2.01-01}
}
\end{figure}

\begin{figure}[h!]
\includegraphics[angle=0, width=0.9\linewidth]{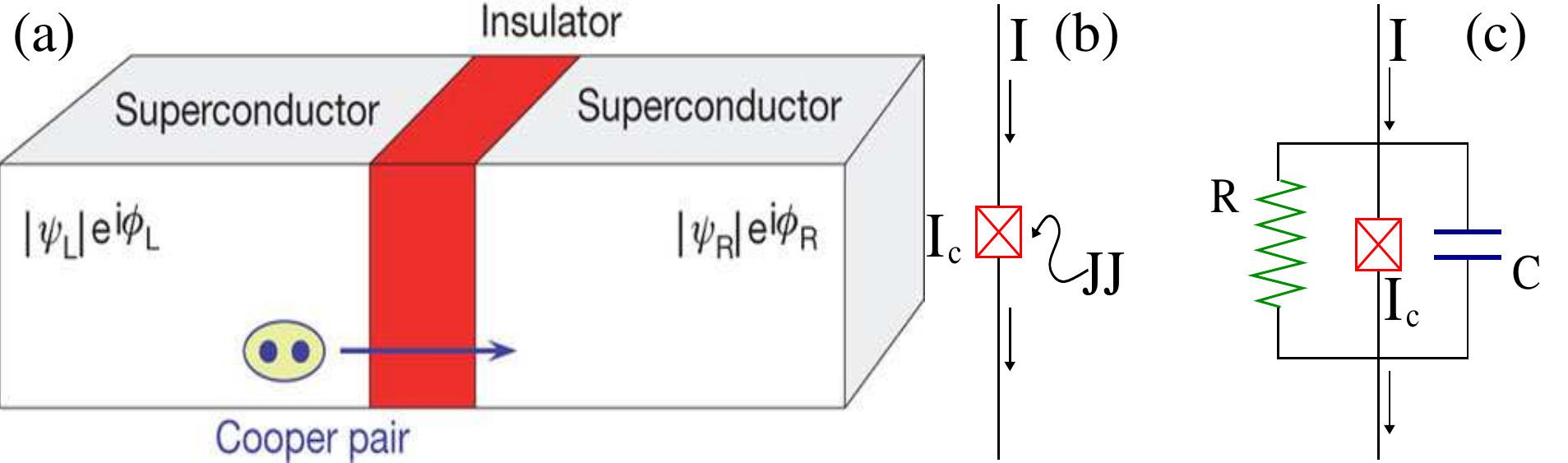}
\caption{
(a) Schematic of a superconductor-insulator-superconductor (SIS) Josephson 
    junction that consists of two superconducting plates separated by a thin 
    insulating layer. Cooper pairs can tunnel through the barrier without loss 
    of energy. The order parameters of the left and right superconductors are 
    $|\Psi_L| \exp(i\Phi_L)$ and $|\Psi_R| \exp(i\Phi_R)$, respectively.
(b) The electrical circuit symbol of a Josephson junction.
(c) The equivalent circuit of a real Josephson junction according to the widely 
    used Resistively and Capacitively Shunted (RCSJ) junction model.
\label{fig2.02-02}
}
\end{figure}

The Superconducting QUantum Interference Device (SQUID) is currently one of the 
most important solid-state circuit elements for superconducting electronics 
\cite{Anders2010}; among many other technological applications \cite{Kleiner2004,
Clarke2004b,Fagaly2006}, SQUIDs are used in devices that provide the most 
sensitive sensors of magnetic fields. Recent advances that led to nano-SQUIDs 
\cite{Wernsdorfer2009} makes the fabrication of SQUID metamaterials at the 
nanoscale an interesting possibility. The radio-frequency (rf) SQUID, in 
particular, shown schematically in Fig. \ref{fig2.01-01}(a), consists of a 
superconducting ring of self-inductance $L$ interrupted by a Josephson junction 
(JJ) \cite{Josephson1962}. A JJ is made by two superconductors connected through 
a "weak link", i.e., through a region of weakened superconductivity. A common type 
of a JJ is usually fabricated by two superconductors separated by a thin 
dielectric oxide layer (insulating barrier) as shown in Fig. \ref{fig2.02-02}(a); 
such a JJ is referred to as a superconductor-insulator-superconductor (SIS) 
junction. The fundamental properties of JJs have been established long ago 
\cite{Barone1982,Likharev1986}, and their usage in applications involving 
superconducting circuits has been thoroughly explored. The observed 
{\em Josephson effect} in such a junction, has been predicted by Brian D. 
Josephson in 1962 and it is of great importance in the field of superconductivity 
as well as in physics. That effect has been exploited in numerous applications 
in superconducting electronics, sensors, and high frequency devices. In an ideal 
JJ, whose electrical circuit symbol is shown in Fig. \ref{fig2.02-02}(b), the 
(super)current (Josephson current) $I_J$ through the JJ and the voltage $V_J$ 
across the JJ are related through the celebrated Josephson relations 
\cite{Josephson1962}
\begin{equation}
\label{Ch5.01}
   V_J (t)=\frac{\Phi_0}{2 \pi} \frac{\partial \phi_J (t)}{\partial t}, \qquad 
   I_J (t) =I_c \, \sin[\phi_J (t)], 
\end{equation}
where $I_c$ is the critical current of the JJ, 
$\Phi_0 =\frac{h}{2 e} \simeq 2.07\times 10^{-15} ~Wb$ is the flux quantum, with
with $h$ and $e$ being the Planck's constant and the electron's charge, 
respectively, and $\phi_J$ is the difference of the phases of the order 
parameters of the superconductors at left ($S_L$) and right ($S_R$) of 
the barrier $\phi_L$ and $\phi_R$, respectively (Fig. \ref{fig2.02-02}(a)), 
i.e., $\phi_J =\phi_L -\phi_R$ (the Josephson phase). In the presence of an 
electromagnetic potential ${\bf A} ({\bf r},t)$, the corresponding 
gauge-invariant Josephson phase is 
\begin{equation}
\label{Ch5.01.777}
   \phi_J (t) =\phi_L (t) -\phi_R (t) 
    -\frac{2\pi}{\Phi_0} \int_{S_L}^{S_R} {\bf A} ({\bf r},t) d{\bf l}.
\end{equation}
In an ideal JJ, in which the (super)current is carried solely by Cooper 
pairs, there is no voltage drop across the barrier of the JJ. In practice, 
however, this can only be true at zero temperature ($T=0$), while at finite 
temperatures there will always be a quasi-partice current. The latter is carried 
single-electron excitations (quasi-electrons) resulting from thermal breaking of 
Cooper pairs due to the non-zero temperature, and it is subjected to losses. In 
superconducting circuits, a real JJ is often considered as the parallel 
combination of an ideal JJ (described by Eqs. (\ref{Ch5.01})), a shunting 
capacitance $C$, due to the thin insulating layer, and a shunting resistance $R$, 
due to quasi-electron tunneling through the insulating barrier, i.e., the 
quasi-particle current (Fig. \ref{fig2.02-02}(c)). This equivalent electrical 
circuit for a real JJ is known as the Resistively and Capacitively Shunted 
Junction (RCSJ) model, and is widely used for modeling SIS Josephson junctions. 
The ideal JJ can be also described as a variable inductor, in a superconducting 
circuit. From the Josephson relations Eqs. (\ref{Ch5.01}) and the current-voltage 
relation for an ordinary inductor $U =L (\partial I/\partial t)$, it is easily 
deduced that the Josephson inductance is
\begin{equation}
\label{Ch5.01.999}
   L \equiv L_J \equiv L_J (\phi_J) =\frac{L_J (0)}{\cos(\phi_J)}, \qquad
   L_J (0) =\frac{\Phi_0}{2\pi I_c}.
\end{equation}
Note that Eq. (\ref{Ch5.01.999}) describes a nonlinear inductance, since $L_J$ 
depends both on the current and the voltage through the Josephson phase $\phi_J$. 

Due to the Josephson element (i.e., the JJ), the rf SQUID is a highly nonlinear 
oscillator that responds in a manner analogous to a magnetic "atom", exhibiting 
strong resonance in a time-varying magnetic field with appropriate polarization. 
Moreover, it exhibits very rich dynamic behavior, including chaotic effects 
\cite{Fesser1983,Ritala1984,Sorensen1985} and tunability of its resonance 
frequency with external fields \cite{Lazarides2012}. The equivalent electrical 
circuit for an rf SQUID in a time-dependent magnetic field ${\bf H}(t)$ threading 
perpendicularly its loop, as shown schematically in Fig. \ref{fig2.01-01}(b), 
comprises a flux source $\Phi_{ext} (t)$ in series with an inductance $L$ and a 
real JJ described by the RCSJ model. 

\begin{figure}[!t]
\includegraphics[angle=0, width=0.75\linewidth]{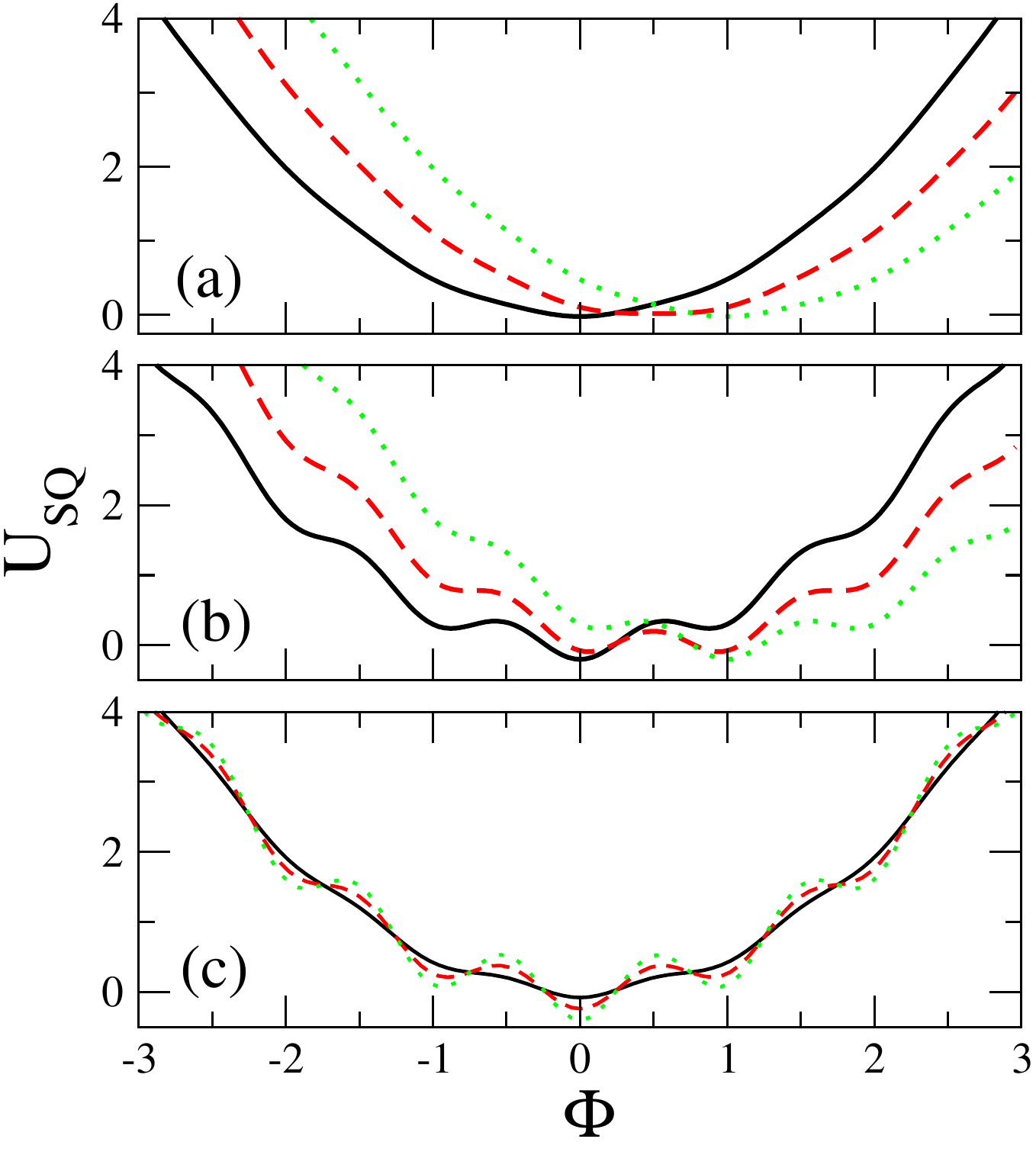}
\caption{
 The rf SQUID potential $U_{SQ}$ from Eq. (\ref{Ch5.06}) as a function of the 
 flux $\Phi$ threading the SQUID ring.
(a) For a non-hysteretic SQUID with $\beta_L \simeq 0.75 <1$ and
    $\phi_{dc}=\Phi_{dc} /\Phi_0 =0$ (black-solid curve);
    $0.5$ (red-dashed curve); $1.0$ (green-dotted curve).
(b) For a hysteretic SQUID with $\beta_L \simeq 8 > 1$ and
    $\phi_{dc}=\Phi_{dc} /\Phi_0 =0$ (black-solid curve);
    $0.5$ (red-dashed curve); $1.0$ (green-dotted curve).
(c) For $\Phi_{dc} =0$ and $\beta_L =0.5 <1$ (black-solid curve);
    $1.5$ (red-dashed curve); $2.5$ (green-dotted curve).
\label{fig2.03-03}
}
\end{figure}

The dynamic equation for the flux $\Phi$ threading the loop of the rf SQUID is 
obtained by direct application of Kirchhoff's laws, as
\begin{equation}
\label{Ch5.01.2}
 C \frac{d^2 \Phi}{dt^2} +\frac{1}{R} \frac{d \Phi}{dt} 
    +I_c \, \sin\left( 2 \pi \frac{\Phi}{\Phi_0} \right) 
    +\frac {\Phi -\Phi_{ext}}{L} =0,
\end{equation}
where $\Phi_0$ is the magnetic flux quantum, $I_c$ is the critical current of 
the JJ, and $t$ is the temporal variable. Eq. (\ref{Ch5.01.2}) is derived from 
the combination of the single-SQUID flux-balance relation 
\begin{equation}
\label{Ch5.02a}
   \Phi =\Phi_{ext} +L\, I, 
\end{equation}
and the expression for the current in the SQUID $I$ provided by the RCSJ model 
\begin{equation}
\label{Ch5.02b}
   -I =+C \frac{d^2 \Phi}{dt^2} +\frac{1}{R} \frac{d \Phi}{dt} 
       +I_c \, \sin\left( 2 \pi \frac{\Phi}{\Phi_0} \right).
\end{equation}
Eq. (\ref{Ch5.01.2}) has been studied extensively for more than three decades, 
usually under an external flux field of the form
\begin{eqnarray}
  \label{Ch5.05}
     \Phi_{ext} = \Phi_{dc}  +\Phi_{ac} \cos(\omega t ) ,
\end{eqnarray}
i.e., in the presence of a time-independent (constant, dc) and/or a 
time-dependent (usually sinusoidal) magnetic field of amplitude $\Phi_{ac}$ and 
frequency $\omega$. The orientation of both fields is such that their flux 
threads the SQUID loop. In the absence of dc flux, and very low amplitude of the 
ac field ($\Phi_{ac} \ll \Phi_0$, linear regime), the SQUID exhibits resonant 
magnetic response at
\begin{equation}
\label{Ch5.03}
  \omega_{SQ} = \omega_{LC} \sqrt{ 1 +\beta_L }, 
\end{equation}
where 
\begin{equation}
\label{Ch5.04}
  \omega_{LC} =\frac{1}{\sqrt{L C}}, \qquad \beta_L = 2\pi \frac{L I_c}{\Phi_0},  
\end{equation}
is the inductive-capacitive ($L C$) SQUID frequency and {\em SQUID parameter}, 
respectively. Eq. (\ref{Ch5.01.2}) is formally equivalent to that of a massive 
particle in a tilted washboard potential
\begin{equation}
\label{Ch5.06}
  U_{SQ} (\Phi) =\frac{1}{C} \left\{ \frac{(\Phi - \Phi_{ext})^2}{2 L} 
          -E_J \, \cos\left(2\pi \frac{\Phi}{\Phi_0}\right) \right\} ,
\end{equation}
with $E_J =\frac{I_c \Phi_0}{2\pi}$ being the Josephson energy. That potential 
has a number of minimums which depends on the value of $\beta_L$, while the 
location of those minimums varies with the applied dc (bias) flux $\Phi_{dc}$. 
For $\beta_L < 1$ (non-hysteretic regime) the potential $U_{SQ} (\Phi)$ is a 
corrugated parabola with a single minimum which moves to the right with 
increasing $\Phi_{ext} =\Phi_{dc}$, as it is shown in Fig. \ref{fig2.03-03}(a). 
For $\beta_L > 1$ (hysteretic regime) there are more than one minimums, while 
their number increases with further increasing $\beta_L$. A dc flux 
$\Phi_{ext} =\Phi_{dc}$ can move all these minimums as well (Fig. 
\ref{fig2.03-03}(b)) The emergence of more and more minimums with increasing 
$\beta_L$ at $\Phi_{ext} =\Phi_{dc}=0$ is illustrated in Fig. 
\ref{fig2.03-03}(c). It should be stressed here that in general the external 
flux $\Phi_{ext}$ is a sum of a dc bias and an ac (time-periodic) term. In that 
case, the potential $U_{SQ} (\Phi)$ as a whole rocks back and forth at the 
frequency of the external driving ac flux, $\Omega$. Then, in order to determine 
the possible stationary states of the SQUID, the principles of minimum energy 
conditions do not apply; instead, the complete nonlinear dynamic problem has to 
be considered. 

{\em Normalization.-} 
For an appropriate normalization of the single SQUID equation (\ref{Ch5.01.2})
and the corresponding dynamic equations for the one- and two-dimensional SQUID 
metamaterials discussed below, the following relations are used
\begin{eqnarray}
\label{Ch5.06.2}
  \phi =\frac{\Phi}{\Phi_0}, ~~~\phi_{ac} =\frac{\Phi_{ac}}{\Phi_0},
  ~~~\phi_{dc} =\frac{\Phi_{dc}}{\Phi_0}, 
  ~~~\tau=\omega_{LC} t , ~~~\Omega=\frac{\omega}{\omega_{LC}} , 
  ~~~i =\frac{I}{I_c} ,  
\end{eqnarray}    
i.e., frequency and time are normalized to $\omega_{LC}$ and its inverse 
$\omega_{LC}^{-1}$, respectively, while all the fluxes and currents are 
normalized to $\Phi_0$ and $I_c$, respectively. Then, Eq. (\ref{Ch5.01.2}) is 
written in normalized form as
\begin{eqnarray}
\label{Ch5.06.3}
   \ddot{\phi} +\gamma \dot{\phi} +\beta \sin\left( 2\pi \phi \right) +\phi 
    =\phi_{ext}, 
\end{eqnarray}
where the overdots denote derivation with respect to the normalized temporal 
variable $\tau$, $\phi_{ext} =\phi_{dc} +\phi_{ac} \cos(\Omega \tau )$ is the 
normalized external flux, and
\begin{equation}
\label{Ch5.06.4}
   \beta=\frac{I_c L}{\Phi_0} =\frac{\beta_L}{2\pi}, \qquad
   \gamma=\frac{1}{R} \sqrt{ \frac{L}{C} } 
\end{equation}
is the rescaled SQUID parameter and dimensionless loss coefficient, respectively. 
The term which is proportional to $\gamma$ in Eq. (\ref{Ch5.06.3}) actually 
represents all of the dissipation coupled to the rf SQUID.

\begin{figure}[h!]
\includegraphics[angle=0, width=0.9\linewidth]{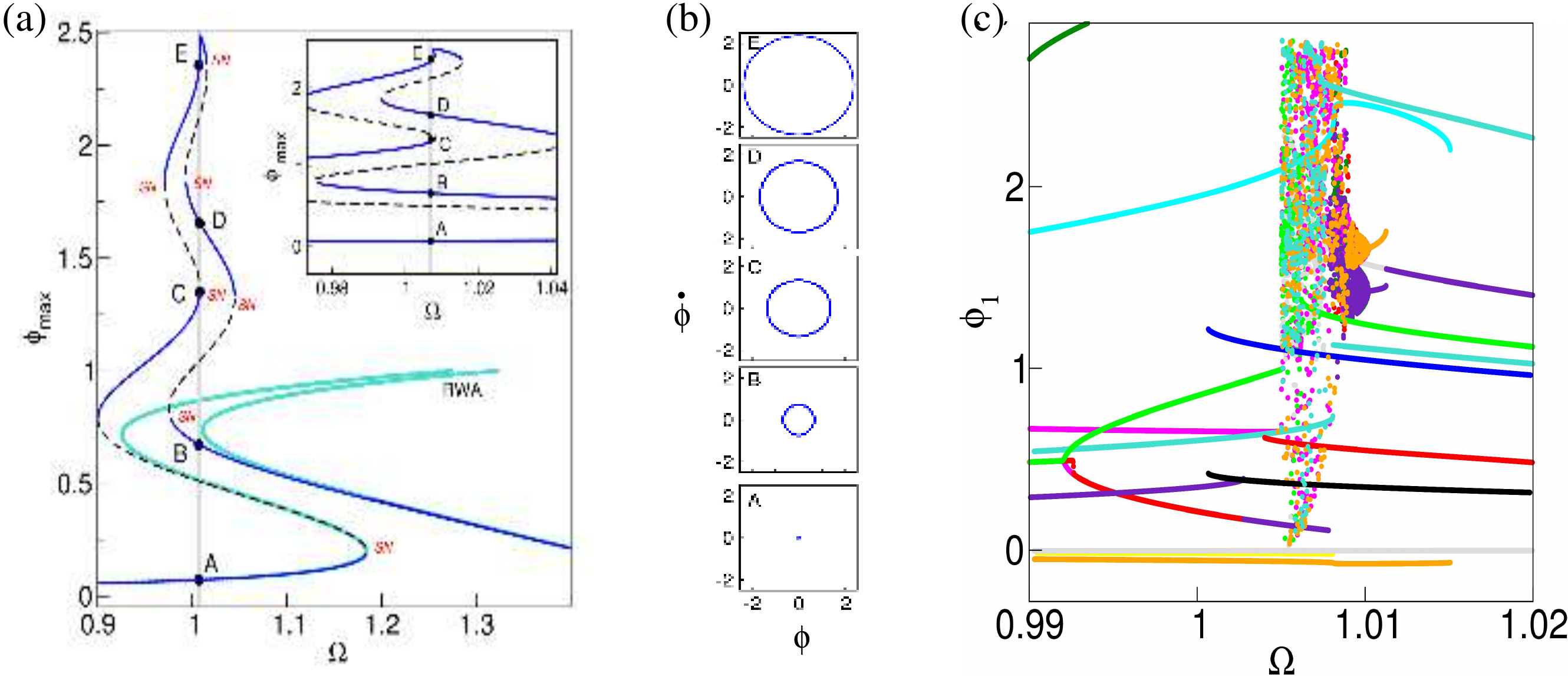}
\caption{
 (a) The snake-like resonance curve for an rf SQUID with $\beta_L=0.86$, 
 $\gamma=0.024$, and ac flux amplitude $\phi_{ac} =0.06$. The solid-blue and 
 dashed-black lines correspond to branches of stable and unstable periodic 
 solutions, respectively. Saddle-node bifurcation points are denoted as SN. The 
 vertical thick gray line corresponds to $\Omega=1.007$, and the turquoise lines 
 are obtained from Eq. (\ref{Ch5.06.5}).
 Inset: enlargement around the maximum multistability frequency $\Omega=1.007$. 
 (b) Phase portraits for the periodic orbits at the points marked as $A - E$
 on the resonance curve and in the inset.
 (c) A bifurcation diagram for two coupled identical SQUIDs around the 
 single-SQUID maximum multistability frequency $\Omega=1.007$. Only stable 
 solutions are shown for clarity. The flux though the loop of say, the first 
 SQUID, $\phi_1$ is plotted at the end of the driving period (see text). 
 The coupling coeffcient between the SQUIDs is $\lambda =-0.025$ and the other 
 parameters as in (a). At least ten (10) stable states, including chaotic ones,
 are visible at $\Omega=1.007$.
\label{fig2.04-04}
}
\end{figure}

The properties of the many variants of the SQUID have been investigated for many 
years, and they can be found in a number of review articles
\cite{Gallop1976,Clarke1986,Jenks1997,Koelle1999,Kleiner2004,Fagaly2006,Clarke2010},
textbooks \cite{Gallop1991}, and a Handbook \cite{Clarke2004a,Clarke2004b}. Here 
we focus on the multistability and the tunability properties of rf SQUIDs, which 
are important for our later discussions on SQUID metamaterials. As it was 
mentioned earlier, the rf SQUID is a nonlinear oscillator which exhibits strong 
resonant response at a particular frequency to a sinusoidal (ac) flux field.
For low amplitudes of the ac flux field, i.e., in the linear regime, the 
single-SQUID resonance frequency is given in Eq. (\ref{Ch5.03}); in units of
the inductive-capacitive ($LC$) SQUID frequency, the single-SQUID resonance 
frequency is   
\begin{equation}
\label{Ch5.06.42}
   \Omega_{SQ} = \frac{\omega_{SQ}}{\omega_{LC}} =\sqrt{ 1 +\beta_L }. 
\end{equation}
It was theoretically demonstrated that the SQUID resonance can be tuned within
a broad band of frequencies either by a dc flux bias $\phi_{dc}$ or by the 
amplitude of an ac flux field $\phi_{ac}$. Soon after these predictions, these
tunability properties were confirmed experimentally 
\cite{Butz2013a,Trepanier2013}. Further experiments showed that the single-SQUID
resonance frequency is also tunable by the power of the ac flux field, as well 
as with temperature \cite{Zhang2015}. In particular, with increasing the 
amplitude of the ac flux field from low to high values, the single-SQUID 
resonance frequency shifts from $\Omega =\Omega_{SQ}$ to $\Omega =1$ (i.e., 
towards lower frquencies). Note that for $\beta_L =0.86$, a typical value for
$\beta_L$ and very close to those obtained in the experiments, the single-SQUID
resonance frequency may vary from $\Omega =\Omega_{SQ} \simeq 1.364$ to 
$\Omega \simeq 1$, that is more than $25 \%$ of variation (broad-band 
tunability). Moreover, the single-SQUID resonance curve, i.e., the oscillation 
amplitude of the flux through the SQUID loop $\phi_{max}$ as a function of the 
driving frequency $\Omega$, changes dramatically its shape. Such a resonance 
curve for relatively high ac flux amplitude $\phi_{ac} =0.06$ is shown in
Fig. \ref{fig2.04-04}(a) \cite{Hizanidis2016a}. Resonance curves like that are 
calculated from Eq. (\ref{Ch5.06.3}). As it can be observed, the curve "snakes" 
back and forth within a narrow frequency band around the geometrical resonance 
frequency $\Omega \sim 1$. The solid-blue branches of the resonance curve 
indicate stable solutions, while the dashed-black branches indicate unstable 
ones (hereafter referred to as stable and unstable branches, respectively). The 
stable and unstable branches merge at particular points (turning points) denoted 
by SN; at all these points, in which $d\Omega / d\phi_{max} =0$, saddle-node 
bifurcations of limit cycles occur. Clearly, by looking at Fig. 
\ref{fig2.04-04}(a) one can identify that there are certain frequency bands in 
which more than one simultaneously stable solutions exist. For the sake of 
illustration, a vertical line has been drawned at $\Omega =1.007$. For that 
frequency, there are five (5) simultaneously stable solutions which are marked 
by the letters $A, B, C, D, E$, and four (4) unstable ones. This can be seen 
more clearly in the inset of Fig. \ref{fig2.04-04}(a) which shows an enlargement 
of the main figure around $\Omega =1.007$. Actually, at this frequency the 
number of possible solutions for the chosen set of simulation parameters is 
maximum. Thus for that set of parameters, $\Omega =1.007$ is the maximum 
multistability frequency. The corresponding trajectories of the stable solutions 
at $A - E$ are shown in the phase portraits $\dot{\phi} (\tau) - {\phi} (\tau)$ 
in Fig. \ref{fig2.04-04}(b). Multistability is enchanced (i.e., more solution 
branches appear) with increasing the ac flux amplitude $\phi_{ac}$ or lowering 
the loss coefficient $\gamma$. 

An approximation to the resonance curve for $\phi_{max} \ll 1$ is given by 
\cite{Hizanidis2016a}
\begin{eqnarray}
\label{Ch5.06.5}
   \Omega^2 =\Omega_{SQ}^2 \pm \frac{\phi_{ac}}{\phi_{max}} 
   -\beta_L \phi_{max}^2 
     \{ a_1 -\phi_{max}^2 [ a_2 -\phi_{max}^2 ( a_3 -a_4 \phi_{max}^2 )] \} ,
\end{eqnarray}
where $a_1 = \pi^2 /2$, $a_2 = \pi^4 /12$, $a_3 = \pi^6 /144$, and 
$a_4 = \pi^8 /2880$, which implicitly provides $\phi_{max}$ as a function of 
$\Omega$. The approximate flux amplitude - driving frequency curves from Eq. 
(\ref{Ch5.06.5}) are shown in Fig. \ref{fig2.04-04}(a) in turquoise color; 
clearly, they show excellent agreement with the numerical snaking resonance 
curve for $\phi_{max} \lesssim 0.6$. From Eq. (\ref{Ch5.06.5}), the frequency
of the first saddle-node bifurcation (the one with the lowest $\phi_{max}$)
can be calculated accurately. For simplicity, set $a_2 =a_3 =a_4 =0$ and 
$a_1 = \pi^2 /2$ into Eq. (\ref{Ch5.06.5}) and then use the condition 
$d\Omega / d\phi_{max} =0$ to obtain 
$\phi_{max,b} =\left( \frac{\mp \phi_{ac}}{\pi^2 \beta_L} \right)^{1/3}$,
where $\phi_{max,b}$ is the flux amplitude at which the first saddle-node
bifurcation occurs. By substitution of $\phi_{max,b}$ into the simplified 
Eq. (\ref{Ch5.06.5}), we get 
$\Omega^2 \equiv \Omega^2_b 
 =\Omega_{SQ}^2 -\frac{3}{2} (\pi^2 \beta_L)^{1/3} \phi_{ac}^{2/3}$, where 
$\Omega_b$ is the frequency at which the first saddle-node bifurcation occurs. 
For the parameters used in Fig. \ref{fig2.04-04}(a), i.e., $\beta_L =0.85$ and 
$\phi_{ac} =0.06$, we get $\phi_{max,b} \simeq 0.192$ and 
$\Omega_b \simeq 1.18$ which agree very well with the numerics. 

The dynamic complexity for frequencies around the single-SQUID resonance 
increases immensely in a SQUID array with a relatively large number of SQUIDS.
Although the coupling between SQUIDs is discussed in detail in the next 
Subsection (Subsection $2.2$), we believe it is appropriate to show here the 
solutions around the single-SQUID multistability frequency for a system of two 
coupled SQUIDs (Fig. \ref{fig2.04-04}(c)). The two SQUIDs, $1$ and $2$, are 
identical, and they are coupled magnetically with strength $\lambda$ through 
their mutual inductances. In Fig. \ref{fig2.04-04}(c), the flux through the loop
of SQUID $1$, $\phi_1$, is plotted as a function of frequency $\Omega$; 
since the presence of chaotic solutions was expected, the value of $\phi_1$ was 
plotted at the end of fifty (50) consecutive driving periods $T$ for each 
$\Omega$ (after the transients have died out). The frequency interval of Fig. 
\ref{fig2.04-04}(c) is the same as that in the inset of Fig. \ref{fig2.04-04}(a). 
Different colors have been used to help distinguishing between different solution 
branches; also, unstable solutions have been omitted for clarity. It can be 
observed that the number of stable states for the two-SQUID system is more than 
two times larger than the stable states of the single-SQUID. Moreover, apart
from the periodic solutions, a number of (coexisting) chaotic solutions has 
emerged. The dynamic complexity increases with increasing the number of SQUIDs
$N$ which are coupled together. This effect, which has been described in the 
past for certain arrays of nonlinear oscillators is named as {\em attractor 
crowding} \cite{Wiesenfeld1989,Tsang1990}. It has been argued that the number 
of stable limit cycles (i.e., periodic solutions) in such systems scale with 
the number of oscillators $N$ as $(N-1)!$. As a result, their basins of 
attraction crowd more and more tightly in phase space with increasing $N$.
The importance of this effect for the emergence of counterintuitive collective
states in SQUID metamaterials is discussed in Subsection $3.4$.

\begin{figure}[h!]
\includegraphics[angle=0, width=0.9\linewidth]{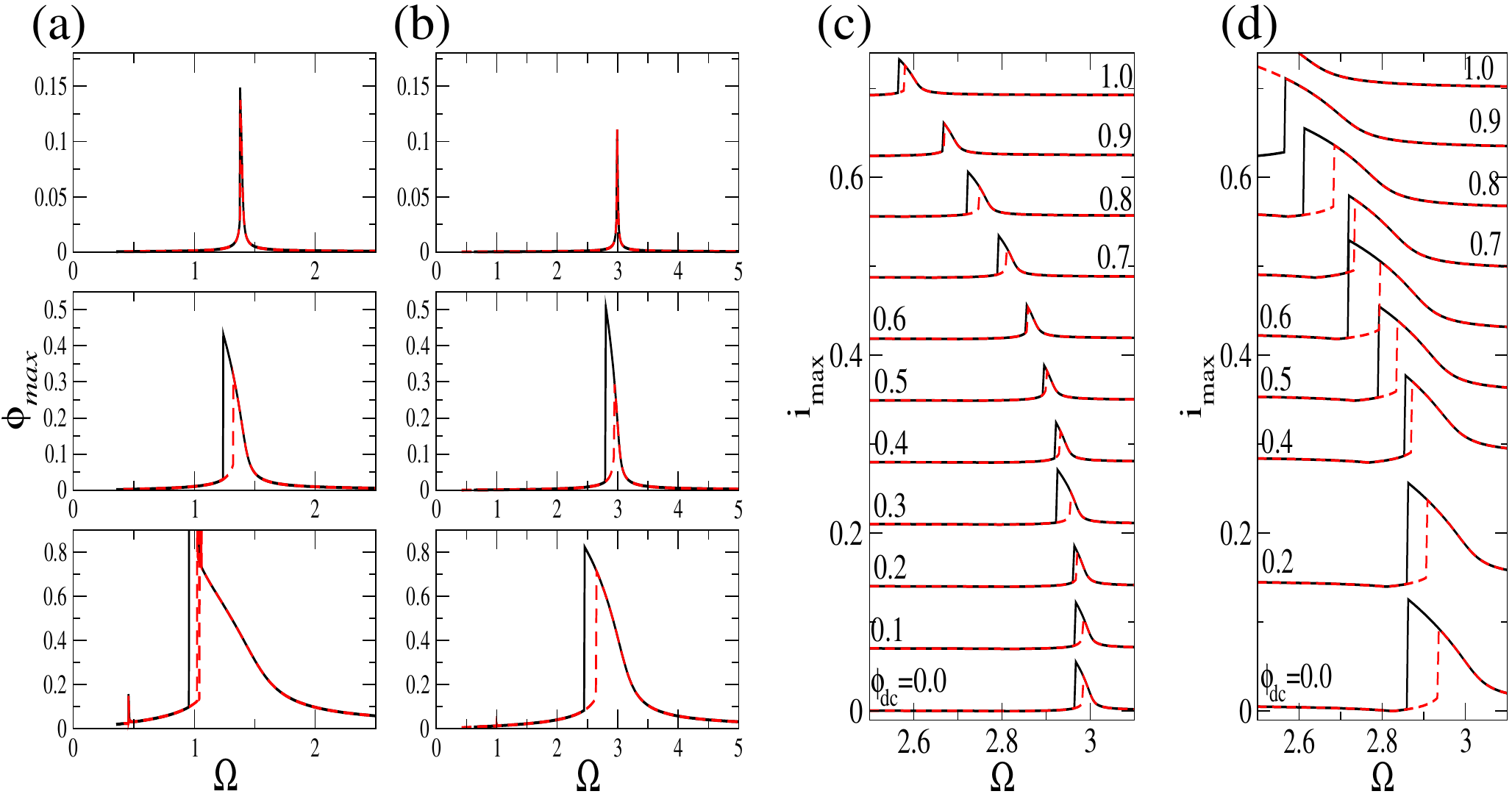}
\caption{
 Flux amplitude - frequency ($\phi_{max} - \Omega$) curves for an rf SQUID with 
 $\phi_{dc}=0$, $\gamma=0.002$, and (a) $\beta =0.15$; (b) $\beta =1.27$, and 
 current amplitude - frequency ($i_{max} - \Omega$) curves for an rf SQUID with 
 $\beta =1.27$, $\gamma=0.002$, and (c) $\phi_{ac}=0.01$; (d) $\phi_{ac}=0.1$.
 In (a) and (b), the ac field amplitude $\phi_{ac}$ increases from top to bottom
 panel: $\phi_{ac}=0.001$ (top); $\phi_{ac}=0.01$ (middle); $\phi_{ac}=0.1$ 
 (bottom). In (c) and (d), the value of the dc flux bias increases from zero in 
 steps of $0.1$ as shown.
\label{fig2.05-05}
}
\end{figure}

In Fig. \ref{fig2.05-05}, a number of flux amplitude - frequency and current 
amplitude-frequency curves are presented to demonstrate the tunability of the 
resonance frequency by varying the amplitude of the ac field $\phi_{ac}$ or by 
varying the dc flux bias $\phi_{dc}$. Since the properties of a SQUID-based 
metamaterial are primarily determined by the corresponding properties of its 
elements (i.e., the individual SQUIDs), the tunability of a single SQUID implies 
the tunability of the metamaterial itself. In Figs. \ref{fig2.05-05}(a) and (b) 
the flux amplitude - frequency curves are shown for a SQUID in the non-hysteretic 
and the hysteretic regime with $\beta_L =0.15$ and $1.27$, respectively. The ac 
field amplitude $\phi_{ac}$ increases from top to bottom panel. In the top 
panels, the SQUID is close to the linear regime and the resonance curves are 
almost symmetric without exhibiting visible hysteresis. A sharp resonance appears 
at $\Omega =\Omega_{SQ}$ as predicted from linear theory. With increasing 
$\phi_{ac}$ the nonlinearity becomes more and more appreciable and the resonance 
moves towards lower frequencies (middle and lower panels). Hysteretic effects 
are clearly visible in this regime. The resonance frequency of an rf SQUID can 
be also tuned by the application of a dc flux bias, $\phi_{dc}$, as shown in 
Figs. \ref{fig2.05-05}(c) and (d). While for $\phi_{dc} =0$ the resonance 
appears close to $\Omega_{SQ}$ (although slightly shifted to lower frequencies 
due to small nonlinear effects), it moves towards lower frequencies for 
increasing $\phi_{dc}$. Importantly, the variation of the resonance frequency 
does not seem to occur continuously but, at least for low $\phi_{dc}$, small 
jumps are clearly observable due to the inherently quantum nature of the SQUID 
which is incorporated to some extent into the phenomenological flux dynamics Eq. 
(\ref{Ch5.06.3}). The shift of the resonance with a dc flux bias in a single 
SQUID has been observed in high critical temperature (high$-T_c$) rf SQUIDs 
\cite{Zeng2000}, as well as in low critical temperature (low$-T_c$) rf SQUIDs 
\cite{Trepanier2013,Zhang2015}. In Fig. \ref{fig2.05-05}, there was no attempt 
to trace all possible branches of the resonance curves for clarity, and also for 
keeping the current in the SQUID to values less than the critical one for the 
JJ, i.e., for $I < I_c$. 

\begin{figure}[h!]
\includegraphics[angle=0, width=0.46\linewidth]{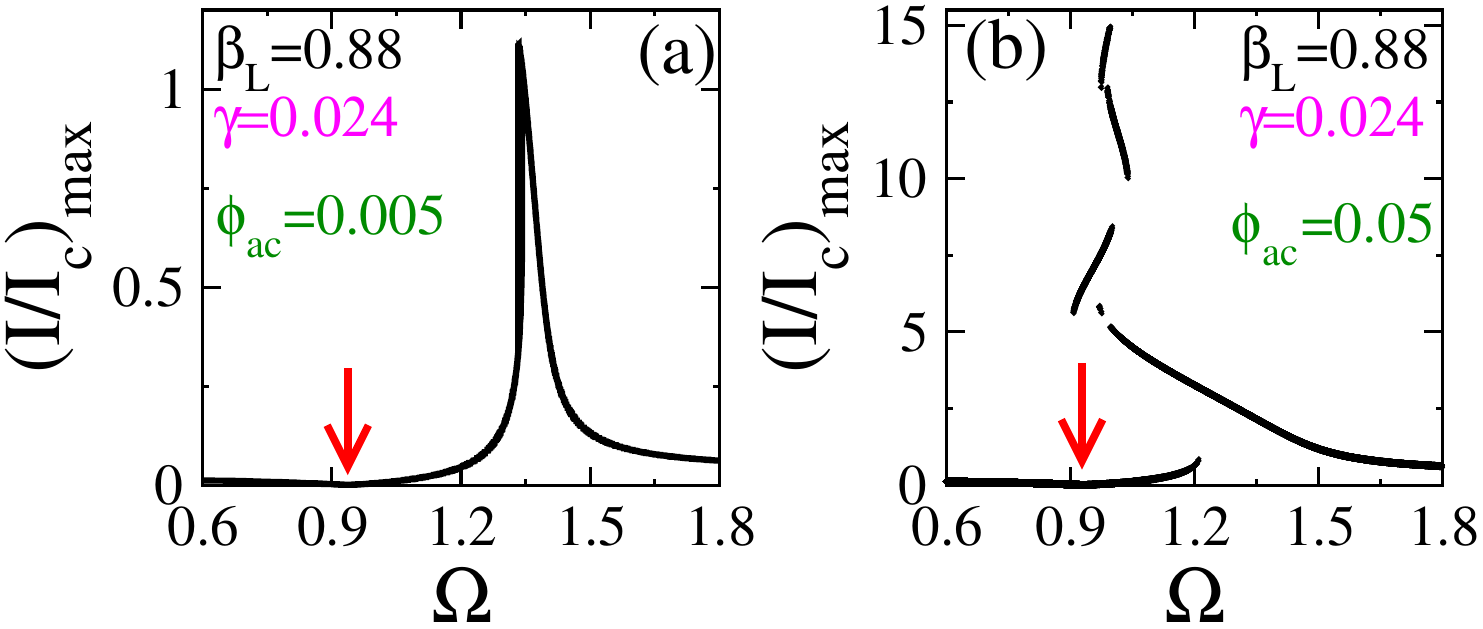}
\includegraphics[angle=0, width=0.46\linewidth]{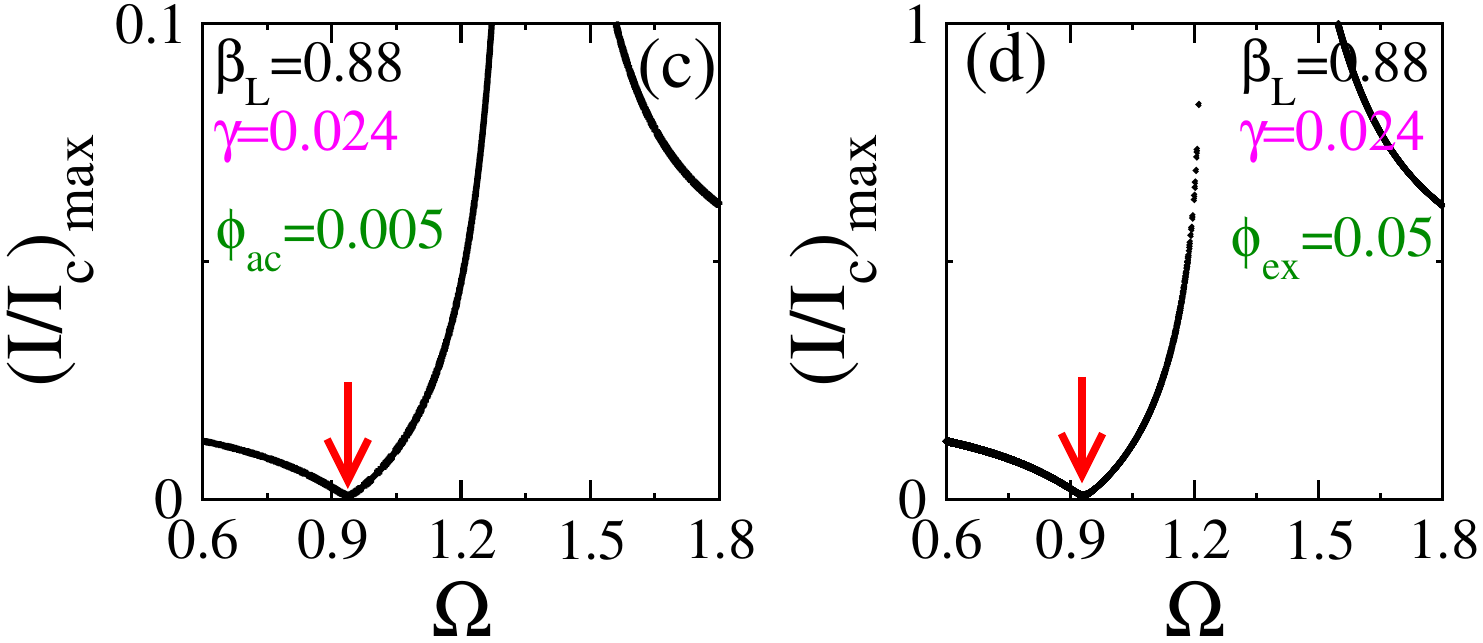}
\caption{
 Current amplitude $i_{max} =I_{max}/I_c$ - driving frequency $\Omega$ 
 characteristics for an rf SQUID with $\beta_L =0.88$, $\gamma=0.024$, 
 $\phi_{dc}=0$, and (a) $\phi_{ac}=0.005$; (b) $\phi_{ac}=0.05$. Enlargements of 
 (a) and (b) around the anti-resonance frequnecy are shown in (c) and (d), 
 respectively. The red arrows point at the location of the anti-resonance.
\label{fig2.06-06}
}
\end{figure}

Another illustration of the multistability in an rf SQUID is shown in Fig. 
\ref{fig2.06-06}, which also reveals an anti-resonance effect. In Figs. 
\ref{fig2.06-06}(a) and (b), the current amplitude - frequency curves are shown 
in two cases; one close to the weakly nonlinear regime and the other in the 
strongly nonlinear regime, respectively. In Fig. \ref{fig2.06-06}(a), the curve 
does not exhibit hysteresis but it is slightly skewed; the resonance frequency 
is $\Omega_R \simeq 1.25$, slightly lower than the SQUID resonance frequency in 
the linear regime, $\Omega_{SQ} \simeq 1.37$ (for $\beta_L =0.88$). In Fig. 
\ref{fig2.06-06}(b), the ac field amplitude $\phi_{ac}$ has been increased by an 
order of magnitude with respect to that in Fig. \ref{fig2.06-06}(a), and thus 
strongly nonlinear effects become readily apparent. Five (5) stable branches can 
be identified in a narrow frequency region around $\Omega \simeq 1$, i.e., 
around the geometrical (inductive-capacitive, $LC$) resonance frequency 
(unstable brances are not shown). The upper branches, which are extremely 
sensitive to perturbations, correspond to high values of the current amplitude 
$i_{max} =I_{max} / I_c$, which leads the JJ of the SQUID to its normal state. 
The red arrows in Figs. \ref{fig2.06-06}, point at the location of an 
{\em anti-resonance} \cite{Hizanidis2016b} in the current amplitude - frequency 
curves. Such an anti-resonance makes itself apparent as a well-defined dip in 
those curves, with a minimum that almost reaches zero. The effect of 
anti-resonance has been observed in nonlinearly coupled oscillators subjected to 
a periodic driving force \cite{Woafo1998} as well as in parametrically driven 
nonlinear oscillators \cite{Chakraborty2013}. However, it has never before been 
observed in a single, periodically driven nonlinear oscillator such as the rf 
SQUID. In Figs. \ref{fig2.06-06}(c) and (d), enlargements of Figs. 
\ref{fig2.06-06}(a) and (b), respectively, are shown around the anti-resonance 
frequency. Although the "resonance" region in the strongly nonlinear case
has been shifted significantly to the left as compared with the weakly nonlinear 
case, the location of the anti-resonance has remained unchanged (eventhough 
$\phi_{ac}$ in Figs. \ref{fig2.06-06}(a) and (b) differ by an order of 
magnitude). The knowledge of the location of anti-resonance(s) as well as the 
resonance(s) of an oscillator or a system of oscillators, beyond their 
theoretical interest, it is of importance in device applications. Certainly 
these resonances and anti-resonances have significant implications for the SQUID 
metamaterials whose properties are determined by those of their elements (i.e., 
the individual SQUIDs). When the SQUIDs are in an anti-resonant state, in which 
the induced current is zero, they do not absorb energy from the applied field 
which can thus transpass the SQUID metamaterial almost unaffected. Thus, in such 
a state, the SQUID metamaterial appears to be transparent to the applied 
magnetic flux as has been already observed in experiments on two-dimensional
SQUID metamaterials \cite{Zhang2015}; the observed effect has been named as 
{\it broadband self-induced transparency}. Moreover, since the anti-resonance 
frequency is not affected by $\phi_{ac}$, the transparency can be observed even 
in the strongly nonlinear regime, for which the anti-resonance frequency lies 
into the multistability region. In that case, the transparency of the 
metamaterial may be turned on and off as it has been already discussed in Ref. 
\cite{Zhang2015}. Thus, the concept of the anti-resonance serves for making a 
connection between an important SQUID metamaterial property and a fundamental 
dynamical property of nonlinear oscillators.  

\begin{figure}[h!]
\includegraphics[angle=0, width=0.95\linewidth]{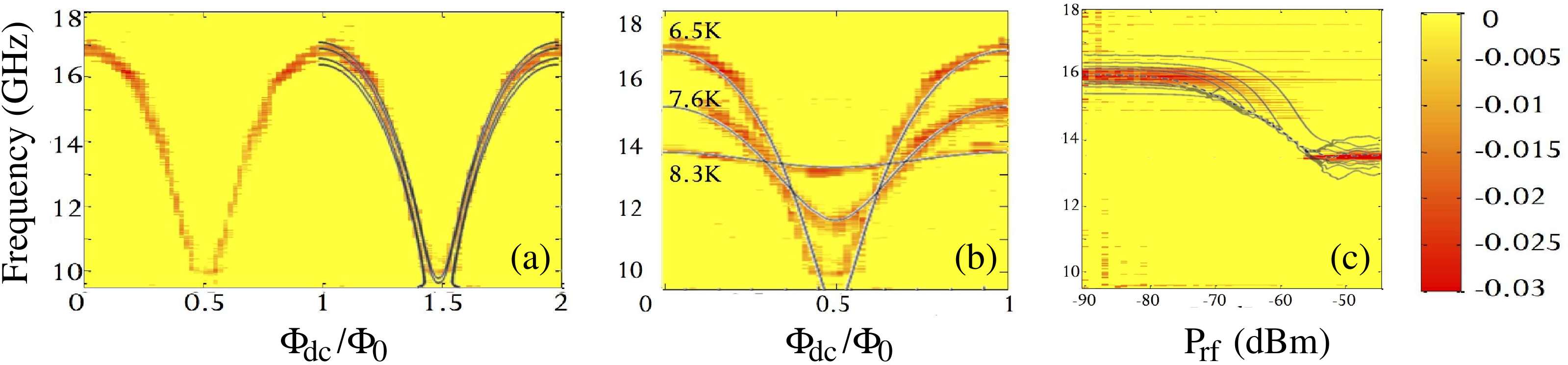}
\caption{
 Experimental measurements of the complex transmission magnitude $|S_{21}|$ of 
 an rf SQUID. The resonant response is identified by the red features. 
 Left: $|S_{21}|$ as a function of frequency $\Omega$ and applied dc flux 
 $\phi_{dc}$ at $-80 ~dBm$ rf power and temperature $T=6.5 ~K$.   
 Middle: $|S_{21}|$ as a function of frequency and applied dc flux at three 
 different temperatures, $T=6.5 ~K$, $7.6 ~K$, and $8.3 ~K$, and $-80 ~dBm$ rf 
 power.
 Right: $|S_{21}|$ as a function of frequency and rf power at fixed dc flux, 
 $\phi_{dc}=1/6$ and temperature $T =6.5 ~K$ \cite{Trepanier2013}.
\label{fig2.07-07}
}
\end{figure}

The tunability of the SQUID resonance with a dc magnetic field and the 
temperature has been investigated in recent experiments 
\cite{Jung2013,Trepanier2013}. Those investigations rely on the measurement of 
the magnitude of the complex transmission $|S_{21}|$ as one or more external 
parameters such as the driving frequency, the dc flux bias, and the temperature 
vary. Very low values of $|S_{21}|$ indicate that the SQUID is at resonance. In 
Fig. \ref{fig2.07-07}, the resonant response is identified by the red features.
In the left panel, it is observed that the resonance vary periodically with the 
applied dc flux, with period $\Phi_0$. In the middle panel of Fig. 
\ref{fig2.07-07}, the effect of the temperature $T$ is revealed; as expected, 
the tunability bandwidth of the resonance decreases with increasing temperature. 
In the right panel of Fig. \ref{fig2.07-07}, the variation of the resonance 
frequency with the applied rf power is shown. Clearly, three different regimes
are observed; for substantial intervals of low and high rf power, the resonance 
frequency is approximatelly constant at $\Omega \sim \Omega_{SQ}$ and 
$\Omega \sim 1$, respectively, while for intermediate rf powers the resonance 
apparently dissapears. The latter effect is related to the broad-band 
self-induced transparency \cite{Zhang2015}.    

\subsection{SQUID Metamaterials Models and Flux Wave Dispersion}
Conventional (metallic) metamaterials comprise regular arrays of split-ring 
resonators (SRRs), which are highly conducting metallic rings with a slit. These 
structures can become nonlinear with the insertion of electronic devices (e.g., 
varactors) into their slits 
\cite{Lapine2003,Shadrivov2006a,Wang2008,Boardman2010,Lapine2014}. SQUID 
metamaterials is the superconducting analogue of those nonlinear conventional 
metamaterials that result from the replacement of the varactor-loaded metallic 
rings by rf SQUIDs as it has been suggested both in the quantum \cite{Du2006} 
and the classical \cite{Lazarides2007} regime. Recently, one- and two-dimensional 
SQUID metamaterials have been constructed from low critical temperature 
superconductors which operate close to liquid Helium temperatures
\cite{Butz2013a,Butz2013b,Trepanier2013,Zhang2015,DZhang2016,Trepanier2017}. The 
experimental investigation of these structures has revealed several novel 
properties such as negative diamagnetic permeability \cite{Jung2013,Butz2013a}, 
broad-band tunability \cite{Butz2013a,Trepanier2013}, self-induced broad-band 
transparency \cite{Zhang2015}, dynamic multistability and switching 
\cite{Jung2014b}, as well as coherent oscillations \cite{Trepanier2017}, among 
others. Some of these properties, i.e., the dynamic multistability effect and 
tunability of the resonance frequency of SQUID metamaterials, have been also 
revealed in numerical simulations \cite{Lazarides2013b,Tsironis2014b}. 
Moreover, nonlinear localization \cite{Lazarides2008a} and the emergence of 
counter-intuitive dynamic states referred to as {\em chimera states} in current 
literature \cite{Lazarides2015b,Hizanidis2016a,Hizanidis2016b} have been 
demonstrated numerically in SQUID metamaterial models. The chimera states have 
been discovered numerically in rings of identical phase oscillators
\cite{Kuramoto2002} (see Ref. \cite{Panaggio2015} for a review).

The applied time-dependent magnetic fields induce (super)currents in the SQUID 
rings through Faraday's induction law, which couple the SQUIDs together through 
dipole-dipole magnetic forces; although weak due to its magnetic nature, that 
interaction couples the SQUIDs non-locally since it falls-off as the inverse 
cube of their center-to-center distance. Consider a one-dimensional linear array 
of $N$ identical SQUIDs coupled together magnetically through dipole-dipole 
forces. The magnetic flux $\Phi_n$ threading the $n-$th SQUID loop is 
\cite{Lazarides2015b}
\begin{equation}
\label{Ch5.07}
  \Phi_n =\Phi_{ext} +L\, I_n +L\, \sum_{m\neq n} \lambda_{|m-n|} I_m ,
\end{equation}
where the indices $n$ and $m$ run from $1$ to $N$, $\Phi_{ext}$ is the external 
flux in each SQUID, $\lambda_{|m-n|} =M_{|m-n|}/L$ is the dimensionless coupling 
coefficient between the SQUIDs at positions $m$ and $n$, with $M_{|m-n|}$ being 
their corresponding mutual inductance, and
\begin{eqnarray}
\label{Ch5.08}
    -I_n =C\frac{d^2\Phi_n}{dt^2} +\frac{1}{R} \frac{d\Phi_n}{dt} 
           +I_c\, \sin\left(2\pi\frac{\Phi_n}{\Phi_0}\right) 
\end{eqnarray}
is the current in each SQUID given by the RCSJ model \cite{Likharev1986}, 
with  $\Phi_0 =h/(2 e)$ and $I_c$ being the flux quantum and the critical 
current of the JJs, respectively. Recall that within the RCSJ framework, $R$, 
$C$, and $L$ are the resistance, capacitance, and self-inductance of the SQUIDs' 
equivalent circuit, respectively. Combination of Eqs.(\ref{Ch5.07}) and 
(\ref{Ch5.08}) gives
\begin{eqnarray}
\label{Ch5.09}
  C \frac{d^2\Phi_n}{dt^2} +\frac{1}{R} \frac{d\Phi_n}{dt}
    +\frac{1}{L} \sum_{m=1}^N  \left( {\bf \hat{\Lambda}}^{-1} \right)_{nm} 
         \left( \Phi_m -\Phi_{ext} \right) 
    +I_c\, \sin\left(2\pi\frac{\Phi_n}{\Phi_0} \right) =0 ,
\end{eqnarray}
where ${\bf \hat{\Lambda}}^{-1}$ is the inverse of the $N\times N$ coupling 
matrix 
\begin{eqnarray}
\label{Ch5.10}
  \left( {\bf \hat{\Lambda}} \right)_{nm} =\left\{ \begin{array}{ll}
      1, & \mbox{if $m= n$}; \\
   \lambda_{|m-n|} =\lambda_1 \, |m-n|^{-3}, & \mbox{if $m\neq n$},\end{array} 
    \right.   
\end{eqnarray}
with $\lambda_1$ being the coupling coefficient betwen nearest-neighboring 
SQUIDs.
\begin{figure}[!h]
\includegraphics[angle=0, width=0.9 \linewidth]{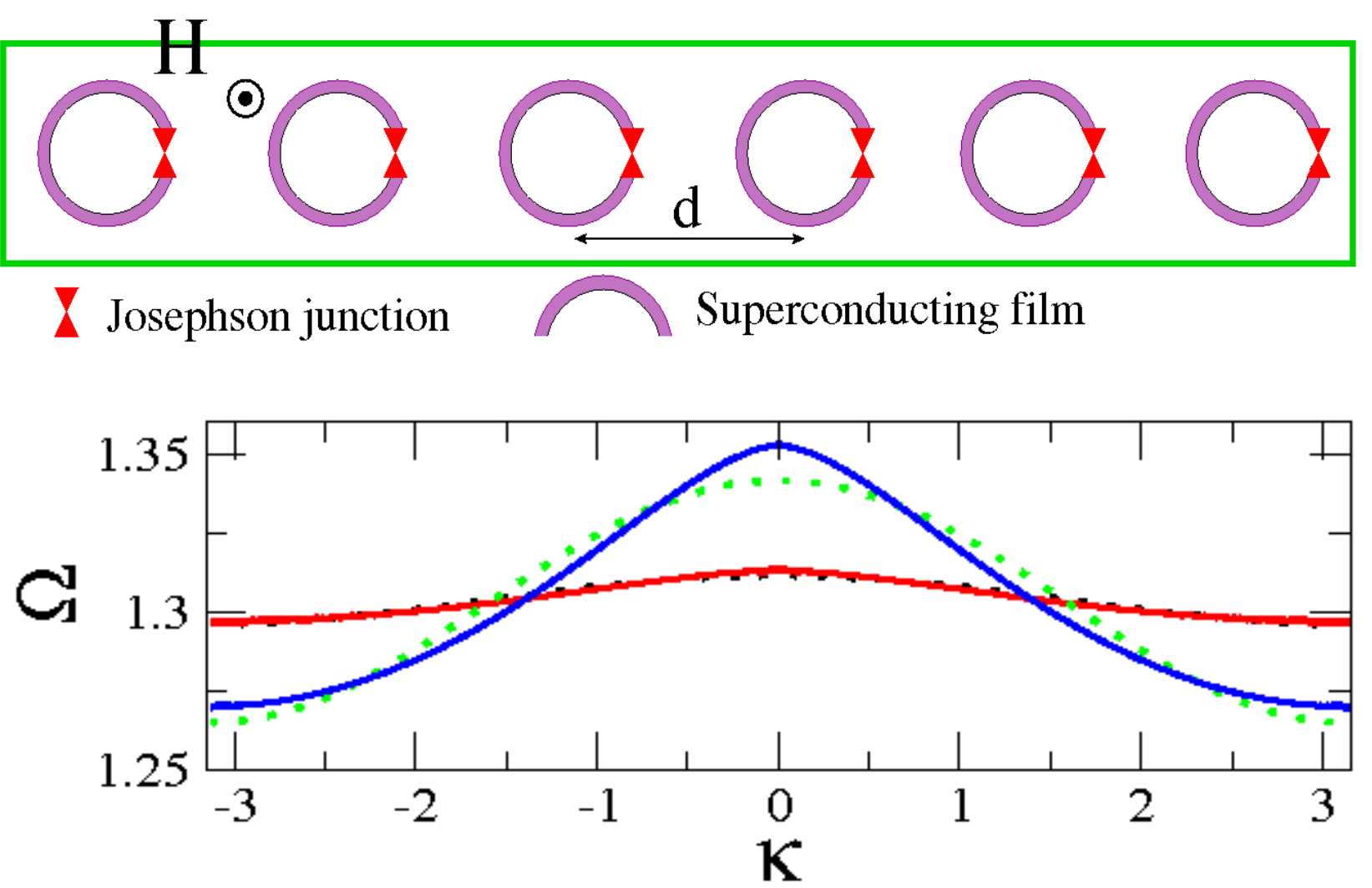}
\caption{
 Upper: Schematic of a one-dimensional SQUID metamaterial. 
 Lower: Frequency dispersion of the SQUID metamaterial with non-local coupling, 
 for $\beta = 0.1114$ ($\beta_L \simeq 0.7$) and $\lambda_0=-0.05$ (blue curve); 
 $-0.01$ (red curve). The corresponding dispersions for nearest-neighbor 
 (local) coupling are shown as green and black dotted curves, respectively. 
\label{fig2.08-08}
}
\end{figure}
The dimensionless coupling strength $\lambda_{|m-n|} =M_{|m-n|}/L$ between the
SQUIDs at site $n$ and $m$ (normalized center-to-center distance $|m-n|$), can
be calculated either analytically or numerically. The self-inductance $L$ of the
SQUIDs, for example, can be either estimated by an empirically derived equation
\cite{Brojeny2003} or it can be calculated using commercially available software
(FastHenry). The mutual inductance $M_{|m-n|}$ can also be obtained numerically
from a FastHenry calculation or it can be approximated using basic expressions
from electromagnetism. The magnetic field generated by a wire loop, which is the
approximate geometry of a SQUID, at a distance greater than its dimensions, is
given by the Biot-Savart law as $B =\frac{\mu_0}{4\pi} \frac{\pi r^2 I_w}{d^3}$,
where $I_w$ is the current in the wire, $r$ is the radius of the loop, and $d$ 
is the distance from the center of the loop. The mutual inductance between two 
such (identical) loops lying on the same plane is given by
\begin{eqnarray}
\label{Ch5.10.999}
   M =\frac{B \pi r^2}{I_w} 
     =\frac{\mu_0}{4\pi} \frac{(\pi r^2)^2}{d^3} \propto d^{-3},
\end{eqnarray}
where it is assumed that the field $B$ is constant over the area of each loop,
$\pi r^2$. For square loops of side $a$, the radius $r$ should be replaced by
$a/\sqrt{\pi}$. Eq. (\ref{Ch5.10.999}) explains qualitatively the inverse cube 
distance-dependence of the coupling strength $\lambda_{|m-n|}$ between SQUIDs.  

In normalized form Eq. (\ref{Ch5.09}) reads ($n=1,...,N$)
\begin{eqnarray}
\label{Ch5.11}
  \ddot{\phi}_n +\gamma \dot{\phi}_n +\beta \sin\left( 2\pi \phi_n \right) 
    =\sum_{m=1}^N \left( {\bf \hat{\Lambda}}^{-1} \right)_{nm} 
         \left( \phi_{ext} -\phi_m \right),
\end{eqnarray}
where the relations given in Eq. (\ref{Ch5.06.2}) have been used. Specifically, 
frequency and time are normalized to $\omega_{LC} =1/\sqrt{LC}$ and its inverse 
$\omega_{LC}^{-1}$, respectively, the fluxes and currents are normalized to 
$\Phi_0$ and $I_c$, respectively, the overdots denote derivation with respect 
to the normalized temporal variable, $\tau$,
$\phi_{ext} =\phi_{dc} +\phi_{ac} \cos(\Omega \tau )$, with 
$\Omega=\omega/\omega_0$ being the normalized driving frequency, and $\beta$, 
$\gamma$ are given in Eq. (\ref{Ch5.06.4}). The (magnetoinductive) coupling 
strength between SQUIDs, which can be estimated from the experimental parameters 
in Ref. \cite{Kirtley2005}, as well as from recent experiments 
\cite{Jung2013,Butz2013a}, is rather weak due to its magnetic nature (of the 
order of $10^{-2}$ in normalized units). Since that strength falls-off 
approximatelly as the inverse-cube of the distance between SQUIDs, a model which 
takes into account nearest-neighbor coupling only is sufficient for making 
reliable predictions. In that case, the coupling matrix assumes the simpler, 
tridiagonal and symmetric form 
\begin{eqnarray}
\label{Ch5.12}
   \left( {\bf \hat{\Lambda}} \right)_{nm} =\left\{ \begin{array}{ll}
        1, & \mbox{if $m= n$};\\
        \lambda_1 & \mbox{if $m =n \pm 1$}; \\
        0 & \mbox{for any other $n$, $m$}. \end{array} \right.   
\end{eqnarray}
For $\lambda_1 \ll 1$, the inverse of the coupling matrix is approximated to order 
${\cal O} (\lambda_1^2)$ by
\begin{eqnarray}
\label{Ch5.13}
   \left( {\bf \hat{\Lambda}}^{-1} \right)_{nm} =\left\{ \begin{array}{ll}
        1, & \mbox{if $m= n$};\\
        -\lambda_1 & \mbox{if $m =n \pm 1$}; \\
        0 & \mbox{for any other $n$, $m$}. \end{array} \right.   
\end{eqnarray}
Substituting Eq. (\ref{Ch5.13}) into Eq. (\ref{Ch5.11}), the corresponding 
dynamic equations for the fluxes through the loops of the SQUIDs of a locally 
coupled SQUID metamaterial are obtained as
\begin{eqnarray}
\label{Ch5.14}
   \ddot{\phi}_n +\gamma \dot{\phi}_n 
   +\phi_n +\beta \sin\left( 2\pi \phi_n \right)
   =\lambda ( \phi_{n-1} +\phi_{n+1} ) +\phi_{eff} , 
\end{eqnarray}
where $\phi_{eff} =(1 -2\lambda)\, \phi_{ext}$ is the "effective" external flux, 
with $\phi_{ext} =\phi_{dc} +\phi_{ac} \cos(\Omega \tau)$ being the normalized 
external flux. The effective flux arises due to the nearest-neighbor 
approximation. For a finite SQUID metamaterial (with $N$ SQUIDs), $\phi_{eff}$ 
is slightly different for the SQUIDs at the end-points of the array; 
specifically, for those SQUIDs $\phi_{eff} =(1 -\lambda)\, \phi_{ext}$ since 
they interact with one nearest-neighbor only.

Linearization of Eq. (\ref{Ch5.11}) around zero flux with $\gamma=0$ and 
$\phi_{ext} =0$ gives for the infinite system
\begin{eqnarray}
\label{Ch5.11.2}
  \ddot{\phi}_n +\left[ \beta_L 
    +\left( {\bf \hat{\Lambda}}^{-1} \right)_{nn} \right] \phi_n
    +\sum_{m\neq n} \left( {\bf \hat{\Lambda}}^{-1} \right)_{nm} \phi_m =0 . 
\end{eqnarray}
By substitution of the plane-wave trial solution 
$\phi_n=A\, \exp[i(\kappa n -\Omega \tau)]$ into Eq. (\ref{Ch5.11.2}), with 
$\kappa$ being the wavevector normalized to $d^{-1}$ ($d$ is the side of the 
unit cell, see Fig. \ref{fig2.01-01}) , and using
\begin{eqnarray}
\label{Ch5.11.3} 
  \sum_{m\neq n} \left( {\bf \hat{\Lambda}}^{-1} \right)_{nm} e^{i\kappa(m-n)} 
       =2 \sum_{m=1}^\infty 
           \left( {\bf \hat{\Lambda}}^{-1} \right)_{m} \cos(\kappa m) ,
\end{eqnarray}
where $m$ is the "distance" from the main diagonal of ${\bf\hat{\Lambda}}^{-1}$, 
we get
\begin{eqnarray}
\label{Ch5.11.4} 
   \Omega =\sqrt{ \Omega_{1}^2 
     +2 \sum_{m=1}^\infty 
         \left( {\bf \hat{\Lambda}}^{-1} \right)_{m} \cos(\kappa m) } ,
\end{eqnarray}
where 
$\Omega_{1}^2 =\beta_L +\left( {\bf \hat{\Lambda}}^{-1} \right)_{nn} \simeq \Omega_{SQ}^2$.
Note that for the infinite system the diagonal elements of the inverse of the 
coupling matrix $\left( {\bf \hat{\Lambda}}^{-1} \right)_{nn}$ have practically 
the same value which is slightly larger than unity. The frequency $\Omega_{1}$ 
is very close to the resonance frequency of individual SQUIDs, $\Omega_{SQ}$. 
Eq. (\ref{Ch5.11.4}) is the nonlocal frequency dispersion. By substitution of 
the same trial solution into Eq. (\ref{Ch5.14}), we get the nearest-neighbor 
frequency dispersion for flux waves, as
\begin{eqnarray}
\label{Ch5.11.5}
   \Omega \equiv \Omega_{\kappa} 
         =\sqrt{ \Omega_{SQ} -2 \lambda \, \cos \kappa } .
\end{eqnarray}
Eqs. (\ref{Ch5.11.4}) and (\ref{Ch5.11.5}) result in slightly different 
frequency dispersion curves as can be observed in the lower panel of Fig. 
\ref{fig2.08-08} for two different values of the coupling coefficient $\lambda$.
\begin{figure}[h!]
\includegraphics[angle=0, width=0.45\linewidth]{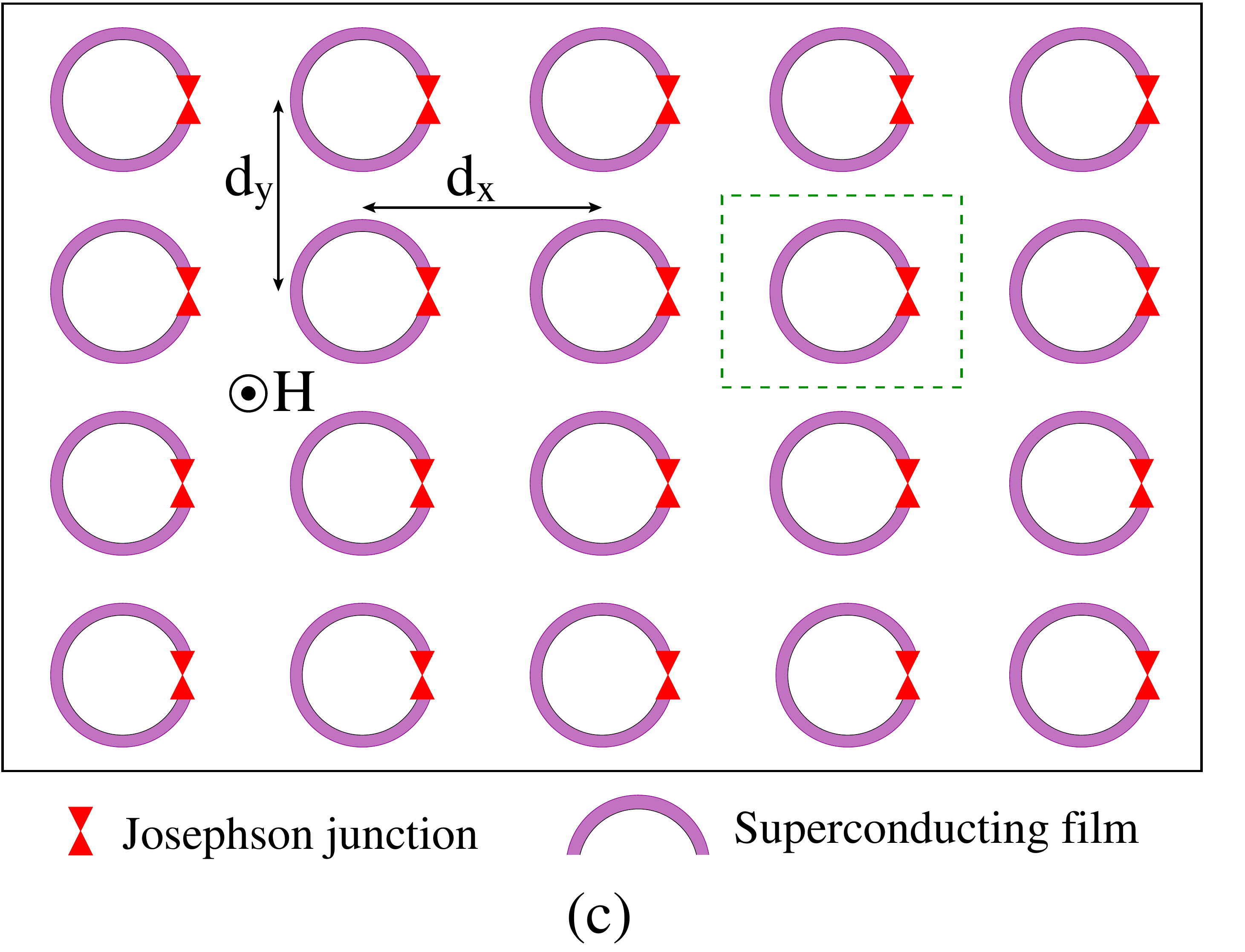}
\includegraphics[angle=0, width=0.45\linewidth]{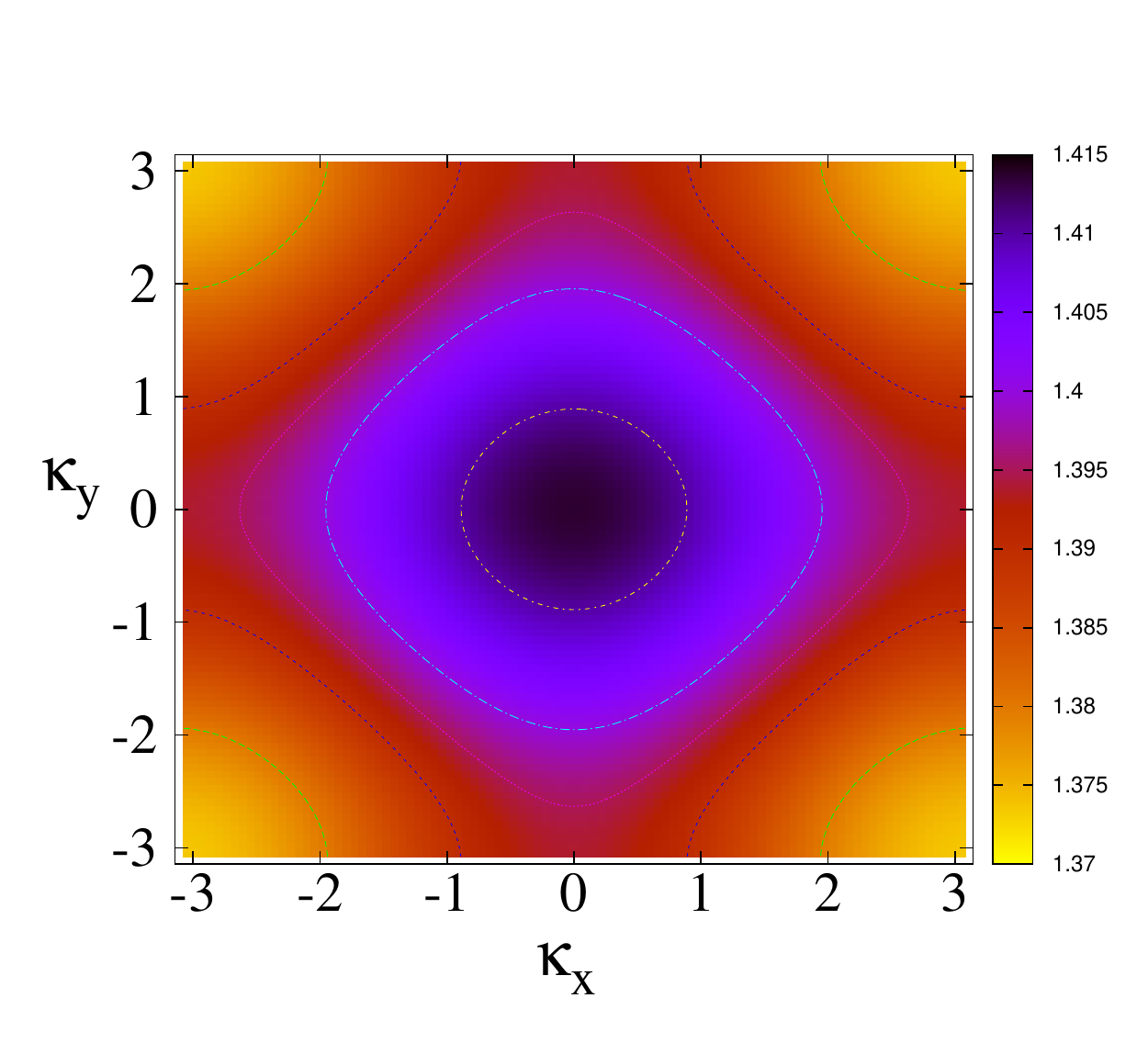}
\caption{
 Left: Schematic drawing of a planar SQUID metamaterial in a time-dependent 
 magnetic field ${\bf H}(t)$.
 Right: Density plot and contours of the linear frequency dispersion 
 $\Omega_{\vec\kappa}$ on the $\kappa_x - \kappa_y$ plane calculated from Eq.
 (\ref{Ch5.15.8}), for a two-dimensional SQUID metamaterial with  
 $\lambda_x = \lambda_y =-0.014$ and $\beta=0.15$ ($\beta_L =0.7$).
\label{fig2.09-09}
}
\end{figure}

In order to increase the dimensionality and obtain the dynamic equations for the 
fluxes through the loops of the SQUIDs arranged in an orthogonal lattice, as 
shown in the left panel of Fig. \ref{fig2.09-09} in which the unit cell is 
enclosed inside the green-dotted line, we first write the corresponding 
flux-balance relations \cite{Lazarides2008a,Tsironis2009,Lazarides2013b}, 
\begin{eqnarray}
\label{Ch5.15}
  \Phi_{n,m} 
  =\Phi_{ext} + L \, \left[ I_{n,m} +\lambda_x ( I_{n-1,m} + I_{n+1,m} ) 
              +\lambda_y ( I_{n,m-1} + I_{n,m+1} ) \right] ,
\end{eqnarray}
where $\Phi_{n,m}$ is the flux threading the $(n,m)-$th SQUID of the metamaterial, 
$I_{n,m}$ is the total current induced in the $(n,m)-$th SQUID of the 
metamaterial, and $\lambda_{x,y} \equiv M_{x,y} / L$ are the magnetic coupling 
coefficients between neighboring SQUIDs, with $M_x$ and $M_y$ being the mutual 
inductances in the $x$ and $y$ directions, ($M_x, M_y <0$). The subscripts $n$ 
and $m$ run from $1$ to $N_x$ and $1$ to $N_y$, respectively. The current 
$I_{n,m}$ is given by the RCSJ model as
\begin{equation}
\label{Ch5.15.2}
  -I_{n,m} = C\frac{d^2 \Phi_{n,m}}{dt^2} +\frac{1}{R} \frac{d \Phi_{n,m}}{dt} 
    + I_c\, \sin\left( 2\pi\frac{\Phi_{n,m}}{\Phi_0} \right) .
\end{equation}
Eq. (\ref{Ch5.15}) can be inverted to provide the currents as a function of the 
fluxes, and then it can be combined with Eq. (\ref{Ch5.15.2}) to give the 
dynamic equations \cite{Lazarides2008a}
\begin{eqnarray}
\label{Ch5.15.3}
   C\frac{d^2 \Phi_{n,m}}{dt^2} +\frac{1}{R} \frac{d \Phi_{n,m}}{dt} 
    +\Phi_{n,m}
    +I_c\, \sin\left( 2\pi\frac{\Phi_{n,m}}{\Phi_0} \right)
    -\lambda_x ( \Phi_{n-1,m} + \Phi_{n+1,m} )
    -\lambda_y ( \Phi_{n,m-1} + \Phi_{n,m+1} )
\nonumber \\
    =[1 -2( \lambda_x +\lambda_y )] \Phi_{ext} .
\end{eqnarray}    
In the absence of losses ($\gamma=0$), the earlier equations can be obtained from 
the Hamiltonian function
\begin{eqnarray}
\label{Ch5.15.4}
   H =\sum_{n,m} \frac{Q_{n,m}^2}{2 C}
      +\sum_{n,m} 
        \left[\frac{1}{2L} (\Phi_{n,m} -\Phi_{ext} )^2
         -E_J \, \cos\left( 2\pi \frac{\Phi_{n,m}}{\Phi_0} \right) \right]
   -\sum_{n,m} \frac{\lambda_x}{L} (\Phi_{n,m} -\Phi_{ext} ) 
                                   (\Phi_{n-1,m} -\Phi_{ext} )
\nonumber \\
   -\sum_{n,m} \frac{\lambda_y}{L} (\Phi_{n,m} -\Phi_{ext} ) 
                                   (\Phi_{n,m-1} -\Phi_{ext} ) , 
\end{eqnarray}
where 
\begin{eqnarray}
  \label{Ch5.15.5}
    Q_{n,m} =C\, \frac{d \Phi_{n,m}}{dt}
\end{eqnarray}
is the canonical variable conjugate to $\Phi_{n,m}$, and represents the charge 
accumulating across the capacitance of the JJ of each rf SQUID. The above 
Hamiltonian function is the weak coupling version of that proposed in the 
context of quantum computation \cite{Roscilde2005}. Eqs. (\ref{Ch5.15.3}) can be 
written in normalized form as 
\begin{eqnarray}
\label{Ch5.15.6}
  \ddot{\phi}_{n,m} +\gamma \dot{\phi}_{n,m} +\phi_{n,m} 
                    +\beta\, \sin( 2 \pi \phi_{n,m} )
  -\lambda_x ( \phi_{n-1,m} +\phi_{n+1,m} ) 
  -\lambda_y ( \phi_{n,m-1} +\phi_{n,m+1} )
   = \phi_{eff} , 
\end{eqnarray}	
where the overdots denote differentiation with respect to the normalized time 
$\tau$, and 
\begin{equation}
\label{Ch5.15.7}
  \phi_{eff} =[1-2(\lambda_x +\lambda_y)] \phi_{ext};
   \qquad
  \phi_{ext} = \phi_{dc}  +\phi_{ac} \cos(\Omega \tau ) .
\end{equation}   
The frequency dispersion of linear flux-waves in two-dimensional SQUID 
metamaterials can be obtained with the standard procedure, by using plane wave
trial solutions into the linearized dynamic equations (\ref{Ch5.15.6}). That 
procudure results in the relation 
\begin{eqnarray}
\label{Ch5.15.8}
   \Omega \equiv \Omega_{\bf \kappa} 
   =\sqrt{\Omega_{SQ}^2 -2( \lambda_x \, \cos \kappa_x
                                +\lambda_y \, \cos \kappa_y ) }  , 
\end{eqnarray}
where ${\bf \kappa}=(\kappa_x, \kappa_y)$ is the normalized wavevector. The 
components of ${\bf \kappa}$ are related to those of the wavevector in physical 
units ${\bf k}=(k_x, k_y)$ through $\kappa_{x,y} = d_{x,y} \, k_{x,y}$ with 
$d_{x}$ and $d_{y}$ being the wavevector component and center-to-center distance 
between neighboring SQUIDs in $x-$ and $y-$direction, respectively. A density 
plot of the frequency dispersion equations (\ref{Ch5.15.8}) on the 
$\kappa_x - \kappa_y$ plane is shown in the right panel of Fig. \ref{fig2.08-08} 
for a tetragonal (i.e., $d_x =d_y$) SQUID metamaterial. Assuming thus that the 
coupling is isotropic, i.e., $\lambda_x =\lambda_y =\lambda$, the maximum and 
minimum values of the linear frequency band are then obtained by substituting 
${\bf \kappa} =(\kappa_x, \kappa_y) =(0, 0)$ and $(\pi,\pi)$, respectively, into 
Eq. (\ref{Ch5.15.8}). Thus we get
\begin{equation}
\label{Ch5.15.81}
  \omega_{max} = \sqrt{ 1 +\beta_L +4 |\lambda| } , \qquad
  \omega_{min} = \sqrt{ 1 +\beta_L -4 |\lambda| } ,
\end{equation}
that give an approximate bandwidth 
$\Delta \Omega \simeq 4 |\lambda| / \Omega_{SQ}$.

Note that the dissipation term $+\frac{1}{R} \frac{d \Phi_{n,m}}{dt}$ appearing 
in Eq. (\ref{Ch5.15.3}) may result from the corresponding Hamilton's equations 
with a time-dependent Hamiltonian \cite{Lazarides2008a}
\begin{eqnarray}
  \label{Ch5.15.9}
   H =e^{-t/\tau_C} \, \sum_{n,m} \frac{Q_{nm}^2}{2 C}
    +e^{+t/\tau_C} \, \sum_{n,m} 
     \left[\frac{1}{2L} (\Phi_{nm} -\Phi_{ext} )^2
      -E_J \, \cos\left( 2\pi \frac{\Phi_{nm}}{\Phi_0} \right) \right]
\nonumber \\
   -\sum_{n,m}  \left[ 
    \frac{\lambda_x}{L} (\Phi_{nm} -\Phi_{ext} ) (\Phi_{n-1,m} -\Phi_{ext} )
   +\frac{\lambda_y}{L} (\Phi_{nm} -\Phi_{ext} ) (\Phi_{n,m-1} -\Phi_{ext} ) 
    \right] ,
\end{eqnarray}
where $E_J \equiv \frac{I_c\, \Phi_0}{2\pi}$ is the Josephson energy, 
$\tau_C=R\, C$, and 
\begin{equation}
    Q_{nm} = e^{+t/\tau_C} \, C\, \frac{d \Phi_{nm}}{dt}
\end{equation}
is the new canonical variable conjugate to $\Phi_{nm}$ which represents the 
generalized charge across the capacitance of the JJ of each rf SQUID. The 
Hamiltonian in Eq. (\ref{Ch5.15.9}) is a generalization in the 
two-dimensional lossy case of that employed in the context of quantum computation 
with rf SQUID qubits \cite{Roscilde2005,Corato2006}.

\subsection{Wide-Band SQUID Metamaterial Tunability with dc Flux}
\begin{figure}[t!]
\includegraphics[angle=0, width=0.9\linewidth]{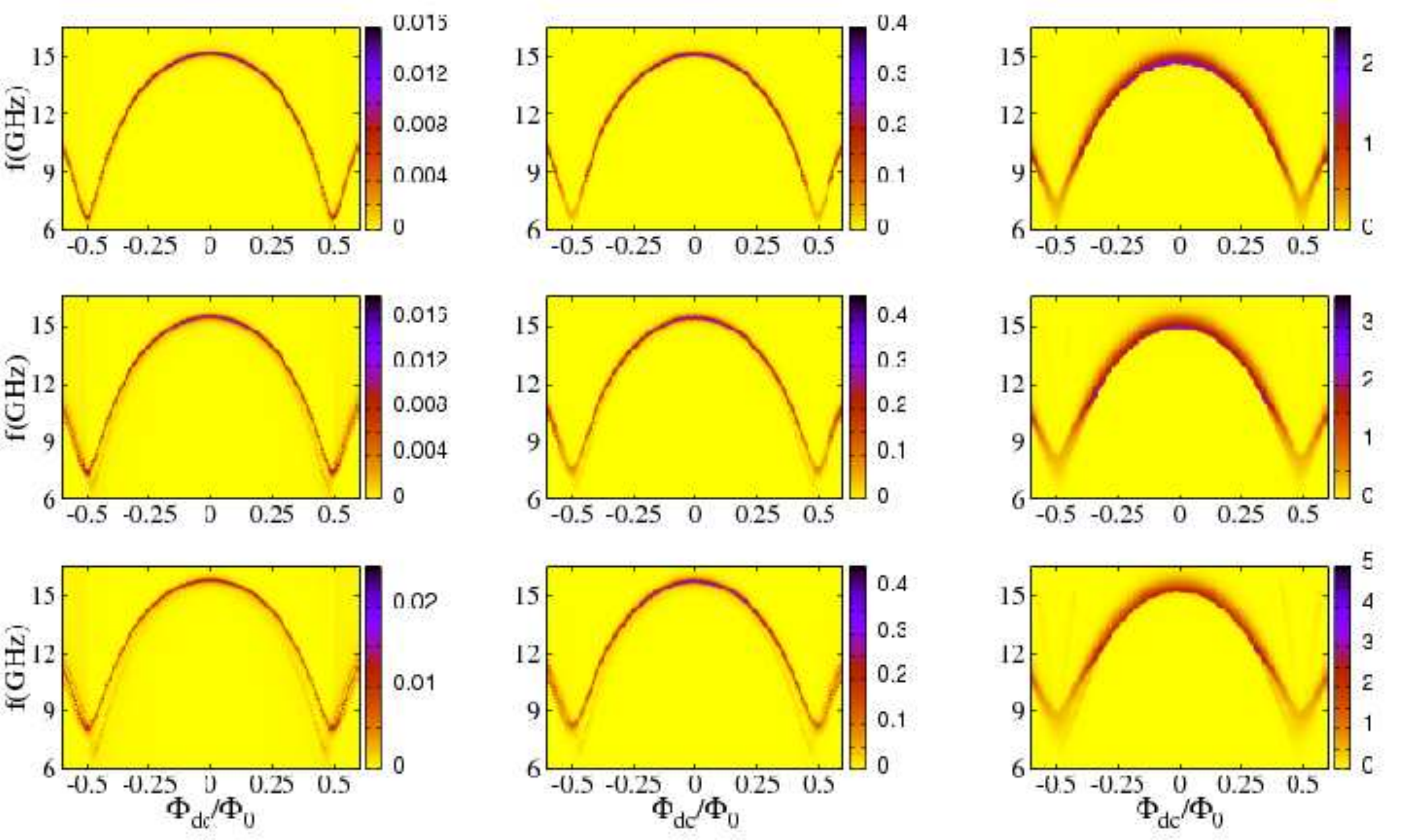}
\caption{ 
 Density plot of the total energy per SQUID, $E_{tot}$, averaged over a period 
 $T$ of temporal evolution, as a function of the dc flux bias 
 $\phi_{dc} =\Phi_{dc}/\Phi_0$ and the driving frequency $f$, for $N_x=N_y=11$, 
 $\beta_L \simeq 0.7$, $\gamma =0.009$, and several combinations of ac flux 
 amplitude $\phi_{ac}$ and coupling coefficients $\lambda_x =\lambda_y$ 
 (tetragonal lattice). The ac flux amplitude $\phi_{ac}$ increases from left to 
 right, while the coupling increases from top bottom.
First row:
 $\lambda_x =\lambda_y =-0.01$, 
and $\phi_{ac}=1/5000$ (left); $1/1000$ (middle); $1/200$ (right).
Second row:
 $\lambda_x =\lambda_y =-0.03$, 
 and $\phi_{ac}=1/5000$ (left); $1/1000$ (middle); $1/200$ (right).
Third row:
 $\lambda_x =\lambda_y =-0.05$, 
 and $\phi_{ac}=1/5000$ (left); $1/1000$ (middle); $1/200$ (right).
 The single-SQUID resonance frequency $f_{SQ}$ used in the calculations is set 
 to $15~GHz$.
\label{fig2.10-10}
}
\end{figure}

An rf SQUID metamaterial is shown to have qualitatively the same behavior as a 
single rf SQUID with regard to dc flux and temperature tuning. Thus, in close 
similarity with conventional, metallic metamaterials, rf SQUID metamaterials 
acquire their electromagnetic properties from the resonant characteristics of 
their constitutive elements, i.e., the individual rf SQUIDs. However, there are 
also properties of SQUID metamaterials that go beyong those of individual rf 
SQUIDs; these emerge through collective interaction of a large number of SQUIDs 
forming a metamaterial. Numerical simulations using the SQUID metamaterial 
models presented in the previous sub-section confirm the experimentally observed 
tunability patterns with applied dc flux in both one and two dimensions. Here, 
numerical results for the two-dimensional model are presented; however, the dc 
flux tunability patterns for either one- or two-dimensional SQUID metamaterials 
are very similar. Due to the weak coupling between SQUIDs, for which the 
coupling coefficient has been estimated to be of the order of $10^{-2}$ in 
normalized units \cite{Zhang2015}, the nearest-neighbor coupling between SQUIDs 
provides reliable results. The normalized equations for the two-dimensional 
model equations (\ref{Ch5.15.6}) are \cite{Lazarides2008a,Lazarides2013b} 
\begin{eqnarray}
\label{Ch5.16}
  \ddot{\phi}_{n,m} +\gamma \dot{\phi}_{n,m} 
   hi_{n,m} +\beta\, \sin( 2 \pi \phi_{n,m} )
  -\lambda_x ( \phi_{n-1,m} +\phi_{n+1,m} ) 
  -\lambda_y ( \phi_{n,m-1} +\phi_{n,m+1} )
   =\phi_{eff},
\end{eqnarray}
where $\phi_{eff} =[1-2(\lambda_x +\lambda_y)] \phi_{ext}$ and
$\phi_{ext} = \phi_{dc}  +\phi_{ac} \cos(\Omega \tau )$. The total (symmetrized) 
energy of the SQUID metamaterial, in units of the Josephson energy $E_J$, is 
then  
\begin{eqnarray}
\label{Ch5.16.2}
   E_{tot} =\sum_{n,m} \left\{ \frac{\pi}{\beta} 
       \left[ \dot{\phi}_{n,m}^2 +( \phi_{n,m} -\phi_{ext} )^2 \right]
           +1 -\cos(2\pi \phi_{n,m}) \right\}
\nonumber \\ 
      -\frac{\pi}{\beta}  \sum_{n,m} \left\{ \left[ 
   \lambda_x ( \phi_{n,m} -\phi_{ext} )( \phi_{n-1,m} -\phi_{ext} )
  +\lambda_x ( \phi_{n+1,m} -\phi_{ext} )( \phi_{n,m} -\phi_{ext} ) \right. \right.
\nonumber \\  
\left. \left. 
  +\lambda_y ( \phi_{n,m-1} -\phi_{ext} )( \phi_{n,m} -\phi_{ext} )
  +\lambda_y ( \phi_{n,m} -\phi_{ext} )( \phi_{n,m+1} -\phi_{ext} ) \right] \right\} . 
\end{eqnarray}
For $\phi_{ext} =\phi_{dc}$, the average of that energy over one period $T$ of 
evolution
\begin{equation}
\label{Ch5.16.3}
   <E_{tot}>_T =\frac{1}{T} \int_0^T d\tau E_{tot} (\tau) ,
\end{equation}
where $T=2\pi /\Omega$ with $\Omega$ being the normalized driving frequency, is 
constant when the obtained solution is locked to that driving frequency. In the 
following, the term "tunability of the resonance" for the SQUID metamaterial is 
used; however, the notion of the resonance is rather appropriate for a single 
SQUID only. The SQUID metamaterial is 
capable of absorbing substantial amount of energy for frequencies within its 
linear frequency band, given by Eq. (\ref{Ch5.15.8}); however, the energy 
absorption in that band is far from being uniform. Thus, the term "tunability of 
the resonance" for the SQUID metamaterial refers to that frequency at which the 
highest absorption of energy occurs (also note that for strong nonlinearities, 
the SQUID metamaterial can absorb significant amount of energies in other 
frequency bands as well, see next Section). Typical resonance tunability 
patterns are shown in Fig. \ref{fig2.10-10} as density plots of $<E_{tot}>_\tau$ 
as a function of the driving frequency $f$ (in nutural units) and the dc flux 
bias $\phi_{dc}$ for several combinations of $\phi_{ac}$ and 
$\lambda_x =\lambda_y$. Note that the energy background in Fig. \ref{fig2.10-10} 
has been removed for clarity. The thick lines with darker color depict the 
regions of the map with high energy absorption. The parameters used in the 
calculations are similar to the experimental ones 
\cite{Butz2013a,Trepanier2013,Zhang2015}, although no attempt was made to 
exactly fit the observed patterns. These parameter values are consistent with 
a single-SQUID resonance frequency $f \simeq 15~GHz$, which is also used in 
the calculations and to express the frequency in natural units. In Fig. 
\ref{fig2.10-10}, the ac flux amplitude $\phi_{ac}$ increases from left to right, 
while the coupling $\lambda_x =\lambda_y$ increases from top bottom. The 
resonance become stronger as we move from the left to the right panels, as the 
nonlinear effects become more and more important with increasing $\phi_{ac}$. 
When going from top to bottom panels, with increasing $|\lambda_x| =|\lambda_y|$, 
a smearing of the resonance is observed, along with the appearence of secondary 
resonances. The latter manifest themselves as thin red (dark) curves that are 
located close to the main shifting pattern, and they are better seen around 
half-intefger values of the applied dc flux. In order to obtain accurate 
tunability patterns, a few hundreds of absorbed energy-frequency curves (one for 
each $\phi_{dc}$) have been calculated. Eqs. (\ref{Ch5.16}) are typically 
integrated with a fourth-order Runge-Kutta algorithm with constant time-step.

\begin{figure}[!t]
\center{
\includegraphics[angle=-0, width=0.9\linewidth]{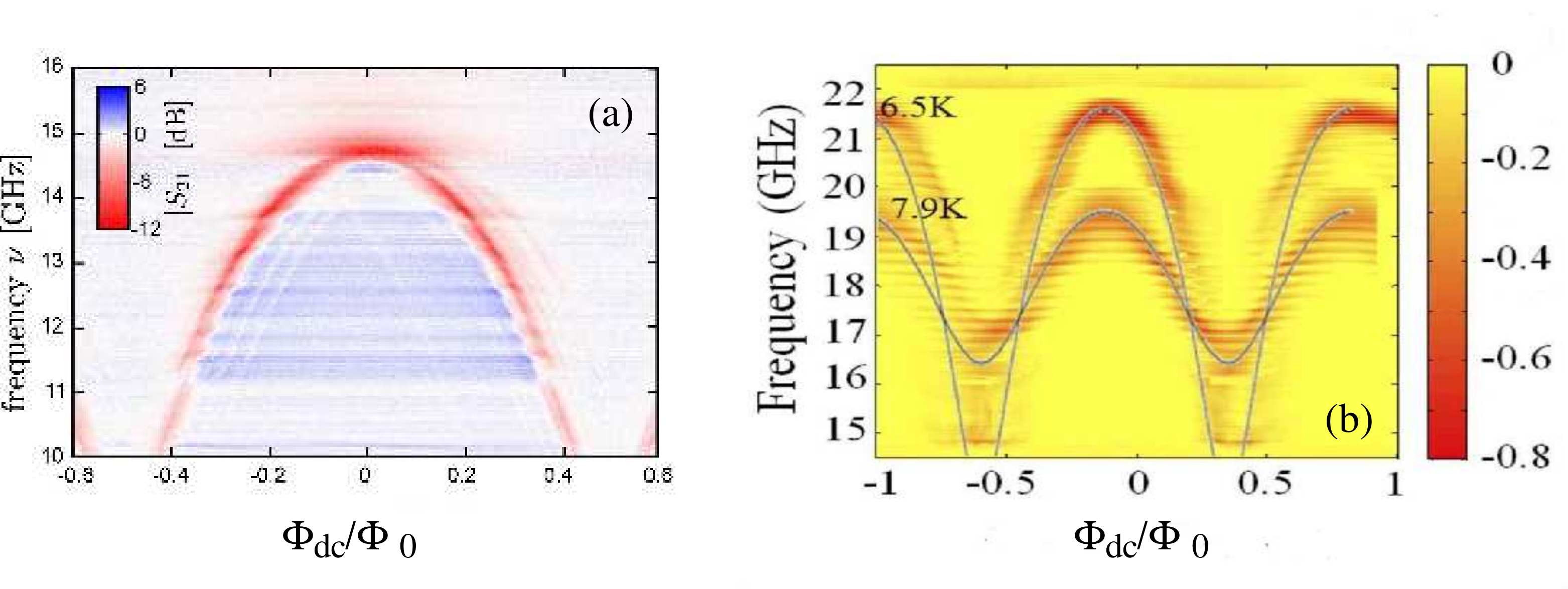}
}
\caption{
 Measured magnitude of the complex transmission $|S_{21}|$ as a function of the 
 driving frequency $f$ of a weak ac flux field and the applied dc flux bias 
 $\phi_{dc} =\Phi_{dc} /\Phi_0$ for 
 (a) a quasi-one-dimensional rf SQUID metamaterials (in a double-chain 
     configuration with each chain comprising $54$ rf SQUIDs) \cite{Butz2013a}; 
 (b) a two-dimensional ($27\times 27$) rf SQUID metamaterial \cite{Trepanier2013}. 
 In both figures the red (dark) features indicate regions of reduced transmission,
 which corresponds to resonant response. The solid-grey lines in (b) are the 
 calculated single-SQUID resonance frequencies. 
\label{fig5.10}
}
\end{figure}

Experimentally, the resonance tunability patterns are obtained by measuring the 
magnitude of the microwave complex transmission $|S_{21}|$ (in $dB$s) of the 
SQUID metamaterials \cite{Butz2013a,Trepanier2013}. The samples, which are either 
quasi-one-dimensional or two-dimensional, comprise nominally identical elements 
and they were placed inside coplanar waveguides. When excited by a weak 
microwave (rf) signal in the presence of a dc flux bias, the resonances of the 
SQUID metamaterials can be detected as dips in the frequency-dependent 
$|S_{21}| (\omega)$ through the waveguide. The resulting wide-band tunability 
patterns, shown in Fig. \ref{fig5.10}(a) and (b) for quasi-one-dimensional and 
two-dimensional SQUID metamaterials, respectively, clearly exhibit a periodicity 
in the dc flux of $\Phi_0$. Note the similarity between these patterns and those 
obtained for a single SQUID in Fig. \ref{fig2.07-07}. Since the coupling between 
SQUIDs is weak, their frequency bands are very narrow; however, the red (dark) 
regions indicating resonant response are actually thinner that the corresponding 
frequency bandwidths. This is because the resonant response is very strong at 
some particular frequencies, seemingly close to the maximum frequency of the 
band. The frequency of highest and lowest response is obtained for dc flux equal 
to integer and half-integer multiples of the flux quantum $\Phi_0$, respectively, 
as it can be seen clearly in Fig. \ref{fig5.10}(b) which actually contains two 
patterns for different temperatures, i.e., for $T \simeq 6.5~K$ and 
$T \simeq 7.9~K$. The resonance frequency for the lowest temperature pattern 
varies with the dc flux from approximately $15~GHz$ to $21.5~GHz$, providing 
nearly $30\%$ tunability! The tunability range reduces with increasing 
temperature, as can be readily inferred by comparing the data for the two 
different temperatures. With an increase of temperature from $T \simeq 6.5~K$ to 
$T \simeq 7.9~K$, the tunability range has been almost halved. Note that similar 
behavior is observed in the corresponding curves for a single SQUID. The 
resonant response of the SQUID metamaterial is stronger close to dc fluxes equal 
to integer multiples of $\Phi_0$ (this is visible in both subfigures of Fig. 
\ref{fig5.10}). Thus, a "useful" frequency range can be identified in which the 
depth of the resonance does not change considerably with $\phi_{dc}$. In Fig. 
\ref{fig5.10}(a), that range lies between $13 ~GHz$ and $14.5 ~GHz$ 
\cite{Butz2013a}.

Another interesting effect which is clearly visible in the fist column of Fig. 
\ref{fig2.10-10}, corresponding to low $\phi_{ac}$ (closer to the linear limit), 
is the slight increase of the resonance frequency at $\phi_{dc} =0$ of the SQUID 
metamaterial with increasing the magnitude of the coupling coefficients 
$\lambda_x =\lambda_y$. This effect, as well as the shape of the resonant 
response for $\phi_{dc}$ between $-1/2$ and $+1/2$ can be understood within an 
approximate treatment which is valid for low ac field amplitudes $\phi_{ac}$.
Assume that $\phi_{n,m} \simeq \phi$ for any $n,m$, i.e., that the SQUIDs are 
synchronized \cite{Lazarides2013b} (note that small deviations from complete 
synchronization always appear for finite size SQUID metamaterials). Then 
substitute $\phi_{n,m} = \phi$, and $\gamma =0$, $\lambda_x =\lambda_y =\lambda$ 
into Eqs. (\ref{Ch5.16}) to get   
\begin{equation}
\label{Ch5.16.4}
   \ddot{\phi} +(1-4\lambda) \phi +\beta\, \sin(2\pi \phi) 
             =(1-4\lambda) ( \phi_{dc} +\phi_{ac} \cos(\Omega t ). 
\end{equation}
In the earlier equation we further use the approximation
$\beta \sin(2\pi \phi) \simeq \beta_L \phi -\frac{2\pi^2}{3} \beta_L \phi^3$,
and the ansatz $\phi =\phi_0 +\phi_1  \cos(\Omega t )$. Substituting into
Eq. (\ref{Ch5.16.4}), using the rotating wave approximation (RWA), and 
separating constant from time-dependent terms, we get
\begin{equation}
\label{Ch5.16.5}
  \frac{2\pi^2}{3} \beta_L \phi_0^3 -(1-4\lambda+\beta_L) \phi_0 
     -\frac{3}{2}\phi_0 \phi_1^2 +(1-4\lambda) \phi_{dc} =0 , \qquad 
  \frac{\pi^2}{2} \beta_L \phi_1^3 
     -\left\{ (1-4\lambda+\beta_L-\Omega^2) -2\pi^2 \beta_L \phi_0^2 \right\} 
      \phi_1 +(1-4\lambda) \phi_{ac} =0 .
\end{equation}
Limiting ourselves to the case $\phi_1 < \phi_0 <<1$, we may simplify Eqs.
(\ref{Ch5.16.5}) by neglecting terms proportional to $\phi_1^3$, $\phi_0^3$, and 
$\phi_0 \phi_1^1$. Note that we keep the term $\propto \phi_0^2 \phi_1$, i.e., 
the lowest order coupling term between the two equations. Then, the resulting 
equations can be easily solved to give
\begin{eqnarray}
\label{Ch5.16.6}
   \phi_0 = \frac{(1-4\lambda) \phi_{dc}}{(1-4\lambda+\beta_L)} ; \qquad
   \phi_1 = \frac{(1-4\lambda) \phi_{ac}}
      {\left\{ (1-4\lambda+\beta_L-\Omega^2) -2\pi^2 \beta_L \phi_0^2 \right\}} .
\end{eqnarray}
Obviously, the ac flux amplitude in the SQUIDs, $\phi_1$, attains its maximum 
value when the expression in the curly brackets in the denominator of Eq. 
(\ref{Ch5.16.6}) is zero. Solving that expression for $\Omega$, we get
\begin{equation}
\label{Ch5.16.7}
   \Omega =\sqrt{ (1-4\lambda+\beta_L) -(2\pi^2 \beta_L)
       \frac{(1-4\lambda)^2 \phi_{dc}^2}{(1-4\lambda+\beta_L)^2} } ,
\end{equation}
or, in natural units
\begin{equation}
\label{Ch5.16.8}
   f =\frac{f_{SQ}}{\sqrt{1+\beta_L}} 
       \sqrt{ (1-4\lambda+\beta_L) -(2\pi^2 \beta_L)
        \frac{(1-4\lambda)^2 \phi_{dc}^2}{(1-4\lambda+\beta_L)^2} } ,
\end{equation}
which corresponds to the "resonance frequency" of the SQUID metamaterial itself,
with $f_{SQ}$ being the single-SQUID resonance frequency. This is exactly the 
frequency for which the resonant response of the SQUID metamaterial is stronger. 
Moreover, the dependence of that frequency on the coupling coefficients, which 
has been experimentally observed and also seen in the numerical simulations, 
implies that at that frequency the SQUIDs in the metamaterials exhibit a high 
degree of synchronization.

\begin{figure}[!t]
\includegraphics[angle=0, width=0.90\linewidth]{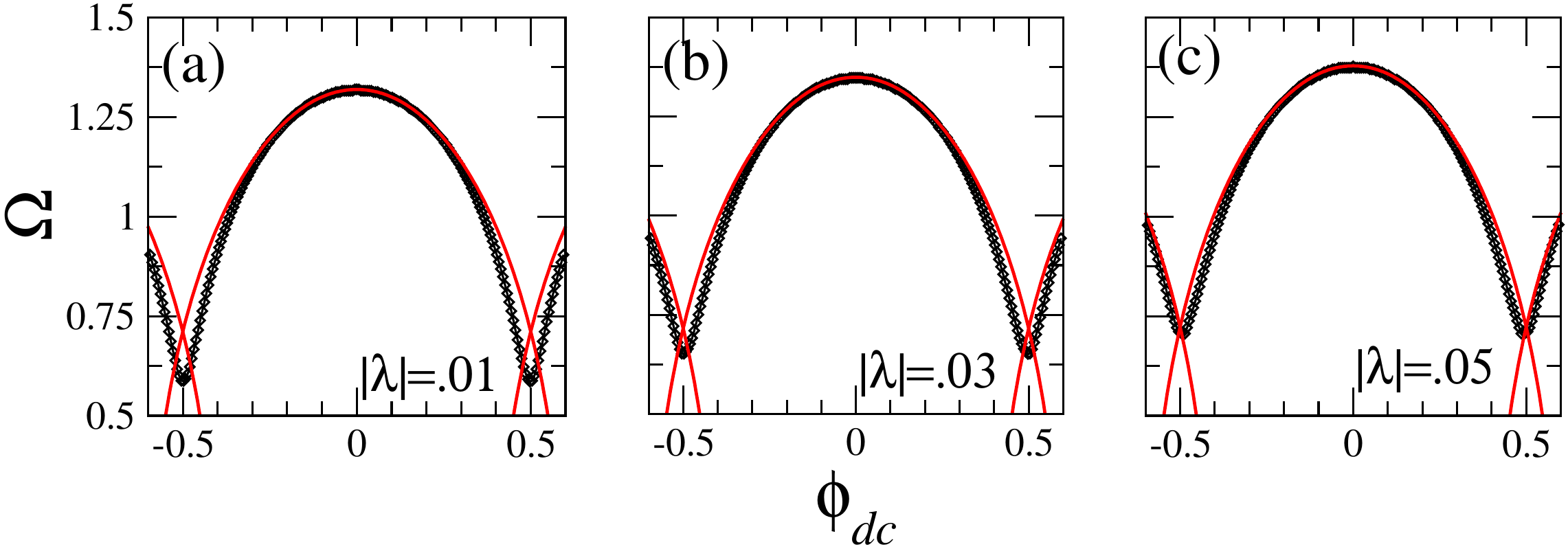}
\caption{
 Normalized frequency at maximum response of the SQUID metamaterial, $\Omega$, 
 as a function of the applied dc flux, $\phi_{dc}$, in the presence of a 
 low-amplitude ac flux $\phi_{ac}$. The black circles have been extracted from 
 the numerical data of Fig. \ref{fig2.10-10}, while the red solid lines are 
 plotted from Eq. (\ref{Ch5.16.7}).
 Parameters: 
 $N_x=N_y=11$, $\phi_{ac}=1/5000$, $\gamma =0.009$, $\beta_L =0.7$, and
  (a) $\lambda_x =\lambda_y =-0.01$; 
  (b) $\lambda_x =\lambda_y =-0.03$; 
  (c) $\lambda_x =\lambda_y =-0.05$.
\label{fig5.10.2}
}
\end{figure}

From the actual numerical data of the resonance tunability patterns presented in 
the left column of Fig. \ref{fig2.10-10} (low $\phi_{ac}$), which are 
calculated for increasing magnitude of the coulping coefficients (from top to 
bottom), the maximum response frequency has been extracted by simply identifying 
that frequency at which $<E_{tot}>_{\tau_i}$ is maximum. These curves, for 
$\lambda_x =\lambda_y =\lambda=-0.01$, $-0.03$, and $-0.05$, are shown in Fig. 
\ref{fig5.10.2}a, b, and c, respectively, along with the corresponding ones 
calculated from Eq. (\ref{Ch5.16.8}). The simple expression (\ref{Ch5.16.8}),
which contains only two parameters, $\lambda$ and $\beta_L$, fairly agrees with 
the simulations for $\phi_{ac} \ll 1$ in a rather wide region of dc fluxes, 
i.e., from $\phi_{dc} \sim -0.3$ to $\sim +0.3$. Within this interval, the 
resonance frequency in Fig. \ref{fig5.10.2}a may change from $\Omega=1.12$ to 
$1.32$ that makes a variation of $\sim 15\%$. Similar tunability ranges are 
observed in Figs. \ref{fig5.10.2}b and c. For larger $\phi_{dc}$, the importance 
of the term $\propto \phi_0^3$ increases and it cannot be neglected for the 
solutions of Eqs. (\ref{Ch5.16.5}). However, the agreement between the two 
curves seems to get better for larger $|\lambda|$. By setting $\phi_{dc} =0$ in 
Eq. (\ref{Ch5.16.8}) we get that 
$f =\frac{f_{SQ}}{\Omega_{SQ}} \sqrt{ (\Omega_{SQ}^2 -4\lambda) }$, which, for 
$f_{SQ} =15~GHz$, $\Omega_{SQ} =1.304$ ($\beta_L=0.7$), 
$\lambda =-0.01, ~-0.03, ~-0.05$ gives respectively, $f=15.2, ~15.5, 15.9 ~GHz$ 
in agreement with the numerical results (Fig. \ref{fig5.10.2}). The 
$\lambda-$dependence of the SQUID metamaterial resonance frequency is weaker in 
the corresponding one-dimensional case. 
The resonance shift due to nonlinearity has been actually observed in a 
Josephson parametric amplifier driven by fields of different power levels 
\cite{Castellanos2007}, while the shift with applied dc flux has been observed 
in high$-T_c$ rf SQUIDs \cite{Zeng2000} and very recently in a low$-T_c$ rf 
SQUID in the linear regime \cite{Jung2013}. Systematic measurements on microwave 
resonators comprising SQUID arrays are presented in Refs. 
\cite{Castellanos2007,Palacios2008}.

\subsection{Energy Transmission in SQUID Metamaterials}
\begin{figure}[t!]
\includegraphics[angle=0, width=0.90\linewidth]{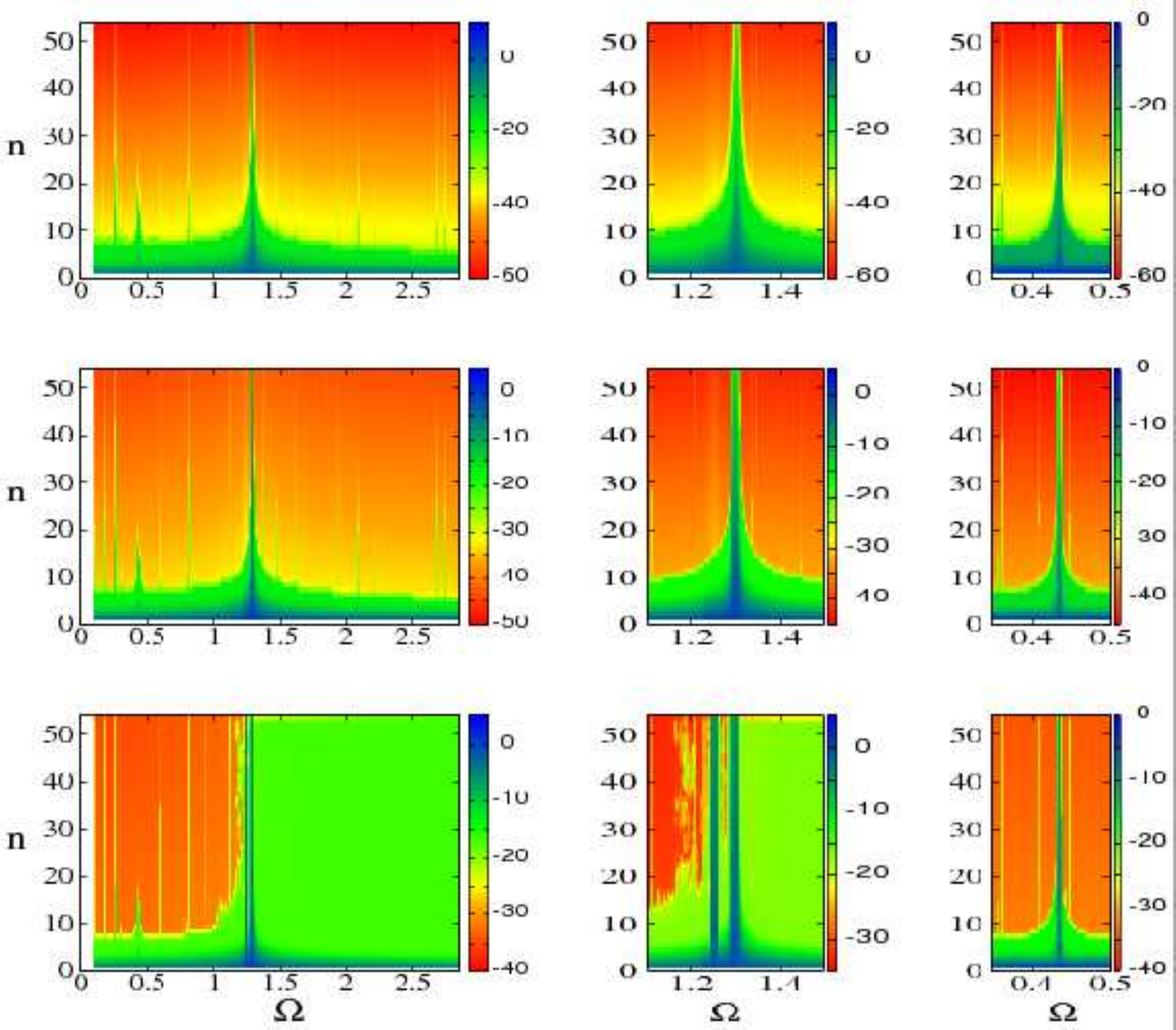}
\caption{ 
 Energy transmission through a one-dimensional SQUID metamaterial with $N=54$ 
 SQUIDs. The decimal logarithm of the energy density averaged over 
 $\tau_i =2000\, T$ time units, $\log_{10} [ <E_n>_{\tau_i} ]$, is mapped on 
 the frequency $\Omega$ - site-number $n$ plane for $\beta_L =0.7$ 
 ($\beta=0.1114$), $\lambda=-0.01$, $\phi{ac}=0.1$, and $\gamma=0.009$ (upper); 
 $\gamma=0.004$ (middle);  $\gamma=0.001$ (lower). The middle and left columns 
 are enlargements of frequency bands around the fundamental and the subharmonic 
 resonance, respectively, at $\Omega \simeq 1.302$ and $0.43$.  
\label{fig5.11}
}
\end{figure}

Conventional SRR-based metamaterials, are capable of transmitting energy through 
the array of resonators (i.e., the SRRs), carried by a new kind of waves, the 
so-called magnetoinductive waves 
\cite{Shamonina2004,Syms2006,Shadrivov2006b,Syms2010,Stevens2010,Lazarides2011a}, 
which have been actually observed in one-dimensional SRR arrays with a 
relatively small number of elements \cite{Wiltshire2003,Shadrivov2007}. Very 
much in the same way, SQUID metamaterials are capable of transmitting energy 
through magnetoinductive flux waves. In order to investigate the transmission of 
energy through a SQUID metamaterial, a one-dimensional array of SQUIDs 
comprising $N=54$ identical elements with $\beta_L =0.7$ ($\beta=0.1114$) 
locally coupled to their nearest neighbors, is considered. That array is driven 
at one end (say the left end, that with $n=1$) by an ac flux field of amplitude 
$\phi_{ac}$ and frequency $\Omega$. Then, Eqs. (\ref{Ch5.14}) become 
\begin{equation}
\label{Ch5.17}
  \ddot{\phi}_n +\gamma \dot{\phi}_n +\phi_n +\beta\, \sin( 2 \pi \phi_n )
  -\lambda ( \phi_{n-1} +\phi_{n+1} ) 
   = (1-2\lambda) \phi_{ext} \, \delta_{n,1} ,
\end{equation}
where the Kroneckers' delta $\delta_{n,1}$ indicates that only the SQUID with 
$n=1$ is driven by the ac field $\phi_{ext} =\phi_{ac}\, \cos(\Omega \tau)$. The 
total energy in this case is obtained by appropriate modification of Eq. 
(\ref{Ch5.16.2}), as
\begin{eqnarray}
\label{Ch5.17.2}
    E_{tot} =\sum_{n=1}^N  \left\{ \frac{\pi}{\beta}
       \left[ \dot{\phi}_{n}^2 +( \phi_{n} -\phi_{ext} \delta_{n,1} )^2 \right] 
           +\left[ 1-\cos(2\pi \phi_n) \right] \right\}
\nonumber \\
  -\frac{\pi}{\beta}  \sum_{n=1}^N \left[
   \lambda ( \phi_{n} -\phi_{ext} \delta_{n,1} )
           ( \phi_{n-1} -\phi_{ext} \delta_{n,1} ) 
  +\lambda ( \phi_{n+1} -\phi_{ext} \delta_{n,1} )
           ( \phi_{n} -\phi_{ext} \delta_{n,1} ) 
   \right] . 
\end{eqnarray}
The dynamic equations (\ref{Ch5.17}) implemented with the boundary conditions 
$\phi_0 =\phi_{N+1}=0$ are integrated in time until transients are eliminated 
and the system reaches a stationary state. Typically, $12000\, T$ time units of 
time-integration, where $T=2\pi / \Omega$, are sufficient for that purpose. 
The energy density of the SQUID metamaterial is calculated as a function of time 
from Eq. (\ref{Ch5.17.2}) for $\tau_i =2000\, T$ time units more; then, the 
decimal logarithm of the energy density averaged over $\tau_i$,
$\log_{10} [ <E_n>_{\tau_i} ]$, is mapped on the frequency $\Omega$ - site 
number $n$ plane and shown in Fig. \ref{fig5.11} (high transmission regions are 
indicated with darker colors). In Fig. \ref{fig5.11}, the quantity 
$\log_{10} [ <E_n>_{\tau_i} ]$ is shown for three different values of the 
dissipation coefficient $\gamma$ for fixed $\lambda=-0.01$ and $\phi{ac}=0.1$. 
For that value of $\phi{ac}$, the nonlinear effects become stronger for lower 
$\gamma$. The $\log_{10} [ <E_n>_{\tau_i} ]$ map for relatively strong 
dissipation ($\gamma=0.009$) is shown in the upper panel of Fig. \ref{fig5.11}; 
apparently, significant energy transmission occurs in a narrow band, of the 
order of $\sim 2\lambda$ around the single SQUID resonance frequency 
$\Omega_{SQ} \simeq 1.3$ (for $\beta_L=0.7$). This band almost coincides with 
the linear band for the one-dimensional SQUID metamaterial. Note that energy 
transmission also occurs at other frequencies; e.g., at $\Omega \sim 0.43$ that 
corresponds to a subharmonic resonance ($1/3$). Subharmonic resonances result 
from nonlinearity; in this case, nonlinear effects are already significant due 
to the relatively high $\phi_{ac}$ for all values of $\gamma$. However, for 
decreasing $\gamma$ (middle panel), more energy is transmitted both at 
frequencies in the linear band and the subharmonic resonance band. In the 
following we refer to the latter as the nonlinear band, since it results from 
purely nonlinear effects. With further decrease of $\gamma$ (lower panel), the 
transmitted energy in these two bands becomes more significant. The comparison 
can be made more clear by looking at the panels in the middle and right columns 
of Fig. \ref{fig5.11}, which show enlarged regions of the corresponding panels 
in the left column. The enlargement around the linear band (middle column) 
clearly reveals the increase of the transmitted energy with decreasing $\gamma$. 
For frequencies in the subharmonic nonlinear band, the SQUID metamaterial 
becomes transparent; that type of self-induced transparency due to nonlinearity 
is a robust effect as can be seen in Fig. \ref{fig5.11} (right column panels), 
in which the loss coefficient has been varied by almost an order of magnitude. 
Moreover, in the case of very low losses ($\gamma=0.001$) the linear band splits 
into two bands, in which significant energy transmission occurs. Of those bands, 
the one at lower frequencies is also a nonlinear band; that phenomenon of energy 
transmission in the gap of the linear band(s) is known as 
{\em supratransmission} \cite{Geniet2002}. In the density plots of Fig. 
\ref{fig5.11}, the upper boundary is a reflecting one, which allows for the 
formation of stationary flux wave states in the SQUID metamaterial. However, 
similar calculations performed with a totally absorbing boundary give 
practically identical results. 

\begin{figure}[h!]
\includegraphics[angle=0, width=0.8\linewidth]{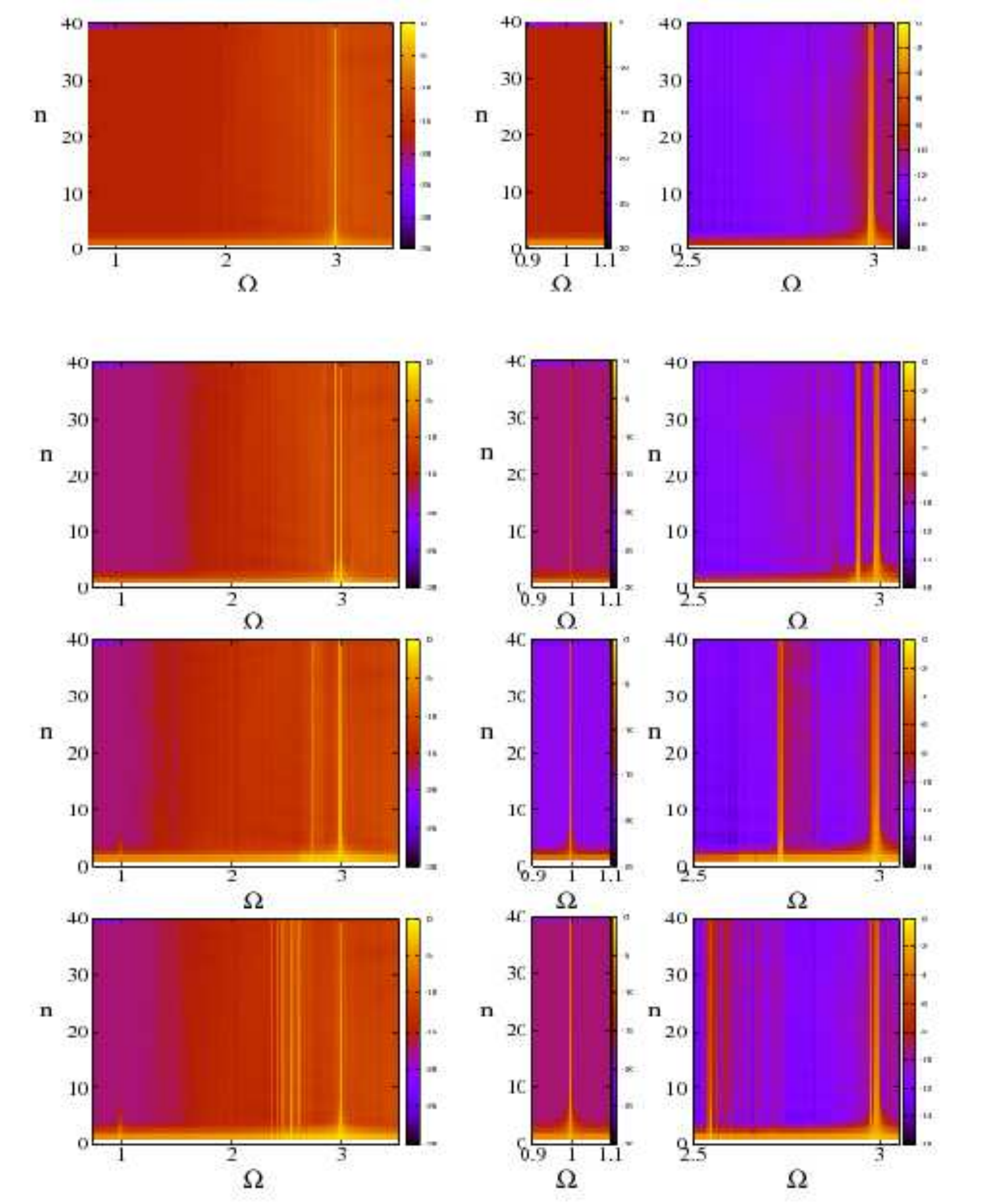}
\caption{ 
 Energy transmission through a one-dimensional SQUID metamaterial with $N=40$ 
 SQUIDs. The decimal logarithm of the energy density averaged over 
 $\tau_i =2000\, T$ time units, $\log_{10} [ <E_n>_{\tau_i} ]$, is mapped on 
 the frequency $\Omega$ - site-number $n$ plane for $\beta=1.27$ 
 ($\beta_L \simeq 8$), $\gamma =0.001$, $\lambda =-0.014$, and 
 (a) $\phi_{ac} =0.001$; (b) $\phi_{ac} =0.01$; (c) $\phi_{ac} =0.1$; 
 (d) $\phi_{ac} =0.2$. From (a) to (d), the panels in the middle and right 
 columns are enlarged regions of the panels in the left column; middle panels 
 enlarge the frequency region around the $L\, C$ resonance, while the right 
 panels enlarge the frequency region around the single SQUID resonance 
 $\Omega_{SQ}$. Red-orange regions indicate the frequency bands in which energy 
 transmission is high.
\label{fig5.12}
}
\end{figure}

In ac driven SQUID metamaterials, the significance of nonlinear effects depends 
both on $\gamma$ and $\phi_{ac}$. For fixed, low $\gamma$ and $\phi_{ac} \ll 1$,
the dynamics is essentially close to be linear, and consequntly the energy 
transmission through a SQUID metamaterial is limited to frequencies within the 
linear band. The strength of nonlinear effects increases, however, with 
increasing $\phi_{ac}$, resulting in the opening of nonlinear energy 
transmission bands just as in the case in which $\gamma$ is varied. In Fig. 
\ref{fig5.12}, maps of $\log_{10} [ <E_n>_{\tau_i} ]$ are shown for a SQUID 
metamaterial with $\beta_L =8$ (note that with that choice of $\beta_L$ the 
SQUIDs are hysteretic). For each of the Figs. \ref{fig5.12}a-d, enlargements of 
the frequency bands around the $L\, C$ frequency ($\Omega =1$) and $\Omega_{SQ}$ 
are shown in the panels of the middle and right columns, respectively. For low 
ac field amplitude $\phi_{ac}$ (Fig. \ref{fig5.12}a), a significant amount of 
energy is transmitted through around $\Omega_{SQ} =3$, as indicated by the 
orange vertical line that is clearly visible in the left panel which corresponds
to the linear band. With increasing $\phi_{ac}$, nonlinearity starts becoming 
important, and the indications of nonlinear transmission are clearly visible in 
Fig. \ref{fig5.12}b. Although on a large scale it appears as a widening of the 
linear band, a closer look (right panel) shows that there are actually two 
distinct bands; the linear band, and a second band which emerges at frequencies 
below it. Both the linear and the nonlinear bands have approximately the same 
width. In the middle panel, a faint orange vertical line indicates that a small 
amount of energy is also transmitted through the array at the $L\, C$ frequency. 
With further increasing $\phi_{ac}$ (Fig. \ref{fig5.12}c), the distance 
separating the nonlinear from the linear band increases; specifically, while the 
linear band remains at frequencies around $\Omega_{SQ}$, the nonlinear band 
shifts to lower frequencies due to nonlinearity. In this case, the energy 
transmitted at the $L\, C$ frequency also becomes significant (middle panel). In 
Fig. \ref{fig5.12}d, the ac field amplitude has been increased to 
$\phi_{ac} =0.2$, where the nonlinearity dominates. While the two main bands 
still persist (with the nonlinear band being shifted to even lower frequencies), 
more, narrower bands seem to appear, while the energy transmission at 
$\Omega \simeq 1$ becomes even more significant.   

For very low $\phi_{ac}$ ($\phi_{dc}=0$), Eqs. (\ref{Ch5.17}) can be linearized 
to 
\begin{eqnarray}
\label{Ch5.17.3}
  \ddot{\phi}_{n} +\gamma \dot{\phi}_{n} +\Omega_{SQ}^2 \phi_{n}
  -\lambda ( \phi_{n-1} +\phi_{n+1} )
  =\bar{\phi}_{ac} \cos(\Omega \tau) \, \delta_{n,1} ,
\end{eqnarray}
where $\bar{\phi}_{ac} =(1-2\lambda) \phi_{ac}$. If we further neglect the loss 
term, Eqs. (\ref{Ch5.17.3}) can be solved exactly in closed form for any 
$\Omega$ and for any finite even $N$, where $N$ is the total number of SQUIDs in 
the one-dimensional metamaterial. By substitution of the trial solution 
$\phi_n = q_n \cos(\Omega \tau)$ into Eqs. (\ref{Ch5.17.3}) and after some 
rearrangement we get
\begin{equation}
\label{Ch5.17.4}
  s q_{n-1} + q_n + s q_{n+1} = \kappa_0 \, \delta_{n,1} ,
\end{equation}
where
\begin{equation}
\label{Ch5.17.5}
  s=-\frac{\lambda}{\Omega_{SQ}^2 -\Omega^2}, \qquad
  \kappa_0 =\frac{\bar{\phi}_{ac}}{\Omega_{SQ}^2 -\Omega^2} ,
\end{equation}
or, in matrix form
\begin{equation}
  \label{Ch5.17.6}
    {\bf q} = \kappa_0 \, \hat{\bf S}^{-1} {\bf E}_1 ,
\end{equation}
where  ${\bf q}$ and ${\bf E}_1$ are $N-$dimensional vectors  with componets 
$q_n$ and $\delta_{n,1}$, respectively, and $\hat{\bf S}^{-1}$ is the inverse of 
the $N\times N$ coupling matrix $\hat{\bf S}$. The latter is a real, symmetric, 
tridiagonal matrix that has its diagonal elements equal to unity, while all the 
other non-zero elements are equal to $s$. The elements of the matrix 
$\hat{\bf S}^{-1}$ can be obtained in closed analytical form 
\cite{Lazarides2010a} using known results for the inversion of more general 
tridiagonal matrices \cite{Huang1997}. Then, the components of the ${\bf q}$ 
vector can be written as
\begin{equation}
  \label{Ch5.17.7}
   q_n = \kappa_0 \left( \hat{\bf S}^{-1} \right)_{n,1} ,
\end{equation}
where $\left( \hat{\bf S}^{-1} \right)_{n,1}$ is the $(n,1)-$element of 
$\hat{\bf S}^{-1}$, whose explicit form is given in reference 
\cite{Lazarides2010a}. Then, the solution of the linear system of Eqs.
(\ref{Ch5.17.4}) with $\gamma=0$ is
\begin{eqnarray}
  \label{Ch5.17.8} 
    \phi_n (\tau) = \kappa_0 \mu \frac{\sin[(N-n+1)\theta']}{\sin[(N+1)\theta']}
       \cos(\Omega \tau) , \qquad
  \theta' = \cos^{-1} \left( \frac{1}{2|s|} \right) , 
\end{eqnarray}
for $s>+1/2$ and $s<-1/2$ (in the linear flux-wave band), and
\begin{eqnarray}
  \label{Ch5.17.82} 
    \phi_n (\tau) = \kappa_ \mu \frac{\sinh[(N-n+1)\theta]}{\sinh[(N+1)\theta]} 
    \cos(\Omega \tau) ,
  \qquad
  \theta = \ln\frac{1+\sqrt{1-(2s)^2}}{2|s|} , 
\end{eqnarray}
for $-1/2 < s < +1/2$ (outside the linear flux-wave band), where
\begin{eqnarray}
  \label{Ch5.17.9}
   \mu= \frac{1}{|s|} \left( -\frac{|s|}{s} \right)^{n-1} .
\end{eqnarray}
The above expressions actually provide the asymptotic solutions, i.e., after the 
transients due to dissipation, etc., have died out. Thus, these driven linear 
modes correspond to the possible stationary states of the linearized system; the 
dissipation however may alter somewhat their amplitude, without affecting very 
much their form. Note also that the $q_n$s are uniquely determined by the 
parameters of the system, and they vanish with vanishing $\phi_{ac}$.

From the analytical solution at frequencies within the linear flux-wave band, 
Eqs. (\ref{Ch5.17.8}) and (\ref{Ch5.17.82}), which correspond to either $s>+1/2$ 
or $s<-1/2$, the resonance frequencies of the array can be obtained by setting 
$\sin[(N+1)\theta'] =0$. Thus we get
\begin{eqnarray}
  \label{Ch5.17.10}
  s\equiv s_m=\frac{1}{2 \cos\left[ \frac{m\pi}{(N+1)} \right]} ,
\end{eqnarray}
where $m$ is an integer ($m=1,...,N$). By solving the first of Eqs. 
(\ref{Ch5.17.5}) with respect to $\Omega$, and substituting the values of 
$s\equiv s_m$ from Eq. (\ref{Ch5.17.10}), we get
\begin{eqnarray}
  \label{}
   \Omega \equiv \Omega_m =
   \sqrt{ \Omega_{SQ}^2 +2\, \lambda \, \cos\left( \frac{m\pi}{N+1} \right) },
\end{eqnarray}
which is the {\em discrete frequency dispersion} for linear flux-waves in a 
one-dimensional SQUID metamaterial, with $m$ being the mode number 
($m=1,...,N$).

\subsection{Multistability and Self-Organization in Disordered SQUID Metamaterials}
The total current of the SQUID metamaterial, devided by the number of SQUIDs and 
normalized to the critical current of the Josephson junctions, $I_c$, is defined 
as
\begin{equation}
\label{Ch5.18}
   i_{tot} (\tau) =\frac{1}{N_x N_y} \sum_{n,m} \frac{I_{n,m} (t)}{I_c} 
                  \equiv \frac{1}{N_x N_y} \sum_{n,m} i_{n,m} (\tau) ,
\end{equation} 
where $i_{n,m} (\tau)$ is the normalized time-dependent current flowing in the 
$(n,m)$th SQUID of the metamaterial. The total current $i_{tot}$ is maximum in a 
homogeneous (synchronized) state of the SQUID metamaterial. Homogeneous states 
are a subset of all possible states which are formed through amplitude and phase 
synchronization of the currents in individual SQUIDs; consequently homogeneous 
states provide a large magnetic response to an ac magnetic flux field, especially 
withing a frequency band around the single SQUID resonance frequency. (Note 
however that homogeneity is never complete in the case of finite size SQUID 
metamaterials). The total current $i_{tot} (\tau)$ is calculated through the 
(normalized) expression
\begin{eqnarray}
\label{Ch5.18.2}
  i_{n,m} =\frac{1}{\beta} \left\{ \phi_{n,m} - \phi_{eff} 
    -\lambda_x ( \phi_{n-1,m} + \phi_{n+1,m} ) 
    -\lambda_y ( \phi_{n,m-1} + \phi_{n,m+1} ) \right\},
\end{eqnarray}
which holds in the weak coupling approximation ($\lambda_x, \lambda_y \ll 1$), 
when the normalized fluxes $\phi_{n,m}$ have been calculated by numerically 
integrating Eqs. (\ref{Ch5.16}). Assuming that the SQUID metamaterial comprises 
identical elements, arranged in a perfect tetragonal lattice, the total current 
amplitude $i_{max}$ is defined as the absolute maximum of the total current 
$i_{tot}$ in one period of temporal evolution $T$, i.e., 
\begin{equation}
\label{Ch5.18.3}
   i_{max}=max_T \left\{\frac{1}{N_x N_y} \sum_{n,m} i_{n,m} (\tau) \right\}.
\end{equation} 
For disordered SQUID metamaterials, the total current amplitude is the average 
of $i_{max}$, over a number of $n_R$ realizations, denoted by 
$< i_{max} >_{n_R}$. In order to account for the termination of the structure 
in finite SQUID metamaterials, Eqs. (\ref{Ch5.16}) are implemented with the 
following boundary conditions  
\begin{eqnarray}
\label{Ch5.18.4}
  \phi_{0,m} (\tau) =\phi_{N_x+1,m} (\tau) = 0, \qquad 
  \phi_{n,0} (\tau) =\phi_{n,N_y+1} (\tau)= 0 .
\end{eqnarray}
In the rest of this Section, two values of the SQUID parameter are used, i.e., 
$\beta_L =0.9$ and $\beta_L =8$ which correspond to the SQUIDs being in the 
non-hysteretic and the hysteretic regime, respectively. For these values of 
$\beta_L$ the corresponding single SQUID resonance frequencies are 
$\Omega_{SQ}=1.4$ and $3$, respectively.
\begin{figure}[h!]
\includegraphics[angle=-0, width=0.45\linewidth]{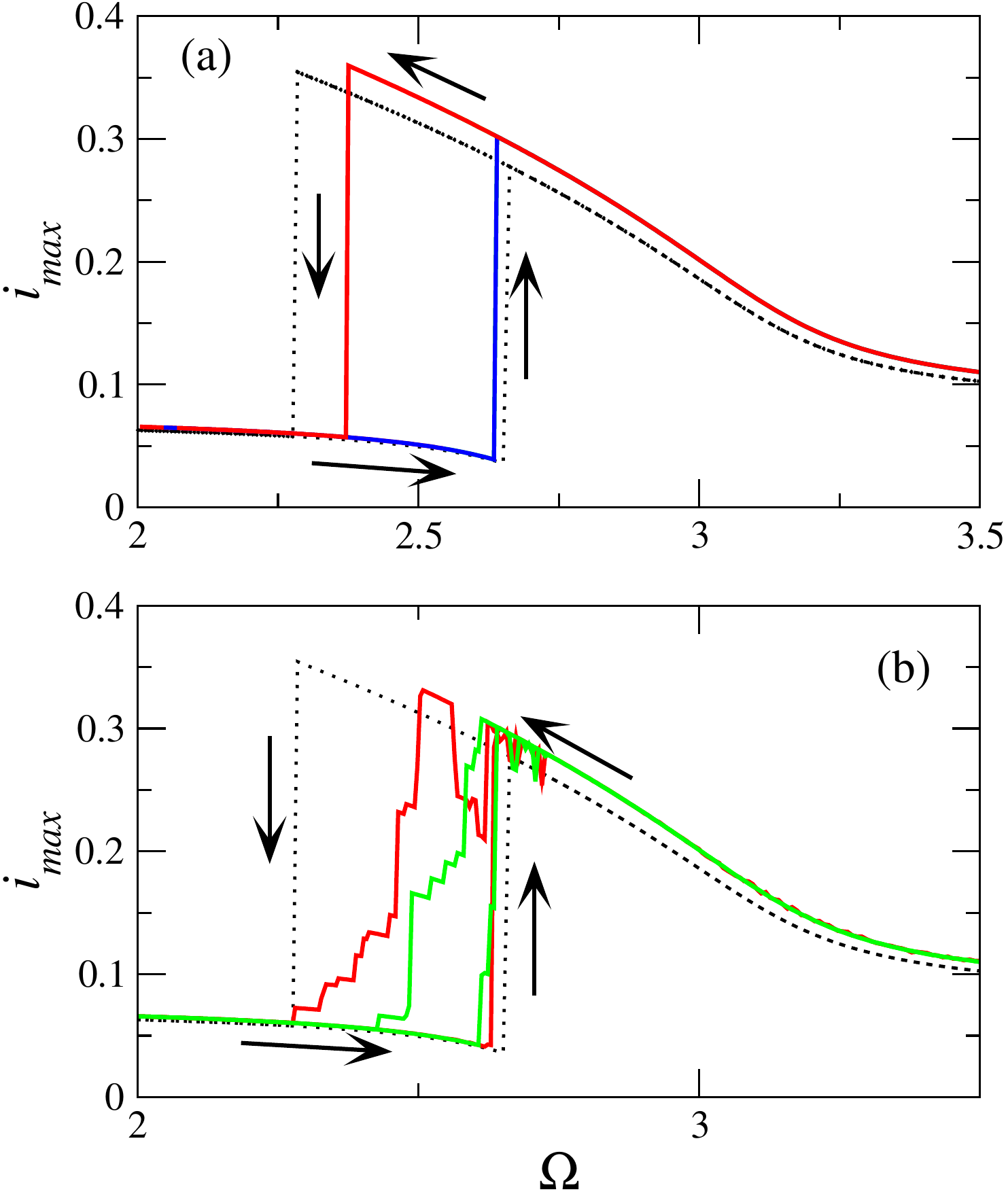}
\includegraphics[angle=-0, width=0.45\linewidth]{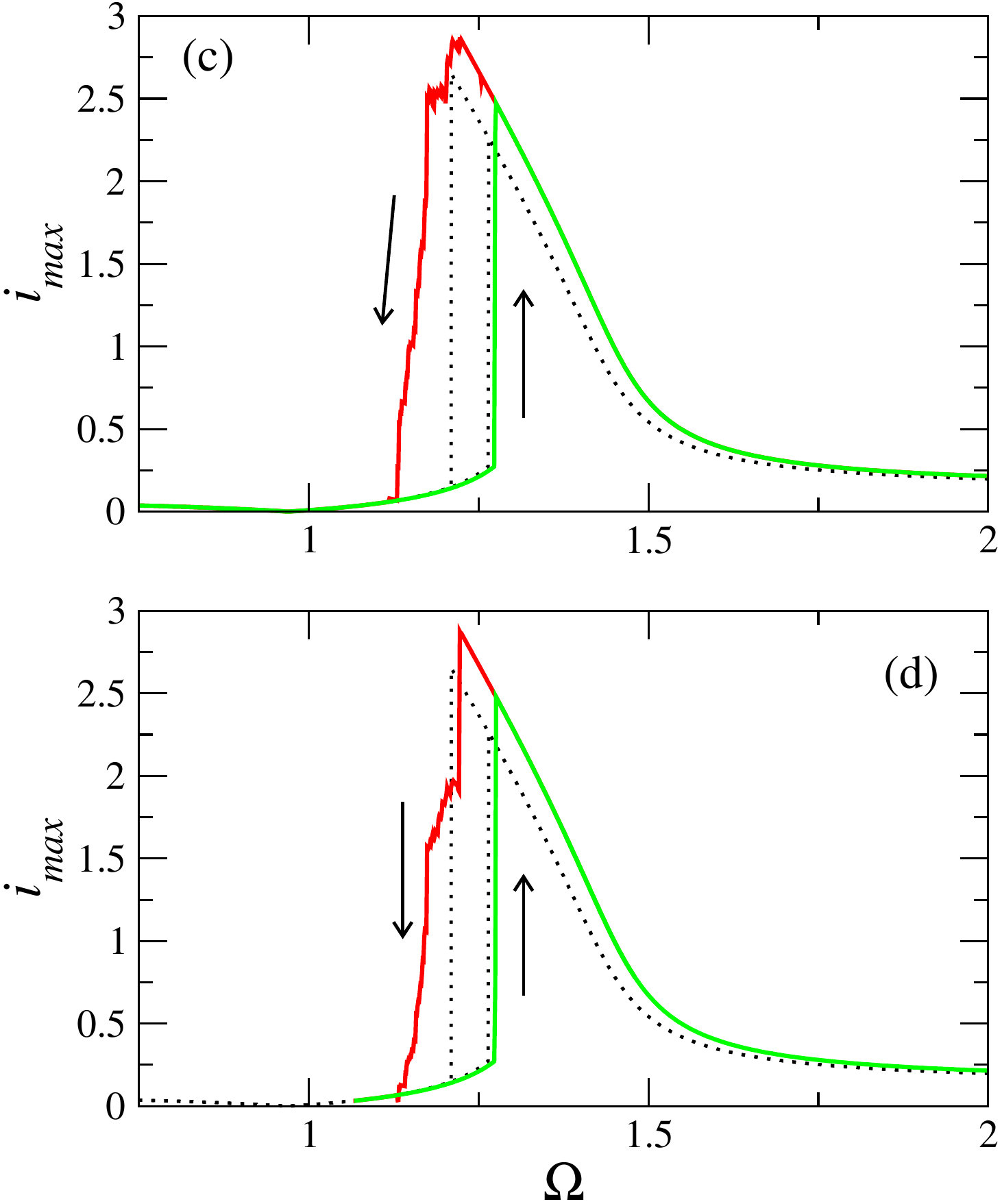}
\caption{ 
 Induced total current amplitude $i_{max}$ as a function of the driving 
 frequency $\Omega$ for two-dimensional $N_x \times N_y$ SQUID metamaterials 
 with $\gamma=0.002$, $\phi_{dc} =0$, and 
(a) $\beta=1.27$, $N_x=N_y=20$, $\phi_{ac} =0.1$, periodic boundary conditions;
(b) $\beta=1.27$, $N_x=N_y=20$, $\phi_{ac} =0.1$, free-end boundary conditions;
(c) $\beta=0.15$, $\phi_{ac} =0.02$, $N_x=N_y=20$, free-end boundary conditions;
(d) $\beta=0.15$, $\phi_{ac} =0.02$, $N_x=N_y=40$, free-end boundary conditions. 
The black dotted lines indicate the corresponding $i_{max}$ vs. $\Omega$ curves 
for a single rf SQUID.
\label{fig5.13}
}
\end{figure}

Typical current amplitude - frequency curves are shown in Fig. \ref{fig5.13}, 
in which the total current amplitude $i_{max}$ is shown as a function of the 
frequency $\Omega$ of the ac flux field (dc flux is set to zero). In this 
figure, the value of $\beta_L$ has been chosen so that the SQUIDs are well into 
the hysteretic regime ($\beta_L \simeq 8$). It is observed that bistability 
appears in a frequency band of substantial width. The corresponding curves 
(black-dotted) for a single SQUID are also shown for comparison. In Fig. 
\ref{fig5.13}a, in which periodic boundary conditions have been employed, the 
bistability region for the SQUID metamaterial is narrower than that for a single 
SQUID, although the total current amplitude is slightly larger than that for a 
single SQUID. For periodic boundary conditions, the size of the metamaterial 
does not affect those results; current amplitude - frequency curves for larger 
arrays with $N_x =N_y=40$ and $N_x =N_y=80$ (not shown here) are practically 
identical to those shown in Fig. \ref{fig5.13}a. In all the other figures till 
the end of this Section, free-end boundary conditions [Eqs. (\ref{Ch5.18.4})] 
which are appropriate for finite-size SQUID metamaterials are assumed. In that 
case, the total current amplitude - frequency curves are very sensitive to the 
initial conditions as well as the model and numerical parameters such as the 
frequency increment, the scanned frequency band, etc. For an illustration, the 
curves shown in Fig. \ref{fig5.13}b in different colors (red and green) have 
been calculated using different initializations of the SQUID metamaterial. The 
parts of the current amplitude - frequency curves for the SQUID metamaterial 
which are close to those for a single SQUID are formed by almost homogeneous 
states, i.e., states in which all the SQUIDs are close to either the 
high-current amplitude or the low-current amplitude single-SQUID states. 
Completely homogeneous states are formed easily in periodic SQUID metamaterials, 
but they are destroyed by perturbations in the finite-size ones; compare Figs.
\ref{fig5.13}a and \ref{fig5.13}b. In the latter figure, the observed 
staircase-like curve with many small steps indicates the existence of many 
different solutions which are formed when a number of SQUIDs are close to the 
high-current single-SQUID state while the others are in the low-current 
single-SQUID state. In Fig. \ref{fig5.13}c and d, the corresponding curves for 
SQUIDs with $\beta_L \lesssim 1$ ($\beta =0.15$) are shown. A comparison with 
the corresponding curves for a single SQUID (black-dotted curves) indicates that 
the bistability regions have nearly the same width. For the values of 
$\phi_{ac}$ used in Figs. \ref{fig5.13}, the nonlinearities are already 
substantial, and thus capable to make the total current amplitude - frequency 
curves significantly hysteretic. The excited nonlinearities are also evident 
from the shifting of the resonance frequency of the SQUID metamaterial, which 
should be close to $\Omega_{SQ}$ in the linear regime, to considerably lower 
values. Indeed, the maximum of the total current amplitude in Figs. 
\ref{fig5.13}c and d is at $\Omega \sim 1.2$, while the corresponding 
$\Omega_{SQ}$ is $\sim 1.39$.
\begin{figure}[h!]
\includegraphics[angle=-0, width=0.45\linewidth]{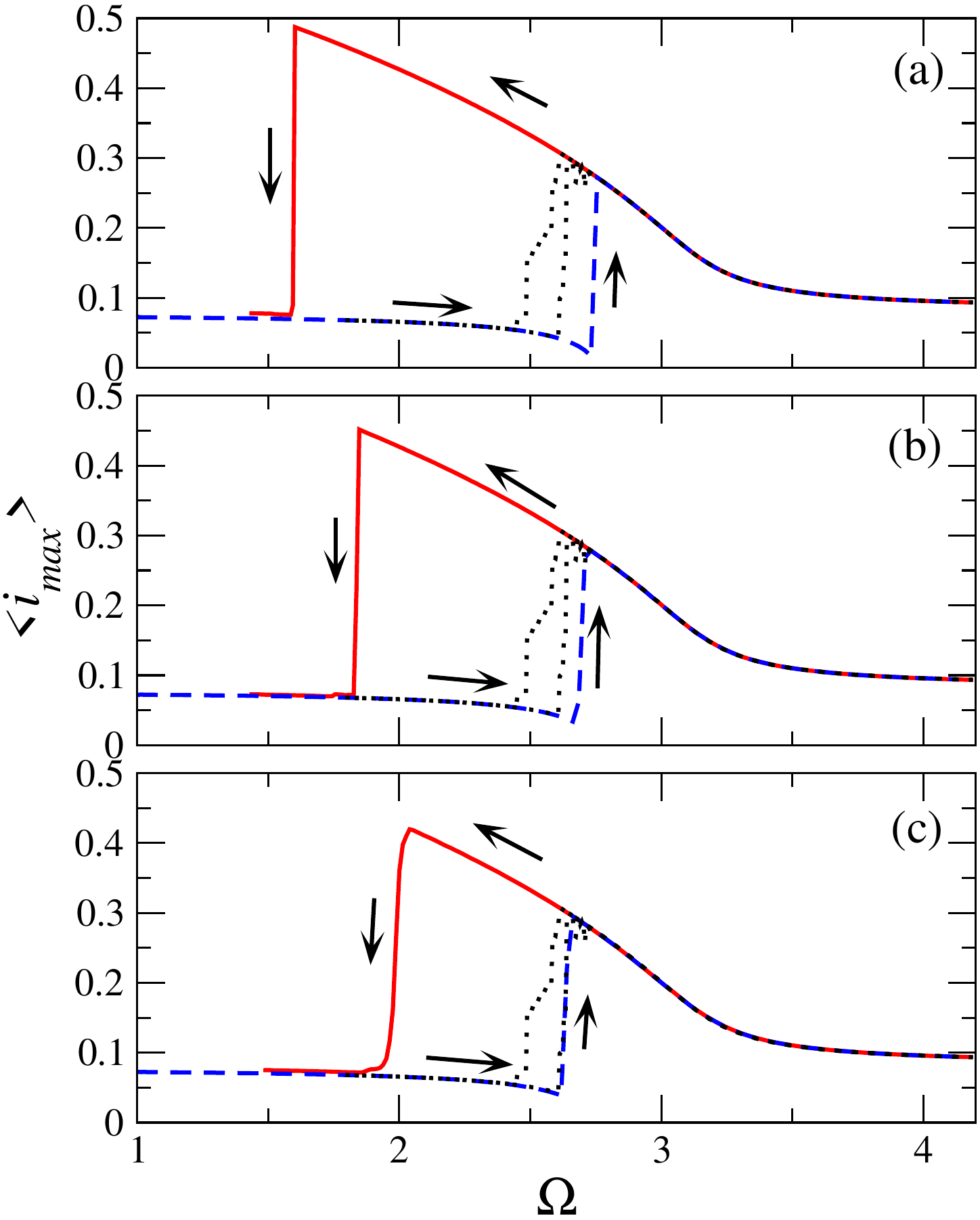}
\includegraphics[angle=-0, width=0.45\linewidth]{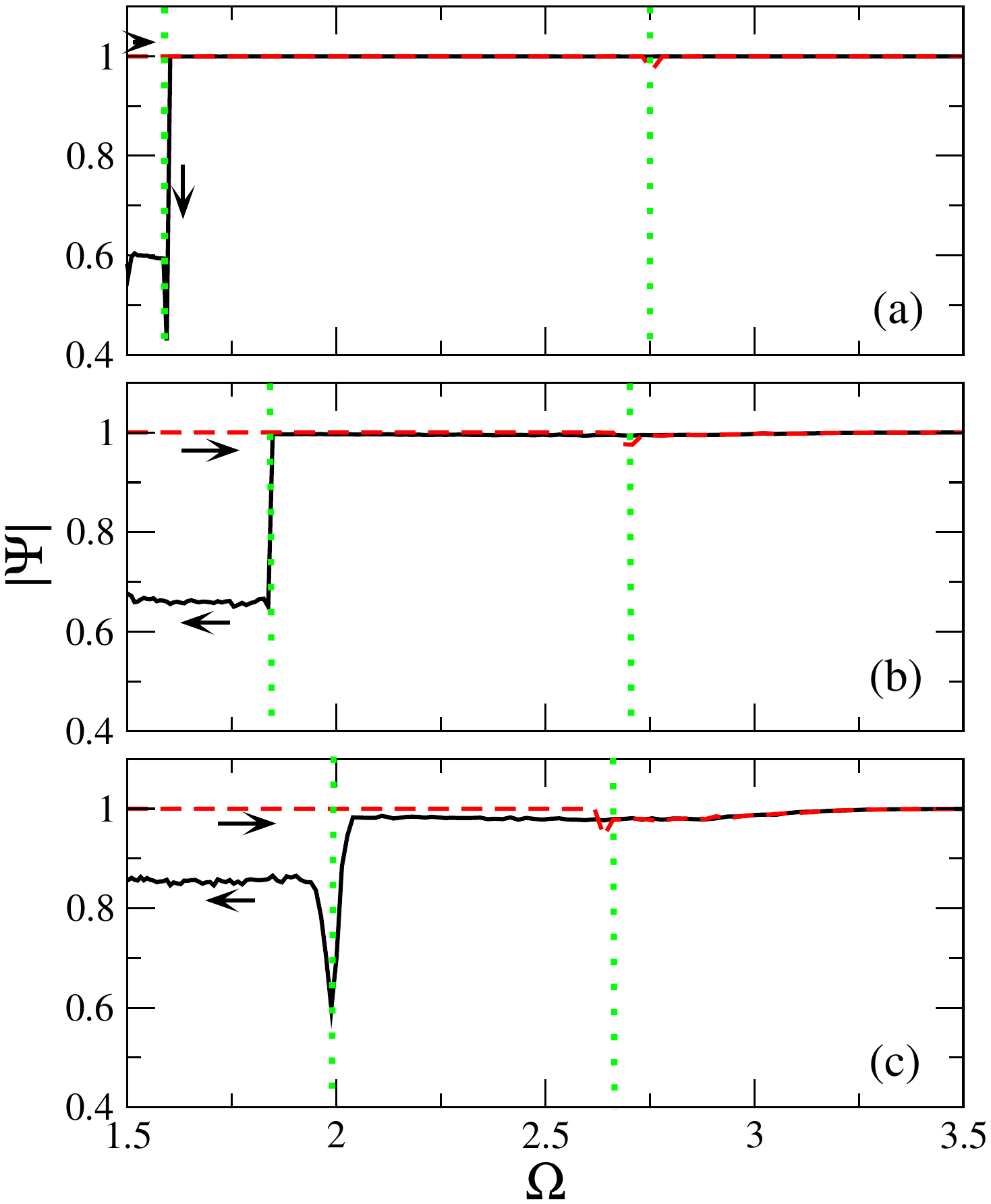}
\caption{
 Left:
 Total current amplitude averaged over $n_R =30$ realizations of disorder, 
 $<i_{max}>_{n_R}$, as a function of the driving frequency $\Omega$ for a SQUID 
 metamaterial with $
 N_x =N_y =20$, $\alpha=0.002$, $\beta=1.27$, $\phi_{ac} =0.1$, $\phi_{dc} =0$, 
 and 
 (a) $\delta \beta =\pm 0.01$; 
 (b) $\delta \beta =\pm 0.05$; 
 (c) $\delta \beta =\pm 0.1$.
 Right:
 The corresponding magnitude of the synchronization parameter averaged over 
 $n_R =30$ realizations of disorder, $<|\Psi|>_{n_R}$, as a function of the 
 driving frequency $\Omega$ in the bistability region for a SQUID metamaterial 
 with $N_x =N_y =20$, $\alpha=0.002$, $\beta=1.27$, $\delta \beta =\pm 0.01$, 
 $\phi_{dc} =0$, $\phi_{ac} =0.1$, and 
 (a) $\delta \beta =\pm 0.01$; 
 (b) $\delta \beta =\pm 0.05$; 
 (c) $\delta \beta =\pm 0.1$.
 The arrows indicate the direction of frequency variation while the green dotted 
 lines the corresponding bistability intervals.
\label{fig5.14}
}
\end{figure}

Up to this point, all the SQUIDs in a metamaterial are assumed to be identical; 
however, slight deviations in the values of the parameters of individual SQUIDs 
may occur in a SQUID metamaterial, due to unavoidable imperfections in the 
fabrication procedure. The existing experience shows that the available 
fabrication technology allows for the fabrication of SQUID metamaterials with 
parameter variation within a few percent from one SQUID to another (typically 
$1-2\%$). Thus, weak quenched disorder is present in all realizable SQUID 
metamaterials, that may affect their collective behavior. One possible source of 
disorder comes through random deviations of the critical currents $I_c$ of the 
Josephson junctions in the SQUIDs from a particular target value. Random 
variation of the critical current from one SQUID to another affects in turn the 
SQUID parameter $\beta_L$ of individual SQUIDs and eventually their nominal 
resonance frequency $\Omega_{SQ}$. There are of course other sources of disorder, 
which are related to the experimental apparatus and/or the procedure of 
measurement \cite{Butz2013b}; e.g., the presence of stray magnetic fields that 
are created either by the magnetic components in the experimental setup or by 
sources from outside, such as the magnetic field of the earth. Stray magnetic 
fields effectively cause inhomogeneities in the applied field, so that different 
SQUIDs are subjected to different bias (dc) flux fields which shift randomly 
their resonance frequency away from its nominal value. Their effect is clearly 
revealed in the spoiled tunability patterns in reference \cite{Butz2013b}. In 
the rest of this Section, $\beta$ (equivalently $\beta_L$) is allowed to vary 
randomly from one SQUID to another around a target (mean) value $\Omega_{SQ}$, 
by a few percent. In order to make statistically meaningful predictions, 
statistical averages of the currents over many realizations, $n_R$, of disorder 
configurations have to be made. Remarkably, the calculations reveal that weak 
disorder does not destroy bistability, but, instead, it stabilizes the system 
against modulational or other instabilities. The robustness of the bistability 
region is important for prospective applications in which SQUID metamaterials 
could replace nonlinear resonators as read-out units for superconducting flux 
qubits \cite{Lupascu2007}, that perform quantum non-demolishion measurements. 

The effect of weak quenched disorder on the total current amplitude - frequency 
curves of SQUID metamaterials is shown in Fig. \ref{fig5.14}, in which the 
$\beta$ parameters of the SQUIDs are drawn from a uniform random distribution of 
zero mean. For obtaining statistically reliable results, statistical averages 
are calculated over $n_R$ realizations of quenched disorder. In the left panels 
of Fig. \ref{fig5.14}a, b, and c, the disorder strength on the parameter $\beta$ 
is $\pm 0.01$, $\pm 0.05$, and $\pm 0.1$, respectively, and $n_R =30$. In the 
left panels of Fig. \ref{fig5.14}, the stability interval of the high-current 
amplitude, almost homogeneous states shrinks with increasing strength of 
disorder. Apparently, weak disorder exhibits wider bistability as compared to 
the corresponding ordered case \cite{Lazarides2013b}. Those results are related 
to earlier work on disordered networks of nonlinear oscillators in which 
moderate disorder may enhance synchronization and stabilize the system against 
chaos \cite{Braiman1995a,Braiman1995b}. Stabilization of Josephson circuits 
against chaos, in particular, has been recently demonstrated by numerical 
simulations in the time domain \cite{Gustavsson2013}. Moreover, experimental 
stabilization of qubit spectral resonance with random pulses has been observed 
\cite{Li2013}. In the context of SQUID metamaterials, synchronization of 
individual SQUIDs in the high or low current amplitude states results in high or 
low total current amplitude for the metamaterial as a whole. This requires that 
(almost) all the SQUIDs are in phase. It could be natural to assume that the 
more nearly identical the elements of a SQUID metamaterial are, the better their 
synchronization will be. However, even in the ideal case of identical elements, 
the earlier assumption may not be true and the in-phase state may be dynamically 
unstable. Then, synchronization is reduced and the SQUID metamaterial cannot 
remain for too long in the high total current amplitude state that is more 
sensitive to instability. This type of disorder-assisted self-organization may 
also occur by introducing local disorder in an array of otherwise identical 
oscillators, i.e., in the form of impurities 
\cite{Gavrielides1998a,Gavrielides1998b}. In this case, the impurities trigger 
a self-organizing process that brings the system to complete synchronization and 
suppression of chaotic behavior.

In order to ensure that the averaged total current amplitude - frequency curves 
in the left panels of Fig. \ref{fig5.14} indeed correspond to (almost) 
homogeneous (uniform) states, an appropriate measure of synchronization has to 
be calculated. Thus, a complex synchronization parameter is defined as 
\begin{equation}
\label{Ch5.18.5}
  \Psi = \left< \frac{1}{N_x \, N_y} 
        \sum_{n,m} e^{2 \pi i \phi_{n,m}} \right>_{\tau,n_R} ,
\end{equation} 
where the brackets $< >$ denote averaging both in time (i.e., in one oscillation 
period, $\tau =T$) and the number of realizations of disorder $n_R$. The 
magnitude of $\Psi$ quantifies the degree of synchronization between the SQUIDs; 
$|\Psi|$ may vary between $0$ and $1$, corresponding to completely asynchronous 
and synchronized states, respectively. The calculated values of  $|\Psi|$ for 
(parts of) the averaged total current amplitude-frequency are shown in the right 
panels of Fig. \ref{fig5.14} for strongly driven SQUID metamaterial and three 
different levels of disorder. The bistability regions shrinks with increasing 
strength of disorder (from top to bottom); the parameter $|\Psi|$ for the low 
current amplitude states remains close to unity for the whole range of 
frequencies shown. That parameter for the high current amplitude states exhibits 
similar behavior within the bistability region; however, as soon as the frequency 
reaches the lower boundary of the bistability region, the high current amplitude 
states start losing their stability and the synchronization breaks down. Thus, 
for frequencies below the left green (dotted) vertical line, the SQUID 
metamaterial has settled to a low current amplitude state which however preserve 
some degree of synchronization.

\begin{figure}[h!]
\includegraphics[angle=0, width=0.45\linewidth]{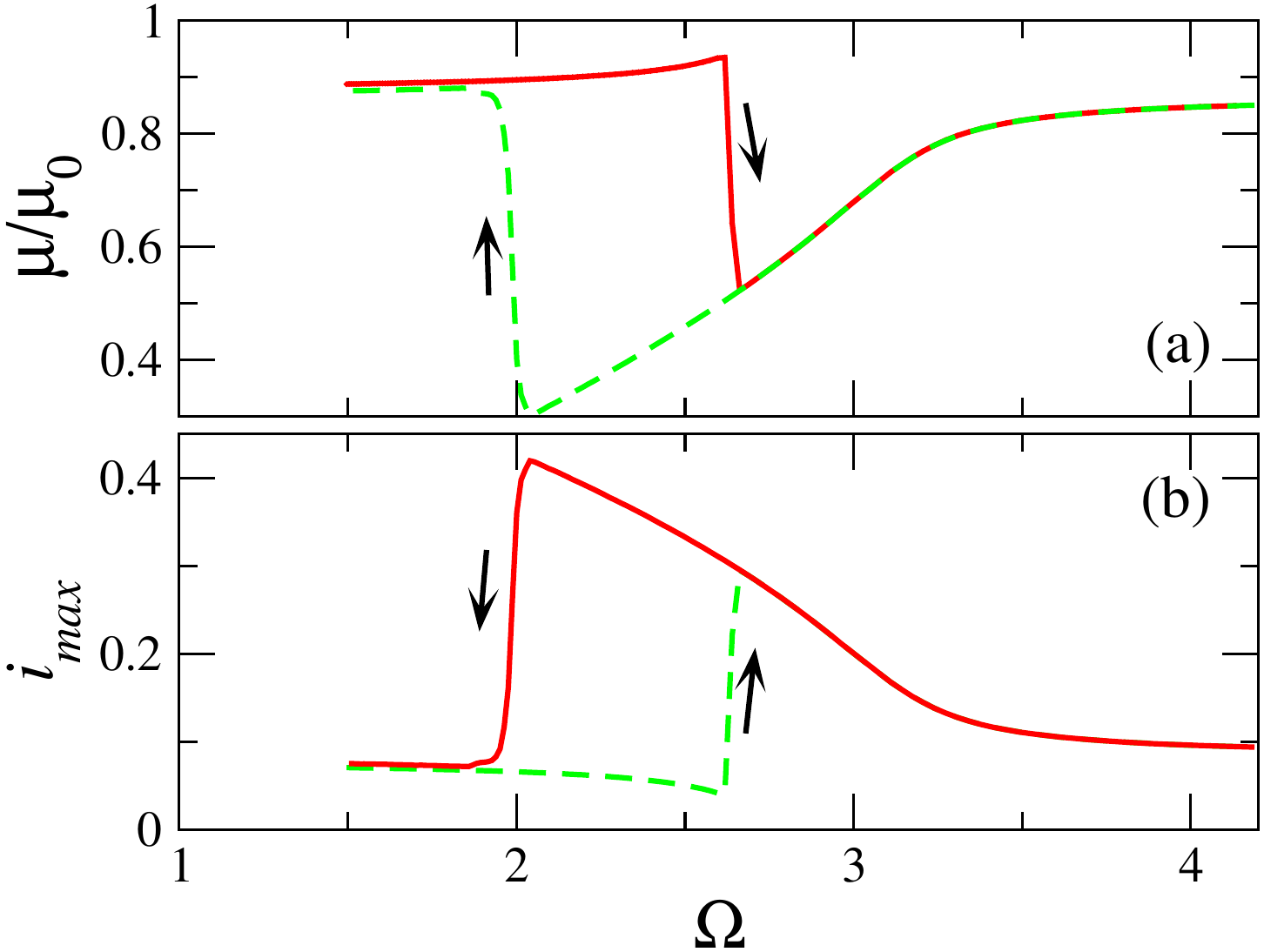}
  \hspace{5mm}
\includegraphics[angle=0, width=0.45\linewidth]{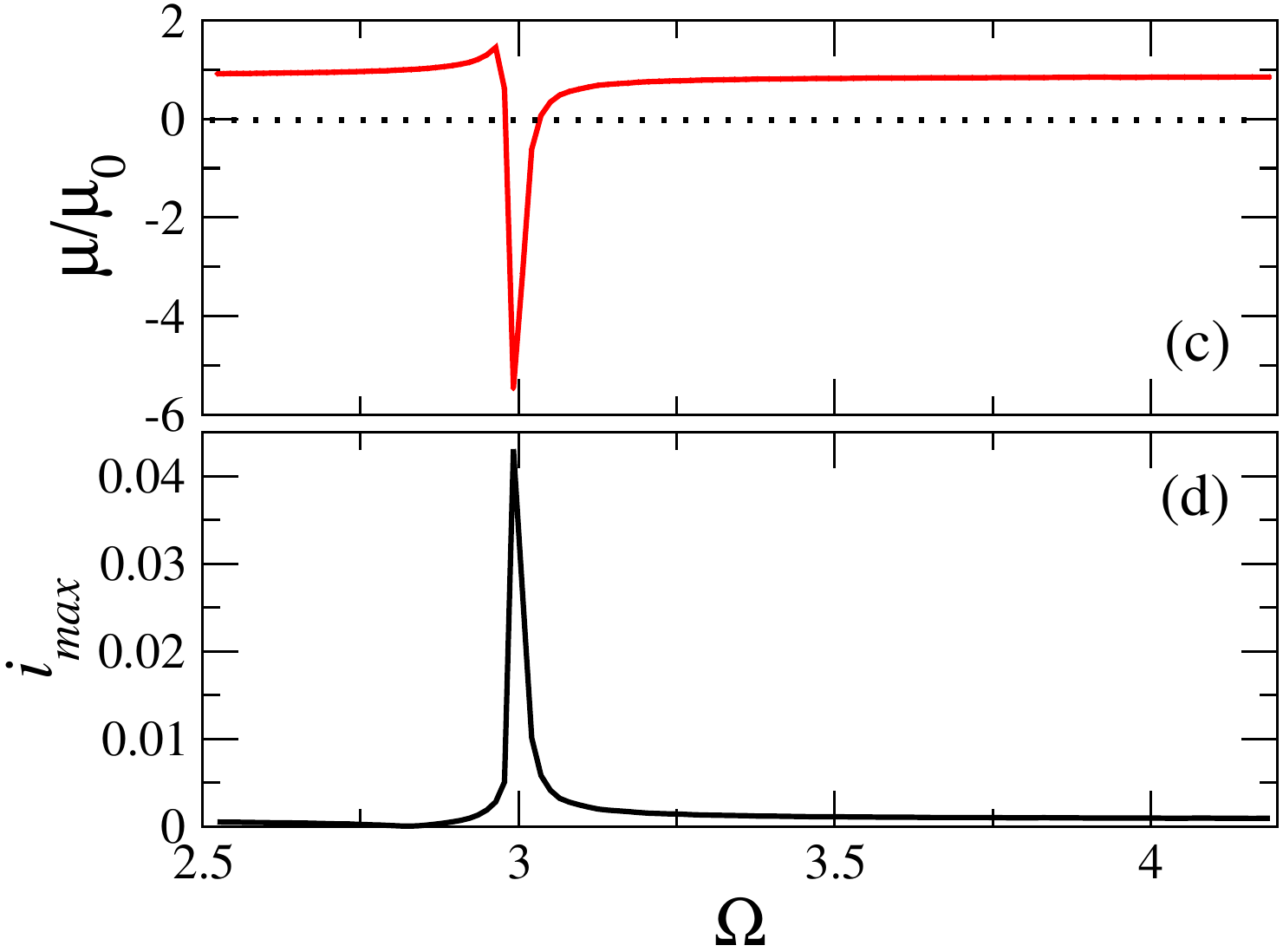}
\caption{ 
 Relative magnetic permeability $\mu_r =\mu/\mu_0$ for the low and high current 
 amplitude states as a function of the driving frequency $\Omega$, for a 
 disordered SQUID metamaterial with 
 $N_x = N_y =20$, $\gamma=0.002$, $\beta=1.27$, $\phi_{dc} =0$, and 
 (a) $\delta \beta=\pm 0.1$, $\phi_{ac} =0.1$;
 (c) $\delta \beta=\pm 0.01$, $\phi_{ac} =0.001$. 
 The corresponding total current amplitude - frequency curves are shown in (b) 
 and (d), respectively.
 Multiple-valued magnetic response is observed in the bistability region of (a). 
 Negative $\mu_r$ is observed in (c) within a narrow frequency band just above 
 the resonance frequency. The parameters of (a) and (b) are the same with those 
 in Fig. \ref{fig5.14}c (left panels).
\label{fig5.15}
}
\end{figure}

The magnetic response of the SQUID metamaterial at any particular state can be 
calculated in terms of the magnetization using simple electrical equivalent 
circuit models \cite{Lazarides2007,Jung2013,Trepanier2013}. Assuming for 
simplicity a tetragonal unit cell ($d_x=d_y=d$) with isotropic interactions 
between neighboring SQUIDs ($\lambda_x =\lambda_y \equiv \lambda$), and a 
squared SQUID area of side $\alpha$, the magnetization is
\begin{equation}
\label{Ch5.18.6}
   M = \frac{\alpha^2 <I>}{d^2 D},
\end{equation} 
where $<I> =I_c \, <i> \equiv I_c \frac{1}{N_x N_y} \sum_{n,m} <i_{n,m}>_\tau$ 
is the spatially and temporally averaged current in the SQUID. Note that SQUID 
metamaterials fabrication technology is currently planar, while the 
magnetization is defined to be inversely proportional to a unit volume. However, 
the experiments on SQUID metamaterials currently involve waveguides, in which 
the samples are placed. Thus, the necessary third dimension, which enters into 
the expression for the magnetization Eq. (\ref{Ch5.18.6}), comes from the length 
of the waveguide cavity in the direction that is perpendicular to the SQUID 
metamaterial plane $D$ \cite{Jung2013,Butz2013a,Butz2013b,Trepanier2013}. 
Using fundamental relations of electromagnetism, the relative magnetic 
permeability can be written as 
\begin{equation}
\label{Ch5.18.7}
  \mu_r = 1 +\frac{M}{H} , 
\end{equation} 
where $H$ is the intensity of a spatially uniform magnetic field applied 
perpendicularly to the SQUID metamaterial plane. The latter is related to the 
external flux applied to the SQUIDs through
\begin{equation}
\label{Ch5.18.8}
  H = \frac{\Phi_0}{\mu_0 \alpha^2} <\phi_{ext}> ,
\end{equation} 
where $\mu_0$ is the magnetic permeability of the vacuum, and the brackets 
denote temporal averaging. Combining Eqs. (\ref{Ch5.18.6})-(\ref{Ch5.18.8}), we 
get
\begin{equation}
\label{Ch5.18.9}
  \mu_r =1 +\kappa \frac{<i>}{<\phi_{ext}>} , 
\end{equation} 
where the coefficient 
$\kappa =\frac{\mu_0 \alpha I_c}{\Phi_0} \frac{\alpha^3}{d^2 D}$ is the analogue 
of the filling factor in the context of conventional metamaterials. For a rough 
estimation of the constant $\kappa$ we assume that $L \sim \mu_0 \alpha$, where 
$L$ is the SQUID self-inductance, and that $D \simeq d$. Then, we have that
$\kappa \sim \beta \left( \frac{\alpha}{d} \right)^3$. Using $\alpha =d/2$ and 
$\beta =1.27$ we get $\kappa \simeq 0.16$.

While the expression for the magnetic permeability is rather simple, there is 
some uncertainty about the value of the factor $\kappa$. However, for a 
reasonable value of $\kappa$ the magnetic permeability can be negative within 
a narrow frequency band above the single-SQUID resonance frequency for weakly 
driven SQUID metamaterials. In that case, increasing disorder results in 
weakening the negative response of the metamaterial; thus, for relatively 
strong disorder the response is not sufficient to provide negative $\mu_r$ even 
for strongly driven SQUID metamaterials, as can be seen in Fig. \ref{fig5.15}a.
In that figure it is also observed that due to the bistability, the relative 
magnetic permeability $\mu_r$ may obtain two different values depending on which 
state the SQUID metamaterial is. The currents $i_{n,m}$ with given $\phi_{ext}$ 
can be calculated from Eq. (\ref{Ch5.18.2}) when the corresponding $\phi_{n,m}$ 
have been calculated from Eq. (\ref{Ch5.16}). Then, Eq. (\ref{Ch5.18.9}) 
provides $\mu_r$ for a particular, parameter-dependent $\kappa$ coefficient. 
Thus, simultaneously stable SQUID metamaterial states exhibit different magnetic 
responses to an external magnetic field and therefore exhibit different values 
of $\mu_r$. Such magnetic multi-response in the presence of disorder is observed 
in Fig. \ref{fig5.15}a, with the corresponding $<i_{max}>_{n_R} - \Omega$ curves 
shown in Fig. \ref{fig5.15}b. The same quantities are shown in Fig. 
\ref{fig5.15}c and d, respectively, for a weakly driven SQUID metamaterial and 
lower strength of disorder. Here, $\beta$ varies randomly by $\pm 0.1$ ($~8\%$) 
around the nominal value $\beta=1.27$. The nonlinear effects become unimportant 
bringing the metamaterial close to the linear limit, and the hysteresis in the 
$<i_{max}>_{n_R} - \Omega$ curve as well as in the $\mu_r - \Omega$ curve 
dissappears. Note that the behavior of $\mu_r$ follows closely that of the 
averaged total current amplitude $<i_{max}>_{n_R}$. Furthermore, at driving 
frequencies below (but very close to) the single-SQUID resonance in the linear 
regime $\Omega_{SQ} \simeq 3$ ($\beta=1.27$), the SQUID metamaterial becomes 
strongly diamagnetic, so that it actually crosses the zero $\mu_r$ line. Such 
extreme diamagnetism corresponds to negative $\mu_r$, which persists within a 
narrow frequency band, just like in conventional metamaterials. Similar 
calculations with a transmission line model fed by experimental transmission 
data, produce qualitatively similar results \cite{Butz2013a}.  
\newpage

\section{SQUID-Based Metamaterials II: Localization and Novel Dynamic States}
\subsection{Intrinsic Localization in Hamiltonian and Dissipative Systems}
Discrete breathers (DBs), also known as intrinsic localized modes (ILMs), are 
spatially localized and time-periodic excitations which appear generically in 
extended periodic discrete systems of weakly coupled {\em nonlinear oscillators} 
\cite{Flach1998,Campbell2004,Flach2008a,Dmitriev2016}. DBs may be generated 
spontaneously as a result of fluctuations \cite{Peyrard1998,Rasmussen2000}, 
disorder \cite{Rasmussen1999}, or by purely deterministic mechanisms 
\cite{Hennig2007a,Hennig2007b,Hennig2015}. Since their discovery 
\cite{Sievers1988}, in a large volume of analytical and numerical studies the 
conditions for their existence and their properties have been explored for a 
variety of nonlinear mathematical models of physical systems. Their very 
existence has been proved rigorously for both energy-conserving (Hamiltonian) 
and dissipative lattices \cite{Mackay1994,Aubry1997}, and several numerical 
algorithms have been designed for their accurate construction 
\cite{Marin1996,Marin2001,Tsironis2002,Bergamin2003}. 
A fundamental requirement for their existence is that their frequency of 
oscillation and its multitudes are outside the linear frequency band. 
Importantly, they have been observed in a variety of physical systems, such as 
solid state mixed-valence transition metal complexes \cite{Swanson1999}, 
quasi-one dimensional antiferromagnetic chains \cite{Schwarz1999}, arrays and 
ladders of Josephson junctions \cite{Binder2000,Trias2000,Mazo2002,Schuster2004}, 
micromechanical cantilever arrays \cite{Sato2003}, optical waveguide systems 
\cite{Eisenberg1998}, layered crystal insulator at $300 K$ \cite{Russell2007}, 
and proteins \cite{Edler2004}. Further experiments concerning breathers in 
crystals are reviewed in Ref. \cite{Dmitriev2016}. Once generated, DBs modify 
system properties such as lattice thermodynamics and introduce the possibility 
of nondispersive energy transport \cite{Tsironis2003,Kopidakis2001}, because of 
their potential for translatory motion (i.e., mobility) along the lattice 
\cite{Flach1999}. In numerical experiments, DB mobility can be achieved by 
applying appropriate perturbations \cite{Chen1996}. From the perspective of 
applications to experimental situations where dissipation is always present, 
dissipative DB excitations (usually driven by a sinusoidal power source) are 
more relevant than their energy-conserved (Hamiltonian) counterparts. 
Dissipative DBs, which possess the character of an attractor for initial 
conditions in the corresponding basin of attraction, are generated whenever 
power balance, instead of the conservation of energy, governs the dynamics of 
the nonlinear lattice. Furthermore, the attractor character of dissipative DBs 
allows for the existence of quasi-periodic and even chaotic DBs 
\cite{Martinez2003,Ikeda2007}. 

\subsection{Dissipative Breathers in SQUID Metamaterials}
The existence of dissipative DBs has been numerically demonstrated in 
conventional (metallic) metamaterials comprising split-ring resonators, both in 
the "bulk" and the "surface" (i.e., in the ends and the edges, respectively, of 
one- and two-dimensional finite systems) 
\cite{Lazarides2006,Eleftheriou2008,Lazarides2008b,Eleftheriou2009}, as well as 
in binary metamaterial models \cite{Molina2009,Lazarides2010b}. In typical 
experimental situations, SQUID metamaterials are driven by an ac (sinusoidal) 
flux field and they are subjected to dissipation, mainly due to quasi-particle
tunneling through the Josephson junction. Their discreteness, along with weak 
coupling between their elements and the (Josephson) nonlinearity, favors the 
appearence of dissipative breathers. Moreover, due to low dissipation in SQUID 
metamaterials, dissipative breathers in those systems could be in principle 
observed experimentally through advanced imaging techniques such as the Laser 
Scanning Microscopy (LSM) \cite{Ustinov2015}. Here, the existence of dissipative 
DBs in SQUID metamaterials is demonstrated in the one-dimensional case, for 
simplicity and ease of presentation; however, it has been demonstrated that 
increasing dimensionality does not destroy breather excitations either in 
conventional or SQUID metamaterials \cite{Lazarides2008a,Tsironis2009}. In SQUID 
metamaterials, dissipative DBs can be generated either by properly designed 
their initial state, or by driving them through a stage of modulational 
instability; the latter method allows for spontaneous formation of dissipative 
DBs. Since the SQUIDs in a metamaterial are weakly coupled (through magnetic 
dipole-dipole forces), the breather structures which are generated with either 
of these two methods are highly localized; thus, a large amount of the energy 
provided initially to the SQUID metamaterial is concentrated in only a few 
SQUIDs. The generation and subsequent evolution of dissipative DBs can thus be 
visualized on three-dimensional plots in which the fluxes through the SQUIDs or 
the currents in the SQUIDs are plotted on the lattice site $n$ - time $\tau$ 
plane. In order to generate DBs by initialization, {\em a trivial breather 
configuration} has to be constructed first, which corresponds to a numerically 
accurate solution in the case of vanishing coupling between SQUIDs 
(anti-continuous limit \cite{Marin1996}). Then, using one of the several 
breather-finding algorithms, a DB family can be obtained by slowly increasing 
the coupling coefficient, say $\lambda$. That family has a member-breather for 
each value of $\lambda$ up to a critical one $\lambda_c$ that depends on the 
other parameters of the system; for $\lambda > \lambda_c$ the DB family 
destabilizes and dissapears. In order to construct a trivial dissipative DB
configuration, sufficiently strong nonlinearity is required for the individual 
SQUIDs to exhibit multistability; for that purpose, one first calculates the 
flux amplitude $\phi_{max}$ - driving frequency $\Omega$ curve of the single 
SQUID. For a particular frequency $\Omega$ in the multistability region, at 
least two simultaneously stable solutions of the single SQUID (all the SQUIDs 
are regarded to be identical) have to be identified, say $(\phi_0, \dot{\phi_0}$ 
($0$) and $(\phi_1, \dot{\phi_1}$ ($1$), with low and high flux amplitude 
$\phi_{max,0}$ and $\phi_{max,1}$, respectively. Then the initial state of the 
SQUID metamaterial is constructed by setting one of the SQUIDs, say that at 
$n=N/2$ in the state-solution $1$, and all the others to the state-solution $0$
($N$ is the number of SQUIDS in the metamaterial). The SQUID at $n=N/2$ is 
hereafter referred to as the central DB site, which also determines its location; 
all the others constitute the "background". That configuration is used as initial 
condition for the integration in time of Eqs. (\ref{Ch5.14}) to numerically 
obtain a stable dissipative DB for a given value of $\lambda$. Note that one 
could start integrating at $\lambda =0$ and then slowly increase the value of 
$\lambda$ up to the desired one ($\lambda < \lambda_c$); however, since 
$\lambda \ll 1$, that continuation procedure may not be necessary. Note also 
that the procedure for obtaining dissipative DBs is easier than the corresponding 
one for obtaining Hamiltonian DBs, for which Newton's method instead of merely 
numerical integration is required. In the obtained dissipative DB state, all the 
SQUIDs are oscillating with frequency $\Omega_b =\Omega$ (period-1 DBs, with 
their frequency being locked to that of the driving flux field $\Omega$), 
although their flux or current amplitudes are generally different. Note however 
that there may also be DBs with more complicated temporal behavior
\cite{Martinez2003}.       

\begin{figure}[h!]
\begin{center}
   \includegraphics[angle=0, width=0.9\linewidth]{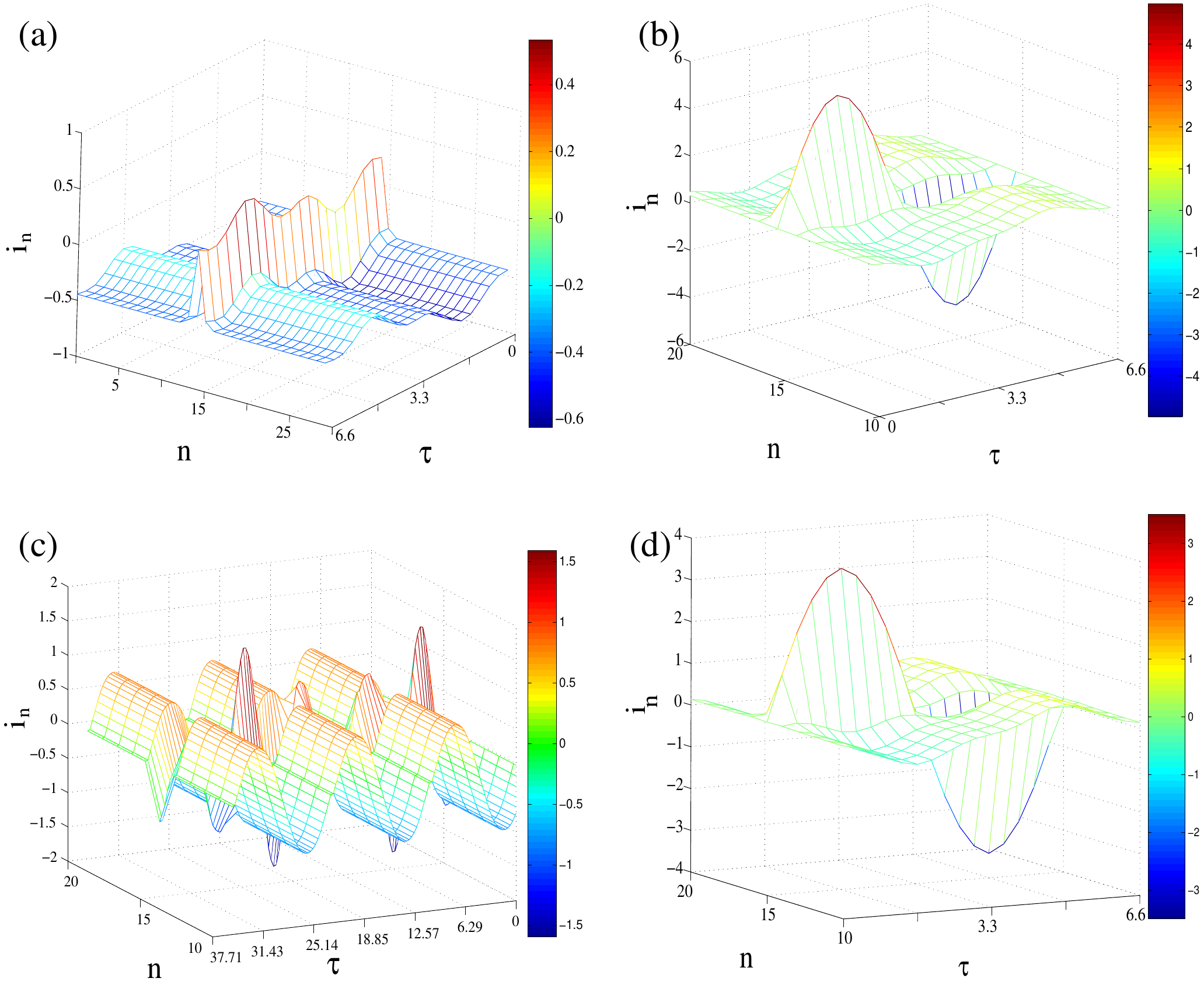} 
\end{center}
\caption{
(a) Temporal evolution of a dissipative discrete breather during one driving 
    period $T_b=2\pi/\Omega_b =6.6$, for $\phi_{dc}=0.5$, $\phi_{ac}=0.2$, 
    $\beta=1.27$, $\gamma=0.001$, $\lambda=-0.1$.
(b) $\&$ (d) Temporal evolution of a dissipative discrete breather during one 
    driving period $T_b=2\pi/\Omega_b =6.6$, for $\phi_{dc}=0$, $\phi_{ac}=0.6$, 
    $\beta=1.27$, $\gamma=0.001$, $\lambda=-0.1$.
(c) Temporal evolution of a dissipative discrete breather during three driving 
    periods $T_b =2\pi/\Omega_b =12.57$, for $\phi_{dc}=0$, $\phi_{ac}=1.2$, 
    $\beta=1.27$, $\gamma=0.001$, $\lambda=-0.0225$.
From (a) to (d), only part of the array ($N=30$) is shown for clarity.
\label{fig55.01}
}
\end{figure}

For a metamaterial comprising hysteretic SQUIDs ($\beta_L > 1$), there are more 
possibilities for constructing trivial DB configurations. Recall that for 
$\beta_L > 1$, the SQUID potential has more than one minimums, with their number 
increasing by further increasing $\beta_L$. Furthermore, a dc flux bias 
$\phi_{dc}$ also affects the SQUID potential at will; for example, by applying 
a dc flux bias $\phi_{dc} =0.5$, the SQUID potential becomes a double-well one. 
Then, there are at least two simultaneously stable states, one with 
$\phi_{max} \sim 0$ and the other with $\phi_{max} \sim 1$, corresponding to the 
left and right minimum of the SQUID potential, respectively. Those states can be 
employed to construct a trivial DB configuration as described earlier. Such a 
double-well dissipative DB is shown in Fig. \ref{fig55.01}(a), in which the 
spatio-temporal evolution of the induced currents $i_n$ ($n=1,2,3,...,N$) is 
plotted during one period of the DB oscillation, $T_b =\frac{2\pi}{\Omega_b}$. 
The currents in both the background SQUIDs and the SQUID on the central site are 
oscillating with the same frequency $\Omega_b =\Omega$, i.e., the frequency of 
the driving flux field. For $\phi_{dc} =0$ but still $\beta_L > 1$, no other DBs 
of that type can be obtained, since the local minimums are highly metastable; 
for sufficiently high ac flux amplitude $\phi_{ac}$ there may be more stable 
states which are generated dynamically due to strong nonlinearity. These states, 
which usually have high flux amplitude $\phi_{max}$ can be used to construct 
trivial DB configurations as described earlier. Two typical examples of such 
dissipative DBs, which are simultaneously stable, are shown in Fig. 
\ref{fig55.01}(b) and (d). This is possible because at that frequency (with 
corresponding period $T_b =6.6$) there are three simultaneously stable 
single-SQUID solutions; one with low flux amplitude $\phi_{max}$ and two with 
high flux amplitude $\phi_{max}$. Each of the high flux amplitude solutions can 
be combined with the low flux amplitude solution so that two trivial DB 
configurations can be constructed, which result in the two different dissipative 
DBs. The DB frequency $\Omega_b$ is again locked to the driving frequency 
$\Omega$ ($\Omega_b=\Omega$). For relatively weak coupling between SQUIDs, 
dissipative DBs can be also obtained which period of oscillation is a multiple 
of that of the external driver $\Omega$ (subharmonic dissipative DBs). Such a 
period-3 dissipative DB is shown in Fig. \ref{fig55.01}(c), in which the current 
in the SQUID of the central DB site apparently oscillates with $T_b =3 T$, with 
$T=2\pi / \Omega$, while the currents in the background SQUIDs oscillate with 
$T_b =T$. Some remarks are here in order: a major difference between Hamiltonian 
and dissipative DBs is that in the former the background oscillators are at rest 
at all time while in the latter the background oscillators oscillate as well, 
although with an amplitude different than that of the oscillator at the central 
DB site. In Fig. \ref{fig55.01}(a), the currents in all the SQUIDs are oscillating 
in phase; to the contrary, in Figs. \ref{fig55.01}(b), (c), and (d), the currents 
in the background SQUIDs are in anti-phase with respect to the current in the 
SQUID of the central DB site. This has significant consequencies for the local 
magnetic response of the SQUID metamaterial, since the local magnetization is 
directly proportional to the induced current. Thus, the observed phase difference 
indicates that the breathers may change locally the magnetic response of the 
system from paramagnetic to diamagnetic or vice versa 
\cite{Lazarides2008a,Tsironis2009}. In Figs. \ref{fig55.01}(a), (b), and (d), 
the time-dependence of the currents in the background SQUIDs is clearly 
non-sinusoidal, due to strong nonlinearities. The time-dependence of the current 
in the SQUID of the central DB site on the other hand seems perfectly sinusoidal, 
which is due to the high flux amplitude and the shape of the SQUID potential.
\begin{figure}[h!]
\begin{center}
   \includegraphics[angle=0, width=0.46\linewidth]{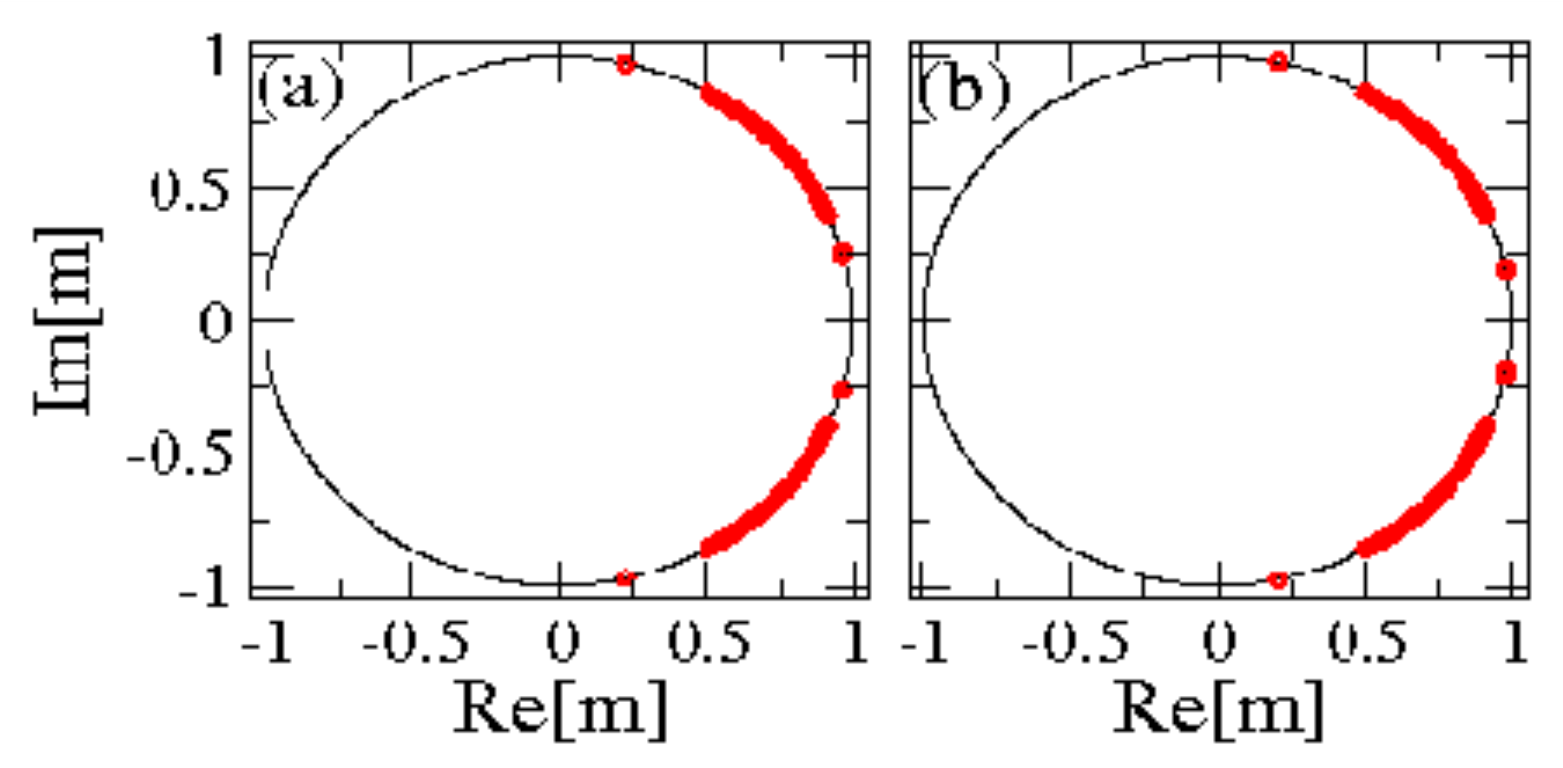} 
   \includegraphics[angle=0, width=0.49\linewidth]{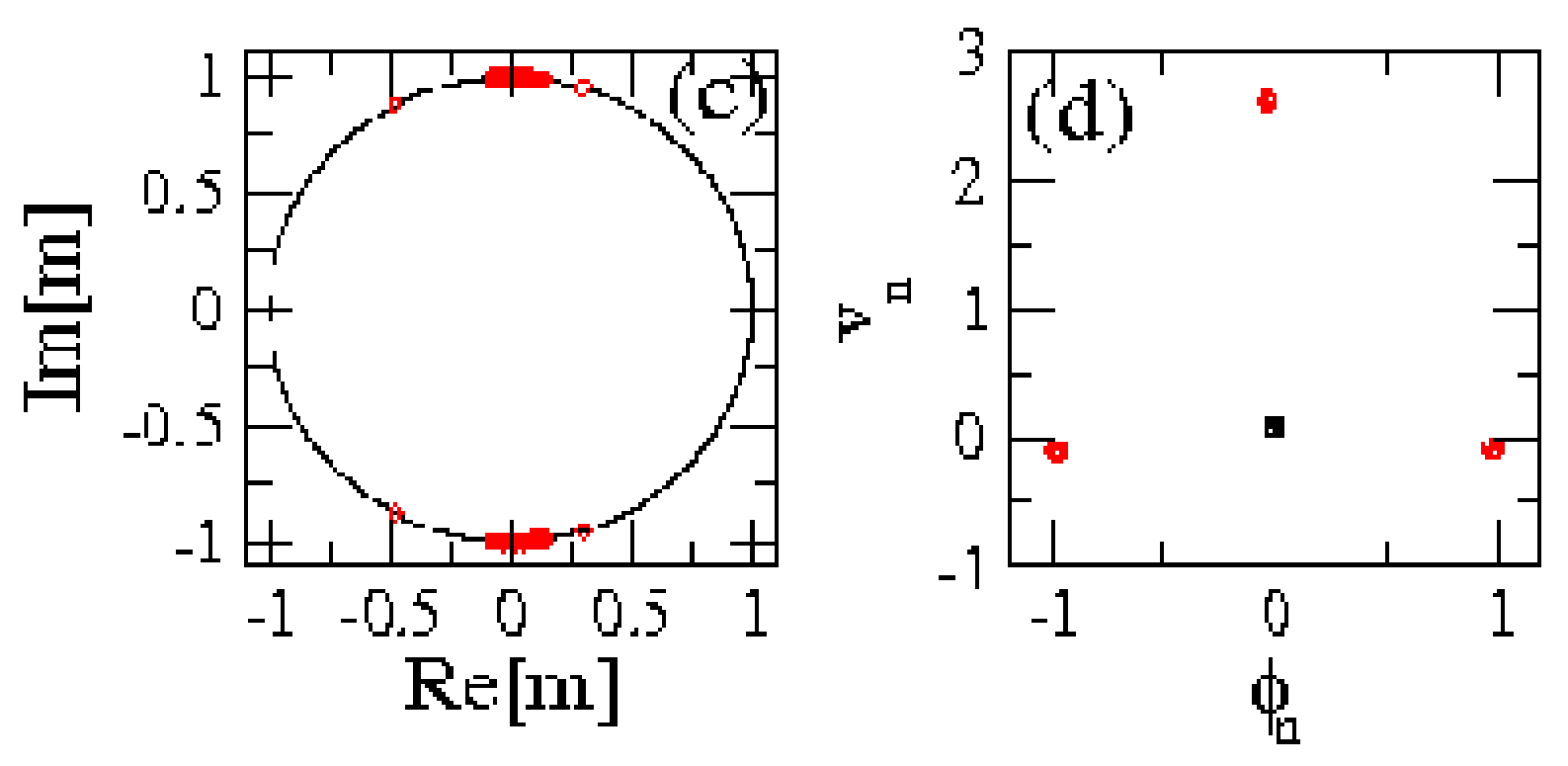} 
\end{center}
\caption{
(a) $\&$ (b) Floquet spectra for the simultaneously stable dissipative discrete 
    breathers shown in Fig. \ref{fig55.01}(b) and (d), respectively; all the 
    eigenvalues lie on a circle of radius 
    $R_e = \exp(-\gamma T_b/2) \simeq 0.996705$ in the complex plane.
(c) Floquet spectra for the period-3 dissipative discrete breather shown in 
    Fig. \ref{fig55.01}(c); all the eigenvalues lie on a circle of radius
    $R_e = \exp(-\gamma T_b/2) \simeq 0.993735$.
(d) Reduced stroboscopic diagrams for the SQUID at the central DB site at 
    $n =n_b =N/2$ (red circles), and the SQUID at $n=7$ in the background 
    (black square),
    for the period-3 dissipative discrete breather shown in 
    Fig. \ref{fig55.01}(c).
\label{fig55.02}
}
\end{figure}

The linear stability of dissipative DBs is addressed through the eigenvalues of 
the Floquet matrix (Floquet multipliers). A dissipative DB is linearly stable 
when all its Floquet multipliers $m_i, ~i=1,...,2N$ lie on a circle of radius 
$R_e = \exp(-\gamma T_b/2)$ in the complex plane \cite{Marin2001}. All the 
dissipative DBs shown in Fig. \ref{fig55.01} are linearly stable. The 
calculated eigenvalues for the two simultaneously stable dissipative DBs in 
Fig. \ref{fig55.01}(b) and (d), are shown respectively in Figs. \ref{fig55.02}(a) 
and (b). The Floquet spectrum for the period-3 dissipative DB in Fig. 
\ref{fig55.01}(c), is shown in Fig. \ref{fig55.02}(c). In Fig. \ref{fig55.02}(d), 
two reduced stroboscopic diagrams are shown on the ${v}_{n} - \phi_{n}$ plane, 
with ${v}_{n} =\dot{\phi}_{n}$ being the normalized instantaneous voltage across 
the Josephson junction of the $n$th SQUID. The one diagram is for the SQUID at 
the central DB site, i.e., at $n =n_b =N/2$. while the other is for a SQUID in 
the background, at $n=7$. Clearly, the trajectory of the SQUID at $n =n_b$ 
crosses the reduced phase space at three points (red circles), while that at 
$n=7$ at only one point (black square).

\begin{figure}[h!]
\begin{center}
\begin{tabular}{c}
  \includegraphics[angle=0, width=0.43\linewidth]{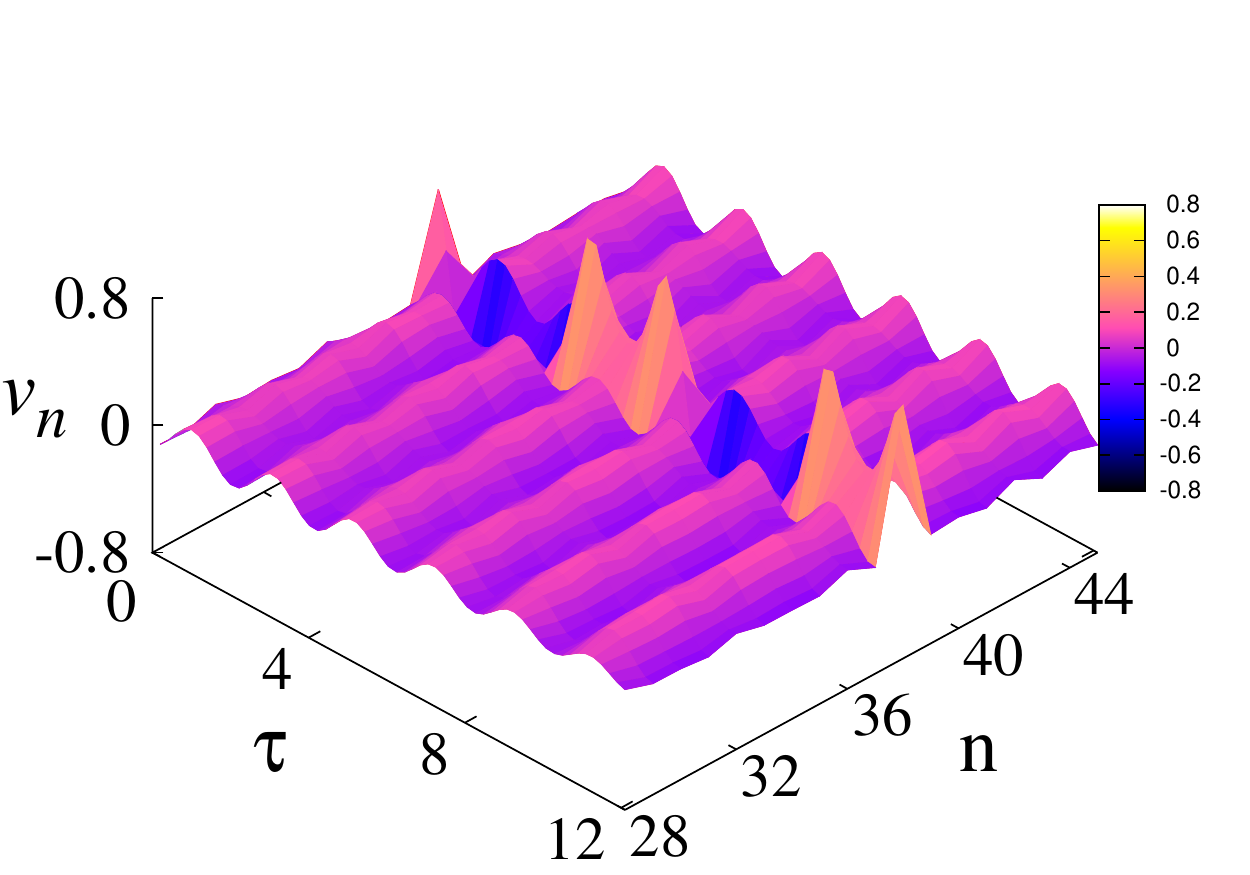} 
  \includegraphics[angle=0, width=0.43\linewidth]{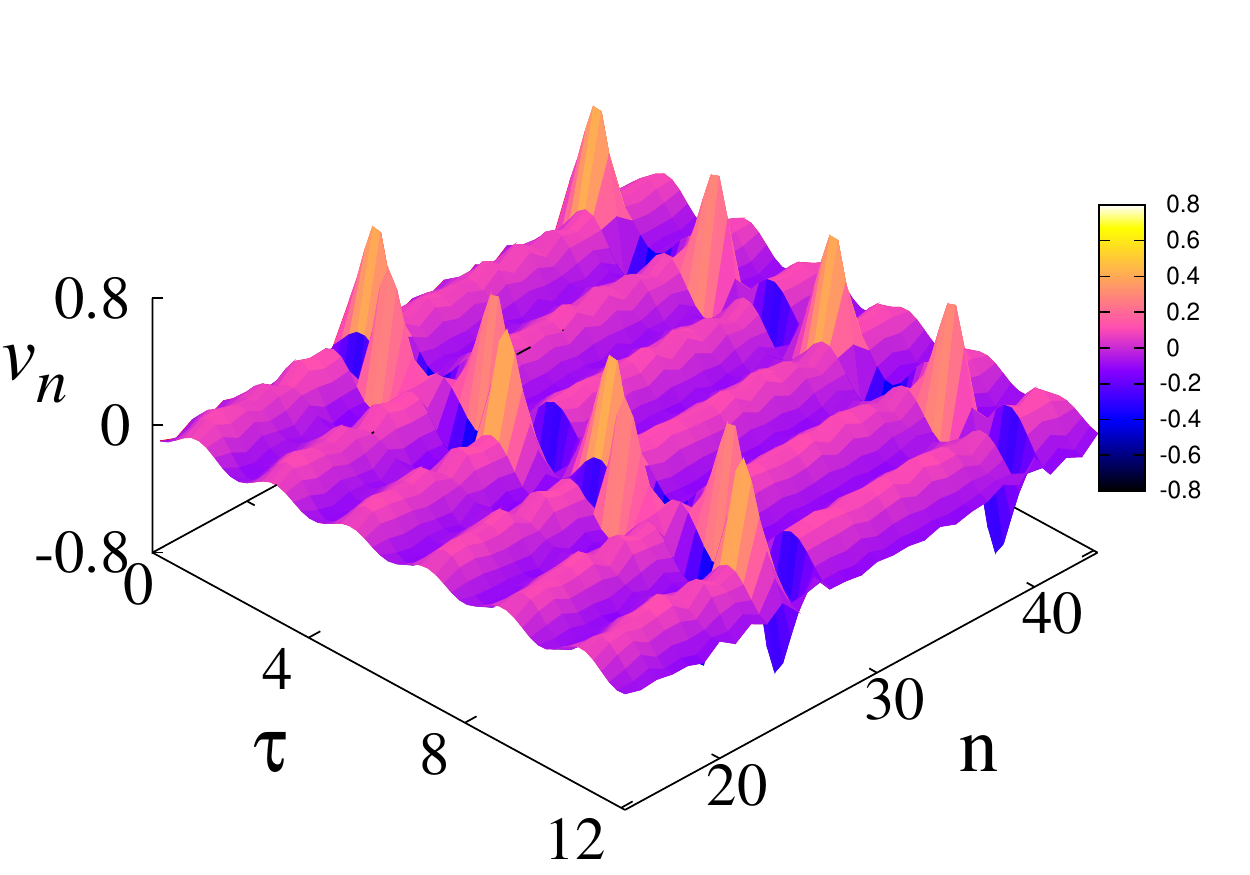} 
\end{tabular}
\end{center}
\caption{
 Spatiotemporal evolution of dissipative discrete breathers excited spontaneously 
 in weakly disordered SQUID metamaterials during six (6) periods of the driving 
 flux field. The voltages $v_n=d\phi/d\tau$ across the Josephson junctions of the 
 SQUIDs in the metamaterial are plotted on the $\tau - n$ plane for $\phi_{dc}=0$, 
 $\phi_{ac}=0.03$, $\beta=1.27$, $\gamma=0.001$, $\lambda=-0.0014$, $\Omega=3.11$, 
 and $N=50$. The left and right panels correspond to different configurations of 
 disorder, which are realized by adding to $\beta$ random numbers from a uniform
 distribution in the interval $[-0.01 \beta, +0.01 \beta]$.
\label{fig55.03}
}
\end{figure}

For generating dissipative DBs experimentally, the approach based on Marin's 
algorithm \cite{Marin2001} is not particularly useful, since it requires from 
the system to be initialized in a rather specific state. However, DB generation 
in SQUID metamaterials may be a relatively easy task whenever weak disorder is 
present, e.g., due to imperfections during fabrication. In a particular 
realization of a SQUID metamaterial, the SQUIDs are not completely identical but 
the values of their parameters slightly fluctuate around a mean nominal value. 
The parameter which is affected the most from those imperfections seems to be 
the critical current of the Josephson element $I_c$ in each SQUID, which varies 
exponentially with the thickness of the insulating barrier. Moreover, $I_c$ is 
proportional to the SQUID parameter $\beta_L$ which multiplies the nonlinear 
term in the SQUID flux equation and essentially determines its characteristic 
behavior. In order to take into account that type of disorder, the parameter 
$\beta =\beta_L /2\pi$ is allowed to vary randomly around its nominal value by 
$\pm 1\%$ of that value. The random numbers in the interval 
$[\beta -1\%, \beta +1\%]$ are drawn from a uniform distribution with mean value 
$\beta$. Numerical simulation of disordered SQUID metamaterials for many 
different configurations of disorder reveal that in most cases spontaneously 
generated dissipative DBs appear. In Fig. \ref{fig55.03}, the spontaneous 
generation of dissipative DBs is illustrated for two different configurations 
of disorder. In that figure, the instantaneous voltage across the Josephson 
junction of each SQUID $v_n =\dot{\phi}_n =\frac{d\phi_n}{d\tau}$ is plotted on 
the $\tau - n$ plane. For different configurations of disorder (while all other 
parameters are kept the same) a different number of dissipative DBs may appear 
at different locations in the metamaterial. As can be observed, the number of 
spontaneously generated DBs is one and three for the left and the right panel 
of Fig. \ref{fig55.03}, respectively. Furthermore, in the left panel, the period 
of voltage oscillations is twice that of the driver and that of the voltages of 
the SQUIDs in the background. Thus, this dissipative DB is a period-2 one.

\subsection{Collective Counter-Intuitive Dynamic States}
The investigation of networks of coupled nonlinear elements pervades all of 
science, from neurobiology to statistical physics, often revealing remarkable 
aspects of collective behavior \cite{Watts1998,Strogatz2001}. The effect of 
non-local interactions, which constitutes the "dark corner" of nonlinear 
dynamics, has been extensively investigated in the last decade and has unveiled 
collective dynamic effects such as synchronization 
\cite{Strogatz2000,Acebron2005}, pattern formation \cite{Battogtokh1999}, and 
Turing instabilities \cite{Viana2011}. Recently, a dynamic state which is 
qualitatively distinct and it has a counter-intuitive structure, referred to in 
current literature as a {\em "chimera state"}, was discovered in numerical 
simulations of non-locally coupled oscillator arrays \cite{Kuramoto2002}. That 
discovery was followed by intense theoretical
\cite{Abrams2004,Omelchenko2008,Abrams2008,Pikovsky2008b,Ott2009,Martens2010,
Omelchenko2011,Yao2013,Omelchenko2013,Hizanidis2014,Zakharova2014,Yeldesbay2014} 
and experimental
\cite{Tinsley2012,Hagerstrom2012,Wickra2013,Nkomo2013,Martens2013,
Schonleber2014,Viktorov2014,Rosin2014,Schmidt2014b,Gambuzza2014,Kapitaniak2014,
Hart2016} 
activity. A chimera state is characterized by the coexistence of synchronous 
and asynchronous clusters (subgroups) of oscillators, even though they are 
coupled symmetrically and they are identical \cite{Smart2012,Panaggio2015}.
Recent works also report on the issue of robustness of chimera states 
\cite{Omelchenko2015} as well as on the emergence of chimera states in systems 
with global \cite{Schmidt2014,Sethia2014,Yeldesbay2014,Bohm2015} and local 
coupling schemes \cite{Laing2015,Hizanidis2016a}. Chimera-like states in modular 
networks \cite{Shanahan2010,Hizanidis2016c} have been also investigated, 
expanding our understanding on the role of topology and dynamics for their 
occurrence. Further research efforts aim to stabilize chimera states by feedback 
schemes \cite{Sieber2014} and to control the localization of the different 
regimes \cite{Bick2015,Isele2016,Omelchenko2016}. Although chimera states are
generally regarded to be metastable \cite{Shanahan2010,Wildie2012}, or even 
chaotic transients \cite{Wolfrum2011}, there are also examples in which they are 
at the global minimum of a system, such as in Ising spins in thermal equilibrium 
\cite{Singh2011}. The level of synchronization and metastability of chimera 
states can be quantified using measures of local and global synchronization 
\cite{Shanahan2010,Wildie2012,Hizanidis2016b}, measures of metastability 
\cite{Shanahan2010,Wildie2012,Lazarides2015b}, the chimera index 
\cite{Gopal2014}, etc. Many different types of non-local interactions between
the oscillators in a given network have been considered in literature, often 
exponentially decaying, that allow a particular system to reach a chimera state. 
The crucial ingredient for the emergence of chimera states is the choise of 
initial conditions. Those states do not actually result from destabilization 
of the more familiar homogeneous (i.e., synchronized) or clustered states, but 
they usually coexist with (some) of them. Thus, without an appropriate choice of 
initial condition, the system will reach one of those instead of a chimeric one. 
SQUID metamaterials seem to be perfect candidates for the observation of chimera 
states, since their constitutive elements are essentially non-locally coupled 
and they are highly nonlinear oscillators. Those elements (i.e., the SQUIDs) may 
also exhibit multistability in a frequency band around the single-SQUID 
resonance. As it has been discussed in the previous Section, SQUIDs are coupled 
magnetically through dipole-dipole forces which fall-off as the inverse cube of 
their center-to-center distance. That coupling, although short-ranged 
\cite{Campa2009} and weak due to its magnetic nature, is clearly non-local. For 
simplicity, the one-dimensional non-local model, Eqs. (\ref{Ch5.11}), with 
appropriate initial conditions are used for obtaining very long-lived chimera 
states in SQUID metamaterials.

\subsection{Chimera States in SQUID Metamaterials}
\subsubsection{\em SQUID metamaterials with non-local coupling}
The one-dimensional SQUID metamaterial is initialized with 
\begin{equation}
\label{Ch55.100} 
   \phi_n (\tau =0) =\phi_R , \qquad \dot{\phi}_n (\tau =0) =0 ,
\end{equation}
where $\phi_R$ is a random number drawn from a flat, zero mean distribution in
$[-\phi_R/2, +\phi_R/2]$. The following boundary conditions
\begin{equation}
\label{Ch55.101} 
   \phi_0 (\tau) =0 , \qquad \phi_{N+1} (\tau) =0 ,
\end{equation}
are used to account for the termination of the structure in a finite SQUID 
metamaterial. The degree of synchronization for the whole SQUID metamaterial or 
just a part of it (e.g., a cluster having $M$ SQUIDs with $M \leq N$, with $N$ 
being the total number of SQUIDs in the metamaterial) is quantified by the 
magnitude of a complex, Kuramoto-type synchronization parameter $\Psi(\tau)$, 
defined as
\begin{equation}
\label{Ch55.102} 
  \Psi(\tau) = \frac{1}{M} \sum_{m=1}^M e^{i [2\pi \phi_m (\tau)]}.
\end{equation}
The magnitude of that synchronization parameter, $r (\tau) =|\Psi(\tau)|$, 
provides a global (for the whole metamaterial) or local (within a cluster) 
measure of spatial coherence at time-instant $\tau$. The value of $r (\tau)$ 
by its definition lies in the interval $[0,1]$, where the extremal values $0$ 
and $1$ correspond to complete desynchronization and synchronization, 
respectively. The mean synchrony level $\bar{r}$, which is an index of the 
global synchronization level, is defined as the average of $r (\tau)$  over the 
total time of integration (excluding transients) \cite{Wildie2012}, while the 
variance of $r (\tau)$, $\sigma_{r}^2$, captures how the degree of synchrony 
fluctuates in time. Fluctuations of the degree of synchronization have been 
associated with metastability and therefore $\sigma_{r}^2$ is indicative of the 
metamaterial's metastability level \cite{Shanahan2010,Wildie2012}.
\begin{figure}[!h]
\includegraphics[angle=0, width=0.95 \linewidth]{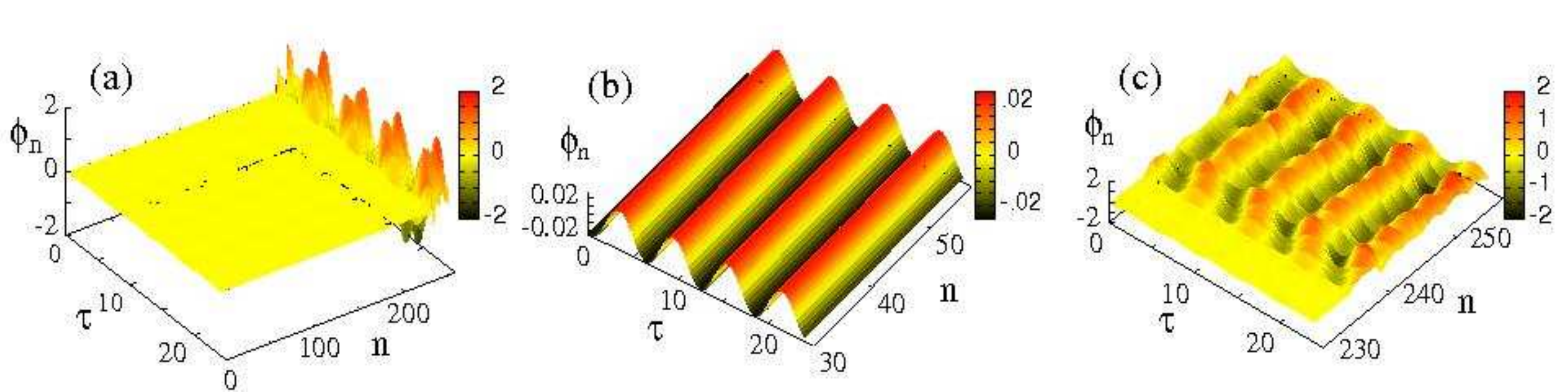} 
\caption{
 Spatio-temporal evolution of the normalized fluxes $\phi_n$ threading the SQUID 
 rings during four driving periods $T=5.9$ for $N=256$, $\gamma=0.0022$, 
 $\lambda_0=-0.05$, $\beta_L\simeq 0.7$, $\phi_{ac}=0.015$, and $\phi_R =0.85$.
(a) for the whole SQUID metamaterial; 
(b) for part of the metamaterial that belongs to the coherent cluster;
(c) for part of the metamaterial that includes the incoherent cluster.
\label{fig55.04}
}
\end{figure}
A typical spatio-temporal flux pattern for the SQUID metamaterial, obtained 
after $10^7$ time units of time-integration, is shown in Fig. \ref{fig55.04}(a), 
in which the evolution of the $\phi_n$s is monitored during four driving periods 
$T=2\pi/\Omega$. In that pattern, two different domains can be distinguished, in 
which the fluxes through the loops of the SQUIDs are oscillating either with low 
or high amplitude. The enlargement of two particular sub-domains shown in Figs. 
\ref{fig55.04}(b) and (c), reveals the unexpected feature which characterizes a 
chimera state; besides the difference in the oscillation amplitudes (i.e., 
low-high), the two groups of SQUIDs exhibit distinctly different dynamic 
behaviors: the low-amplitude oscillations are completely synchronized (Fig. 
\ref{fig55.04}(b)) while the high-amplitude ones are desynchronized both in 
phase and amplitude (Fig. \ref{fig55.04}(c)). Note that since the SQUID 
metamaterial is driven at a particular frequency $\Omega$, there can be no net 
frequency drift as in phase oscillators \cite{Kuramoto2002}; instead, the period 
of each SQUID in the asynchronous cluster fluctuates around that of the driver, 
$T$.

The chimera states are very sensitive to slight changes of the model parameters, 
the parameters of the applied flux field(s), as well as the integration 
parameters such as the the time-step $\Delta \tau$ of the integration algorithm. 
The latter is chosen to be $0.02$, which provides reliable results for systems 
of nonlinear oscillators. Decreasing of the time-step (i.e., to 
$\Delta \tau=0.01$) leads the SQUID metamaterial to a different chimera state 
due to metastability effects; that state may be either more or less synchronized 
than the previous one, depending on the other parameters. For the parameters 
used in this Section, the SQUID metamaterial reaches spontaneously a chimera 
state for most of the initial flux density configurations with 
$\phi_R \sim \Phi_0$.

\begin{figure}[!h]
\includegraphics[angle=0, width=0.52 \linewidth]{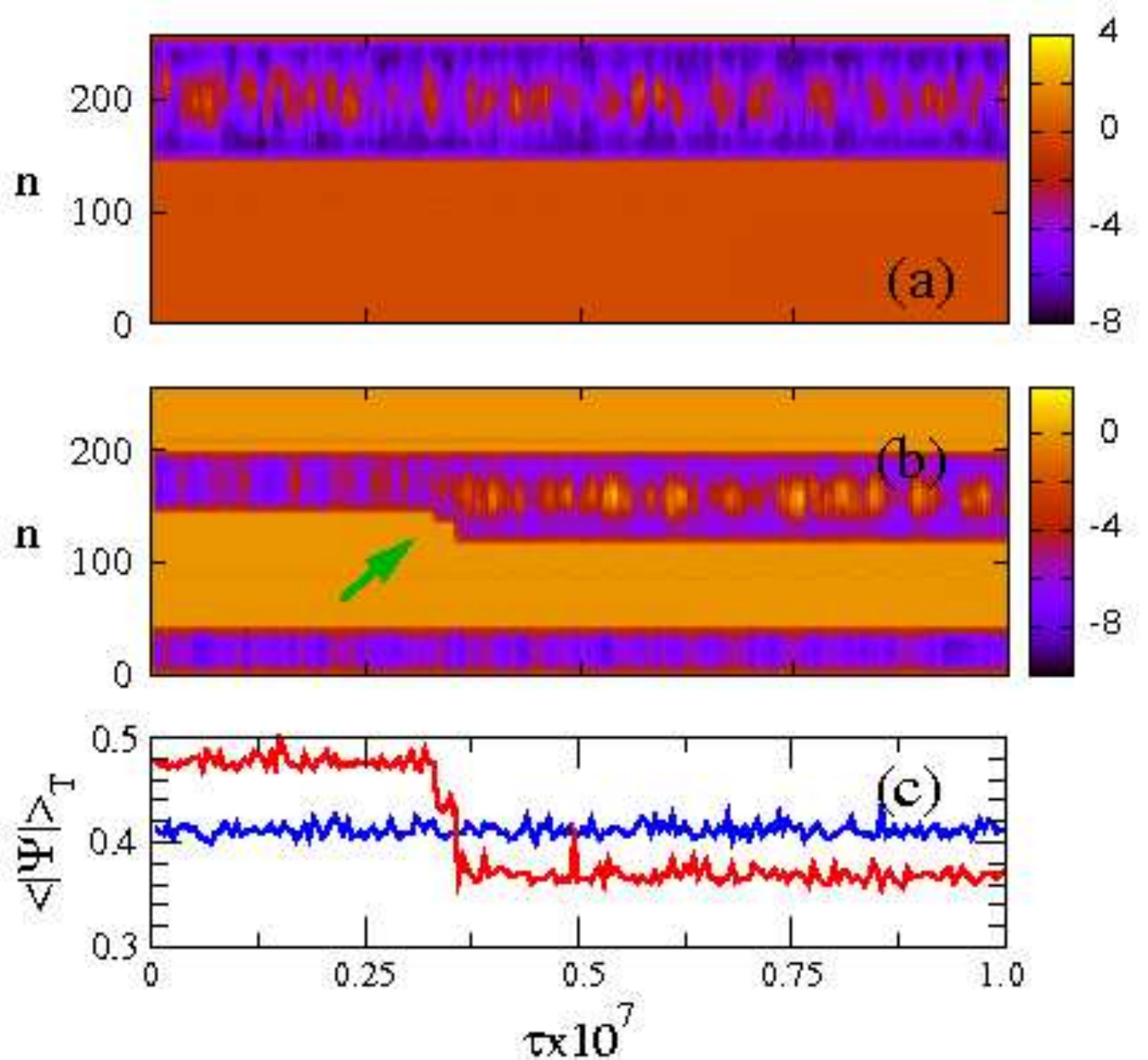} 
\includegraphics[angle=0, width=0.38 \linewidth]{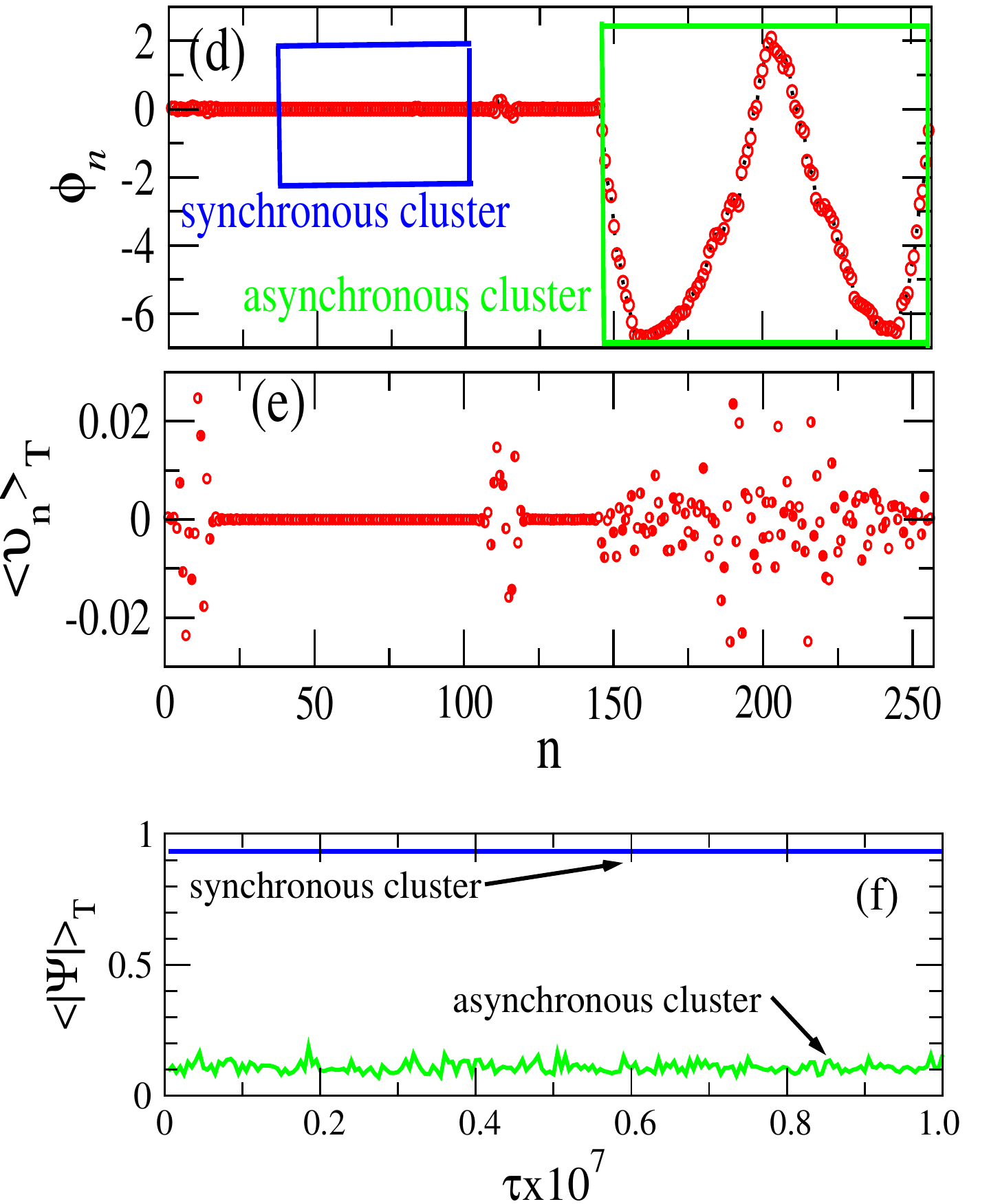} 
\caption{
(a) Flux density $\phi_n$ as a function of site number $n$ and normalized time 
    $\tau$ for a non-locally coupled SQUID metamaterial with $N=256$, 
    $\gamma=0.0021$, $\lambda_0=-0.05$, $\beta_L\simeq 0.7$, and $\phi_R =0.9$, 
    driven by an ac flux field of amplitude $\phi_{ac}=0.015$ and period $T=5.9$.
(b) Same as (a) with $\phi_R =0.8$. The green arrow indicates sudden expansions 
    of the corresponding asynchronous cluster.
(c) The corresponding magnitude of the synchronization parameter averaged over 
    the driving period $T$, $<r (\tau)>_T$ as a function of $\tau$; the blue and 
    red curves are obtained for the chimera state shown in (a) and (b), 
    respectively.
(d) Spatial profile of the fluxes $\phi_n$ threading the SQUID rings at 
    $\tau=10^7$ time units for the parameters of (a) and (b).
(e) The corresponding averaged voltage profile 
    $<v_n (\tau)>_T =<\dot{\phi}_n (\tau)>_T$.
(f) The magnitude of the synchronization parameter averaged over $T$, 
    $<r (\tau)>_T$, as a function of $\tau$, calculated for the coherent cluster 
    in the small-blue box (blue curve) and for the incoherent cluster in the 
    large-green box (green curve) in (d).
\label{fig55.05}
}
\end{figure}

In Figs. \ref{fig55.05}(a) and b, the long-term spatio-temporal evolution for 
the fluxes $\phi_n$ is mapped on the $n - \tau$ plane for two different initial 
flux configurations (i.e., different $\phi_R$); the values of the $\phi_n$s are 
obtained at time-instants that are multiples of the driving period $T$, so that 
uniform (non-uniform) colorization indicates synchronous (asynchronous) dynamics. 
In Fig. \ref{fig55.05}(a), the spontaneous formation of two large clusters of 
SQUIDs, one with synchronized and the other with desunchronized dynamics, can be 
observed. More clusters of SQUIDs, two with synchronized and two with 
desynchronized dynamics, can be observed in Fig. \ref{fig55.05}(b), in which the 
effect of metastability is reflected in the sudden expansions of the upper 
asynchronous cluster at around $\tau \sim 0.35\times 10^7$ t.u. (green arrow). 
In the corresponding time-dependent magnitudes of the synchronization parameter 
averaged over the driving period $T$, $<r (\tau)>_T =<|\Psi (\tau)|>_T$, those 
sudden expansions correspond to jumps towards lower synchronization levels 
(Fig. \ref{fig55.05}(c)). Note that the same calculations, when performed using 
nearest-neighbor (local) coupling, result not in chimera states but instead in 
{\em clustered states}. The latter are also non-homogeneous states, in which two 
or more groups of SQUIDs are spontaneously formed; the SQUID dynamics is 
synchronized within each cluster, however, the clusters are not synchronized to 
each other. Thus, in a clustered state, $<r (\tau)>_T$ can be significantly 
lower than unity (indicating a relatively low degree of synchronization). For 
zero initial conditions, both the non-locally and locally coupled SQUID 
metamaterials result in homogeneous, competely synchronized states with 
$<r(\tau)|>_T$ practically equal to unity at all times. Typical spatial profiles 
of $\phi_n$ and the time-derivatives of the fluxes averaged over $T$, 
$<\dot{\phi}_n (\tau)>_T \equiv <v_n (\tau)>_T$, at the end of the integration 
time (at $\sim 10^7$ time units) of Fig. \ref{fig55.05}(a) are shown in Figs. 
\ref{fig55.05}(d) and (e), respectively. Note that $v_n (\tau)$ is the 
instantaneous voltage across the Josephson junction of the $n-$th SQUID, and it 
is the analogue of the time-derivative of the phases of the oscillators in 
Kuramoto-type phase oscillator models. The pattern of $<v_n (\tau)>_T$ (Fig. 
\ref{fig55.05}(e)) is distinctly different from the standard one for chimera 
states in phase oscillator models \cite{Kuramoto2002}, while it resembles the 
corresponding one for globally coupled, complex Ginzburg-Landau oscillators 
\cite{Sethia2014}. In Figs. \ref{fig55.05}(d) and (e), synchronized clusters of 
SQUIDs are indicated by horizontal segments; it can be observed that besides the 
large incoherent cluster extending from $n=143$ to $256$, there are actually two 
small ones (at around $n\sim 5$ and $n\sim 112$, more clearly seen in Fig. 
\ref{fig55.05}(e) which are not visible in Fig. \ref{fig55.05}(d). The measure 
$<r (\tau)>_T$ as a function of $\tau$ for two different clusters of SQUIDs 
enclosed into the blue (small) and green (large) boxes, which exhibit 
synchronized and desynchronized dynamics, respectively, is shown in 
Fig. \ref{fig55.05}(f). The measure $<r (\tau)>_T$ for the synchronized cluster 
which extends from $n=36$ to $100$, is close to unity for all times (blue curve 
in Fig. \ref{fig55.05}(f)), while that for the desynchronized cluster has a 
significantly lower average and exhibits strong fluctuations which do not 
decrease with time.

In order to determine the metastability levels of the chimera states presented 
in Figs. \ref{fig55.05}(a) and (b), the distributions of the values of 
$x \equiv <|\Psi (\tau)|>_T$s, $pdf(x)$, at all time-steps taken during the 
simulation period (excluding transients, see below) can be calculated (Fig. 
\ref{fig55.06}(a)). A transient period of $100 T$ ($\sim 5900$ time units) was 
allowed, for which the data were discarded. Consider first the black-solid 
curves in the figure, which are actually not symmetric but they fit well to an 
empirical skewed Gaussian function of the form \cite{Rusch1973} 
\begin{equation}
\label{Ch55.103}
   pdf(x) =pdf_m \exp\left\{-\ln(2) 
       \left[ \frac{1}{b} \ln \left( 1 +\frac{2 b (x-x_m)}{D} \right) \right]^2 
        \right\} , 
\end{equation}
where $pdf_m =pdf(x_m)$ is the maximum of the distribution, $x_m$ is the value 
of $x$ at which the maximum of the distribution occurs, $b$ is the asymmetry 
parameter, and $D$ is related to the full-width half-maximum (FWHM) of the 
distribution, $W$, by
\begin{equation}
\label{Ch55.104} 
   W=D\frac{\sinh(b)}{b} .
\end{equation}
The green-dotted curve in Fig. \ref{fig55.06}(a) is a fit of the black-solid 
distribution with $b=0.37$ and $D=0.0116$, while $pdf_m$ and $x_m$ are taken 
from the calculated distribution. That fit gives $W \simeq 0.012$ for the 
non-locally coupled SQUID metamaterial. For the quantification of the 
metastability level, the FWHM of the distribution is used here (which for a 
symmetric Gaussian distribution is directly proportional to the standard 
deviation $\sigma_{r}$, and thus $W$ is proportional to the variance 
$\sigma_{r}^2$ which is a measure of the metastability level). The corresponding 
$pdf_m$ for non-homogeneous (clustered) states in locally coupled SQUID 
metamaterials are effectivelly $\delta-$functions, and the corresponding $W$s 
are smaller by more than two orders of magnitude, indicating the high 
metastability level of the chimera states compared to that of the clustered 
states. The difference in the dynamic behavior between SQUIDs in synchronized 
and desynchronized clusters is also revealed in the power spectra of 
$\phi_n (\tau)$. Two such spectra for frequencies around the fundamental 
(driving) one are shown in semi-logarithmic scale in Fig. \ref{fig55.06}(b), the 
one for a SQUID in the synchronized cluster ($n=40$) and the other in the 
desynchronized cluster ($n=190$). Note that for the chosen parameters, the 
resonance frequency of individual SQUIDs is at $\Omega_{SQ} \simeq 1.3$, while 
the linear band of the SQUID metamaterial extends from $\Omega_{min} \simeq 1.27$ 
to $\Omega_{max} \simeq 1.35$. The driving frequency is $\Omega \simeq 1.06$, 
well below the lower bound of the linear spectrum, $\Omega_{min}$. The spectrum 
for the SQUID at $n=40$ (black curve) exhibits very low noise levels and a 
strong peak at the driving frequency $\Omega$. The smaller peaks in the spectrum 
of the $n=40$ SQUID are also part of it, and they are located at frequencies 
within the linear band of the SQUID metamaterial, i.e., within the range 
$[\Omega_{min}, \Omega_{max}]$. The longer arrow at right points at the 
resonance frequency (the eigenfrequency) of individual SQUIDs. Note that only a 
small number of the eigenfrequencies of the SQUID metamaterial are excited in 
that spectrum, which seem to be selected by random processes. To the contrary, 
the spectrum for the SQUID at $n=190$ exhibits significant fluctuations, the 
peak at the driving frequency, and in addition to that a frequency region around 
$\Omega \sim 0.9-1.05$ in which the average fluctuation level remains 
approximatelly constant, forming a shoulder that often appears in such spectra 
for SQUIDs in desynchronized clusters of chimera states.
\begin{figure}[!h]
\includegraphics[angle=0, width=0.95 \linewidth]{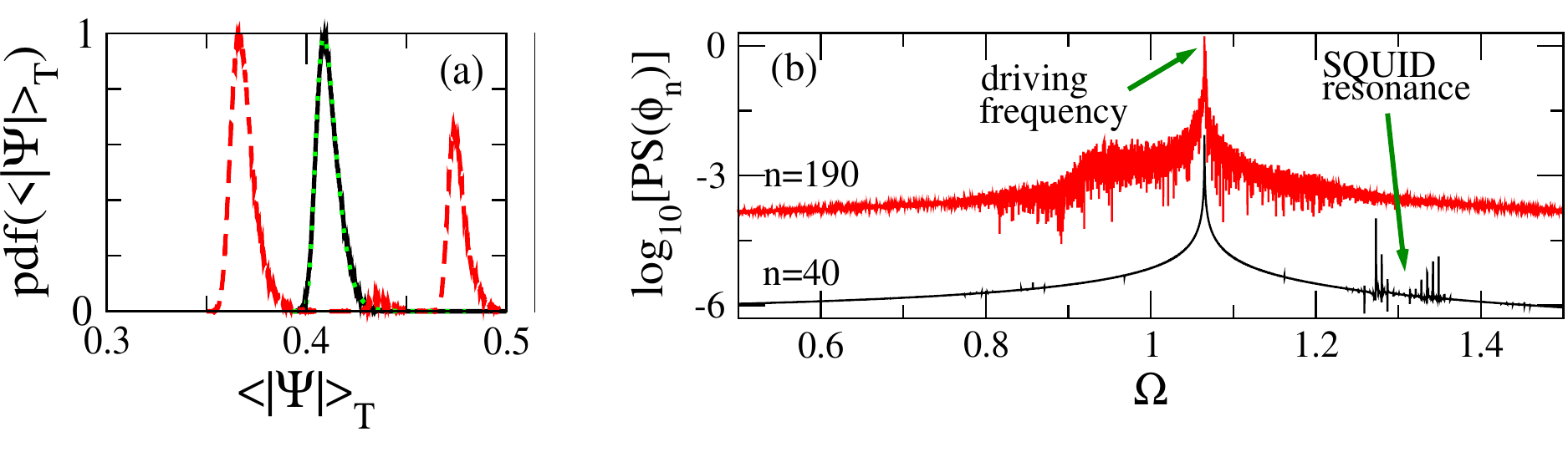}
\caption{
(a) The distributions (divided by their maximum value) of $<|\Psi (\tau)|>_T$s 
    at all instants of the simulation period ($\sim 10^7$ time units with 
    time-step $\Delta t=0.02$) for the states shown in Fig. \ref{fig55.05}(a) 
    and b (black-solid and red-dashed curve, respectivelly). The green-dotted 
    curve is a fit with Eq. (\ref{Ch55.103}).
(b) The {\em power spectrums} of $\phi_n (\tau)$ in semi-logarithmic scale for 
    the SQUIDs with $n=40$ and $n=190$ that belong to the synchronized (black 
    curve) and the desynchronized (red curve) cluster, respectively, of Fig. 
    \ref{fig55.05}(a). The arrow at right points at the eigenfrequency of 
    individual SQUIDs, $\Omega_{SQ} \simeq 1.3$.
\label{fig55.06}
}
\end{figure}

\begin{figure}[!t]
  \includegraphics[angle=0, width=1 \linewidth]{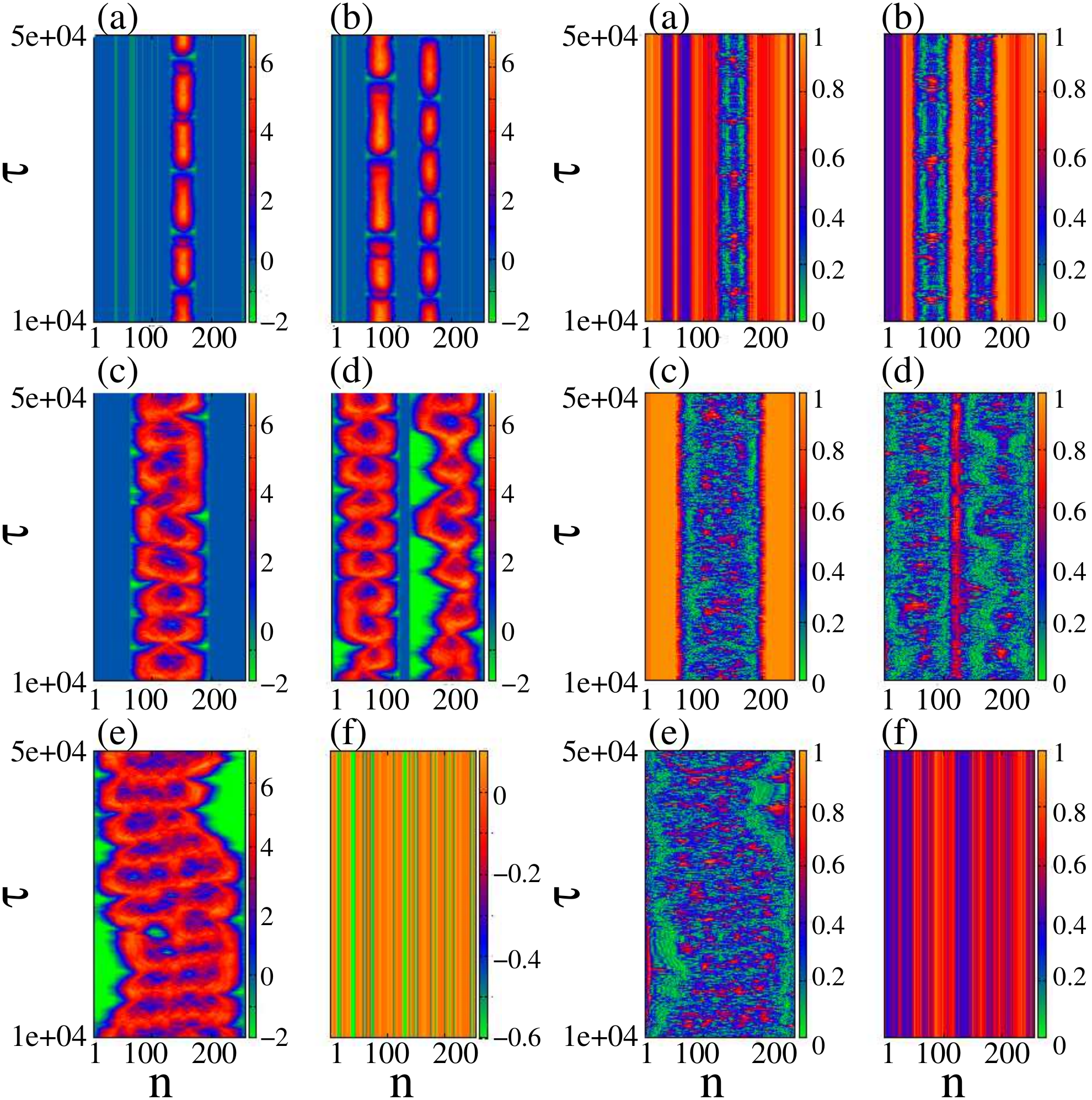}
\caption{
 Left:
 Space-time plots for the flux density $\phi_n$ of the SQUID metamaterial for 
 different initial conditions. Panels (a) and (c) show chimera states with one 
 desynchronized region, panels (b) and (d) show chimera states with two 
 desynchronized regions, while panel (e) show a state with a drifting 
 desynchronized domain, and panel (f) shows a pattern with solitary states.
 Right:
 The corresponding space-time plots for the local order parameter $|Z_n|$ for 
 the states shown in the left panels.
 Parameter values are: $T=5.9$, $N=256$, $\gamma=0.0021$, $\lambda_0=-0.05$, 
 $\beta_L\simeq 0.7$, $\phi_{ac}=0.015$, $\phi_{dc}=0.0$.
\label{fig55.07}
}
\end{figure}

More chimera states, emerging from a variety of initial conditions along with 
the corresponding {\em local synchronization parameter} $|Z_n|$ are shown in 
the left and right panels, respectively, of Fig. \ref{fig55.07}. The real-valued
local synchronization parameter $|Z_n|$ is a measure that can be calculated for 
a group of $2\delta$ coupled oscillators (a sub-system of a larger system of $N$
coupled oscillators) at every instant of time. Its value indicates the 
instantaneous degree of spatial coherence, i.e., the instantaneous degree of 
synchronization, of that group. Is is defined as the magnitude of the complex 
parameter \cite{Omelchenko2011} 
\begin{equation}
\label{Ch55.105}
   Z_n=\frac{1}{2 \delta} \sum_{|m-n| \le \delta} e^{2\pi i \phi_m}, 
       \quad n=\delta+1,\dots, N-\delta.
\end{equation}
A value of the local order parameter $|Z_n| =1$ ($|Z_n| <1$) indicates that the 
$n-$th oscillator belongs to a synchronized (desynchronized) cluster of the 
system of $N$ oscillators. For a finite system of $N$ coupled oscillators such 
as the SQUID metamaterial considered here (i.e., whose boundary conditions are 
not periodic), Eq. (\ref{Ch55.105}) holds for the oscillators whose indices run 
from $n=\delta+1$ to $N-\delta$. For the oscillators close to the boundaries of 
the structure, the calculation of the local order parameter has to be modified 
as follows
\begin{equation}
\label{Ch55.106}
   Z_n=\frac{1}{\delta} \sum\limits_{m=n}^{n+\delta} e^{i2\pi \phi_m} , 
   ~~~{\rm for}~~~
   n=1,\dots,\delta,
\end{equation}
and 
\begin{equation}
\label{Ch55.107}
   Z_n=\frac{1}{\delta} \sum\limits_{m=n-\delta}^{n} e^{i2\pi \phi_m} ,
   ~~~{\rm for}~~~
   N-\delta+1,\dots,N.
\end{equation}

Here the local order parameter is employed to quantufy locally the degree of
synchronization of the collective states obtained for the SQUID metamaterial.  
In the spatiotemporal flux patterns shown in the left panels of Fig. 
\ref{fig55.07}, the values of the $\phi_n$s are obtained at time-instants that 
are multiples of the driving period $T=2\pi/\Omega$ of the ac flux field. 
In particular, Figs. \ref{fig55.07}(a) and (c) correspond to typical chimera 
patterns which exhibit a cluster of desynchronized SQUIDs. In Fig. 
\ref{fig55.07}(a) this cluster is small and it is located around $n=150$, while
in Fig. \ref{fig55.07}(c) it is much larger, spanning the region from 
$n \simeq 70$ to $n \simeq 190$. The SQUIDs which do not belong to these 
clusters are not all synchronized to each other. Instead, small sub-clusters 
of SQUIDs are apparent as stripes with uniform colorization. The SQUIDs 
that belong to such a stripe are synchronized; however, the stripes are not 
synchronized to each other. Furthermore, the flux oscillations in the SQUIDs
that belong to the desynchronized clusters are much stronger than those in the
other SQUIDs. In Figs. \ref{fig55.07}(b) and (d), chimera states exhibitng
two desynchronized clusters are shown. In Fig. \ref{fig55.07}(b) these two
clusters are small and they are located around $n \simeq 80$ and $n \simeq 160$.  
In Fig. \ref{fig55.07}(d), on the other hand, the two desynchronized clusters
are so large that do not leave any space for any synchronized cluster to exist.
A drifting pattern can be observed in Fig. \ref{fig55.07}(e), in which the 
largest part of the SQUID metamaterial forms a desynchronized cluster which size
and position vary in time. Finally, Fig. \ref{fig55.07}(f) demonstrates a 
pattern of low-amplitude flux oscillations with multiple so-called solitary 
states \cite{Maistrenko2014}, where many SQUIDs have escaped from the main 
synchronized cluster and perform oscillations of higher amplitudes (depicted 
by the light green stripes in the otherwise orange background). 
The degree of synchronization within the aforementioned states is visualized 
through the corresponding space-time plots of the local synchronization 
parameter, Eqs. (\ref{Ch55.105}) - (\ref{Ch55.107}), which are shown in the 
right panels of Fig. \ref{fig55.07}. Red-orange colors denote the synchronized 
or coherent regions and blue-green colors the desynchronized or incoherent ones. 
These plots reveal the complexity of the synchronization levels in the SQUID 
metamaterial: For example in Fig. \ref{fig55.07}(a) (right panels) it can be 
seen that the incoherent region located in the center of the metamaterial 
achieves {\em periodically} high values of synchronization demonstrated by the 
orange `"islands" within the cluster. This is related to metastability, which 
was also investigated in reference \cite{Lazarides2015b}. In the coherent 
cluster, on the other hand, blue stripes of low synchronization are observed, 
indicating solitary states. Note that {\em periodic synchronization}, 
characterized by periodic variation of the synchronization parameter, has been 
previously observed in phase oscillator models with external periodic driving 
both with and without an inertial term \cite{Choi1994,Hong1999}.

\begin{figure}[!h]
  \includegraphics[angle=0, width=0.45 \linewidth]{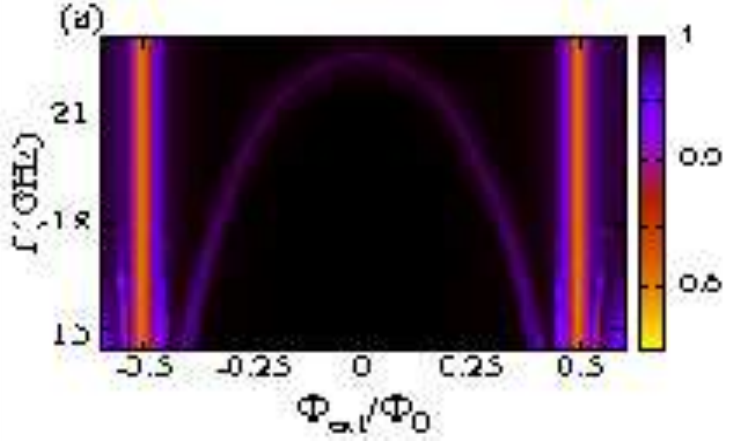}
  \includegraphics[angle=0, width=0.45 \linewidth]{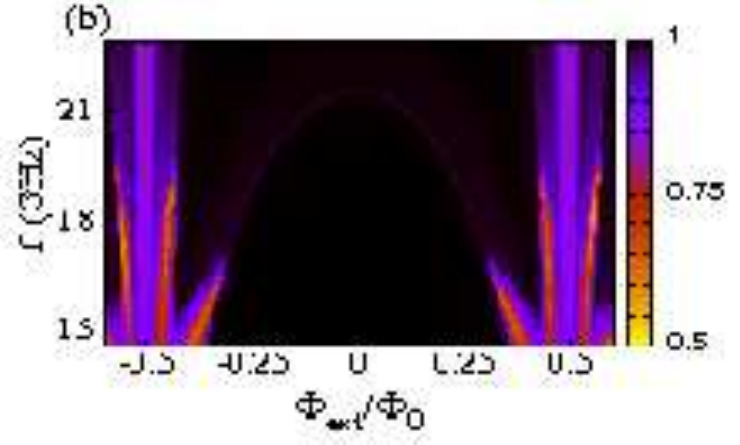} \\
  \includegraphics[angle=0, width=0.45 \linewidth]{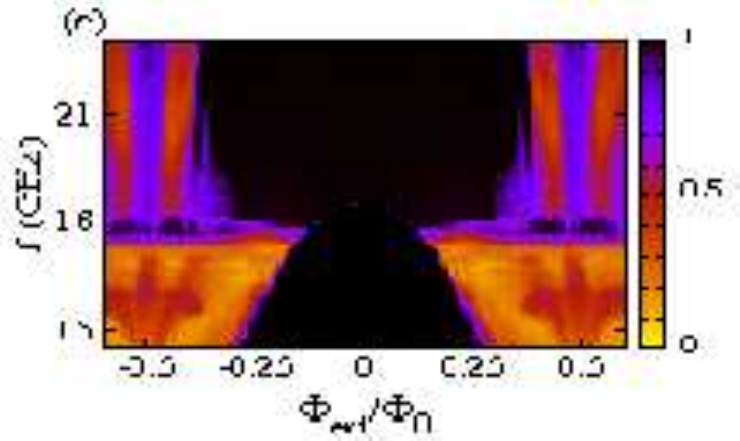}
  \includegraphics[angle=0, width=0.40 \linewidth]{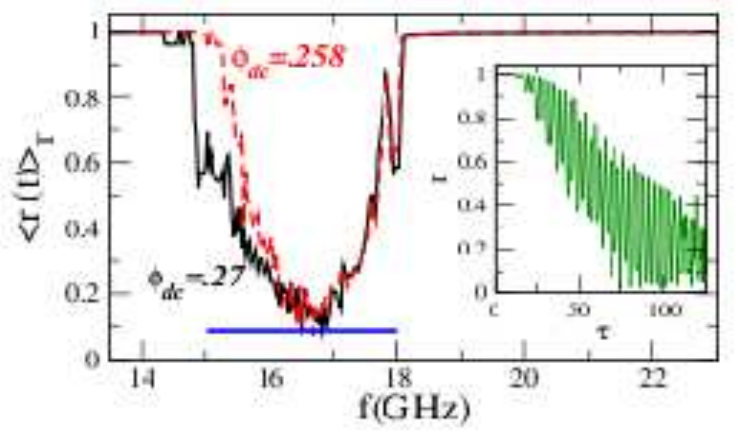}
\caption{
 The magnitude of the global synchronization parameter averaged over the 
 steady-state integration time $<r>_{\Delta \tau}$ mapped as a function of the 
 driving frequency $f$ and the dc flux bias $\phi_{ext} =\phi_{dc}$, for a 
 $27 \times 27$ SQUID metamaterial with $\gamma=0.024$, $\beta_L =0.88$, 
 $\phi_{ac}=0.05$, and (a) $\lambda_0=-0.01$; (b) $\lambda_0=-0.03$; (c) 
 $\lambda_0=-0.05$. In (d), $<r>_{\Delta \tau}$ is plotted as a function of $f$ 
 for $\phi_{dc}=0.27$ (black curve) and $\phi_{dc}=0.258$ (red curve).
 Inset: the instantaneous value of $r$ is shown as a function of time $\tau$ for 
 $\phi_{dc}=0.258$ and $f =16.8 ~GHz$.
\label{fig55.08}
}
\end{figure}

The calculation of the global synchronization parameter averaged over the 
steady-state integration time, $<r>_{\Delta \tau}$ in a physically relevant 
region of the parameter space of driving frequency $f$ and dc bias flux 
$\phi_{ext} =\phi_{dc}$, reveals the possibility of 
synchronization-desynchronization transitions in SQUID metamaterials with 
non-local coupling. Note that the SQUID metamaterial is initialized with zero 
$\phi_n$ and $\dot{\phi}_n$ at each point of the parameter plane. In Fig. 
\ref{fig55.08}(a)-(c), three maps of $<r>_{\Delta \tau}$ are shown on the 
$f - \phi_{dc}$ plane for relatively strong ac driving flux amplitude 
$\phi_{ac}=0.05$ and three values of the coupling coefficient $\lambda=-0.01$ (a), 
$-0.03$ (b), and $-0.05$ (c), for a two-dimensional tetragonal
$27 \times 27$ SQUID metamaterial. The frequency $f$ is given in natural units 
(GHz), while the single SQUID resonance frequency is $f =f_{SQ} =22.6 ~GHz$ 
\cite{Trepanier2013}. As can be seen in the corresponding colorbars, for that 
value of $\phi_{ac}$ and strong coupling coefficient ($\lambda=-0.05$, Fig. 
\ref{fig55.08}(c)), the synchronization parameter $<r>_{\Delta \tau}$ assumes very 
low values in some regions of the parameter plane. Those regions are located 
around $\phi_{dc} =\pm 1/4$ in units of $\Phi_0$. From Fig. \ref{fig55.08}(c), 
two values of $\phi_{dc}$ close to $1/4$ are selected, and the corresponding 
$<r>_{\Delta \tau}$ as a function of the driving frequency $f$ are plotted in 
Fig. \ref{fig55.08}(d). Both curves (black and red) exhibit similar behavior;
for low $f$ the metamaterial is completely synchronized with 
$<r>_{\Delta \tau} =1$. At $f \simeq 14.8 ~GHz$ and $f \simeq 15.2 ~GHz$ (black 
and red curve, respectively), $<r>_{\Delta \tau}$ begins dropping to lower 
values until it reaches a minimum at around $\sim 0.1$. That drop signifies a 
synchronization to desynchronization transition of the SQUID metamaterial with 
the most desynchronized state being observed at $f \sim 16.8 ~GHz$ for both 
curves. For further increasing frequencies, the value of $<r>_{\Delta \tau}$ 
increases gradually until reaching the value of unity ($<r>_{\Delta \tau} =1$) 
where the SQUID metamaterial has return to a completely synchronized state. Note 
that in that desynchronization to synchronization transition, the two curves 
follow closely each other. Similar synchronization to desynchronization 
transitions have been observed in arrays of Josephson junctions 
\cite{Wiesenfeld1996}. In the inset of Fig. \ref{fig55.08}(d), the time-dependence 
of the instantaneous value of the synchronization parameter $r (\tau)$ is shown 
for $f =16.8 ~GHz$, i.e., the value of $f$ for which $<r>_{\Delta \tau}$ is the 
lowest. It is observed that the SQUID metamaterial remains synchronized for only 
about $25$ time units, and then $<r>_{\Delta \tau}$ starts oscillating strongly 
while its average value falls rapidly. In fact, in about $100$ time units, 
$<r>_{\Delta \tau}$ has already reached its lowest value. Note that the minimum 
value of $<r>_{\Delta \tau}$ cannot be zero since the system is finite. As shown 
in the inset, however, the instantaneous value of $r (\tau)$ can actually reach 
values close to zero. The minimum value of $<r>_{\Delta \tau}$ decreases with 
increasing system size, while it vanishes in the "thermodynamic limit", i.e., 
for a very large SQUID metamaterial. 

\subsubsection{\em SQUID metamaterials with local coupling}
\begin{figure}[!h]
  \includegraphics[angle=0, width=0.45 \linewidth]{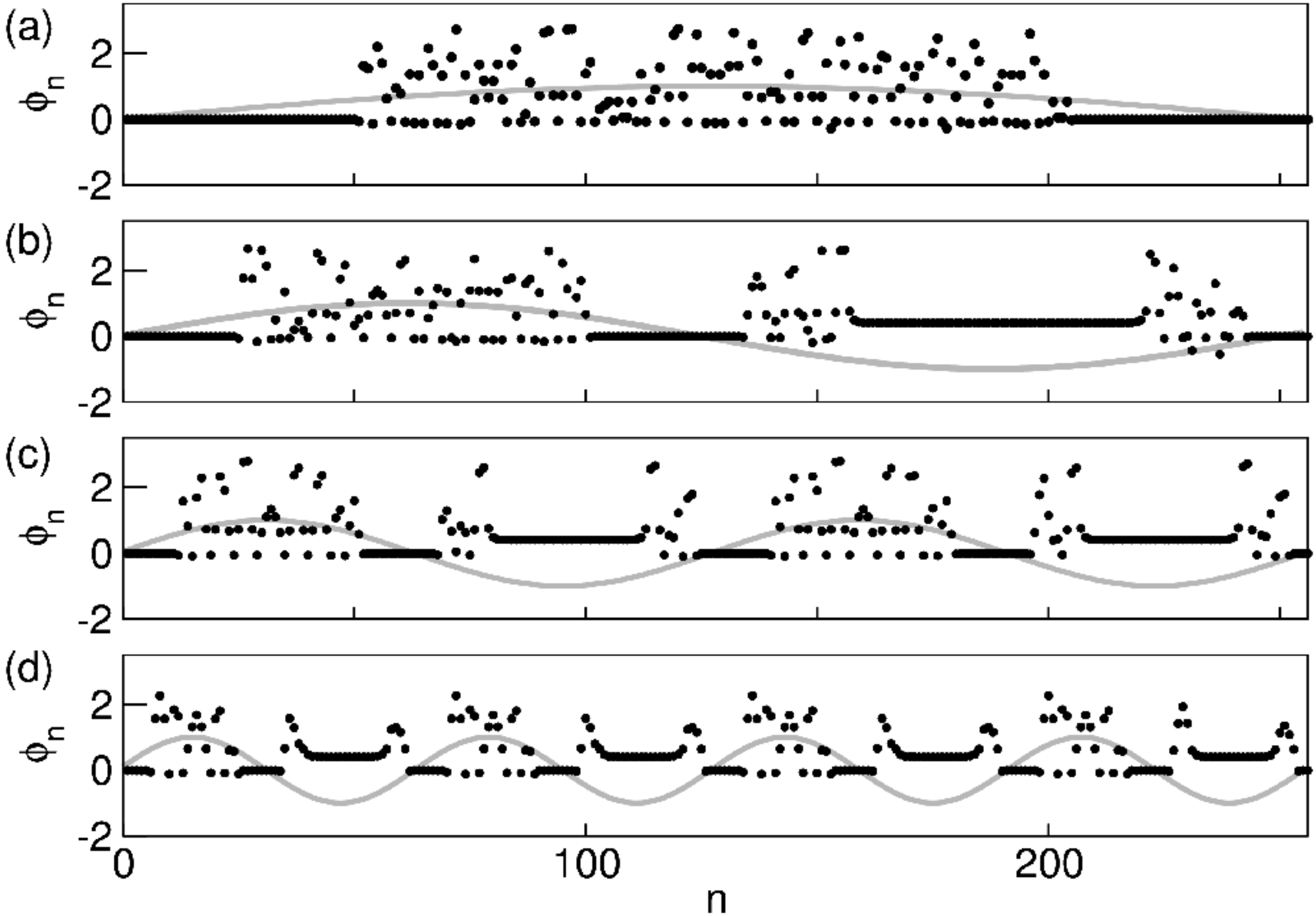}
  \includegraphics[angle=0, width=0.45 \linewidth]{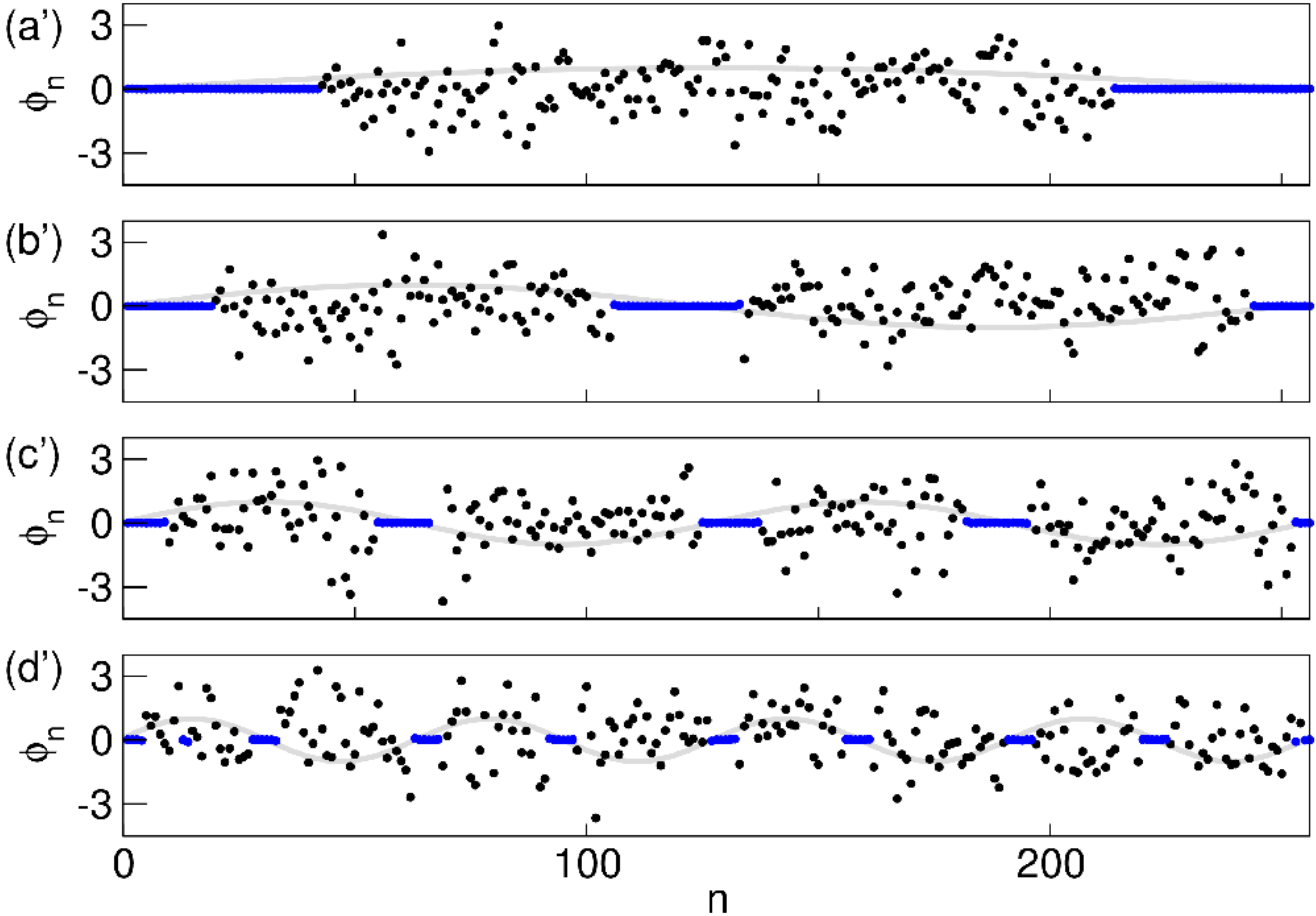}
\caption{
 Snapshots of the magnetic flux density $\phi_n$ at time $\tau =5000$ time units 
 for two different values of the loss coefficient: $\gamma=0.024$ in (a)-(d) and 
 $\gamma=0.0024$ in (a')-(d'). Grey solid lines mark the initial magnetic flux 
 distribution used in the simulations. Blue solid lines in the right panel 
 emphasize the coherent clusters of the chimera states. The other parameters are 
 $T=6.24$ ($\Omega\simeq 1.007$), $\beta_L = 0.86$,and $\phi_{ac}=0.06$.
\label{fig55.09}
}
\end{figure}

\begin{figure}[!h]
  \includegraphics[angle=0, width=0.45 \linewidth]{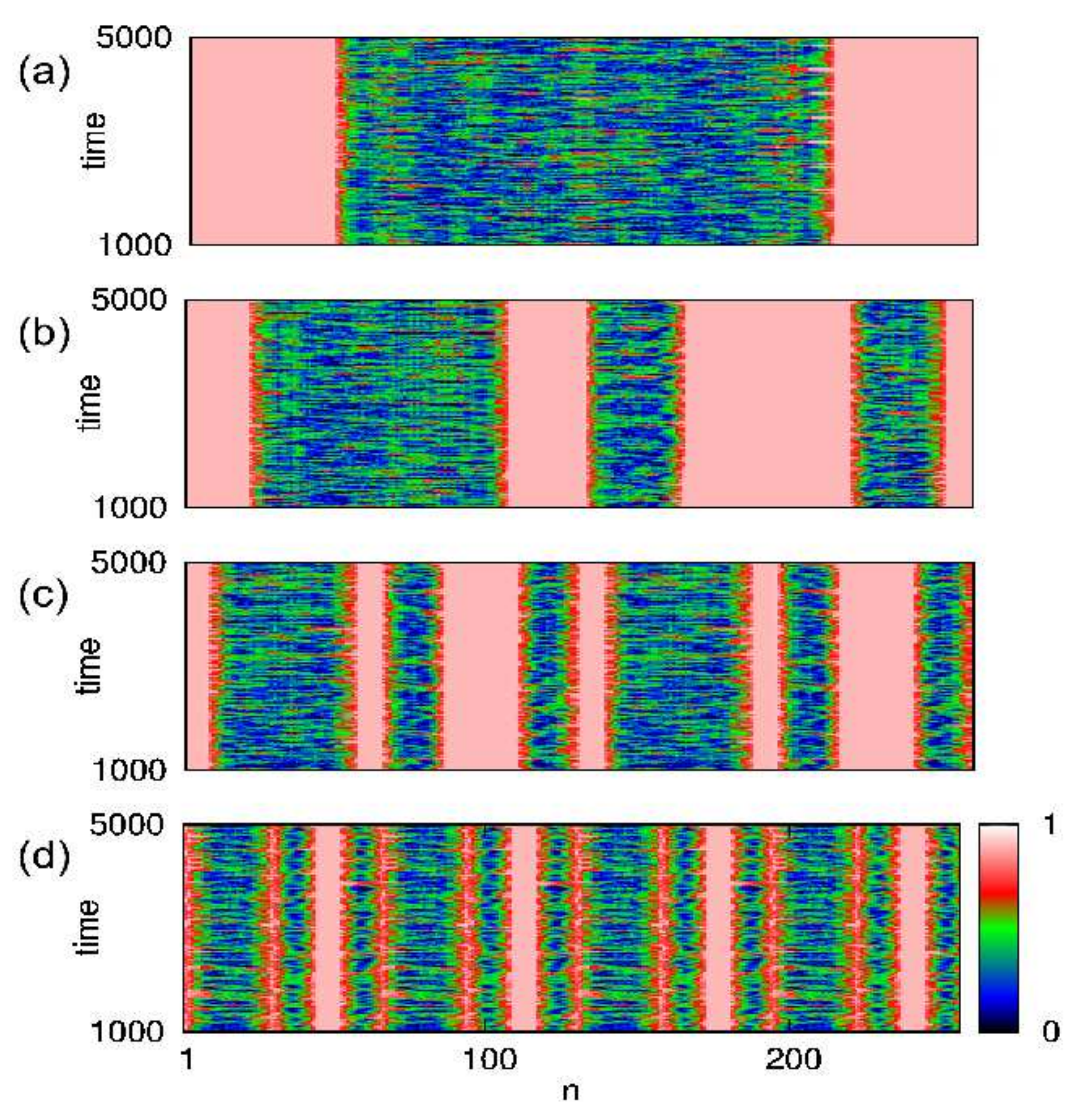}
  \includegraphics[angle=0, width=0.45 \linewidth]{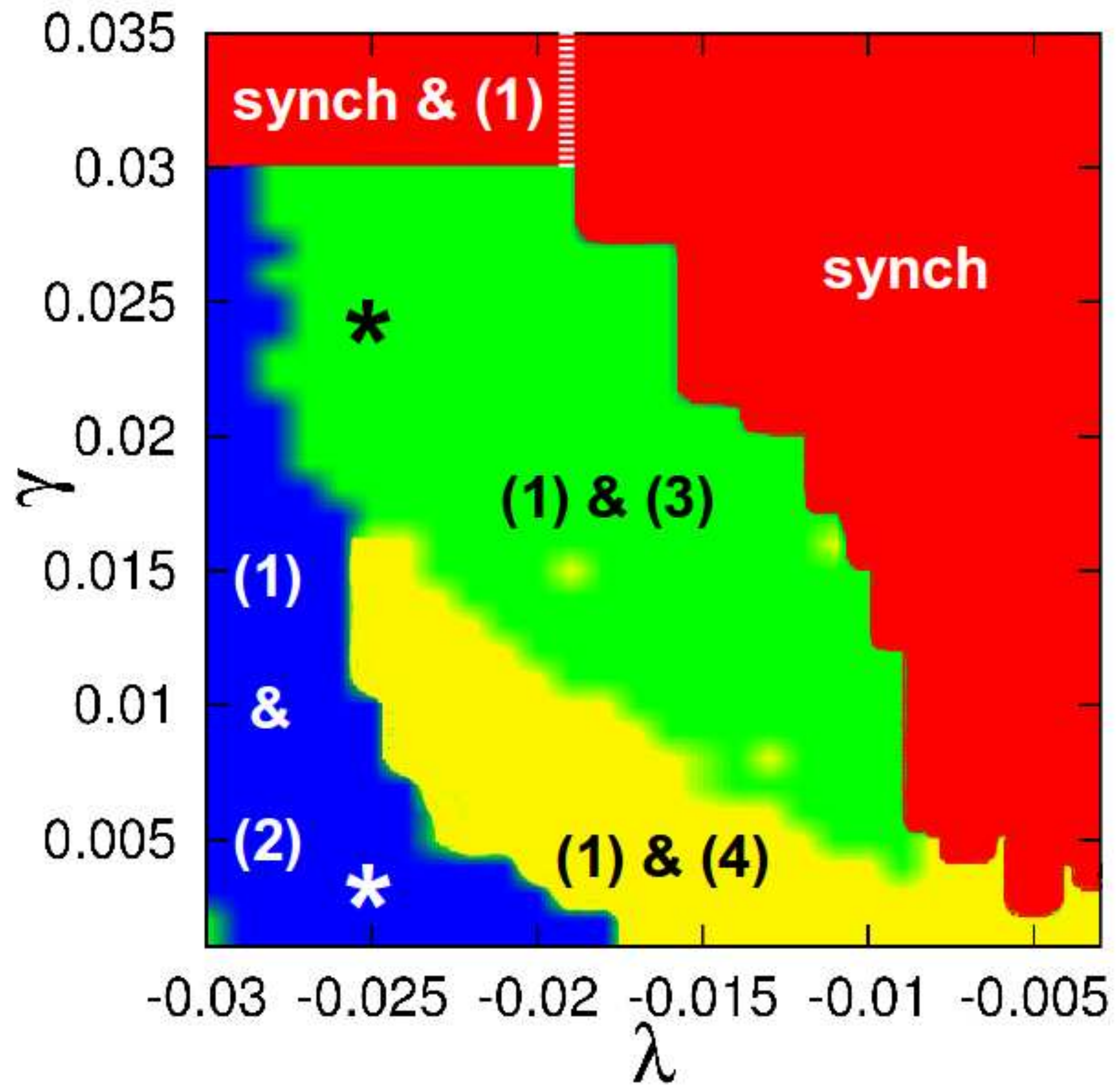}
\caption{
 Left panel: Space-time plots for the magnitude of the local synchronization 
 parameter $|Z_n|$ of the chimera states corresponding to Fig. 
 \ref{fig55.09}(a), (b), (c), and (d).
 Right panel: Map of dynamic regimes in the $(\gamma,\lambda)$ parameter space 
 for the initial conditions of Fig. \ref{fig55.09}(a) and (b). Numbers in 
 brackets denote the multiplicity of the chimera state while "synch" stands for 
 synchronization. The other parameters are as in Fig. \ref{fig55.09}.
\label{fig55.10}
}
\end{figure}

Chimera states have mostly been found for non-local coupling between the coupled 
oscillators. This fact has given rise to a general notion that non-local 
coupling, is an essential ingredient for their existence. However, recently, it 
has been demonstrated that chimeras can be achieved for global coupling as well 
\cite{Schmidt2014,Sethia2014,Yeldesbay2014,Bohm2015}. The case of local coupling 
(i.e., only nearest-neighbor interactions between the oscillators) has been 
studied less. In reference \cite{Laing2015}, chimera states were found in 
locally coupled networks, but the oscillators in the systems under consideration 
were not completely identical. Very recently, the emergence of single- and 
double-headed chimera states in neural oscillator networks with local coupling 
has been reported \cite{Bera2016}. That system, however, is known to exhibit 
high metastability, which renders the chimera state non-stationary when tracked 
in long time intervals \cite{Hizanidis2016c}. The emergence of multi-clustered 
robust chimera states in locally coupled SQUID metamaterials can be demonstrated 
in a relevant parameter region which has been determined experimentally 
\cite{Trepanier2013,Zhang2015}. Fig. \ref{fig55.09} shows time-snapshots of 
the fluxes $\phi_n$ for different initial conditions and for two values of the 
loss coefficient $\gamma$ which differ by an order of magnitude. The left panel 
is for $\gamma=0.024$. The initial "sine wave" flux distribution for each 
simulation is shown by the gray solid line. The SQUIDs that are prepared at 
lower values form the coherent clusters of the chimera state, while those that 
are initially set at higher flux values oscillate incoherently. Moreover, as the 
"wavelength" of the initial flux distribution increases, so does the chimera 
state multiplicity, i.e., the number of coherent and/or incoherent regions. 
Similar behavior is observed for lower values of the loss coefficient 
$\gamma=0.0024$ as shown in the right panel of Fig. \ref{fig55.09}. Here, the 
incoherent clusters are better illustrated since they are approximately of equal 
size and do not contain oscillators that "escape" from the incoherent cluster 
abiding around low magnetic flux values, something which is visible in the left 
panel. Furthermore, the coherent clusters (emphasized by the blue solid lines) 
are fixed around $\phi=0$, unlike in the left panel where additional clusters 
located at slightly higher values also form. Here we must recall that the 
snaking resonance curve of a single SQUID increases significantly its winding
with decreasing values of $\gamma$ (right panel), creating thus new branches of 
stable (and equally unstable) periodic (period-1) solutions.
These branches are larger in number and smaller in size compared to those of 
higher $\gamma$ values (left panel). The lower amplitude branches which are the 
biggest ones attract the SQUIDs that eventually form the coherent clusters. The 
other SQUIDs have a plethora of higher flux amplitude states to choose from 
and, therefore, create a more chaotic incoherent cluster than in the case of 
higher $\gamma$ values. The observed chimera states can be quantified again 
through the local synchronization parameter $|Z_n|$ \cite{Hizanidis2016b}, which 
is a measure for local synchronization. A spatial average with a window size of 
$\delta=5$ elements, can be employed. In the left panel of Fig. \ref{fig55.10} 
the space-time plots of $|Z_n|$ corresponding to the chimera states of Figs. 
\ref{fig55.09}(a)-(d) are shown. The number of (in)coherent regions 
increases according to the number of half-wavelengths in the initial conditions 
and the size and location of the clusters is constant in time. Previous works on 
SQUID metamaterials demonstrated that for nonlocal coupling, single- and 
double-headed chimera states coexist with solitary states \cite{Jaros2015} and 
metastable states of drifting (in)coherence, in a dynamical area of the SQUID 
metamaterial in which the driving frequency lied outside the multistability 
regime \cite{Lazarides2015b,Hizanidis2016b}. For a suitable choice of the 
driving frequency $\Omega$, stable chimera states can be achieved for non-local 
coupling as well. However, those chimera states exist only for low coupling 
strengths $\lambda$; the threshold value of the coupling strength in the case 
of local coupling is much higher. Local coupling is therefore crucial for the 
emergence of \emph{robust} chimera states, both in structure and in lifetime, 
for large areas of parameter space.

In the previous paragraphs, the importance of multistability and the impact of 
the dissipation coefficient $\gamma$ in the formation of chimera states in SQUID 
metamaterials was stressed. In addition to that, it is important to note the 
role of the network topology which is defined through the local nature of 
interactions and the coupling strength $\lambda$. As already shown in Fig.
\ref{fig55.09}, SQUID metamaterials exhibit a variety of coexisting 
multi-clustered chimera states. A systematic study in the $(\lambda, \gamma)$ 
parameter space is depicted in the right panel of Fig. \ref{fig55.10}, in which 
the observed patterns for the initial conditions of Figs. \ref{fig55.09}(a)/(a') 
and b/b' are mapped out. The numbers in the brackets correspond to the 
multiplicity of the respective chimera states and "synch" denotes the 
synchronized states. The black and white asterisk mark the $(\lambda, \gamma)$ 
values used in the left and right panel of Fig. \ref{fig55.09}, respectively. 
For low coupling strengths, single- and four-headed chimera states exist but 
only for low values of $\gamma$. As $\gamma$ increases, the effect of 
multistability diminishes and the system enters the synchronized state. As the 
coupling $\lambda$ becomes stronger, the synchronization threshold for $\gamma$ 
is shifted to higher values, below which three-headed chimeras coexist with 
single-headed ones. The latter persist for even higher $\lambda$ and $\gamma$ 
values and at the same time double-headed chimeras appear as well. For initial 
conditions with a larger modification in space (like in Figs. 
\ref{fig55.09}(c)/(c') and (d)/(d')), chimera states with higher multiplicity 
may emerge, but the mechanism towards synchronization is the same: By increasing 
$\gamma$, the multiplicity of the chimera state decreases and eventually the 
fully coherent state is reached through the appearance of solitary states 
\cite{Jaros2015,Hizanidis2016b}.

\begin{figure}[!h]
  \includegraphics[angle=0, width=0.95 \linewidth]{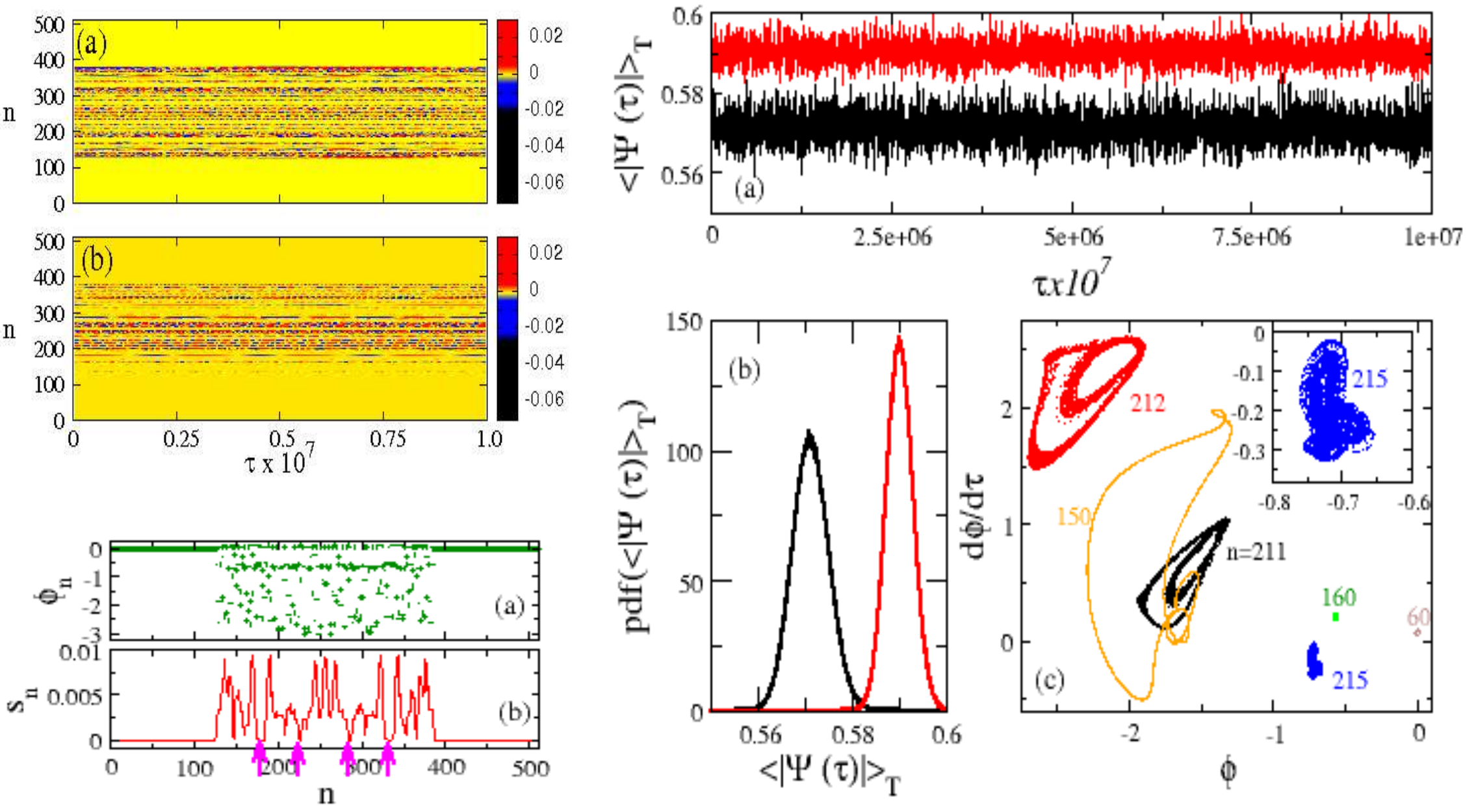}
\caption{
(a) $\&$ (b) Density plot of the fluxes $\phi_n$ through the SQUIDs on the 
    $n - \tau$ plane for a SQUID metamaterial with $N=512$, $\beta_L = 0.86$, 
    $\gamma=0.024$, $\lambda_0=-0.025$, $\phi_{ac}=0.06$, and $T=6.24$ 
    ($\Omega\simeq 1.007$), and different initial conditions (see text).
(c) The flux density profile $\phi_n$ at $\tau=10^7$ time units 
    ($\sim 1.6 \times 10^6 ~T$ for the chimera state shown in (a).
(d) The local standard deviation $\sigma_n$ as a function of the SQUID number 
    $n$ for the chimera state shown in (a), for $\delta=10^{-4}$. 
    The arrows indicate the location of small synchronized clusters.
(e) The magnitude of the synchronization parameter averaged over the driving 
    period $T$, $<r (\tau)>_T$, as a function of time $\tau$; the black 
    (lower) and red (upper) curves which average to $\sim 0.571$ and $\sim 0.59$ 
    correspond to the chimera states shown in (a) and (b), respectively.
(f) The distributions of $<r (\tau)>_T$ with full-width half-maximum 
    $\sim 0.0091$ and $\sim 0.0066$ for the chimera states in (a) and (b), 
    respectively.
(g) Phase portraits in the reduced, single-SQUID phase space for several SQUIDs, 
    which number is indicated on the figure, for the chimera state in (a).
    Inset: Expanded phase portrait for the $n=215$ SQUID.
\label{fig55.11}
}
\end{figure}

Single-headed chimera states with very long life-times can be also obtained 
generically in SQUID metamaterials with nearest-neighbor coupling using initial 
conditions of the form $\phi_n (\tau=0) =-1.7$ and $\dot{\phi}_n (\tau=0) =+1$ 
for $n$ in $[n_\ell,n_r]=[128,384]$ and zero otherwise, or as 
$\phi_n (\tau=0) =3+\phi_R$ and $\dot{\phi}_n (\tau=0) =\phi_R$ for $n$ in 
$[n_\ell,n_r]=[128,384]$ and zero otherwise, with $\phi_R$ drawn from a flat, 
zero mean distribution in $[-4,+4]$. There is nothing special about those 
particular initial conditions; however, different sets of initial conditions 
result in different chimera states. The obtained chimeric patterns are shown in 
Fig. \ref{fig55.11}(a) and (b), respectively, for those initial conditions, 
respectively. The average over a driving period $T$ of the voltages in the 
Josephson junctions of the SQUIDs, $<\dot{\phi}_n>_T$, are mapped onto the 
$n-\tau$ plane so that uniform colorization indicates synchronized dynamics (for 
which the fluxes execute low amplitude oscillations). In the region of 
desynchronized dynamics in the interval $[n_\ell,n_r]$, one can still distinguish 
a few small synchronized clusters that break it into several subclusters. The 
voltage oscillations in the desynchronized clusters differ both in amplitude and 
phase, since the SQUIDs there are close to or in a chaotic state. The profile of 
the fluxes $\phi_n$ threading the SQUID loops for the chimera state in Fig. 
\ref{fig55.11}(a) is shown in Fig. \ref{fig55.11}(c) at the end of the integration 
time. The desynchronized region is indicated by the seemingly randomly scattered 
points in the interval $[n_\ell, n_r]$. The emergence of chimera states in SQUID 
metamaterials can be clearly attributed to the extreme multistability around the 
geometrical resonance frequency of individual SQUIDs, which leads to 
{\em attractor crowding} \cite{Wiesenfeld1989} accompanied by the generation of 
several chaotic states. Thus, with proper choise of initial conditions, a large 
number of SQUIDs may find themselves in a chaotic state forming thus one (or 
more) desynchronized cluster(s). Moreover, there are also periodic states in 
this frequency region which are highly metastable due to attractor crowding that 
shrinks their basins of attraction. Thus, the flux in some of the SQUID 
oscillators may jump irregularly from one periodic state to another resulting in 
effectively random dynamics.  

For the characterization of the chimera states in Figs. \ref{fig55.11}(a) and 
(b), except the global measure of synchronization $<r>_T =<|\Psi (\tau)|>_T$, a 
{\em measure of incoherence} and a {\em chimera index} are defined as follows 
\cite{Gopal2014}. First, we define $v_n (\tau) \equiv <\dot{\phi}_n>_T (\tau)$, 
where the angular brackets indicate average over the driving period $T$, and 
$\bar{v}_n (\tau) \equiv \frac{1}{n_0 +1} \sum_{n=-n_0/2}^{+n_0/2} v_n (\tau)$ 
which is the local spatial average of $v_n (\tau)$ in a region of length $n_0+1$ 
around the site $n$ at time $\tau$ ($n_0 <N$ is an integer).  Then, the local 
standard deviation of $v_n (\tau)$ is defined as 
\begin{equation}
\label{Ch55.108}
   \sigma_n (\tau) \equiv \left< \sqrt{ \frac{1}{n_0 +1} 
   \sum_{n=-n_0/2}^{+n_0/2} \left( v_n -\bar{v}_n \right)^2 } \right>_{n_T} , 
\end{equation}
where the large angular brackets denote averaging over the number of driving 
periods of time-integration (excluding transients). The index of incoherence is 
then defined as $S=1 -\frac{1}{N} \sum_{n=1}^N s_n$, where 
$s_n=\Theta(\delta -\sigma_n)$ with $\Theta$ being the Theta function. The index 
$S$ takes its values in $[0,1]$, with $0$ and $1$ corresponding to synchronized 
and desynchronized states, respectively, while all other values in between them 
indicate a chimera or multi-chimera state. Finally, the chimera index is defined 
as $\eta =\sum_{n=1}^N |s_n -s_{n+1}| / 2$ and equals to unity (an integer 
greater than unity) for a chimera (a multi-chimera) state. The local standard 
deviation $\sigma_n$ as a function of $n$ is shown in Fig. \ref{fig55.11}(d). 
Its value is practically zero in the synchronized regions, while it fluctuates 
between zero and unity in the desynchronized region. However, there are four 
small clusters indicated by the arrows in which the dynamics is synchronized. In 
order for these features to be visible (also apparent in Fig. \ref{fig55.11}(a)), 
the integer $n_0$ has to be close to the number of the SQUIDs that belong to the 
small synchronized clusters (here $n_0=4$). The small synchronized clusters 
divide the central region of the SQUID metamaterial in a number of 
desynchronized clusters. The indices of incoherence for the chimera states shown 
in Figs. \ref{fig55.11}(a) and (b), is $S=0.46$ and $S=0.44$, respectively, very 
close (within $1\%$) to $1-<r (\tau)>_T$ in both cases for a threshold value 
$\delta=10^{-4}$. The choice of both $n_0$ and $\delta$ is rather subjective, 
but they have to be such that the resulting indices agree with what we get by 
inspection. When properly chosen, however, they are very useful for comparing 
chimera states resulting from different initial conditions. The chimera index 
for the states in Figs. \ref{fig55.11}(a) and (b), is $\eta=5$ and $\eta=7$, 
respectively, roughly corresponding to the number of desynchronized clusters of 
a multi-headed chimera state. The global synchronization measure 
$<r (\tau)>_T$ as a function of $\tau$ is shown in Fig. \ref{fig55.11}(e) 
as black (lower) and red (upper) curve for the chimera state in Figs. 
\ref{fig55.11}(a) and (b), respectively. The average over all integration times 
(initial transients have been excluded) gives, respectively, $0.571$ and $0.59$. 
The strong fluctuations of these curves are a distinguishing feature of both
single-headed and multi-headed chimera states; when the SQUID metamaterial is in 
a homogeneous or clusteres state, the size of fluctuations practically vanishes. 
For the obtained values of both the index of incoherence and $<r (\tau)>_T$ it 
can be concluded that the chimera state in Fig. \ref{fig55.11}(b) is slightly 
more synchronized than that in Fig. \ref{fig55.11}(a).

Moreover, their level of metastability can be estimated from the full-width 
half-maximum (FWHM) of the distributions of the values of $<r (\tau)>_T$ shown 
in Fig. \ref{fig55.11}(f). These distributions are well fitted to a Gaussian shape 
while their maximums are located at the long time averages of $<r (\tau)>_T$. 
The FWHM for the black (lower hight) and the red (higher hight) distributions 
turn out to be $\sim 0.57$ and $\sim 0.59$, respectively. Thus, it can be 
concluded that the less synchronized chimera state in Fig. \ref{fig55.11}(a) is 
at a higher metastability level. In Fig. \ref{fig55.11}(g), the phase portraits 
for several SQUIDs on the reduced single SQUID phase space $\phi_n -\dot{\phi}_n$ 
are shown for $n=60,150,160,211,212,215$. Those SQUIDs have been chosen because
they exhibit different dynamical behaviors which ranges from periodic (i.e., the 
SQUID is phase locked to the driver as for e.g., $n=60$) to chaotic (e.g. for 
$n=212$), in which the trajectory explores a significant part of the reduced 
phase space.       
\newpage

\section{SQUID Metamaterials on Lieb lattices}
\label{SQUIDS-S5}
\begin{figure}[!h]
\includegraphics[angle=0, width=0.95 \linewidth]{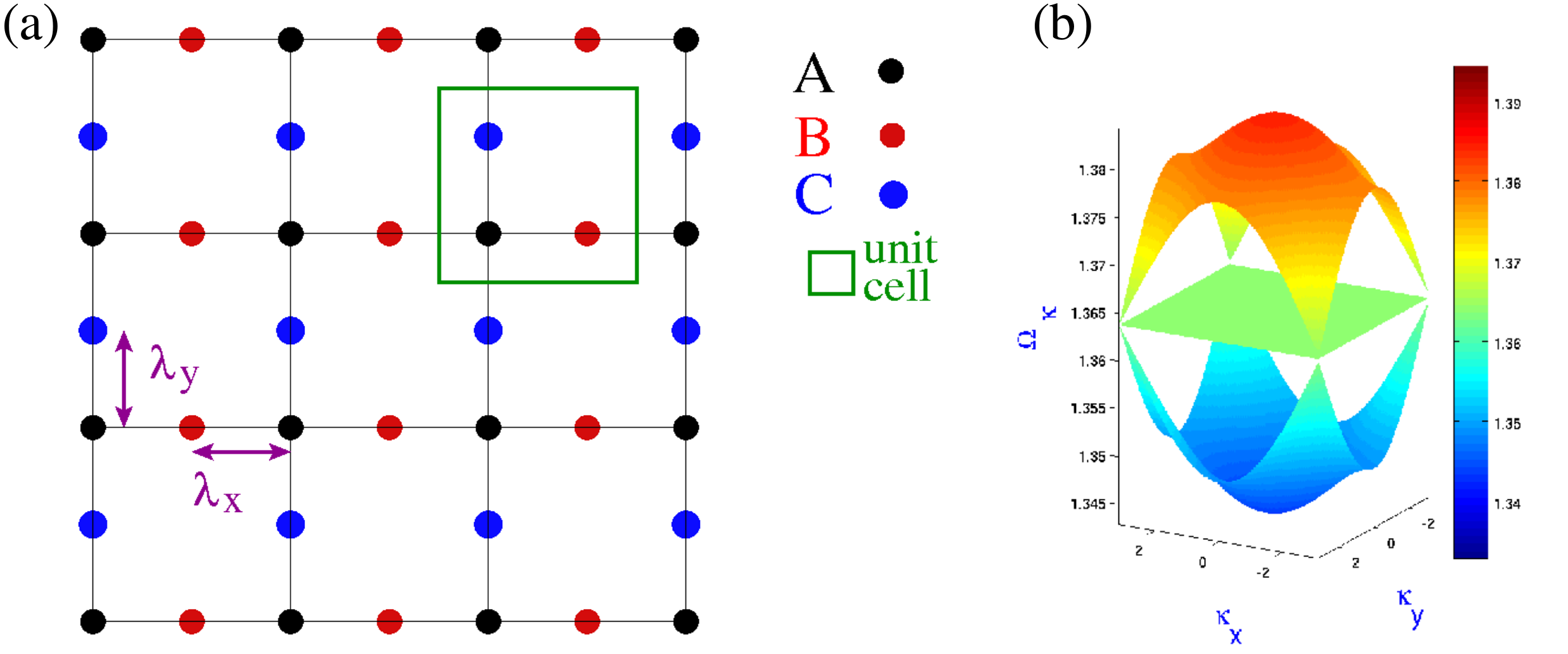}
\caption{
 (a) Schematic of a Lieb lattice. The three sublattices are indicated in black 
 (corner SQUIDs), red (edge SQUIDs), and blue (edge SQUIDs) color. The 
 nearest-neighbor couplings $\lambda_x$ and $\lambda_y$  and the unit cell 
 (green box) are also indicated.   
 (b) The linear frequency spectrum $\Omega_{\bf \kappa} ({\bf \kappa})$ of the 
 SQUID Lieb metamaterial for $\beta=0.86$, and $\lambda_x =\lambda_y =-0.02$. 
 The flat-band frequency is $\Omega_{FB} =\Omega_{SQ} \simeq 1.364$. 
\label{fig9.01}
}
\end{figure}

Besides the freedom of engineering the properties of the individual "particles" 
or devices such as the SQUIDs which play the role of "atoms" in a metamaterial, 
one also has the freedom to choose the arrangement of those "particles" in space. 
That means that one has the freedom to choose a particular type of lattice which 
sites will be occupied by those "particles". Interestingly, there are some 
specific lattice geometries which give rise to novel and potentially useful 
frequency spectrums. Such an example is the so-called {\em Lieb} lattice. The
latter is actually a square-depleted (line-centered tetragonal) lattice, 
described by three sites in a square unit cell as illustrated in Fig. 
\ref{fig9.01}(a); it is characterized by a band structure featuring a Dirac 
cone intersected by a topological flat band \cite{LiuZheng2014,Leykam2018}, 
shown in Fig. \ref{fig9.01}(b). As it is well-known, systems with a flat-band 
in their frequency spectrum support localized eigenmodes also called 
{\em localized flat-band modes}; such states have been recently successfully 
excited and subsequently observed in photonic Lieb lattices 
\cite{Vicencio2015,Mukherjee2015a}.

\subsection{Nearest-Neighbor Model and Frequency Spectrum}
Consider the Lieb lattice shown schematically in Fig. \ref{fig9.01}(a), in which 
each site is occupied by a SQUID. That SQUID Lieb Metamaterial (SLiMM) can be 
regarded as the combination of three sublattices indicated by different colors 
(black, red, blue). The SQUIDs are assumed to be identical, and they are coupled 
magnetically through their mutual inductances to their nearest neighbors. The 
dynamic equations for the fluxes through the loops of the SQUIDs are obtained 
from the combination of the flux-balance relations
\begin{eqnarray}
  \Phi_{n,m}^A = \Phi_{ext} +L \left\{
  I_{n,m}^A +\lambda_x \left[ I_{n-1,m}^B +I_{n,m}^B \right] 
     +\lambda_y \left[ I_{n,m-1}^C +I_{n,m}^C \right] \right\} , 
\nonumber \\
\label{Ch9.01}
  \Phi_{n,m}^B = \Phi_{ext} +L \left\{
  I_{n,m}^B +\lambda_x \left[ I_{n,m}^A +I_{n+1,m}^A \right] \right\} , \\
  \Phi_{n,m}^C = \Phi_{ext} +L \left\{
  I_{n,m}^C +\lambda_y \left[ I_{n,m}^A +I_{n,m+1}^A \right] \right\} , 
\nonumber
\end{eqnarray}
where $\phi_{n,m}^k$ with $k=A, B, C$ is the flux through the SQUID of the 
$(n,m)$th unit cell of kind $k$, $I_{n,m}^k$ is the current in the SQUID of the 
$(n,m)$th unit cell of kind $k$, $\Phi_{ext}$ is the applied (external) flux, 
and $\lambda_x =M_x /L$ ($\lambda_y =M_y /L$) is the dimensionless coupling 
coefficient along the horizontal (vertical) direction, with $M_x$ ($M_y$) being 
the corresponding mutual inductance between neighboring SQUIDs and $L$ the 
self-inductance of each SQUID. The currents $I_{n,m}^k$ in the SQUIDS are 
provided by the resistively and capacitively shunted junction (RCSJ) model to be
\cite{Likharev1986}
\begin{equation}
\label{Ch9.02}
  -I_{n,m}^k =C\frac{d^2 \Phi_{n,m}^k}{dt^2} +\frac{1}{R} \frac{d \Phi_{n,m}^k}{dt}
   +I_c \sin\left( 2\pi \frac{\Phi_{n,m}^k}{\Phi_0} \right) ,
\end{equation}
where $R$ is the subgap resistance through the Josephson junction of each SQUID,
$C$ is the capacitance of the Josephson junction of each SQUID, and $I_c$ is the 
critical current of the Josephson junction of each SQUID. From Eqs. 
(\ref{Ch9.01}) and (\ref{Ch9.02}), assuming that all the terms proportional to 
$\lambda_x^a \lambda_y^b$ with $a+b >1$ are negligible, we get 
\cite{Lazarides2017} 
\begin{eqnarray}
  L C \frac{d^2 \Phi_{n,m}^A}{dt^2} +\frac{L}{R} \frac{d \Phi_{n,m}^A}{dt}
   +L I_c \sin\left( 2\pi \frac{\Phi_{n,m}^A}{\Phi_0} \right) +\Phi_{n,m}^A 
  = \lambda_x \left( \Phi_{n,m}^B +\Phi_{n-1,m}^B \right)
   +\lambda_y \left( \Phi_{n,m}^C +\Phi_{n,m-1}^C \right) 
   +\Phi_{eff}^A , 
\nonumber \\
\label{Ch9.03}
  L C \frac{d^2 \Phi_{n,m}^B}{dt^2} +\frac{L}{R} \frac{d \Phi_{n,m}^B}{dt}
   +L I_c \sin\left( 2\pi \frac{\Phi_{n,m}^B}{\Phi_0} \right) +\Phi_{n,m}^B 
  = \lambda_x \left( \Phi_{n,m}^A +\Phi_{n+1,m}^A \right) +\Phi_{eff}^B , 
 \\ 
  L C \frac{d^2 \Phi_{n,m}^C}{dt^2} +\frac{L}{R} \frac{d \Phi_{n,m}^C}{dt}
   +L I_c \sin\left( 2\pi \frac{\Phi_{n,m}^C}{\Phi_0} \right) +\Phi_{n,m}^C 
  = \lambda_y \left( \Phi_{n,m}^A +\Phi_{n,m+1}^A \right) +\Phi_{eff}^C . 
\nonumber
\end{eqnarray}
where $\Phi_{eff}^A =[1-2(\lambda_x +\lambda_y)] \Phi_{ext}$, 
$\Phi_{eff}^B =( 1-2 \lambda_x ) \Phi_{ext}$, and 
$\Phi_{eff}^C =( 1-2 \lambda_y ) \Phi_{ext}$ are the "effective" external fluxes.

Using the relations $\tau =\omega_{LC} t$, 
$\phi_{n,m}^k=\frac{\Phi_{n,m}^k}{\Phi_0}$,
and $\phi_{ext}=\frac{\Phi_{ext}}{\Phi_0}$, where 
$\omega_{LC} = {1}/{\sqrt{LC}}$ is the inductive-capacitive ($LC$) SQUID 
frequency, the dynamic equations for the fluxes through the SQUIDs can be 
written in the normalized form
\begin{eqnarray}
  \ddot{\phi}_{n,m}^A+ \gamma \dot{\phi}_{n,m}^A
   +\beta \sin\left( 2\pi \phi_{n,m}^A \right) +\phi_{n,m}^A 
  = \lambda_x \left( \phi_{n,m}^B +\phi_{n-1,m}^B \right) 
   +\lambda_y \left( \phi_{n,m}^C +\phi_{n,m-1}^C \right) 
   +[1-2(\lambda_x +\lambda_y)] \phi_{ext} ,
\nonumber \\
\label{Ch9.04}
  \ddot{\phi}_{n,m}^B +\gamma \dot{\phi}_{n,m}^B
   +\beta \sin\left( 2\pi \phi_{n,m}^B \right) +\phi_{n,m}^B 
  = \lambda_x \left( \phi_{n,m}^A +\phi_{n+1,m}^A \right) 
   +( 1-2 \lambda_x ) \phi_{ext} ,
\\
  \ddot{\phi}_{n,m}^C +\gamma \phi_{n,m}^C
   +\beta \sin\left( 2\pi \phi_{n,m}^C \right) +\phi_{n,m}^C 
  = \lambda_y \left( \phi_{n,m}^A +\phi_{n,m+1}^A \right) 
   +( 1-2 \lambda_y ) \phi_{ext} ,
\nonumber
\end{eqnarray}
where $\beta$ and $\gamma$ is the SQUID parameter and the loss coefficient, 
respectively, given by Eq. (\ref{Ch5.06.4}), and the overdots on $\phi_{n,m}^k$ 
denote differentiation with respect to the normalized temporal variable $\tau$.

In order to obtain the frequency spectrum of the SLiMM, we set $\gamma =0$ and 
$\phi_{ext} =0$ into Eqs. (\ref{Ch9.04}), and then we linearize them using the 
relation 
$\beta \, \sin \left( 2\pi \phi_{n,m}^k \right) \simeq \beta_L \, \phi_{n,m}^k$, 
where $\beta_L =2 \pi \beta$, and substitute the trial solution
\begin{equation}
\label{Ch9.05}
  \phi_{n,m}^k = {\cal F}_k \exp[i (\Omega \tau -\kappa_x n -\kappa_y m)],
\end{equation}
where $\kappa_x$ and $\kappa_y$ are the $x$ and $y$ components of the 
two-dimensional, normalized wavevector $\bf \kappa$, and 
$\Omega =\omega / \omega_{LC}$ is the normalized frequency. Then, the condition 
of vanishing determinant for the resulting algebraic system for the amplitudes 
${\cal F}_k$ gives
\begin{eqnarray}
\label{Ch9.06}
  \Omega_{\bf \kappa} =\Omega_{SQ}, \qquad
  \Omega_{\bf \kappa} =\sqrt{ \Omega_{SQ}^2 \pm 2 \sqrt{ 
                    \lambda_x^2 \cos^2 \left(\frac{\kappa_x}{2} \right)
                   +\lambda_y^2 \cos^2 \left(\frac{\kappa_y}{2} \right) } } ,  
\end{eqnarray} 
where only positive frequencies are considered. Eqs. (\ref{Ch9.06}) provide the 
{\em linear frequency spectrum of the SLiMM}. Thus, the frequency band 
structure, as it is shown in the right panel of Fig. \ref{fig9.01}, exhibits two 
dispersive bands forming a Dirac cone at the corners of the first Brillouin 
zone, and  a flat band crossing the Dirac points. Note that the flat-band 
frequency $\Omega_{FB}$ is equal to the resonance frequency of individual SQUIDs 
in the linear limit $\Omega_{SQ}$, i.e., $\Omega_{FB} =\Omega_{SQ}$. We also 
note that the flat band is an intrinsic property of this lattice in the 
nearest-neighbor coupling limit and thus it is not destroyed by any anisotropy 
(i.e., when $\lambda_x \neq \lambda_y$).

\subsection{From flat-Band to Nonlinear Localization}
\begin{figure*}[!t] 
\includegraphics[angle=0, width=0.9 \linewidth]{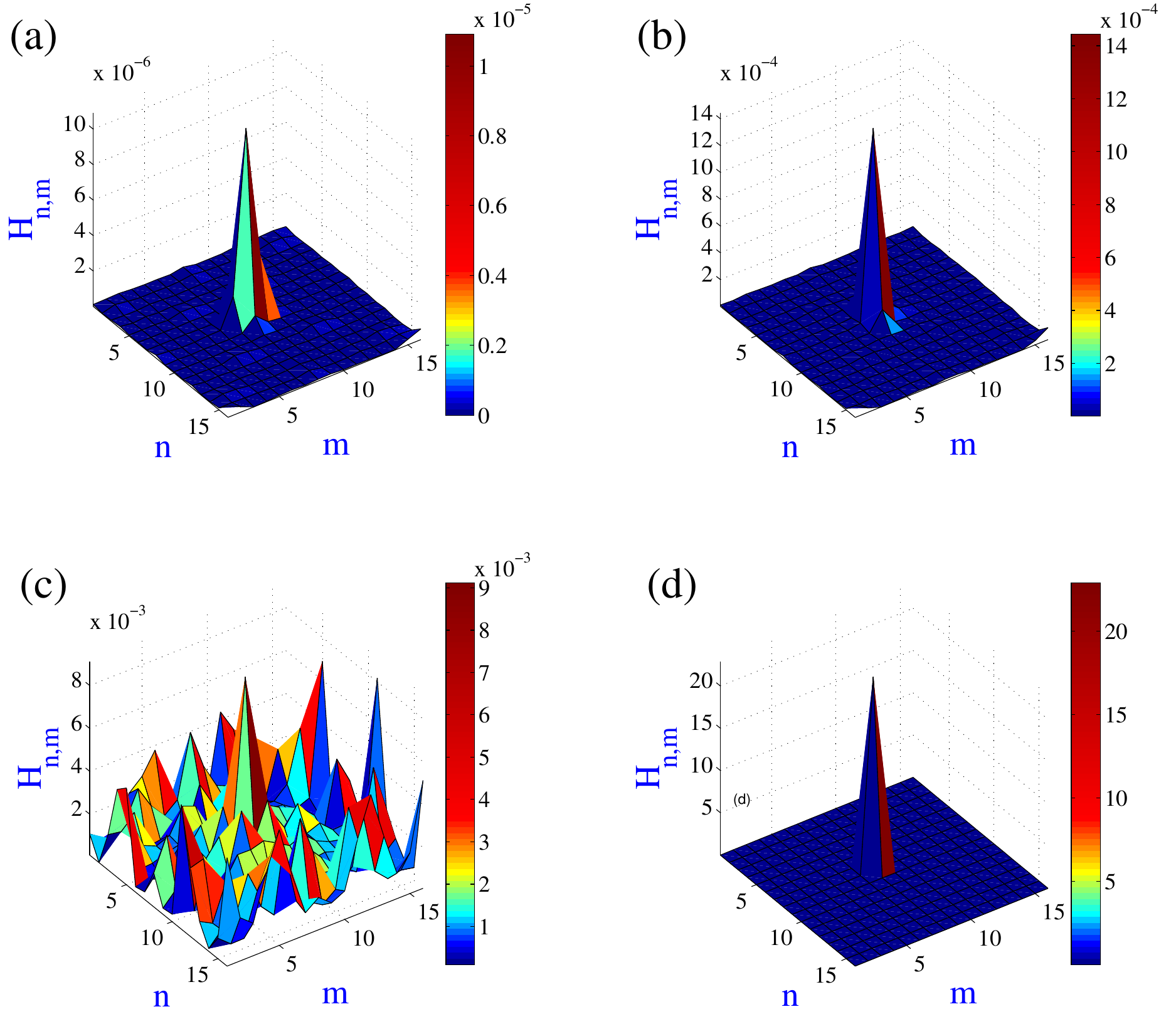} 
\caption{
 Energy density profiles $H_{n,m}$ plotted as a function of $n$ and $m$ at 
 $\tau =10^5~T_{SQ}$ time units for $N_x =N_y =16$, 
 $\lambda_x =\lambda_y =-0.02$, and $\beta_L =0.86$. An edge ($C$) SQUID is 
 initially excited with amplitude 
 (a) $A_m =0.001$; (b) $0.01$; (c) $0.1$; (d) $1$.
\label{fig9.02}
}
\end{figure*}

Eqs. (\ref{Ch9.04}) for the fluxes through the SQUIDs for the "free" SLiMM, i.e.,
that with $\gamma=0$ and $\phi_{ext} =0$, can be derived as the Hamilton's 
equations from the Hamiltonian
\begin{equation}
\label{Ch9.07}
  H =\sum_{n,m} H_{n,m} ,
\end{equation}
where the Hamiltonian density $H_{n,m}$, defined per unit cell, is given by 
\begin{eqnarray}
\label{Ch9.08}
  H_{n,m} =\sum_{k} \left\{ \frac{\pi}{\beta} \left[ 
           \left( q_{n,m}^k \right)^2 +\left( \phi_{n,m}^k \right)^2 \right]
  -\cos\left( 2\pi \phi_{n,m}^k \right) \right\}
\nonumber \\
  -\frac{\pi}{\beta} \{
   \lambda_x [  \phi_{n,m}^A \phi_{n-1,m}^B  +2 \phi_{n,m}^A \phi_{n,m}^B 
               +\phi_{n,m}^B \phi_{n+1,m}^A ]
   +\lambda_y [ \phi_{n,m}^A \phi_{n,m-1}^C 
               +2 \phi_{n,m}^A \phi_{n,m}^C +\phi_{n,m}^C \phi_{n,m+1}^A ] \} ,
\end{eqnarray}
where $q_{n,m}^k =\frac{d\phi_{n,m}^k }{d \tau}$ is the normalized instantaneous 
voltage across the Josephson junction of the SQUID in the $(n,m)$th unit cell of 
kind $k$. Both $H$ and $ H_{n,m}$ are normalized to the Josephson energy, $E_J$. 
The total energy $H$ remains constant in time. For the numerical integration of 
Eqs. (\ref{Ch9.04}) with $\gamma=0$ and $\phi_{ext} =0$, an algorithm that 
preserves the symplectic structure of a Hamiltonian system should be selected. 
In the present case, a second order symplectic St{\"o}rmer-Verlet scheme 
\cite{Hairer2003}, which preserves the total energy $H$ to a prescribed accuracy 
which is a function of the time-step $h$ can be safely used. {\em Periodic 
boundary conditions} are used for simplicity, and the SLiMM is initialized with 
a single-site excitation of the form 
\begin{eqnarray}
\label{Ch9.09}
  \phi_{n,m}^k (\tau=0) =\left\{ \begin{array}{ll}
         A_m,  & \mbox{if $n=n_e$ and $m=m_e$};    \\
         0,    & \mbox{otherwise ,} \end{array} \right.
 \qquad 
 \dot{\phi}_{n, m}^k (\tau=0) =0 , \mbox{for any $n$, $m$} , 
\end{eqnarray}
where $A_m$ is the amplitude, $n_e =N_x/2$, $m_e =N_y/2$, and $k=A$, $B$ or $C$. 
The magnitude of $A_m$ determines the strength of the nonlinear effects. Four 
profiles of the Hamiltonian (energy) density $H_{n,m}$ on the $n-m$ plane for 
$A_m$ spanning four orders of magnitude are shown in Fig. \ref{fig9.02}. For all 
those profiles the initially excited SQUID is of type $C$; the initialization 
with an excitation of a type $B$ SQUID gives the same results due to the 
isotropic coupling ($\lambda_x =\lambda_y$). Apparently, for both low and high 
$A_m$, those profiles exhibit localization; however, for an intermediate value
of $A_m$ the corresponding profile seems to be disordered and no localization 
takes place. Localization for low and high $A_m$ takes place due to different 
mechanisms; for low $A_m$, the SLiMM is in the linear regime in which the 
localized state is due to the flat-band, while for high $A_m$ the localized 
state is due to nonlinear effects. Importantly, no flat-band localization occurs 
for initially exciting an $A$ type SQUID, in agreement with the experiments on 
photonic Lieb lattices. 

\begin{figure}[!h]
\includegraphics[angle=0, width=0.7 \linewidth]{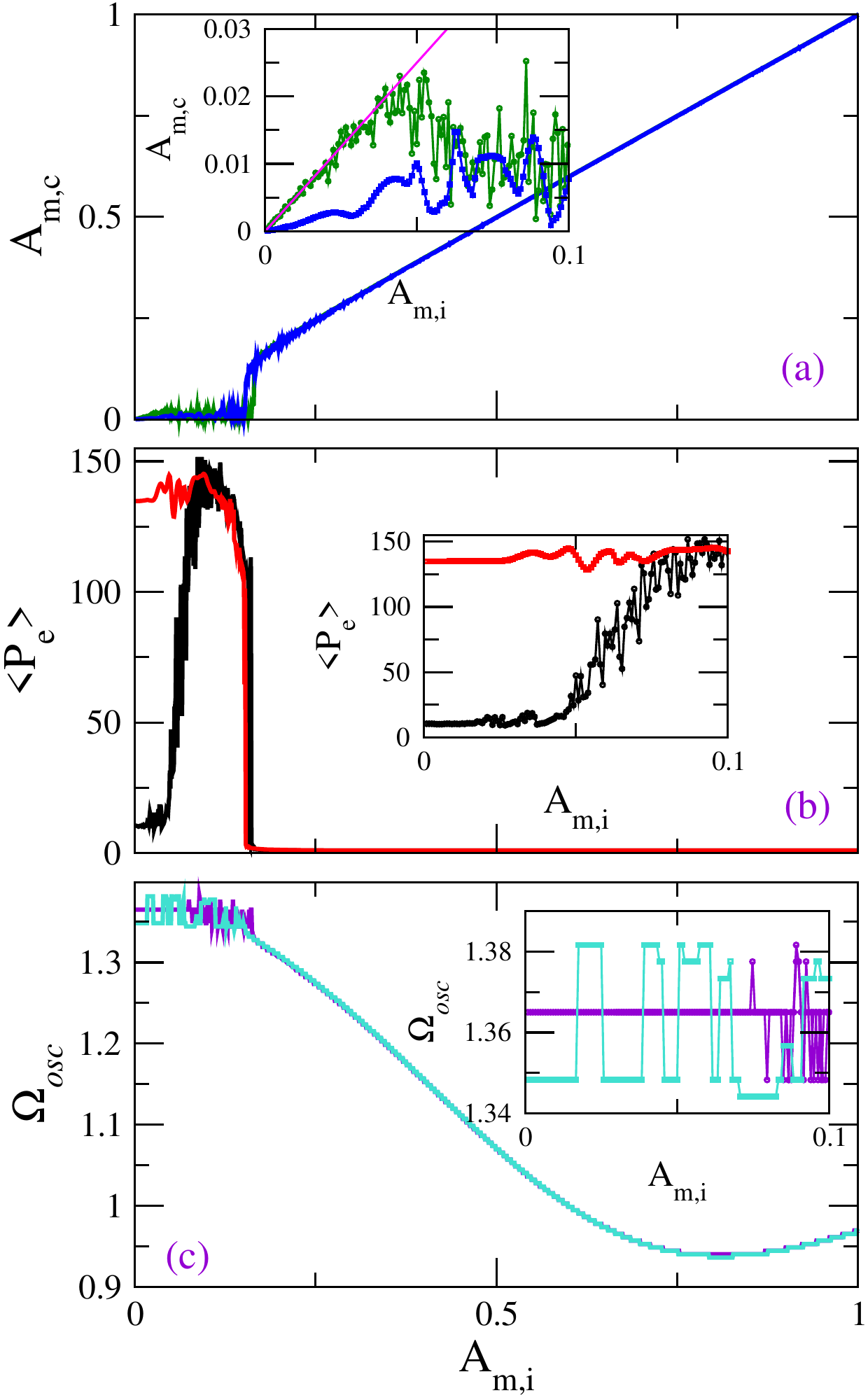} 
\caption{(
(a) The amplitude $A_{m,c}$ of the flux $\phi_{n_e,m_e}^k$ of the initially 
    excited SQUIDs with $k=A$ (blue curve) and $k=C$ (green curve) calculated at 
    the end of the integration time as a function of the initial excitation 
    amplitude $A_{m,i}$. 
    Inset: Enlargement around low $A_{m,i}$. The line $A_{m,i} /2$ is shown in 
    magenta color. 
(b) The energetic participation ratio averaged over the steady-state integration 
    time, $<P_e>$, 
    (transients were discarded) for the SQUID Lieb metamaterial when a corner 
    ($A$) SQUID (red curve) and an edge ($C$) SQUID (black curve) is initially 
    excited, as a function of the initial excitation amplitude $A_{m,i}$. 
(c) The oscillation frequency $\Omega_{osc}$ of the flux $\phi_{n_e,m_e}^k$ of 
    the initially excited $k=A$ (violet curve) and $k=C$ (turquoise curve) SQUID 
    as a function of the initial excitation amplitude $A_{m,i}$.
    Parameters:  
    $N_x =N_y =16$, $\lambda_x =\lambda_y =-0.02$, and $\beta_L =0.86$.
\label{fig9.03}
}
\end{figure}

In order to roughly determine the boundaries between these three regimes, i.e., 
the linear, the intermediate, and the nonlinear one, the {\em energetic 
participation ratio}
\begin{equation}
\label{Ch9.10}
   P_e =\frac{ 1}{ \sum_{n,m} \epsilon_{n,m}^2 } , \qquad
   \epsilon_{n,m} =\frac{H_{n,m}}{H}
\end{equation}
is used to quantify the degree of localization of the resulting flux states. 
Note that $P_e$ measures roughly the number of excited cells in the SLiMM; its 
values range from $P_e =1$ (strong localization, all the energy in a single cell) 
to $P_e =N$, with $N=N_x N_y$ (the total energy is equally shared between the 
$N$ SQUIDs). In Fig. \ref{fig9.03}, the equations for the SLiMM are integrated 
numerically in time (for $\gamma=0$ and $\phi_{ext} =0$) with an initial 
single-site excitations of amplitude $A_{m,i}$; the integration time is long 
enough for the transients to die out, and for the system to be in the 
steady-state for a substantial time-interval. Then, the amplitude of oscillation 
of the initially excited SQUID $A_{m,c}$ is retrieved, along with the frequency 
$\Omega_{osc}$ of that oscillation (the oscillation of the flux through the 
initially excited SQUID), and the $P_e$ averaged over the steady-state 
integration time (denoted as $<P_e>$). The initial amplitude $A_{m,i}$ is 
increased in small steps and the calculations are repeated. That procedure is 
performed for both a $C$ type and an $A$ type SQUID.

The results shown in Fig. \ref{fig9.03}, reveal clearly how the linear, 
intermediate, and nonlinear regimes are separated according to the values of 
$A_{m,i}$ for the set of parameters used in the calculations. As it can be 
observed in Fig. \ref{fig9.03}(a), $A_{m,c}$ remains low for low initial 
amplitudes $A_{m,i} < 0.15$, while for $A_{m,i} > 0.15$ the calculated amplitude 
$A_{m,c}$ increases linearly with increasing $A_{m,i}$ 
($A_{m,c} \simeq A_{m,i}$). The behavior for $A_{m,i} > 0.15$ is a result of the 
strong localization due to nonlinearities and it does not depend on which kind 
of SQUID (edge or corner, $A$ or $B, C$) is initially excited. However, a closer 
look to the two curves for $A_{m,i} < 0.15$ (inset), reveals significant 
differences, especially for $A_{m,i} < 0.05$. Here, the calculated amplitude 
$A_{m,c}$ for $k=C$ follows the relation $A_{m,c} \simeq A_{m,i}/2$, indicating 
localization due to the flat band. This conclusion is also supported by Figs.
\ref{fig9.03}(b) and \ref{fig9.03}(c). In Fig. \ref{fig9.03}(b), the energetic 
participation ratio averaged over the steady-state integration time $<P_e>$, for 
low values of $A_{m,i}$ attains very different values depending on which kind of 
SQUID is initially excited, $A$ or $C$ ($<P_e> \sim 140$ and $<P_e> \sim 10.5$, 
respectively). That large difference is due to delocalization in the former case 
and flat-band localization in the latter case. In the inset of Fig. 
\ref{fig9.03}(b), it can be observed that $<P_e>$ for an initially excited $C$ 
SQUID starts increasing for $A_{m,i} > 0.05$ indicating gradual degradation of 
flat-band localization and meets the $<P_e>$ curve for an initially excited $A$ 
SQUID at $A_{m,i} \sim 0.1$. In Fig. \ref{fig9.03}(c), for $A_{m,i} < 0.15$, the 
oscillation frequency  $\Omega_{osc}$ (either of an $A$ kind or a $C$ kind SQUID) 
is around that of the linear resonance frequency of a single SQUID, 
$\Omega_{SQ}$. As it can be seen in the inset of Fig. \ref{fig9.03}(c), when a 
$C$ SQUID is initially excited, then $\Omega_{osc} =\Omega_{SQ}$ for 
$A_{m,i} \lesssim 0.075$. However, when an $A$ SQUID is initially excited, the 
frequency $\Omega_{osc}$ jumps slightly above and below $\Omega_{SQ}$ 
irregularly, but it remains within the bandwidth of the linear frequency 
spectrum. For $A_{m,i} > 0.15$, the frequency $\Omega_{osc}$ decreases with 
increasing $A_{m,i}$, although it starts increasing again with increasing 
$A_{m,i}$ at $A_{m,i} \sim 0.8$. In this regime, nonlinear localized modes of 
the discrete breather type are formed, which frequency and its multitudes lie 
outside the linear frequency spectrum and depends on its amplitude, as it should 
be. Thus, for the parameter set used in these calculations, it can be inferred 
that flat-band localization occurs for initial amplitudes up to 
$A_{m,i} \simeq 0.05$ (linear regime), while delocalization occurs in the 
interval $0.05 < A_{m,i} < 0.15$ (intermediate regime). For larger $A_{m,i}$, 
strong nonlinear localization with $<P_e> \sim 1$ occurs (nonlinear regime). 

Remarkably, flat-band localization occurs only when an edge SQUID ($B$ or $C$) 
is initially excited. The excitation of a corner ($A$) SQUID does not lead to 
excitation of flat-band modes and thus such an initial state delocalizes rapidly. 
Note that the observed flat-band localization is not very strong as compared to 
the nonlinear localization because single-site excitations of a $B$ or $C$ SQUID 
do not correspond to exact localized flat-band eigenmodes.   
\newpage

\section{Quantum Superconducting Metamaterials}
\label{Quantum-S6}
\subsection{Introduction}
In the 1980's, A. Leggett envisioned the possibility of achieving quantum 
coherence in macroscopic circuits comprising Josephson junctions; since then, 
the macroscopic quantum effects which are present in low-capacitance Josephson 
junction circuits allowed for the realization of several kinds of 
superconducting, effectivelly two-level quantum systems capable of storing 
information in binary form, i.e., {\em superconducting quantum bits or qubits}. 
These solid-state qubit devices are currently at the heart of quantum 
information processing schemes, since they seem to satisfy the requirements for 
being the building blocks of viable quantum computers 
\cite{Devoret2013,Paraoanu2014,Georgescu2014}. Indeed, they found to exhibit 
relatively long coherence times, extremely low dissipation, and scalability 
\cite{Wendin2007}. Several variants of superconducting qubits which relay on the 
Josephson effect \cite{Josephson1962} and utilize either charge or flux or phase 
degrees of freedom have been proposed for implementing a working quantum 
computer; the recently anounced, commercially available quantum computer with 
more than $1000$ superconducting qubit CPU, known as D-Wave 2X$^{TM}$ (the 
upgrade of D-Wave Two$^{TM}$ with $512$ qubits CPU), is clearly a major 
advancement in this direction. A single superconducting charge qubit (SCQ) 
\cite{Pashkin2009} at milikelvin temperatures can be regarded under certain 
conditions as an artificial two-level "atom" in which two states, the ground and 
the first excited ones, are coherently superposed by Josephson coupling. When 
coupled to an electromagnetic (EM) vector potential, a single SCQ does behave, 
with respect to the scattering of EM waves, as an atom in space. This has been 
confirmed for a "single-atom laser" consisted of a superconducting charge qubit 
coupled to a transmission line resonator playing the role of a "cavity") 
\cite{Astafiev2007}. Thus, it would be anticipated that a periodic arrangements 
of such qubits would  demonstrate the properties of a transparent material, at 
least in a particular frequency band. The idea of building materials comprising 
artificial "atoms" with engineered properties, i.e., {\em metamaterials}, and in 
particular superconducting ones, is currently under active development. 
{\em Superconducting quantum metamaterials} (SCQMMs) comprising a large number 
of qubits could hopefully maintain quantum coherence for times long enough to 
reveal new, exotic collective properties. 

The first SCQMM which was implemented recently, comprises $20$ flux qubits 
arranged in a double chain geometry \cite{Macha2014}. Furthermore, lasing in the 
microwave range has been demonstrated theoretically to be triggered in an SCQMM 
initialized in an easily reachable factorized state \cite{Asai2015}. The 
considered system comprised a large number of SCQs which do not interact 
directly, placed in a one-dimensional (1D) superconducting waveguide. In this 
SCQMM, the lasing, i.e., a coherent transition of qubits to the lower-energy 
(ground) state was triggered by an initial field pulse traveling through the 
system. That type of dynamics is associated with the induced qubit-qubit 
coupling via their interaction with the EM field. The decoherence time of 
realistic SCQs as well as the relaxation times of 1D superconducting resonators 
and superconducting transmission lines (STLs) exceed significantly the 
characteristic times of energy transfer between the SCQs and the EM field, even 
for weak SCQ - EM field interaction. Thus, decoherence and leakage can be 
neglected \cite{Rakhmanov2008}. The lasing dynamics of that SCQMM is also 
accompanied by two peculiar phenomena, i.e., the appearance of higher harmonics 
of the EM field and the chaotization of SCQ subsystem dynamics. The lasing 
process in SCQMMs has been found to be quite robust against disorder arising 
from unavoidable variations of the parameters of the SCQs in the fabrication 
process \cite{Koppenhofer2016}. That disorder makes the level-splittings of the 
SCQs as well as other SCQMM parameters to vary locally in a random way. For the 
investigation of that process, a model Tavis-Cummings Hamiltonian was employed, 
which holds not only for real atoms but also for artificial ones, such as 
superconducting flux and charge qubits. Disordered SCQMMs comprising 
superconducting flux qubits with randomly varying excitation frequencies in a 
microwave resonator have been also investigated with a model Tavis-Cummings 
Hamiltonian which contained a qubit-qubit interaction term \cite{Shapiro2015}. 
It is demonstrated that photon phase-shift measurements allow to distinguish 
individual resonances in that flux qubit metamaterial with up to a hundred 
qubits. Numerical simulations of the phase-shift as a function of external flux 
(which modifies the qubit excitation energies), using exact diagonalization of 
the Hamiltonian in a single excitation basis, are in agreement with recent 
experimental results \cite{Macha2014}. Further theoretical studies have revealed 
the emergence of {\em collective quantum coherent phenomena} using an approach 
borrowed from mesoscopic physics \cite{Volkov2014}. It is demonstrated that the 
chain of $N$ qubits, incorporated into a low-dissipation resonant cavity, 
exhibits synchronized dynamics, even though the energy splittings $\Delta_i$ and 
thus the excitation frequencies $\omega_i =\Delta_i / \hbar$ are different from 
one qubit to the other. Those quantum coherent oscillations are characterized by 
two frequencies, 
$\omega_1 =\bar{\Delta} / \hbar$ and $\omega_2 =\tilde{\omega}_R$ where 
$\bar{\Delta} =(1/N) \sum_{i=1}^N \Delta_i$ with $\tilde{\omega}_R$ being the 
resonator frequency "dressed" by the interaction. In a similar SCQMM, i.e., a 
charge qubit array embedded in a low-dissipative resonator, various equilibrium 
photon states were investigated \cite{Iontsev2016}. When the photon energy of 
the resonator $\hbar \omega_0$ is much smaller than the energy splitting of 
qubits $\Delta$ (identical qubits have been considered), a second order phase 
transition is obtained in the state of photons. Specifically, at $T >T^\star$ 
(high temperatures, with $T^\star$ being the transition temperature) the photon 
state is incoherent. At $T < T^\star$ (low temperatures), however, coherent 
states of photons with two different polarizations occur in such a SCQMM. 
Interestingly, these two macroscopic coherent states of photons have equal 
energies, but they are separated by a barrier. Different photon states 
manifest themselves as resonant drops in the frequency-dependent transmission
coefficient $D( \omega)$; thus, incoherent and coherent photon states display a 
single drop and three drops, respectively, in their $D( \omega)$. Moreover, the 
resonant structure of $D( \omega)$ in the latter case provides direct evidence 
of macroscopic quantum oscillations between two different coherent states of 
photons. Quantum synchronization has been also demonstrated theoretically for a 
disordered SCQMM (in which the energy splitting of the $i-$th qubit is 
$\Delta_i$) comprising an array of flux qubits (3-Josephson junction SQUIDs) 
which is coupled to a transmission line \cite{Fistul2017}.

Also, remarkable quantum coherent optical phenomena, such as self-induced 
transparency \cite{McCall1967} (SIT) and Dicke-type superradiance 
\cite{Dicke1954} (collective spontaneous emission, SRD), occur during light-pulse 
propagation in SCQMMs comprising SCQs \cite{Ivic2016}. The occurence of the 
former or the latter effect solely depends on the initial state of the SCQ 
subsystem. Specifically, in self-induced transparency (superradiance) all the 
SCQs are initially in their ground (excited) state; such an extended system 
exhibiting SIT or SRD effects is often called a coherent amplifier or attenuator, 
respectively. These fundamental quantum coherent prosesses have been investigated 
extensively in connection to one- and two-photon resonant two-level systems.
It is demonstrated that SIT or SRD electomagnetic pulses propagating in the SCQMM 
induce to that quantum coherence in the form of {\em "population inversion"} 
pulses, which move together with the SIT or SRD EM pulses at the same speed. The 
experimental confirmation of such quantum coherence effects in SCQMMs may open a 
new pathway to potentially powerful quantum computing. Superradiant effects have 
been recently observed in quantum dot arrays \cite{Scheibner2007} and spin-orbit 
coupled Bose-Einstein condensates \cite{Hamner2014}. These findings suggest that 
these systems can radiatively interact over long distances.

\subsection{Superconducting Qubits}
In the past twenty years, impressive progress has been achieved both 
experimentally and theoretically in superconducting quantum bits (qubits), which 
comprise Josephson junctions. Those superconducting qubits have opened a new 
research area with many potential applications in quantum-information processing. 
The Josephson junctions \cite{Josephson1962}, which are equivalent to nonlinear 
inductors, provide strong nonlinearity to the superconducting qubits; this is a 
desired property for designing effectively two-level systems, as we discuss 
below. The superconducting qubits are essentially macro-mesoscopic devices which 
enter into the fully quantum regime at milli-Kelvin temperatures; such low 
temperatures are needed for the superconducting qubits to maintain their quantum 
states. In close analogy to natural atoms, the superconducting qubits have 
discrete energy levels and therefore can be regarded as {\em artificial atoms}. 
In contrast to natural atoms, however, their properties (e.g., their energy 
levels) as well as the coupling between them can be engineered and/or adjusted 
by external fields. Here, we briefly present the basics for superconducting 
qubits. A basic requirement for the superconducting qubits to function as 
artificial two-level systems (i.e., bits) is the nonlinearity, which 
differentiates the energy spacing between sequencial energy levels. As far as 
that spacing is concerned, the Josephson junctions play an important role as 
highly nonlinear elements. Moreover, Josephson junctions have negligibly small 
energy dissipation, which is yet another desired property for a superconducting 
qubit component. From the two celebrated Josephon relations discussed in Section 
$2.1$ (Eq. (\ref{Ch5.01})) it can be easily deduced that an ideal Josephson 
juction acts as a nonlinear inductance   
\begin{equation}
\label{Ch6.900}
  L_J =\frac{\hbar}{2 e I_c \cos\phi_J} ,
\end{equation}
whose value may even become negative. In Eq. (\ref{Ch6.900}), $\hbar$ is the 
Planck's constant devided by $2\pi$, $e$ is the eletron's charge, $I_c$ is the
critical current which characterizes the Josephson junction, and $\phi_J$ is the
gauge-invariant Josephon phase which has been discussed in Subsection $2.1$.

In an electrical equivalent circuit consideration of superconducting qubits, it 
is exactly that equivalent Josephson inductance $L_J$ that provides the desired 
nonlinearity. Thus, in a given superconducting qubit, the two lowest energy 
levels can be selected to form an effectively two-level system (a bit), 
appropriate for quantum information processing. There are three basic types of 
superconducting qubits comprising Josephson junctions, which are usually 
operating at frequencies in the microwave regime, which rely on different 
"degrees of freedom", i.e., either on charge, or flux, or phase. They are 
classified by the ratio of the Josephson energy to the charging (capacitive) 
energy 
\begin{equation}
\label{Ch6.901}
  \varepsilon_q =\frac{E_J}{E_C}, ~~~{\rm where}~~~ 
  E_J=\frac{\hbar I_c}{2 e}, ~~~ E_C =\frac{e^2}{2 C_J}. 
\end{equation}
In Eq. (\ref{Ch6.901}), $E_J$ and $E_C$ denote the Josephson and charging energy,
respectively, $C_J$ denotes the capacitance of the Josephson junction. The 
quantized superconducting qubits are described by the canonically conjugate 
variables $\phi_J$, i.e., the gauge-invariant Josephson phase, and the number of 
Cooper pairs $n$. Those variables  satisfy the commutation relation 
$[\phi_J, n] =i$ and obey the Heisenberg uncertainty principle 
$\Delta \phi_J \Delta n \geq 1$. It is important to note that the operator $n$ 
has integer eigenvalues whereas $\phi_J$ is an operator corresponding to the 
position of a point on the unit circle (an angle modulo $2 \pi$). For large 
enough systems with $n >> 1$, the number operator $n$ can be replaced by 
$-i \partial / \partial \phi_J$. The three basic types of superconducting qubits, 
i.e., phase, flux, and charge qubits are distinguished by the relations between 
the parameters $E_C$, $E_J$, and the energy difference between the two levels 
$\hbar \omega_0$, with $\omega_0 =\sqrt{ 2 E_C E_J } / \hbar$.
The three basic types of qubits have been described in great detail in several 
excellent review articles \cite{Wendin2003,Devoret2004,Zagoskin2007,Wendin2007,
Martinis2009}. Therefore, here we only briefly refer to them.

{\em Charge qubits.} The prototypical charge qubit (also called Cooper pair box) 
was the first to be described theoretically. Superconducting charge qubits (SCQs) 
are usually formed by small superconducting islands with $n$ Cooper pairs 
grounded through a Josephson junction. A gate voltage $V_g$ can be applied to 
that island through a (gate) capacitance $C_g$, in order to control the spacing 
between the energy-levels of the SCQ. Then, for a non-zero $V_g$, the charging 
energy is $E_C =\frac{e^2}{2 (C_J +C_g)}$, with $C_J$ being the capacitance of 
the Josephson junction, and $n_g =C_g V_g /(2e)$ is the gate-charge number. It 
can be shown that the Hamiltonian of that device is   
\begin{equation}
\label{Ch6.902}
   H =E_C \left( n -n_g \right)^2 -E_J \cos(\phi_J).
\end{equation} 
The eigenenergies and eigenfunctions of the Hamiltonian Eq. (\ref{Ch6.902}) can 
be calculated in terms of special functions which are known with arbitrary 
precision. Note that the eigenspectrum can be modified either by varying $n_g$ 
or $E_J$. Let us now limit ourselves to the two lowest levels of the box. Then, 
near the degeneracy point (optimal point) $n_g =1/2$, where the two charge 
states $|n=0>$ and $|n=1>$ (which differ by a single Cooper pair) have equal 
electrostatic energy, the Hamiltonian Eq. (\ref{Ch6.902}) can be reduced to 
\begin{equation}
\label{Ch6.903}
   H_q =-E_z \left( \sigma_z + X_c \sigma_x \right) ,
\end{equation} 
where $\sigma_x$ and $\sigma_z$ are the Pauli spin operators. The eigenstates of 
Hamiltonian (\ref{Ch6.903}) are coherent superpositions of the states $|n=0>$ 
and $|n=1>$, i.e., they are of the form $( |n=0> \pm |n=1> ) / \sqrt{2}$. In the 
limit $E_C >> E_J$, in which the charging behavior of the capacitance dominates, 
we have that $E_z =E_J /2$ and $X_c =2 (E_C / E_J ) [ (1/2)−n_g ]$. The main 
disadvantage of the charge qubit is its very strong sensitivity to charge noise, 
which can be mitigated to some extent by operating the qubit in the intermediate 
regime $E_J \lesssim E_C$.

{\em Phase qubits.-}
Phase qubits operate in the "phase regime" in which the Josephson term dominates 
the Hamiltonian (\ref{Ch6.902}), i.e., when $E_C \lesssim E_J$. They consist of 
a single Josephson junction which is biased by an external current $I_b$. The 
Hamiltonian of the superconducting phase qubit can be written as
\begin{equation}
\label{Ch6.904}
   H =-E_C \partial_{\phi_J}^2 -E_J \cos(\phi_J) -\frac{I_b \Phi_0}{2\pi} \phi_J 
     \equiv -E_C \partial_{\phi_J}^2 -E_J \left( \cos(\phi_J) +\frac{I_b}{I_c} \right) ,
\end{equation} 
which is the Hamiltonian of a quantum particle in a tilted washboard potential. 
The phase qubit operates typically in the subcritical regime (in practice, when 
$I_b \simeq 0.95 I_c - 0.98 I_c$), so that only a few quantized levels remain in 
each local minimum of the Josephson potential 
$U_J =-E_J \left( \cos(\phi_J) +\frac{I_c}{I_b} \right)$. The tunneling 
probability out of the lowest two levels is very small, and thus these can be 
taken as qubit states $|0>$ and $|1>$. For $I_b \simeq I_c$ we have that 
$\phi_J \simeq \pi/2$ and the Josephson potential can be approximated by 
\begin{equation}
\label{Ch6.905}
  U_J =E_J \left[ \left( 1- \frac{I_c}{I_b} \right) \phi_J -\frac{1}{6} \phi_J^3 \right].
\end{equation} 
The classical oscillation frequency at the bottom of the well (so-called plasma 
oscillation) is given by
\begin{equation}
\label{Ch6.906}
  \omega_p =\omega_0 \left[ \left( 1- \frac{I_b}{I_c} \right)^2 \right]^{1/4} .
\end{equation} 
Quantum-mechanically, energy levels can be found for the potential in Eq. 
(\ref{Ch6.905} with non-degenerate spacings. The first two levels, which have a 
transition frequency $\omega_{01} \simeq 0.95 \omega_p$, can be used for qubit 
states. In practive, $\omega_{01} / (2 \pi)$ falls in the $5-20 ~GHz$ range. 
Defining $\Delta I \equiv I_b -Ic$, the phase qubit Hamiltonian is given by
\begin{equation}
\label{Ch6.907}
  H_q =\frac{\hbar \omega_{01}}{2} \sigma_z +\sqrt{ \frac{\hbar}{2\omega_{01} C} }
          \Delta I \left( \sigma_x +\chi \sigma_z \right) ,
\end{equation} 
where $\chi =\sqrt{ \frac{\hbar \omega_{01}}{3 \Delta U} } \simeq 1/4$ for typical 
operating parameters, with 
$\Delta U =\frac{2\sqrt{2}}{3} I_c \left( 1- \frac{I_c}{I_b} \right)^{3/2}$.

{\em Flux qubits.-}
Another possibility to realize a qubit in the limit $E_J >> E_C$ is to take 
advantage of the degeneracy between two current-carrying states of an rf SQUID. 
The Hamiltonian for this system can be written as 
\begin{equation}
\label{Ch6.908}
  H =-E_C \partial_{\phi_J}^2 -E_J \cos(\phi_J) 
     +\frac{1}{2} E_L \left( \phi_J -\phi_{ext} \right)^2 ,
\end{equation} 
where $\phi_{ext} =2\pi \frac{\Phi_{ext}}{\Phi_0}$ is the reduced flux through 
the loop of the rf SQUID due to an external magnetic field, and 
$E_L =\frac{\Phi_0^2}{2 \pi L}$ is the inductive energy due to the 
self-inductance $L$ of the rf SQUID. The potential energy is in this case formed 
by the last two terms in the Hamiltonian Eq. (\ref{Ch6.908}). For 
$\phi_{ext} \simeq \pi$, the potential of the rf SQUID has two almost degenerate 
minimums. The lowest energy states correspond to a persistent current 
circulating in the loop of the rf SQUID in opposite directions, and they can be 
conveniently used as the the $|0>$ and $|1>$ states of the flux qubit. Tunneling 
between the two potential wells is enabled by the first term in Eq. 
(\ref{Ch6.908}). Then, in the $\{ |0>, |1> \}$ subspace, the effective flux 
qubit Hamiltonian is 
\begin{equation}
\label{Ch6.909}
  H =-\frac{\varepsilon}{2} \sigma_z + \frac{\Delta}{2} \sigma_x ,
\end{equation} 
where $\varepsilon \propto <1|H_q|1> -<0|H_q|0>$ is the energy bias between the 
two levels (level splitting or "gap"), and $\Delta \propto <0|H_q|1>$ is the 
tunneling amplitude. For $\phi_{ext} =\pi$, i.e., exactly on the degeneracy 
point, we have that $\varepsilon =0$ and the eigenstates of the Hamiltonian 
Eq. (\ref{Ch6.909}) are again of the form 
$\left( |0> \pm |1> \right) / \sqrt{2}$.    

\subsection{Self-Induced Transparency, Superradiance, and Induced Quantum Coherence}
\subsubsection{Description of the Model System}
\begin{figure}[!t]
   \centering
   \includegraphics[width=0.78 \linewidth]{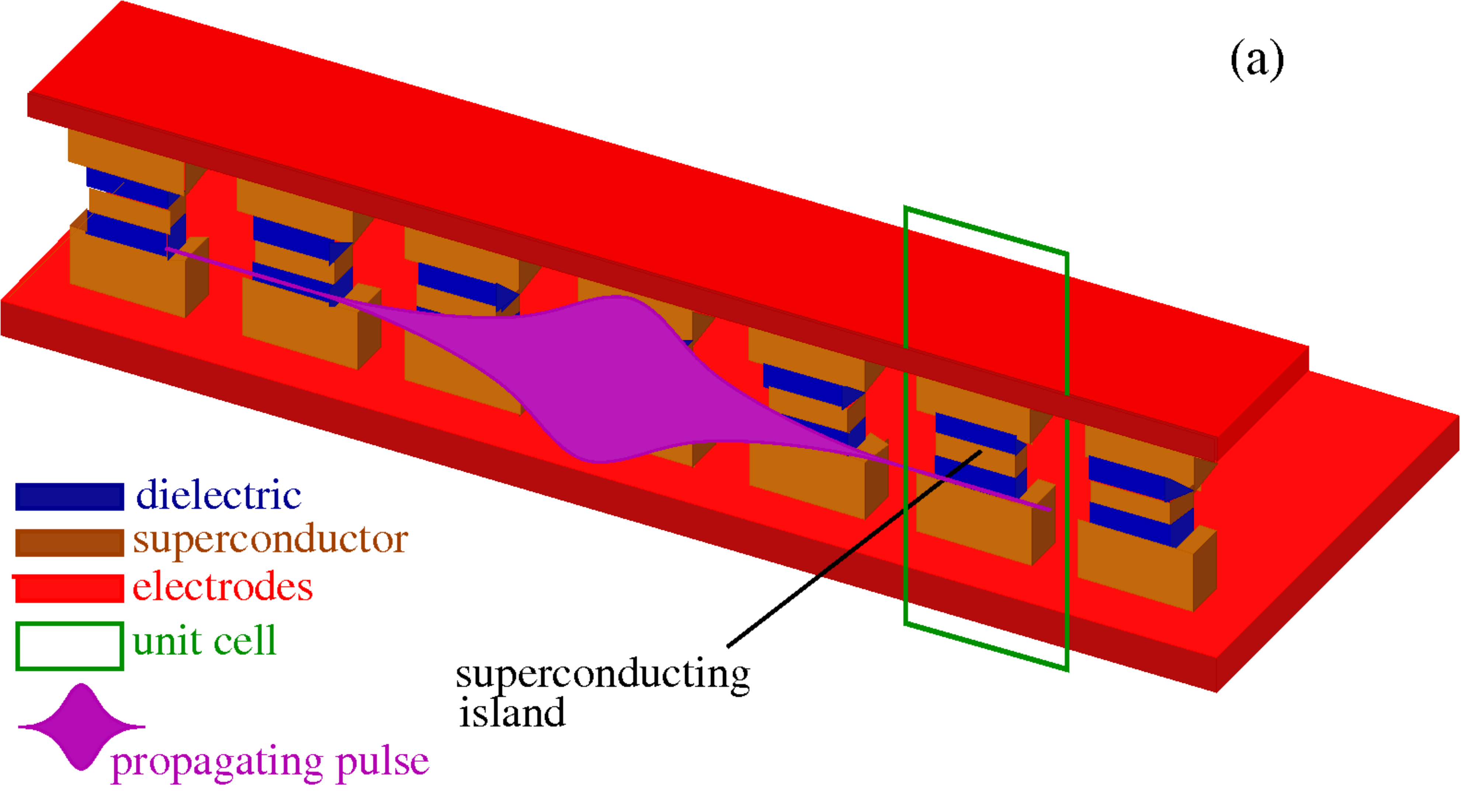}  
   \includegraphics[width=0.78 \linewidth]{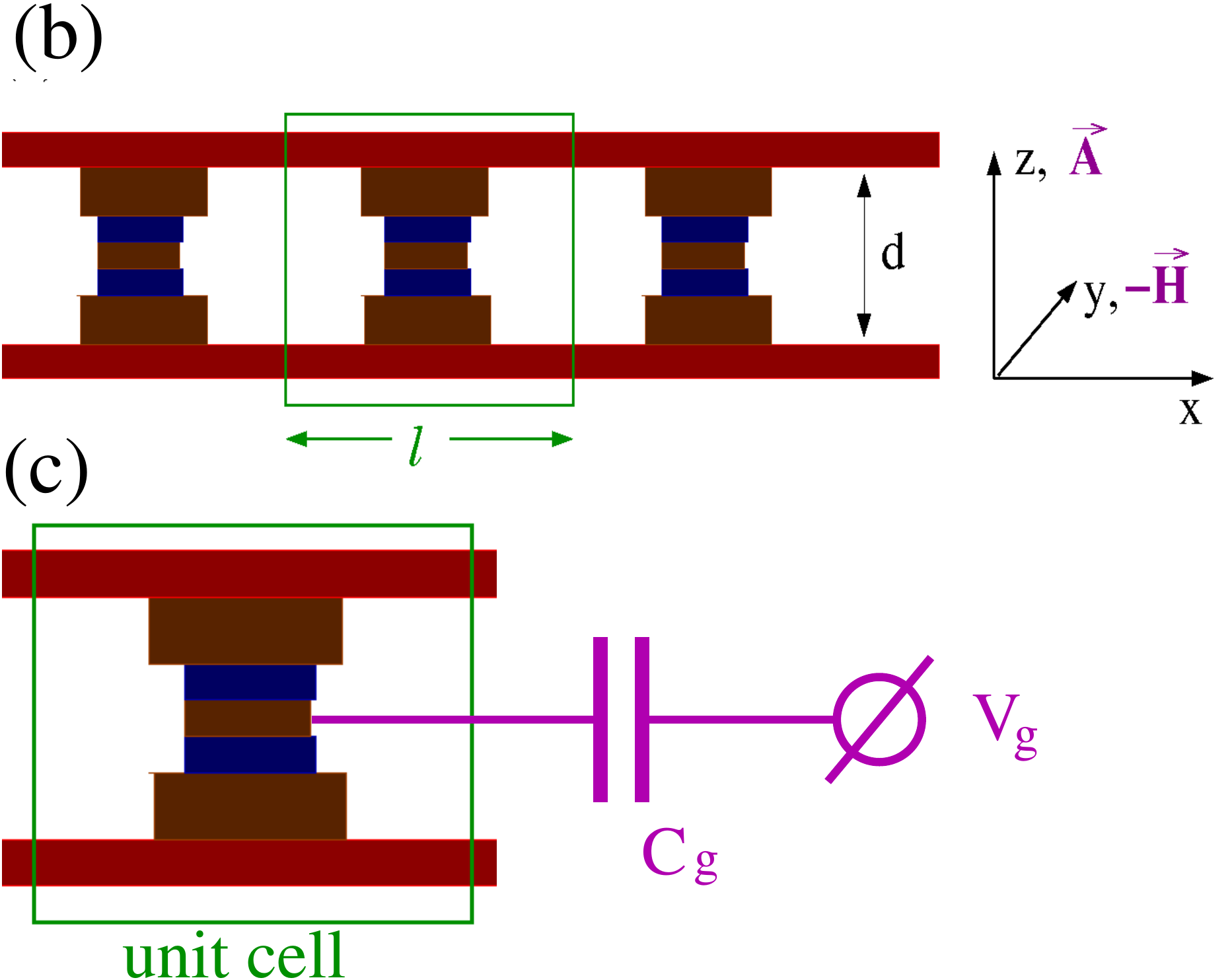}
\caption{
 Schematic drawing of a superconducting quantum metamaterial comprising charge 
 qubits.
(a) An infinite chain of identical charge qubits periodically loaded in a 
    superconducting transmission line (STL). Each qubit is a tiny superconducting 
    island that is connected to the electrodes of the STL through two Josephson 
    junctions, formed in the regions of the insulating (dielectric) layers 
    (blue). The propagating electromagnetic vector potential pulse is also shown 
    schematically and out of scale.
(b) Side view of the SCQMM in which the relevant geometrical parameters and the 
    field orientations are indicated.
(c) A single unit cell of the SCQMM, in which the control circuitry for each 
    individual charge qubit is shown. The gate potential $V_g$ is applied to a 
    superconducting charge qubit through the gate capacitor $C_g$. 
}
\label{fig6.01}
\end{figure}

Consider an infinite, one-dimensional (1D) periodic array comprising 
Superconducting Charge Qubits (SCQs). That array is placed in a Superconducting 
Transmission Line (STL) consisting of two superconducting strips of infinite 
length \cite{Rakhmanov2008,Shvetsov2013}, as shown in Fig. \ref{fig6.01}(a); 
each of the SCQs, in the form of a tiny superconducting island, is connected to 
each bank of the STL by a Josephson junction (JJ). Control circuitry can be 
added to that structure, so that each individual SCQ is coupled to a gate 
voltage source $V_g$ through a gate capacitor $C_g$ (Fig. \ref{fig6.01}(c). Thus, 
local control of the SCQMM can be achieved by altering independently the state 
of each SCQ \cite{Zagoskin2011}. The SCQs exploit the nonlinearity of the 
Josephson effect \cite{Josephson1962} and the large charging energy resulting 
from nanofabrication to create artificial mesoscopic two-level systems. A 
propagating EM field in the STL gives rise to indirect interactions between the 
SCQs, which are mediated by its photons \cite{vanLoo2013}. Those interactions 
are of fundamental importance in quantum optics, quantum simulations, and 
quantum information processing, as well. Since the qubits can be in a coherent 
superposition of quantum states, such a system demonstrates interesting effects, 
e.g., it may behave as a "breathing" photonic crystal with an oscillating band 
gap \cite{Rakhmanov2008}. That gap depends on the quantum state of the qubits, 
that makes this system a {\em quantum photonic crystal}. Thus, a variation of 
the microscopic quantum state of the qubits will change the macroscopic EM
response of the system. The key ingredient of these effects is that the optical 
properties of the Josephson transmission line are controlled by the quantum 
coherent state of the qubits. The progress on the emerging field of microwave 
photonics with superconducting quantum circuits has been reviewed in Ref. 
\cite{XiuGu2017}. Below we discuss two remarkable quantum coherent optical 
phenomena, i.e., self-induced transparency and Dicke-type superradiance, 
which may occur during light-pulse propagation in the above mentioned SCQMM. 
Moreover, it appears that the propagating pulses induce quantum coherence in the 
chain of SCQs, in the form of "population inversion" pulses.

In the following, the essential building blocks of the SCQMM model are 
summarized in a self-contained manner, yet omitting unnecessary calculational 
details which are presented in the Appendix. The system of units in Refs. 
\cite{Rakhmanov2008,Shvetsov2013} is used from here to the end of the 
article. An EM vector potential pulse $\vec{A}=A_z (x,t) \hat{z}$ can propagate 
in the SCQMM structure, which extends over the $x-$direction, and couples to the 
SCQs. Then, the energy per unit cell of the SCQMM - EM vector potential pulse 
can be readily written in units of the Josephson energy 
$E_J=\frac{\Phi_0 I_c}{2\pi C}$, with $\Phi_0$, $I_c$ and $C$ being the magnetic 
flux quantum, the critical current of the JJ, and the capacitance of the JJ, 
respectively, as \cite{Rakhmanov2008,Shvetsov2013}
\begin{eqnarray}
\label{Ch6.01}
   H= \sum_n \left\{ \left[ \dot\varphi^2_n-2\cos \varphi_n \right]
            +\left[ \dot\alpha^2_n+\beta^2(\alpha_{n+1}-\alpha_n)^2 \right] 
            +\left[ 2\cos \varphi_n (1-\cos \alpha_n) \right] \right\},
\end{eqnarray}
where $\varphi_n$ is the superconducting phase on the $n$th island, 
$\alpha_n =2\pi d A_{x,n} /\Phi_0$ is the normalized and discretized EM vector 
potential in the center of the $n$th unit cell, with $d$ being the separation 
between the electrodes of the STL, $\beta =\frac{1}{\sqrt{8\pi l d E_J}} 
\frac{\Phi_0}{2\pi}$, 
and the overdots denote differentiation with respect to the temporal variable 
$t$. The three terms in the square brackets in Eq. (\ref{Ch6.01}) represent the 
energy of the SCQs (equivalently the energy of the two JJs in an EM vector 
potential having the orientation shown in Fig. \ref{fig6.01}(b)), the energy of 
the EM field, and their interaction energy, respectively.

For the discretization of the EM vector potential, it is assumed that the 
wavelength $\lambda$ of the carrier EM field with vector potential $A_z (x,t)$ 
is much larger that all the other characteristic lengths of the SCQMM structure, 
i.e., that $\lambda >> \ell, d$, with $\ell$ being the distance between 
neighboring SCQs. Then, the EM potential can be regarded to be approximately 
constant within a unit cell, so that in the center of the $n$th unit cell 
$A_z (x,t) \simeq A_{z,n} (t)$ or equivalently 
$\alpha (x,t) \simeq \alpha_n (t)$. Note that the coupling between the SCQs and 
the EM field is realized from the requirement of having gauge-invariant 
Josephson phase in each of the junctions.

\subsubsection{Second Quantization and Reduction to Maxwell-Bloch Equations}
The quantization of the SCQ subsystem requires the replacement of the classical 
variables $\varphi_n$ and $\dot\varphi_n$ by the corresponding quantum operators 
$\hat{\varphi}_n$ and $-i({\partial}/{\partial \hat{\varphi}_n})$, respectively. 
While the EM field is treated classically, the SCQs are regarded as two-level 
systems, so that only the two lowest energy states are retained; under these 
considerations, the second-quantized Hamiltonian corresponding to Eq. 
(\ref{Ch6.01}) in the semi-classical approximation is
\begin{eqnarray}
\label{Ch6.02}
  H=\sum_n \sum_{p} E_{p}(n) a^{\dagger}_{n,p} a_{n,p} 
   +\sum_n \left[ \dot\alpha^2_n+\beta^2(\alpha_{n+1}-\alpha_n)^2 \right] 
   +4\sum_n \sum_{p,p'} 
    V_{p,p'}(n) a^{\dagger}_{n,p} a_{n,p'} \sin^2\frac{\alpha_n}{2},
\end{eqnarray}
where $p, p'=0, 1$, $E_0$ and $E_1$ are the energy eigenvalues of the ground and 
the excited state, respectively, the operator $a^{\dagger}_{n,p}$ ($a_{n,p}$) 
excites (de-excites) the $n$th SCQ from the ground to the excited (from the 
excited to the ground) state, and 
$V_{p,p'}=\int d\varphi \Xi^*_p(\varphi) \cos \varphi \Xi_p(\varphi)$ are the 
matrix elements of the effective SCQ - EM field interaction. The basis states 
$\Xi_{p}$ can be obtained by solving the single-SCQ Schr{\"o}dinger equation
$(-{\partial^2}/{\partial \varphi^2} -E_p +2\cos\varphi ) \Xi_p=0$.
In general, each SCQ is in a superposition state of the form
$|\Psi_n\rangle=\sum_p \Psi_{n,p}(t) a^{\dagger}_{n,p} |0\rangle$. The 
substitution of $|\Psi_n\rangle$ into the Schr{\"o}dinger equation with the 
second-quantized Hamiltonian Eq. (\ref{Ch6.02}), and the introduction of the 
Bloch variables
\begin{equation}
\label{Ch6.022}
   R_x (n)=\Psi^\star_{n,1} \Psi_{n,0} +\Psi^\star_{n,0} \Psi_{n,1},
   \qquad
   R_y (n)=i(\Psi^\star_{n,0} \Psi_{n,1} -\Psi^\star_{n,1} \Psi_{n,0}),
   \qquad
   R_z (n)=|\Psi_{n,1}|^2 -|\Psi_{n,0}|^2,
\end{equation}
provides the re-formulation of the problem into the Maxwell-Bloch equations
\begin{eqnarray}
\label{Ch6.03}
  \dot{R}_x(n)=-\left[ \Delta +8 D \sin^2 \frac{\alpha_n}{2} \right] R_y (n), 
 ~ 
  \dot{R}_y(n)=\left[ \Delta +8 D \sin^2 \frac{\alpha_n}{2} \right] R_x (n)
                                     -8\mu \sin^2 \frac{\alpha_n}{2} R_z (n) , 
~
  \dot{R}_z(n)=+8\mu \sin^2 \frac{\alpha_n}{2} R_y (n) , 
\end{eqnarray}
that are {\em nonlinearly coupled} to the resulting equation for the normalized 
EM vector potential
\begin{equation}
\label{Ch6.04}
  \ddot\alpha_n 
   +\left\{ \Omega^2 + \chi \left[ \mu R_x (n) +D R_z (n) \right] \right\} 
     \sin \alpha_n =\beta^2 \delta a_n ,
\end{equation}
where 
\begin{equation}
\label{Ch6.042}
   \delta \alpha_n =\alpha_{n-1} -2\alpha_n +\alpha_{n+1}, ~~
   D=\frac{(V_{11} -V_{00})}{2\chi}, ~~
   \Omega^2 =\frac{( V_{00} +V_{11} )}{2}, ~~ 
   \mu=\frac{V_{10}}{\chi} =\frac{V_{01}}{\chi}, ~~
   \Delta=\epsilon_1 -\epsilon_0 \equiv \frac{(E_1-E_0)}{\chi} ,
\end{equation}
with $\chi=\hbar \frac{\omega_J}{E_J}$. In Eqs. (\ref{Ch6.03}) and 
(\ref{Ch6.04}), the overdots denote differentiation with respect to the 
normalized time $t \rightarrow \omega_J t$, in which 
$\omega_J = \frac{e I_c}{\hbar C}$ is the Josephson frequency and $e$, $\hbar$ 
are the electron charge and the Planck's constant devided by $2\pi$, 
respectively.

\subsubsection{Approximations and Analytical Solutions}
For weak EM fields, $|\alpha_n| \ll 1$ for any $n$, the sine term can be 
linearized as $\sin\alpha_n \simeq \alpha_n$. Then, by taking the continuum 
limit of Eqs. (\ref{Ch6.03}) and (\ref{Ch6.04}), using the relations 
$\alpha_n (t) \rightarrow \alpha(x,t)$ and 
$R_i (n; t) \rightarrow R_i (x; t)$ $(i=x,y,z)$, a set of simplified, yet still 
nonlinearly coupled equations is obtained, similar to those encountered in 
{\em two-photon self-induced transparency} (SIT) in resonant media 
\cite{Belenov1969}. Further simplification can be achieved with the slowly 
varying envelope approximation (SVEA), using the ansatz 
$\alpha(x,t)=\varepsilon(x,t)\cos \Psi(x,t)$ for the EM vector potential, where 
$\Psi(x,t)=k x-\omega t+\phi(x,t)$ and $\varepsilon(x,t)$, $\phi(x,t)$ are the 
slowly varying pulse envelope and phase, respectively, with $\omega$ and $k$
being the frequency of the carrier wave of the EM pulse and its wavenumber in 
the STL, respectively. The dispersion relation (see the Appendix for the 
derivation)
\begin{equation}
\label{Ch6.0444}
   k =\pm \frac{ \sqrt{\omega^2 -\Omega^2} }{\beta},
\end{equation}
provides the dependence of $k$ on $\omega$ or vice versa. In the absence of 
the SCQ chain, the EM pulse propagates in the STL with speed $\beta$. At the 
same time, Eqs. (\ref{Ch6.03}) for the Bloch vector components are transformed 
according to
\begin{equation}
\label{Ch6.043}
   R_x = r_x \cos (2\Psi) +r_y \sin(2\Psi), \qquad
   R_y= r_y\cos(2\Psi) -r_x \sin(2\Psi), \qquad
   R_z =r_z .
\end{equation}
Then, collecting the coefficients of $\sin\Psi$ and $\cos\Psi$ while neglecting 
the rapidly varying terms, and averaging over the phase $\Psi$, results in a set 
of truncated equations. Further manipulation of the resulting equations and the 
enforcement of the {\em two-photon resonance condition} $\Delta =2 \omega$, 
results in
\begin{eqnarray}
\label{Ch6.05}
  \dot{\varepsilon} +c \varepsilon_x =-\chi \frac{\mu}{\Delta} \varepsilon r_y ,
  \qquad 
  \dot{\phi} +c \phi_x =-\chi \frac{2 D}{\Delta} r_z ,
\end{eqnarray}
where $c=\frac{\beta^2 k}{\omega} =2 \frac{\beta^2 k}{\Delta}$, and the 
truncated Maxwell-Bloch equations
\begin{eqnarray}
\label{Ch6.052}
   \dot{r}_x=-2 D \varepsilon^2 r_y, 
   \qquad
   \dot{r}_y=+2 D \varepsilon^2 r_x -\frac{\mu\varepsilon^2}{2} R_z,
   \qquad
   \dot{r}_z=+\frac{\mu\varepsilon^2}{2} r_y,
\end{eqnarray}
in which the $n-$dependence of the $r_i$ $(i=x,y,z)$ is suppressed, in 
accordance with common practices in quantum optics. Also, from Eqs. 
(\ref{Ch6.052}), the conservation law $r_x^2 +r_y^2 +r_z^2 =1$ can be obtained. 

The $r_i$ can be written in terms of new Bloch vector components $S_i$ using the 
unitary transformation 
\begin{eqnarray}
\label{Ch6.053}
   r_x =S_x \cos\Phi -S_z \sin\Phi, \qquad r_y =S_y, \qquad 
   r_z =S_z \cos\Phi +S_x \sin\Phi,
\end{eqnarray}
where $\Phi$ is a constant angle which will be determined later. Using a 
procedure similar to that for obtaining the $r_i$, we get
\begin{eqnarray}
\label{Ch6.054}
   \dot S_x =0, \qquad \dot S_y =-\frac{1}{2} W \varepsilon^2 S_z, \qquad 
   \dot S_z =+\frac{1}{2} W \varepsilon^2 S_y,
\end{eqnarray}
where $W=\sqrt{(4D)^2 +\mu^2}$ and $\tan \Phi \equiv \gamma =\frac{4 D}{\mu}$.
The combined system of Eqs. (\ref{Ch6.054}) and  (\ref{Ch6.05}) admits exact 
solutions of the form $\varepsilon=\varepsilon(\tau=t-x/v)$ and 
$S_i =S_i (\tau=t-x/v)$, where $v$ is the pulse speed. For the slowly varying 
pulse envelop, we obtain
\begin{eqnarray}
\label{Ch6.07}
  \varepsilon (\tau) =\varepsilon_0 
   \left[ 1 +\left( \frac{\tau -\tau_0}{\tau_p} \right)^2 \right]^{-\frac{1}{2}} ,    
\end{eqnarray}
where the pulse amplitude an its duration are given respectively by
\begin{eqnarray}
\label{Ch6.072}
   \varepsilon_0 =\sqrt{ \frac{8\sigma^2}{\Delta} \frac{v}{(c-v)} },
   \qquad
   \tau_p =\left\{ 2\chi \frac{\sigma \mu}{\Delta} \frac{v}{(c-v)} \right\}^{-1} ,
\end{eqnarray}
with $\sigma =\frac{\mu}{W} =\frac{1}{\sqrt{1+\gamma^2}}$. The decoherence 
factor $\gamma$ can be expressed as a function of the matrix elements of the 
effective interaction between the SCQ subsystem and the EM field, $V_{ij}$, as 
$\gamma=2 \frac{(V_{11} -V_{00})}{V_{10}}$ that can be calculated when the 
latter are known. Lorentzian propagating pulses of the form of Eq. 
(\ref{Ch6.07}) have been obtained before in two-photon resonant media
\cite{Tan-no1975a,Nayfeh1978}; however, SIT in quantum systems has only been 
demonstrated in one-photon (absorbing) frequency gap media, in which solitonic 
pulses can propagate without dissipation \cite{John1999}. The corresponding 
solution for the population inversion, $R_z$, reads
\begin{equation}
\label{Ch6.073}
  R_z (\tau) =\pm\left[-1 
     +\left( \frac{\varepsilon (\tau)}{\varepsilon_M} \right)^2 \right] ,
\end{equation}
where $\varepsilon_M =\sqrt{ \frac{8}{\Delta} \frac{v}{(c-v)} }$, and the plus 
(minus) sign corresponds to absorbing (amplifying) SCQMMs; these are specified 
through the initial conditions as
\begin{equation}
\label{Ch6.074}
   R_x (-\infty) =R_y (-\infty)=0, \qquad R_z(-\infty)=-1, \qquad 
   \varepsilon(-\infty)=0,
\end{equation}
for absorbing SCQMMs, and 
\begin{equation}
\label{Ch6.075}
   R_x (-\infty) =R_y (-\infty)=0, \qquad R_z(-\infty)=+1, \qquad 
   \varepsilon(-\infty)=0,
\end{equation}
for amplifying SCQMMs. The initial conditions specified by Eqs. (\ref{Ch6.074}) 
(Eqs. (\ref{Ch6.075})) ensure that before the arrival of the EM pulse, all the 
SCQs are in their ground (excited) state in order to achieve absorption 
(amplification). Since the frequency $\omega$ has been chosen to match the 
two-photon resonance, the wavevector $k$ has been uniquely determined through
the dispersion relation Eq. (\ref{Ch6.0444}) to be 
$k =\pm \frac{\sqrt{\Delta^2 -4 \Omega^2}}{2 \beta}$, or, after replacements of
$\Omega$ and $\Delta$ from Eq. (\ref{Ch6.042}),
\begin{equation}
\label{Ch6.0777}
  k = k_r =\pm 
   \frac{\sqrt{ \left( \frac{E_1 -E_0}{\chi} \right)^2 -4 \frac{V_{00} +V_{11}}{2}}}
        {2 \beta}
\end{equation}
Obviously, the propagation of EM pulses in the SCQMM is only possible for real
$k$. The requirement for the wavenumber $k$ to be real results in the relation 
\begin{equation}
\label{Ch6.076}
    2 \chi^2 (V_{11}+V_{00}) < (E_1 -E_0)^2,
\end{equation}
which provides a necessary condition for pulse propagation in the SCQMM. Note
that the {\em propagation condition} Eq. (\ref{Ch6.076}) contains only 
qubit-related parameters, i.e., the diagonal matrix elements $V_{00}$, $V_{11}$
and the energy levels $E_0$ and $E_1$, whose values can be in principle tailored 
during fabrication or tuned in real time by the gate voltages $V_g$. 

The corresponding velocity-amplitude relation of a propagating pulse in the 
SCQMM under the two-photon resonance condition reads
\begin{eqnarray}
\label{Ch6.08}
   v= c \left[1 \pm { \chi \frac{8 \sigma^2}{\Delta \varepsilon_0^2}} \right]^{-1}, 
\end{eqnarray}
where the value of $c$ is that at the two-photon resonance $\omega=\Delta/2$, 
i.e., $c =c_r =2 \beta^2 k_r /\Delta$. A relation between the pulse amplitude and 
its duration can be obtained by combining Eqs. (\ref{Ch6.072}); the resulting 
relation is then $\varepsilon_0^2 \tau_p =4/(\chi W)$, which can be used to
transform Eq. (\ref{Ch6.08}) into a velocity-duration expression. 
From Eq. (\ref{Ch6.08}), it is obvious that $c_r =2 \beta^2 k_r /\Delta$ plays 
the role of a limiting velocity, since 
$v \rightarrow c$ for $\varepsilon_0 \rightarrow \infty$. Thus, the velocity 
$c_r$ sets an upper (lower) bound on the pulse velocity in absorbing (amplifying) 
SCQMM structures. It is generally lower than the corresponding one for two-photon
SIT or SRD in ordinary media, $\beta$. Moreover, $c_r$ depends only on the qubit
parameters, so that its value can be also tailored during fabrication or tuned
by the gate voltages $V_g$.

\begin{figure}[!t]
   \centering
   \includegraphics[width=0.9\linewidth]{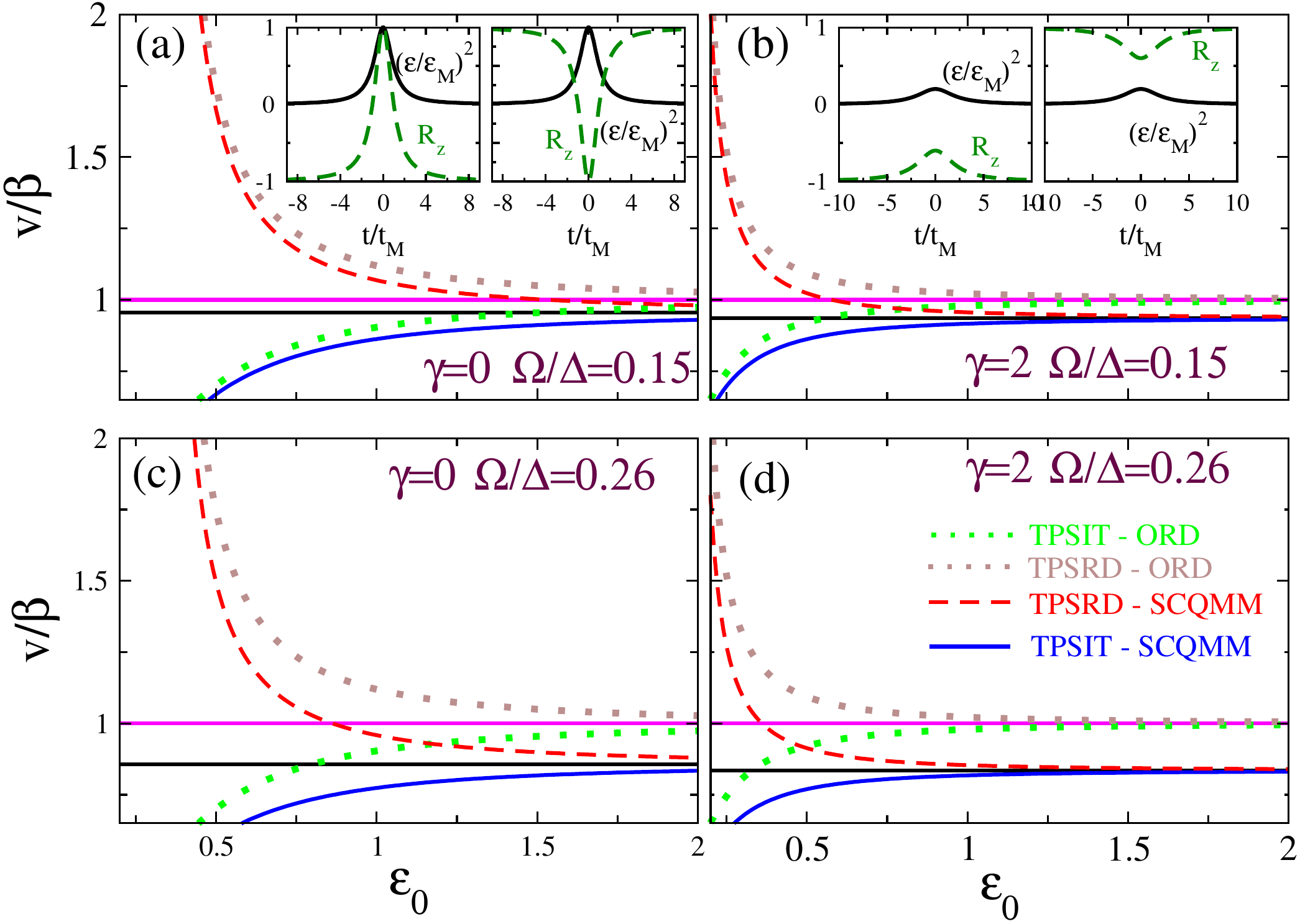}
\caption{
 The normalized pulse velocity $v / \beta$ - amplitude $\varepsilon_0$ relation 
 in two-photon superradiant (TPSRD, amplifying) and two-photon self-induced 
 transparent (TPSIT, absorbing) superconducting quantum metamaterials (SCQMMs) 
 and pulse envelops. In (a)-(d), the normalized pulse velocity $v / \beta$ is 
 plotted as a function of the electromagnetic (EM) vector potential pulse 
 amplitude $\varepsilon_0$ and compared with the corresponding curves for 
 ordinary amplifying (brown-dotted curves) and absorbing (green-dotted curves)
 mediums. The horizontal magenta-solid (resp. black-solid) lines indicate the 
 limiting velocity in ordinary amplifying and absorbing mediums, $v/\beta=1$ 
 (resp. amplifying and absorbing SCQMMs, $v =c_r < \beta$).
(a) $V_{00} =V_{11} =1$, $V_{01} =V_{10} =0.8$, $\chi=1/5$, $E_1 -E_0=3$
   ($\gamma=0$ and $\Omega/\Delta=0.15$).
   Left Inset: The envelop of the EM vector potential pulse
   $(\varepsilon/\varepsilon_M)^2$ and the population inversion $R_z (n)$
   profiles as a function of $\tau/\tau_M$ for TPSIT (absorbing) SCQMMs.
   Right Inset: Same as in the left inset for TPSRD (amplifying) SCQMMs.
(b) $V_{00}=0.6$, $V_{11} =1.4$, $V_{01} =V_{10} =0.8$, $\chi=1/5$, $E_1 -E_0=3$
   ($\gamma=2$ and $\Omega/\Delta=0.15$).
   Left Inset: The envelop of the EM vector potential pulse
   $(\varepsilon/\varepsilon_M)^2$ and the population inversion $R_z (n)$
   profiles as a function of $\tau/\tau_M$ for TPSIT (absorbing) SCQMMs.
   Right Inset: Same as in the left inset for TPSRD (amplifying) SCQMMs.
(c) $V_{00} =V_{11} =3$, $V_{01} =V_{10} =0.8$, $\chi=1/5$, $E_1 -E_0=3$
   ($\gamma=0$ and $\Omega/\Delta=0.26$).
(d) $V_{00} =3$, $V_{11} =3.8$, $V_{01} =V_{10} =0.8$, $\chi=1/5$, $E_1 -E_0=3$
   ($\gamma=2$ and $\Omega/\Delta=0.26$).
}
\label{fig6.02}
\end{figure}

In Fig. \ref{fig6.02}, several velocity-amplitude curves $v/\beta$ as a function 
of $\varepsilon_0$ are shown along with profiles of the envelops of the EM 
vector potential pulse $(\varepsilon/\varepsilon_M)^2$ and the population 
inversion $R_z (n)$ as functions of the slow variable $(\tau/\tau_M)$, 
$\tau_M =\frac{\Delta}{2\chi \mu} \frac{c-v}{v}$, in a frame of reference which 
is moving with velocity $v$, both for absorbing and amplifying SCQMMs. In all 
subfigures, the horizontal magenta-solid lines indicate the limiting velocity in 
ordinary amplifying and absorbing mediums, $v=\beta$, while the black-solid 
lines indicate the limiting velocity in amplifying and absorbing SCQMMs, 
$v =c_r < \beta$. All the curves exhibit a 
hyperbolic dependence; moreover, the corresponding curves for ordinary mediums 
and SCQMMs are close to each other, especially for low $\varepsilon_0$. 
The major difference is that the limiting velocity in SCQMMs is always lower 
than the corresponding one in ordinary mediums. Moreover, that limiting velocity 
in SCQMMs is parameter-dependent as mentioned above; this becomes clear by 
comparing curves of the same color (e.g., the red-dashed curves) in Figs. 
\ref{fig6.02}a and c, for which the ratio $\Omega/\Delta$ is $0.15$ and $0.26$, 
respectivelly, and provides $v/\beta \simeq 0.95$ and $\simeq 0.88$.
The limiting velocity $c_r$ for SCQMMs can be reduced further with increasing 
further the ratio $\Omega/\Delta$. Thus, parameter engineering for the SCQMM can 
slow down the speed of the pulses $v$ at the desired level for high enough 
amplitudes $\varepsilon_0$. Effective control of $v$ in SCQMMs could also be 
achieved in principle by an external field \cite{Park2001} or by real time 
tuning of the qubit parameters through the gate voltages $V_g$, as mentioned 
above. That ability to control the flow of "optical", in the broad sense, 
information may have technological relevance to quantum computing 
\cite{Cornell2001}. The effect of non-zero $\gamma$ factor become apparent by 
comparing again curves of the same color in Figs. \ref{fig6.02}a and b, for 
which $\gamma=0$ and $2$, respectively (the rest of the parameters are the same). 
The velocity-amplitude curves approach their limiting value for lower 
$\varepsilon_0$ with increasing $\gamma$. The same conclusion can be drawn by 
comparing curves of the same color in Figs. \ref{fig6.02}c and d. The effect of 
non-zero $\gamma$ is also revealed in the insets of Figs. \ref{fig6.02}a and b; 
in each inset, the left (right) panel is for absorbing (amplifying) SCQMMs. 
In both cases, the amplitudes of the pulse envelops for $\gamma=2$ are about 
four times smaller than those for $\gamma=0$. Thus, only for $\gamma=0$, i.e., 
for $V_{00}=V_{11}$, can the envelops of the EM vector potential pulses and the 
population inversion pulses (either for absorbing or for amplifying SCQMMs) 
attain their maximum amplitude value. 

Note that we have not been concerned with decoherence effects due to dephasing 
and energy relaxation in the SCQs. This is clearly an idealization which is 
partly justified as long as the coherence time exceeds the wave propagation time 
across a relatively large number of unit cell periods (i.e., a large number of 
SCQs). In a recent experiment \cite{Gambetta2006}, a charge qubit coupled to a 
strip line had a dephasing time in excess of $200 ~ns$, i.e., a dephasing rate 
of $5~MHz$, and a photon loss rate from the cavity of $0.57 ~MHz$. Those 
frequencies are very small compared with the transition frequency of the 
considered SCQs which is of the order of the Josephson energy (i.e., a few $GHz$) 
\cite{Rakhmanov2008,Shvetsov2013}. Therefore, we have neglected such decoherence 
effects here. The decoherence factor $\gamma$, which in Figs. \ref{fig6.02}b and 
d has been chosen according to the parameter values in \cite{Rakhmanov2008}, is 
not related to either dephasing or energy relaxation. That factor attains a 
non-zero value whenever the matrix elements of the effective SCQ-EM field 
interaction, $V_{11}$ and $V_{00}$, are not equal.

\subsubsection{Numerical Simulations}
\begin{figure}[!t]
   \centering
   \includegraphics[width=0.9 \linewidth]{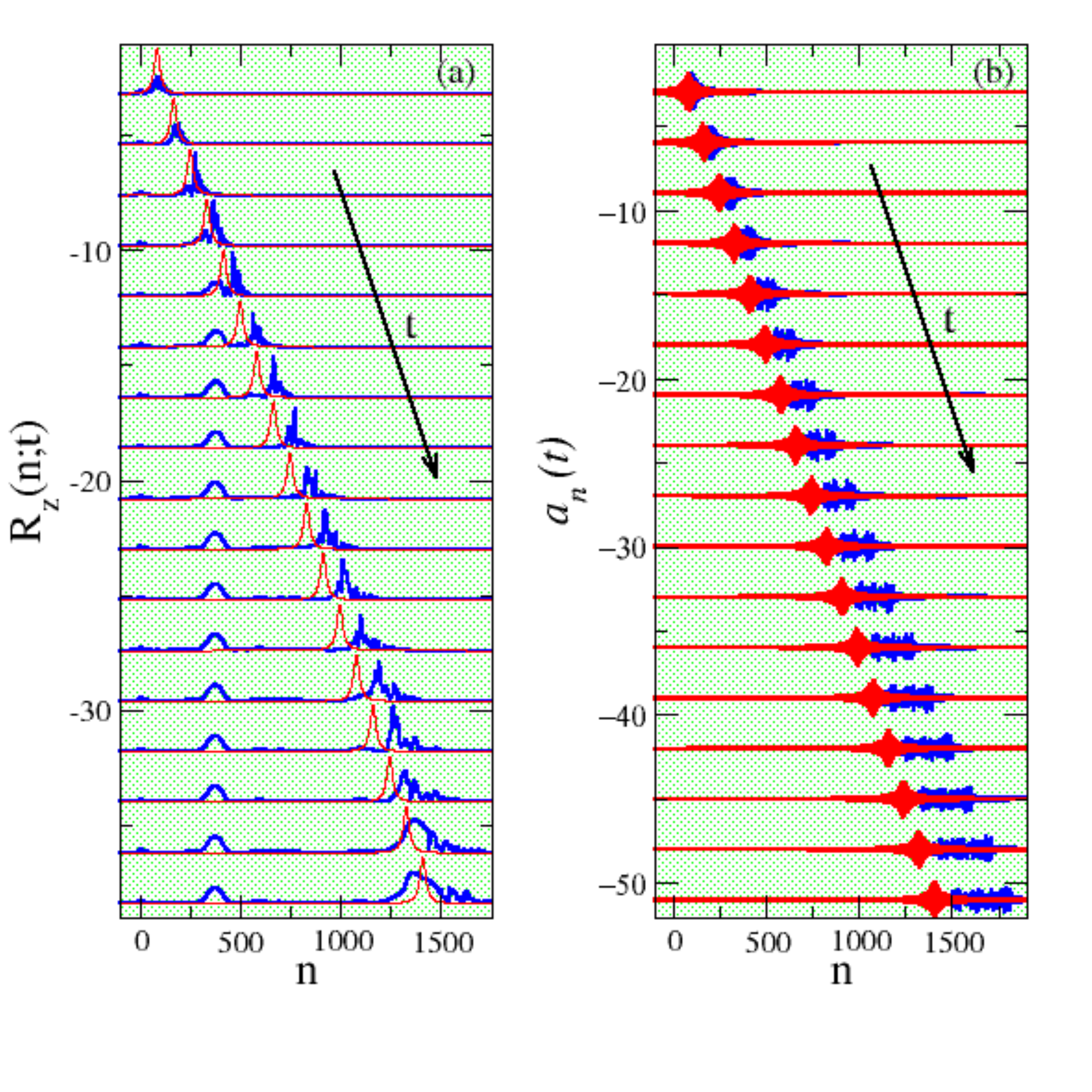}
\caption{
(a) Snapshots of the induced population inversion pulse $R_z (n;t)$, excited by 
    the electromagnetic vector potential pulse $a_n (t)$, whose corresponding 
    snapshots are shown in (b), for two-photon self-induced transparent (TPSIT, 
    absorbing) superconducting quantum metamaterials. 
(b) Snapshots of the corresponding electromagnetic vector potential pulse 
    $a_n (t)$.
    In both (a) and (b), the numerically obtained pulses are shown in blue color, 
    while the analytical solutions are shown in red color.  
 Parameters: 
 $\chi=1/5$, $\beta=6$, $V_{00}=V_{11}=1$, $V_{01}=V_{10}=0.8$, $E_1 -E_0=3$, 
 and $v/c=0.7$.
 Only a small part of the simulated array of SCQs is shown for clarity. 
}
\label{fig6.03}
\end{figure}

The above analytical predictions should be confirmed numerically, by integrating 
Eqs. (\ref{Ch6.03}) and (\ref{Ch6.04}). Any integration algorithm such as a 
fourth order Runge-Kutta scheme with constant time-step can be used for that 
purpose. Using that scheme, small time-steps, e.g., $\Delta t =10^{-3}$, or even 
smaller are required to conserve up to high accuracy the total and partial 
probabilities as the compound system of the SCQs and the EM vector potential 
evolve in time. For the numerical results presented below, periodic boundary 
conditions have been assumed. Due to the Lorentzian shape (Lorentzian) of the EM 
vector potential pulse and the population inversion pulse in the SCQ subsystem, 
very large systems with $N=2^{13} =8192$ and $N=2^{14} =16384$ have been simulated 
to diminish as much as possible the effects from the ends (i.e., to avoid the 
interaction of the pulse tail with itself). In some cases, it is necessary to 
simulate even larger systems, with $N=50,000$. In order to observe two-photon 
self-induced transparent (TPSIT) pulses $a_n (t)$ and the induced population 
inversion pulses $R_z (n;t)$, the following initial coditions are implemented: 
for the former, the analytically obtained solution resulting for the given set 
of parameters, while for the latter all the SCQs are set to their ground state, 
i.e., the state with eigenenergy $E_0$. In terms of the Bloch variables $R_i$, 
$i=x,y,z$, that initial condition is specified as:
\begin{equation}
\label{Ch6.09}
  R_x (t=0) =R_y (t=0) =0, \qquad R_z(t=0)=-1 , 
\end{equation}
for any $n=1,...,N$. Then, the TPSIT pulses $a_n (t)$ and $R_z (n;t)$ exist for 
velocities less than the corresponding limiting velocity for TPSIT media, i.e., 
\begin{equation}
\label{Ch6.092}
   v < c_r =2 \beta^2 \frac{k_r}{\Delta}.
\end{equation}  
Recall that the last equation is valid only when {\em the two-photon resonance 
condition} $\omega=\Delta/2$ has been imposed. In Eq. (\ref{Ch6.092}), the 
frequency $\omega$ of the carrier wave of the EM vector potential is fixed by
the two-photon resonance condition (and thus it is eliminated from the equation).
On the other hand, the wavenumber of the carrier wave of the EM vector potential
may vary in an interval which is restricted by the condition Eq. (\ref{Ch6.076}) 
which ensures that $k$ is real. The integration of Eqs. (\ref{Ch6.03}) and 
(\ref{Ch6.04}) in time and the inspection of the evolving profiles indeed reveal 
that the $a_n (t)$ can propagate in the SCQMM structure, and that at the same 
time, it is capable of exciting an $R_z (n;t)$ pulse of similar shape which also 
propagates at the same speed $v$ for a substantial temporal window. In Figs. 
\ref{fig6.03}a and b, several snapshots of the two-photon self-induced 
transparent (TPSIT) propagating pulses $R_z (n;t)$ and $a_n (t)$, respectively, 
are shown, at instants differing by $20$ time-units (the first snapshot is taken 
at $t=20$). Note that the snapshots are displaced vertically (to avoid 
overlapping) and that time increases downwards. The numerical (analytical) 
results are shown in blue (red) color. In Fig. \ref{fig6.03}a, the amplitude of 
the $R_z (n;t)$ pulse gradually grow to the expected maximum around unity in 
approximately $60$ time-units; then, the pulse continues its course with 
fluctuating amplitude for $\sim 160$ more time-units, during which it moves at 
the same speed as the EM vector potential pulse $a_n (t)$ (Fig. \ref{fig6.03}b). 
However, due to the inherent discreteness in the SCQ chain and the lack of 
direct coupling between the SCQs, the induced population inversion pulse 
$R_z (n;t)$ splits at certain instants leaving behind small "probability bumps" 
which are pinned at particular SCQs. After the end of the almost coherent 
propagation regime, the $R_z (n;t)$ pulse broadens and slows-down until it stops 
completely. At the same time, the width of the $a_n (t)$ pulse increases in the 
course of time due to discreteness-induced dispersion. A comparison with the 
corresponding analytical expressions reveals fair agreement during the almost 
coherent propagation regime, although both the $R_z (n;t)$ and $a_n (t)$ pulses 
travel slightly faster than expected from the analytical predictions. The 
temporal variable here is normalized to the inverse of the Josephson frequency 
$\omega_J$ which for typical parameter values is of the order of a few $GHz$ 
\cite{Rakhmanov2008}. Then, the almost coherent pulse propagation regime in the 
particular case shown in Fig. \ref{fig6.03} lasts for $\sim 160 \times 10^{-9}~s$, 
or $\sim 160~ns$, which is of the same order as the reported decoherence time 
for a charge qubit in reference \cite{Gambetta2006} (i.e., $200~ns$).   

\begin{figure}[!t]
   \centering
   \includegraphics[width=0.9 \linewidth]{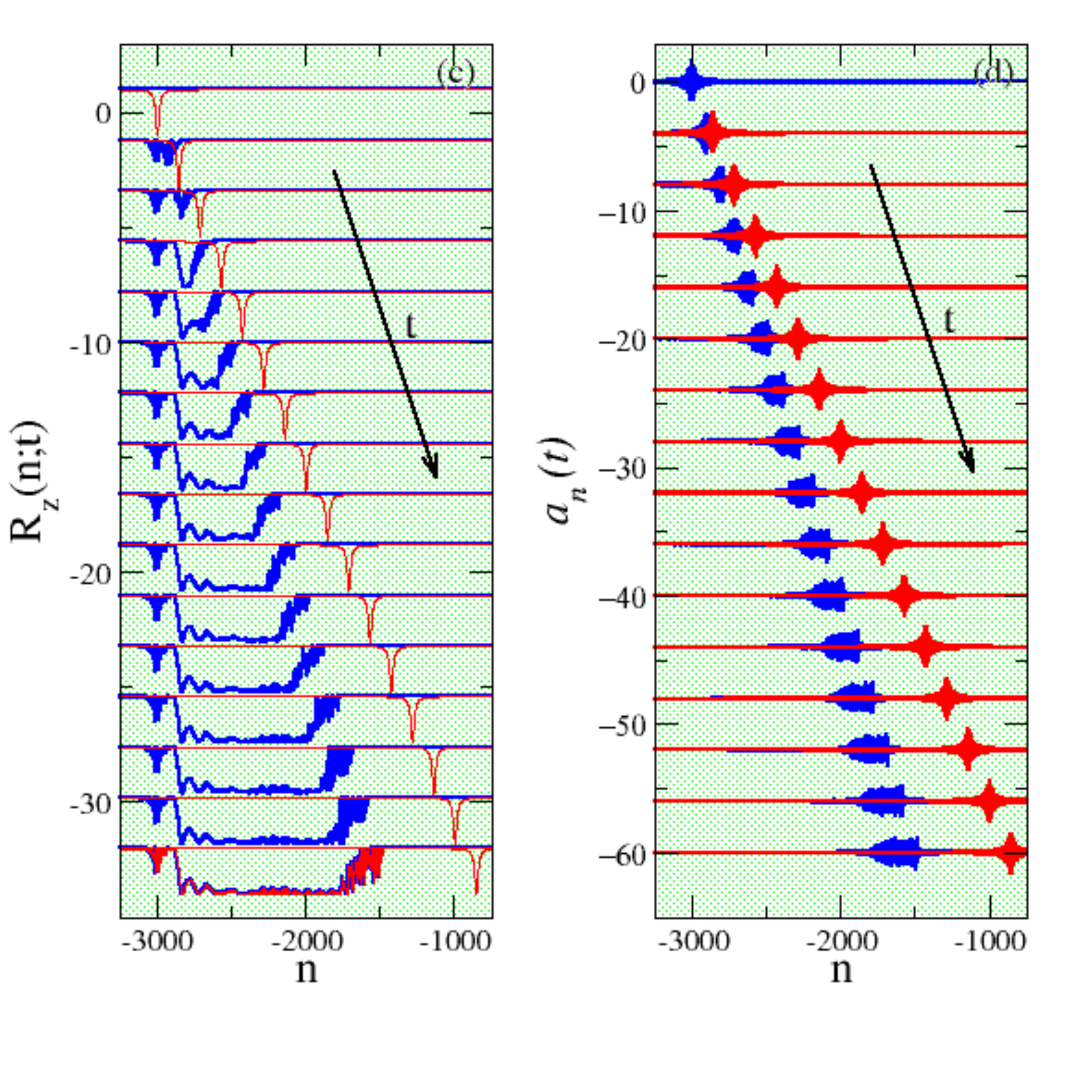}
\caption{
(a) Snapshots of the induced population inversion pulse $R_z (n;t)$, excited by 
    the electromagnetic vector potential pulse $a_n (t)$, whose corresponding 
    snapshots are shown in (b), for two-photon superradiant (TPSRD, amplifying) 
    superconducting quantum metamaterials. 
(b) Snapshots of the electromagnetic vector potential pulse $a_n (t)$.
 In both (a) and (b), the numerically obtained pulses are shown in blue color, 
 while the analytical solutions are shown in red color.  
 Parameters: 
 $\chi=1/5$, $\beta=6$, $V_{00}=V_{11}=1$, $V_{01}=V_{10}=0.8$, $E_1 -E_0=3$, 
 and $v/c=0.7$.
 Only a small part of the simulated array of SCQs is shown for clarity. 
}
\label{fig6.04}
\end{figure}

\begin{figure}[!t]
   \centering
   \includegraphics[width=0.45 \linewidth]{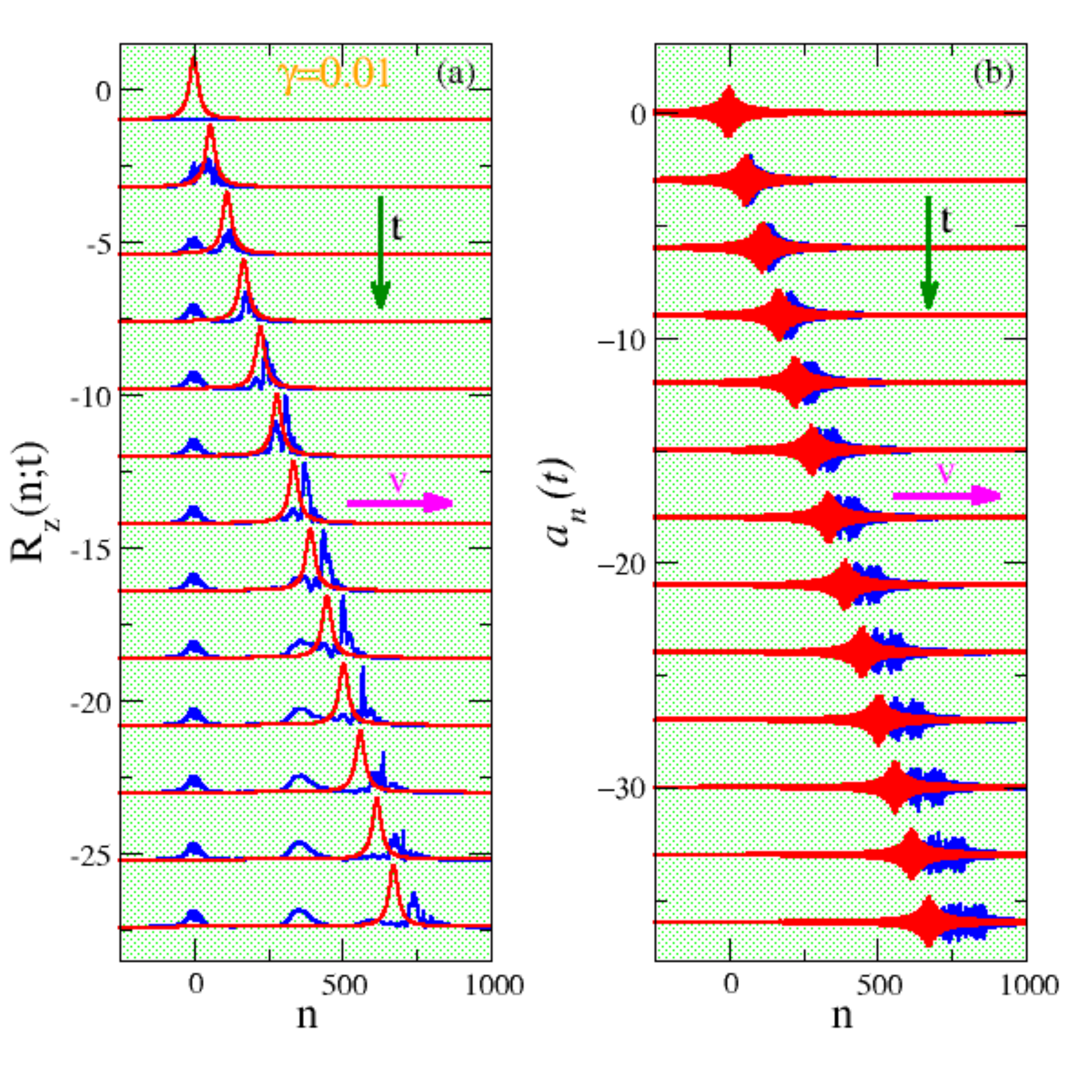}
   \includegraphics[width=0.45 \linewidth]{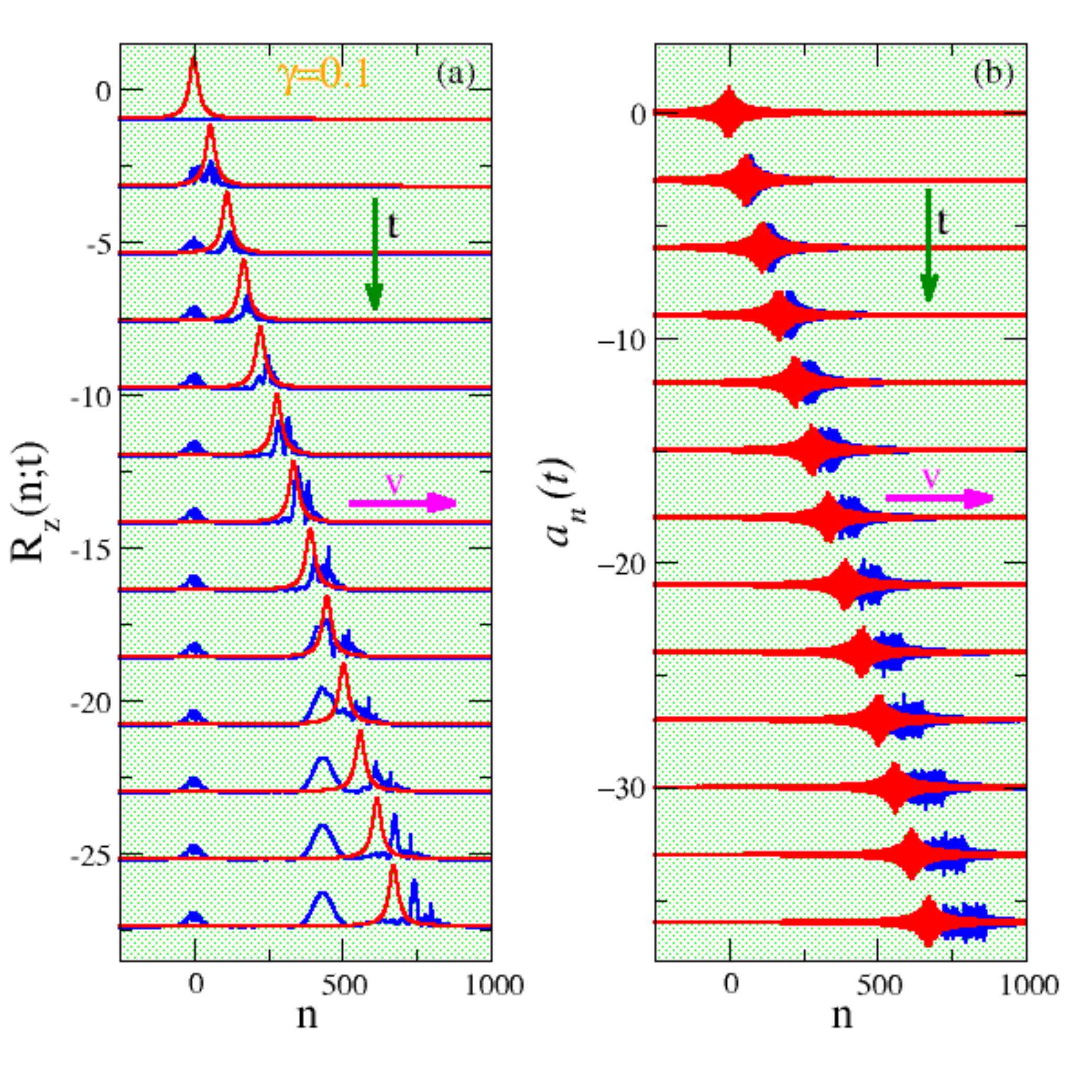}
\caption{
 Snapshots of the induced population inversion pulse $R_z (n;t)$ and the 
 electromagnetic vector potential pulse $a_n (t)$ in two-photon self-induced 
 transparent (TPSIT, absorbing) superconducting quantum metamaterials for a 
 non-zero $\gamma$ factor.
(a) Snapshots of $R_z (n;t)$ for $\gamma=0.01$ ($V_{00}=0.998$, $V_{11}=1.002$).
(b) Snapshots of $a_n (t)$ which excite the corresponding $R_z (n;t)$ pulses in (a).
(c) Snapshots of $R_z (n;t)$ for $\gamma=0.1$ ($V_{00}=0.98$, $V_{11}=1.02$).
(d) Snapshots of $a_n (t)$ which excite the corresponding $R_z (n;t)$ pulses in (c).
 The other parameters are: 
 $\chi=1/4.9$, $\beta=6$, $V_{01}=V_{10}=0.7$, $E_1 -E_0=3$, and $v/c=0.7$.
 In (a)-(d), the numerically obtained pulses are shown in blue color, while the 
 analytical solutions are shown in red color.  
 Only a small part of the simulated array of SCQs is shown for clarity. 
}
\label{fig6.05}
\end{figure}

The situation seems to be different, however, in the case of two-photon 
superradiant (SRD) pulses, as can be observed in the snapshots shown in Figs. 
\ref{fig6.04}a and b, for $R_z (n;t)$ and $a_n (t)$, respectively. Here, the 
lack of the direct interaction between SCQs is crucial, since the SCQs that make 
a transition from the excited to their ground state as the peak of the $a_n (t)$ 
pulse passes by their location, cannot return to their excited states after 
the $a_n (t)$ pulse has gone away. It seems, thus, that the $a_n (t)$ pulse 
excites a kink-like $R_z (n;t)$ front that propagates along with it at the same 
velocity. It should be noted that the common velocity of the $R_z (n;t)$ kink
and the $a_n (t)$ pulse is considerably lower than the analytically predicted 
one, as it can be observed in Figs. \ref{fig6.04}a and b. Even more complicated 
behavioral patterns of two-photon SRD propagating pulses have been also obtained
\cite{Ivic2016}. 
The effect of non-zero $\gamma$ factor on the $R_z (n;t)$ and 
$a_n (t)$ pulses is clearly revealed in Fig. \ref{fig6.05}. These snapshots are
taken at instants eparated by $14$ time-units, from $t=0$ to $t=168$. They are
shifted downwards to avoid overlapping, and only part of the array is shown for 
clarity. A small value of the factor $\gamma$ ($\gamma =0.01$) has practically 
negligible effect on the $R_z (n;t)$ and $a_n (t)$ pulses (Figs. 
\ref{fig6.05}a and b). However, when $\gamma$ increases (e.g., to $\gamma =0.1$, 
as in Figs. \ref{fig6.05}c and d), the amplitude of the envelop of the 
$R_z (n;t)$ pulse decreases significantly. For even higher values of $\gamma$, 
the excitation of $R_z (n;t)$ pulses becomes impossible. 

Thus, even for non-zero $\gamma$, an $a_n (t)$ pulse is able to excite or induce
$R_z (n;t)$ pulse whose amplitude gradually increases until it attains its 
maximum value; that value is close to but less than unity for a small but 
non-zero $\gamma$ (e.g., for $\gamma =0.01$ as in Fig. \ref{fig6.05}a and b).  
The induced $R_z (n;t)$ pulse propagates along with the $a_n (t)$ pulse with 
velocity $v'$. In Fig. \ref{fig6.05}a, the amplitude of the $R_z (n;t)$ pulse 
reaches its maximum at $t\simeq 70$ time units; subsequently it evolves in time 
while it keeps its amplitude almost constant for at least the next $56$ 
time-units. After that, its amplitude starts decreasing until it is completely 
smeared (not shown). During the time interval in which the amplitude of the 
$R_z (n;t)$ pulse is close to (but less than) unity, the SCQMM is considered to 
be in an {\em almost coherent} regime. Note that at about $t=84$ a little bump 
starts to appear which grows to a little larger one as time advances 
({\em "probability bump"}). This probability bump, as well as that observed also 
in Fig. \ref{fig6.03}a for a different parameter set which gives $\gamma =0$, is 
immobile and its peak is located on a site around $n \sim 300$. 
A second such bump appears at $n=0$ due to the initial "shock" of the qubit 
subsystem because of the sudden onset of the $a_n (t)$ pulse. A comparison of 
the numerical $R_z (n;t)$ profiles with the analytical ones reveals that the 
velocity of propagation $v'$, the same for the numerically onbtained $R_z (n;t)$ 
and $a_n (t)$ pulses, (Figs. \ref{fig6.05}a and b), is slightly larger than the 
analytically obtained one $v$ ($v' > v$).
In Figs. \ref{fig6.05}c and d, the diagonal effective matrix interaction elements 
$V_{ij}$ ($i, j=0, 1$) have been chosen so that the factor $\gamma$ has the value 
of $0.1$. That value is obtained by choosing, specifically, $V_{00}=0.98$, 
$V_{11}=1.02$, and $V_{01}=V_{10}=0.7$, and it is already high enough to 
Stark-shift considerably the energy levels of the SCQs. The effect of 
$\gamma =0.1$ becomes apparent by comparing Figs. \ref{fig6.05}a and b, with Figs. 
\ref{fig6.05}c and d, respectively. Remarkably, the EM vector potential pulse 
$a_n (t)$ does not seem to be affected significantly. However, the numerically 
obtained, induced population inversion pulse $R_z (n;t)$ has much lower amplitude 
compared with that of the analytically predicted form which is only slightly 
affected by the high value of $\gamma$. Note that the speed of the $R_z (n;t)$ 
pulse is the same as that in the case of lower value of $\gamma$ 
($\gamma =0.01$). Even the unwanted "probability bumps" in Figs. \ref{fig6.05}a 
and c, for $\gamma =0.01$ and $0.1$, respectively, appear at about the same 
locations with almost the same amplitude and shape. 
\newpage

\section{Summary}
\label{Summary-S7}
SQUIDs metamaterials exploit geometry, superconductivity, and the Josephson 
effect, to exhibit extraordinary metamaterial properties and very rich dynamic 
behavior. Many aspects of their behavior have been investigated both 
theoretically and experimentally; properties such as negative diamagnetic 
permeability, wide-band tunability of the resonance either by a magnetic 
field or the temperature, fast switching between multistable states, 
broad-band self-induced transparency, and coherent oscillations, have been 
experimentally observed. Moreover, some of these properties have been 
numerically confirmed. Further theoretical results, which relay primarily on 
numerical simulations, have predicted the existence of nonlinearly localized 
states in the form of discrete breathers, the existence the counter-intuitive 
chimera states, the nonlinear band opening, and the existence of flat-band 
localized modes, the latter in SQUID metamaterials on Lieb lattices. All the
above results presented in this review have been obtained in the classical
regime, although even those phenomenological, equivalent circuit models used 
to simulate SQUID metamaterials, encompass macroscopic quantum effects.
In the quantum regime, a particular paradigmatic model describing an array
of superconducting charge qubits periodically loaded in a superconducting 
transmission line, has been reviewed. That system could be also regarded
as an array of SQUIDs, which SQUIDs however are strongly coupled to each 
other through direct conducting paths. For that superconducting quantum 
metamaterial, the possibility for the opening of an oscillating photonic
band gap, the propagation of self-induced transparent and superradiant pulses,
and the induction of quantum coherence in the qubit chain by the propagating 
pulses, has been theoretically demonstrated.

The presented review cannot be exhaustive on SQUID metamaterials and quantum
superconducting metamaterials, since that area of the field of metamaterial
research evolves very fast; experiments on both classes of systems are going
on that will perhaps reveal further surprising results. There was an attempt 
to guide the reader through the main results in this area, and to stress that 
the presented properties of the SQUID metamaterials are to a large extent a
result of nonlinearity of their elementary units, i.e., the individual SQUIDs.
Indeed, substantial nonlinearity leads to multistability in a single SQUID
for driving frequencies close to its resonance. That resonance can be tuned
by a dc and/or a harmonic field; moreover, the amplitude of the latter also
determines the strength of the nonlinearity. The SQUID metamaterials, inherit
to a large degree the properties of their elements; moreover, the complexity 
of their dynamics increases immensely in the multistability region with 
increasing the number of SQUIDs. That makes possible the appearance of 
collective states such as the chimera states. The nonlinearity of individual
SQUIDs is also responsible for the localization of energy leading to 
the generation of discrete breathers which exist due to a delicate balance 
between intrinsic dissipation and incoming power from the external field
(dissipative breathers). The flat-band localized modes, on the other hand, 
may exist due to the particular lattice geometry (Lieb lattice), in the 
linear regime (very low field intensities).

For the superconducting quantum metamaterials, the Josephson nonlinearity 
is again crucial in order to form a particular type of a superconducting 
{\em qubit}, i.e., an effectively two-level system (by neglecting all the other,
higher energy states). Those qubits, couple through their Josephson junction(s) 
to the magnetic component of the electromagnetic field of a pulse launched 
form one end; that interaction allows for the generation of important
quantum optical effects, such as self-induced transparency and superradiance.
Moreover, the propagating pulses, which are shaped to acquire Lorentzian 
profiles, are able to induce quantum coherence in the form of population 
inversion pulses of the same shape in the qubit chain. Also, proper design
of the qubit parameters allows for controlling the (common) speed of the 
propagating pulses in the superconducting quantum metamaterial, which is not
possible in a natural material.   
\newpage

\section*{Acknowledgments}
This work is partially supported by 
the Ministry of Education and Science of the Russian Federation in the framework 
of the Increase Competitiveness Program of NUST "MISiS" (No. K2-2017-006), 
and by
the European Union under project NHQWAVE (MSCA-RISE 691209).
The authors thank, in no particular order: Alexey V. Ustinov, Steven M. Anlage,
Alexandre M. Zagoskin, and Zoran Ivi{\'c} for helpful discussions, comments, and
suggestions.
NL gratefully acknowledges the Laboratory for Superconducting Metamaterials, 
NUST "MISiS" for its warm hospitality during visits.
\newpage

\appendix
\section{Derivation of the Maxwell-Bloch-sine-Gordon equations}
\subsection{Quantization of the Qubit Subsystem}
Consider an infinite number of superconducting charge qubits (SCQs) of the form 
of mesoscopic superconducting islands, periodically loaded in a transmission 
line (TL) that consists of two superconducting plates separated by distance $d$. 
The center-to-center distance between the qubits, $\ell$, is of the same order 
of magnitude as $d$. The SCQs are connected to each electrode of the TL with a 
Josephson junction (JJ). Assume that an electromagnetic (EM) wave corresponding 
to a vector potential $\vec{A}=A_z (x,t) \hat{z}$ which propagates along that 
superconducting TL, in a direction parallel to the electrodes and perpendicular 
to the direction of propagation of the EM wave. In the following it is assumed 
that the wavelength $\lambda$ of the EM field is much larger than the other 
length scales such as the separation of the electrodes $d$ and the distance 
between SCQs $\ell$, i.e., $\lambda >> \ell, d$. Then, the magnitude of the EM 
vector potential component $A_z (x,t)$ is approximatelly constant within a unit 
cell, so that $A_z (x,t) \simeq A_{z,n} (t)$. The Hamiltonian of the compound 
qubit array - EM field system shown in Fig. \ref{fig6.01} of the paper is then
\begin{equation}
\label{A4.101}
   H=\sum_n \left\{ \dot\varphi^2_n -2\cos \alpha_n\cos\varphi_n +\dot\alpha^2_n 
                   +\beta^2(\alpha_{n+1}-\alpha_n)^2 \right\} ,
\end{equation}
where $\varphi_n$ is the superconducting phase on $n-$th island, 
$\beta^2=(8\pi d E_J)^{-1}  \left( \Phi_0/(2\pi) \right)^2$, 
$a_n (t) =(2\pi d/\Phi_0) A_{n} (t)$ is the normalized and discretized EM vector 
potential at the $n-$th unit cell, and the overdots denote derivation with 
respect to time $t$. The Hamiltonian Eq. (\ref{A4.101}) is given in units of the 
Josephson energy $E_J =(\Phi_0 I_c)/(2\pi C)$, where $I_c$ and $C$ is the 
critical current and capacitance, respectively, of the JJs, and $\Phi_0 =h/(2 e)$ 
is the flux quantum, with $h$ and $e$ being the Planck's constant and the 
electron charge, respectively. By adding and subtracting $2\cos \phi_n$ to the 
Hamiltonian $H$ and subsequently rearranging, we get the more transparent form
\begin{equation}
\label{A4.102}
   H =H_{qub} +H_{emf} +H_{int} ,
\end{equation}
where the qubit subsystem energy $H_{qub}$, the EM field energy $H_{emf}$, and 
their interaction energy $H_{int}$, take respectively the form
\begin{equation}
\label{A4.103}
        H_{qub}=\sum_n \{ \dot\varphi^2_n-2\cos \varphi_n\} , 
 \qquad H_{emf}=\sum_n \{ \dot\alpha^2_n+\beta^2(\alpha_{n+1}-\alpha_n)^2\} ,
 \qquad H_{int}=\sum_n \{ 2\cos \varphi_n (1-\cos \alpha_n)\} .
\end{equation}

In order to quantize the qubit subsystem, the classical variables $\varphi_n$ 
and $\dot{\varphi}_n$ are replaced by the quantum operators $\hat{\varphi_n}$ 
and $\dot{\varphi}_n\rightarrow-i\frac{\partial}{\partial \varphi_n}$, 
respectively, in $H_{qub}$ and $H_{int}$. The exact energy spectrum $E_p (n)$ 
and the corresponding wavefunctions $\Xi_p (n)$ of the $n$th qubit may then be 
obtained by mapping the Schr{\"o}dinger equation with its Hamiltonian 
$H_{qub,n}=\dot\varphi_n^2-2\cos \varphi_n$, onto the Mathieu equation
\begin{equation}
\label{A4.110}
  \bigg(\frac{\partial^2}{\partial \varphi^2_n}+E_{p,n}-2\cos\varphi_n\bigg)
   \Xi_{p,n}=0.
\end{equation}
Since the qubits are identical and non-interacting directly, the second 
quantization of the qubit subsystem proceeds as follows (the subscript $n$ is 
dropped): First $H_{qub}$ is written as
\begin{equation}
\label{A4.199}
  H_{qub} =-\int d\varphi_n \hat{\Psi}^{\dagger}(\varphi)
            \bigg(\frac{\partial^2}{\partial \varphi^2}+2\cos\varphi\bigg) 
            \hat{\Psi}(\varphi) ,
\end{equation}
where $\hat \Psi^{\dagger}$ and  $\hat \Psi$ are field operators. Then, using 
the expansion $\hat\Psi(\varphi)=\sum_p a_p \Xi_p(\varphi)$, where the operators 
$a^{\dagger}_p$ ($a_p$) create (annihilate) qubit excitations of energy $E_p$, 
the Hamiltonian Eq. (\ref{A4.199}) is transformed into 
\begin{equation}
\label{A4.177}
   H_{qub}=\sum_{p=0,1,...} E_{p} a^{\dagger}_{p} a_{p} .
\end{equation}
We hereafter restrict $H_{qub}$ to the Hilbert subspace of its two lowest levels, 
i.e., those with $p=0, 1$, so that in second quantized form the Hamiltonian Eqs.
(\ref{A4.102}) and (\ref{A4.103}) read
\begin{equation}
\label{A4.107}
   H= \sum_n \sum_{p} E_{p}(n)a^{\dagger}_{n,p}a_{n,p} 
     +\sum_{p,p'} V_{p,p'}(n)a^{\dagger}_{n,p} a_{n,p'}\sin^2\frac{\alpha_n}{2}
     +\sum_n \{ \dot\alpha^2_n+\beta^2(\alpha_{n+1}-\alpha_n)^2\} ,
\end{equation}
where $p, p'=0,1$ and 
\begin{equation}
\label{A4.108}
   V_{p,p'}(n) \equiv V_{p',p}(n) 
       =\int d\varphi_n \Xi^*_p(\varphi_n) \cos \varphi_n\Xi_{p,n} (\varphi_n) ,
\end{equation}
are the matrix elements of the $n$th qubit - EM field interaction. In the 
reduced state space, in which a single qubit can be either in the ground ($p=0$) 
or in the excited ($p=1$) state, the normalization condition 
$\sum_p a^{\dagger}_{n,p} a_{n,p}=1$ holds for any $n$.

\subsection{Maxwell-Bloch Formulation of the Dynamic Equations} 
In accordance with the semiclassical approach adopted here, the time-dependent 
Schr{\"o}dinger equation 
\begin{equation}
\label{A4.198}
   i \hbar\frac{\partial}{\partial t}|\Psi\rangle =\bar{H} |\Psi\rangle ,
\end{equation}
in which $\bar{H}$ is the Hamiltonian from Eq. (\ref{A4.107}) in physical units, 
i.e., $\bar{H} =H E_J$, is employed for the description of the qubit subsystem. 
The state of each qubit is a superposition of the form
\begin{equation}
\label{A4.111}
   |\Psi_n\rangle=\sum_p\Psi_{n,p}(t)a^{\dagger}_{n,p}|0\rangle ,
\end{equation}
whose coefficients $\Psi_{n,p}$  satisfy the normalization conditions
\begin{equation}
\label{A4.112}
   \sum_p|\Psi_{n,p} (t)|^2 =1 , \qquad \sum_{n,p} |\Psi_{n,p}(t)|^2=N ,
\end{equation}
in which a finite $N-$qubit subsystem is implied. Assuming that the pulse power 
is not very strong, the approximation 
$[1-\cos(\alpha_n)] \simeq (1/2) \alpha^2_n$ can be safely applied in the 
interaction part of the Hamiltonian $H_{int}$. Then, the substitution of 
$|\Psi\rangle = |\Psi_n\rangle$ from Eq. (\ref{A4.111}) into the Schr\"odinger 
equation (\ref{A4.198}), and the derivation of the classical Hamilton's equation 
for the normalized EM vector potential $\alpha_n$, yields
\begin{eqnarray}
\label{A4.114}
&& i\dot \Psi_{n,p}=\epsilon_p\Psi_{n,p}+\frac{1}{\chi}
                     \sum_{p'}V_{p,p'}(n){\Psi}_{n,p'}\alpha^2_n, \\
\label{A4.114.2}
&& \ddot\alpha_n-\beta^2(\alpha_{n+1}+\alpha_{n-1}-2\alpha_n)
                   +\sum_{p,p'}V_{p,p'}\Psi^{*}_{n,p}\Psi_{n,p'} \alpha_n=0 ,
\end{eqnarray}
where $\chi= \hbar\omega_J / E_J$. In Eqs. (\ref{A4.114}) and (\ref{A4.114.2}),
the temporal variable is renormalized according to $t \rightarrow \omega_J t$ 
and thus the dimensionless energy of the qubit excitations is redefined 
according to $E_P \rightarrow \epsilon_p =E_p / \chi$.

The evolution Eqs. (\ref{A4.114}) and (\ref{A4.114.2}) can be rewritten in terms 
of the $n-$dependent Bloch vector components through the transformation
\begin{equation}
\label{A4.115}
   R_z(n)=|\Psi_{n,1}|^2-|\Psi_{n,0}|^2 , \qquad
   R_y(n)=i(\Psi^*_{n,0}\Psi_{n,1}-\Psi^*_{n,1}\Psi_{n,0}) , \qquad
   R_x(n)=\Psi^*_{n,1}\Psi_{n,0}+\Psi^*_{n,0}\Psi_{n,1} , 
\end{equation}
in which the variables $R_i$ ($i=x,y,z$) apply to each single qubit, as
\begin{eqnarray}
\label{A4.119.2}
   \dot R_x (n) =-(\Delta +2 D \alpha_n^2) R_y (n), \qquad
   \dot R_y (n) =+(\Delta +2 D \alpha_n^2) R_x (n) -2\mu \alpha_n^2 R_z (n), \qquad
   \dot R_z (n) =+2 \mu \alpha_n^2 R_y (n), \\
\label{A4.120.2}
   \ddot\alpha_n + \chi [ \Omega^2 +\mu R_x (n)+D R_z (n) ] \alpha_n
                =\beta^2 ( \alpha_{n-1}-2\alpha_n+\alpha_{n+1} ) ,
\end{eqnarray}
where $D=\frac{(V_{11}-V_{00})}{2\chi}$, $\Omega^2 =\frac{(V_{11}+V_{00})}{2}$,
$\mu =\frac{V_{10}}{\chi}$, and 
$\Delta=\epsilon_1-\epsilon_0\equiv\frac{(E_1-E_0)}{\chi}$. By taking the 
continuous limit of Eqs. (\ref{A4.119.2}) and (\ref{A4.120.2}), we obtain 
\begin{eqnarray}
\label{A4.119}
   \dot R_x=-(\Delta +2 D \alpha^2) R_y , \qquad 
   \dot R_y=+(\Delta +2 D \alpha^2) R_x -2\mu \alpha^2 R_z , \qquad
   \dot R_z=+2\mu\alpha^2 R_y , \\ 
\label{A4.120}
   \ddot\alpha-\beta^2 \alpha_{xx}+\Omega^2\alpha
                           =-\chi (D R_z +\mu R_x) \alpha ,
\end{eqnarray}
where $R_x$, $R_y$, $R_z$, and $\alpha$ are functions of the spatial variable 
$x$ and the normalized temporal variable $t$, while the overdots denote partial 
derivation with respect to the latter. The Bloch equations (\ref{A4.119}) 
possess the dynamic invariant $\sum_i  R^2_i =1$.

\subsection{Slowly Varying Envelope Approximation and Reduced Dynamic Equations}
The Slowly Varying Envelope Approximation (SVEA) relies on the assumption that 
the envelop of a travelling pulse in a nonlinear medium varies slowly in both 
time and space compared with the period of the carrier wave, which makes 
possible to introduce slow and fast variables. According to the SVEA, the EM 
vector potential can be approximated as
\begin{equation}
\label{A4.121}
   \alpha(x,t) =\varepsilon(x,t) \cos \psi(x,t),
\end{equation}
where $\psi(x,t) =k x-\omega t +\phi(x,t)$, with $k$ and $\omega$ being the 
wavenumber and frequency of the carrier wave, respectively, which depend on each 
other through the dispersion relation, and $\varepsilon(x,t)$ and $\phi(x,t)$ are 
the slowly varying envelop and phase, respectively. Using fast and slow variables, 
the $x$ and $y$ Bloch vector components, $R_x (n)$ and $R_y (n)$, can be 
expressed as a function of new, in-phase and out-of-phase Bloch vector components 
$r_x$ and $r_y$ as
\begin{equation}
\label{A4.122}
  R_x=r_x \cos (2\psi) +r_y \sin(2\psi) , \qquad 
  R_y= r_y \cos(2\psi) -r_x \sin(2\psi) ,
  \qquad R_z=r_z .
\end{equation}

From Eqs. (\ref{A4.121}) and (\ref{A4.122}), the second temporal and spatial 
derivative of $\alpha(x,t)$ can be approximated by
\begin{equation}
\label{A4.300}
   \ddot{\alpha}
   \approx 2\omega \dot\varepsilon \sin\psi 
       +(2\omega \dot\phi -\omega^2) \varepsilon \cos\psi ,
   \qquad
    \alpha_{xx} \approx 
     -2k \varepsilon_x \sin\psi-(2k \phi_x -k^2)\varepsilon \cos\psi ,
\end{equation}
in which the rapidly varying terms of the form 
$\ddot\varepsilon$, $\varepsilon_{xx}$, $\phi^2$, $\phi_{xx}$, $\ddot\phi$, 
$\phi_x\varepsilon_x$, etc., have been neglected. Substitution of Eqs. 
(\ref{A4.300}) and (\ref{A4.122}), into Eq. (\ref{A4.120}) gives 
\begin{equation}
\label{A4.400}
   2(\omega \dot\varepsilon +k\beta^2 \varepsilon_x) \sin\psi 
    +[ 2(\dot\phi \omega +k \phi_x) -\omega^2 +\Omega^2 +\beta^2 k^2 ]
         \varepsilon \cos\psi=
   -\chi \{D r_z +\mu [ r_x \cos(2\psi) +r_y \sin(2\psi) ] \} \varepsilon \cos\psi .
\end{equation}
Equating the coefficients of $\sin\psi$ and $\cos\psi$ in the earlier equation 
yields
\begin{equation}
\label{A4.301}
   \omega \dot\varepsilon +k \beta^2 \varepsilon_x 
  =-\chi \mu r_y \varepsilon \cos^2\psi ,
\end{equation}
and
\begin{equation}
\label{A4.302}
   2(\dot\phi \omega +k \phi_x)
   -\{ \omega^2 -\Omega^2 -\beta^2 k^2 \}=-\chi [D r_z +\mu r_x \cos(2\psi) ] .
\end{equation}
The dispersion relation $\omega=\omega(k)$ is obtained from Eq. (\ref{A4.302}) 
by zeroing the expression in the curly brackets as 
\begin{equation}
\label{A4.125}
   k =\pm \frac{ \sqrt{ \omega^2 -\Omega^2 } }{ \beta } .
\end{equation}
Thus, EM waves propagate through the superconducting quantum metamaterial 
(SCQMM) only when their frequency exceeds a critical one,
$\omega_c =\Omega =\sqrt{ (V_{00} +V_{11}) / 2 }$. Finally, Eqs. (\ref{A4.301}) 
and (\ref{A4.302}) are averaged in time over the period of the fast time-scale 
$T=2\pi/\omega$ of the phase $\psi (x,t)$. Due to the assumed time-dependence of 
$\psi (x,t)$ within the SVEA framework, that averaging requires the calculation 
of integrals of the form
\[ \langle {\cal F}(\sin f(\psi),\cos g(\psi) )\rangle 
=\frac{1}{2\pi}\int^{2\pi}_0 {\cal F}(\sin f(\psi),\cos g(\psi) ) d\psi.\]
This procedure, when it is applied to the two evolution Eqs. (\ref{A4.301}) and 
(\ref{A4.302}) provides the truncated equations for slow amplitude and phase
\begin{equation}
\label{A4.123}
   \dot\varepsilon +c \varepsilon_x =-\chi \frac{\mu}{2\omega} \varepsilon r_y ,
   \qquad  \dot\phi +c \phi_x=-\chi \frac{D}{\omega}R_z ,
\end{equation}
where $c=\beta^2 k / \omega$ is a critical velocity.

Substituting Eq. (\ref{A4.121}) and (\ref{A4.122}) into Eqs. (\ref{A4.119}) for 
the original Bloch vector components, we get
\begin{eqnarray}
\label{A4.303}
   (\dot r_x+2\dot{\psi}r_y) \cos(2\psi) +(\dot r_y-2\dot{\psi} r_x)\sin(2\psi)
           =-(\Delta +2D\varepsilon^2 \cos^2\psi) [r_y\cos(2\psi) -r_x\sin(2\psi)]i \\
\label{A4.304}
   (\dot r_y -2\dot{\psi} r_x) \cos(2\psi) -(\dot r_x +2\dot{\psi} r_y) \sin(2\psi)
              =(\Delta +2D\varepsilon^2 \cos^2\psi) [r_x\cos(2\psi) +r_y\sin(2\psi)]
                -2\mu\varepsilon^2 \cos^2\psi r_z \\
\label{A4.305}
   \dot r_z=2\mu\varepsilon^2 \cos^2\psi [r_y\cos(2\psi) -r_x\sin(2\psi)] .
\end{eqnarray}
By multiplication of Eqs. (\ref{A4.303}) and (\ref{A4.304}) by $\cos(2\psi)$ and 
$\sin(2\psi)$, respectively, and subsequent subtraction of the one equation from 
the other, we get 
\begin{equation}
\label{A4.306}
  \dot r_x +2\dot{\psi}r_y
  =-(\Delta +2D\varepsilon^2 \cos^2\psi) r_y 
     +2\mu\varepsilon^2 \cos^2\psi \cos(2\psi) r_z .
\end{equation}
Similarly, by multiplication of Eqs. (\ref{A4.303}) and (\ref{A4.304}) by 
$\sin(2\psi)$ and $\cos(2\psi)$, respectively, and subsequent addition of the 
resulting equations, we get
\begin{equation}
\label{A4.307}
  \dot r_y -2\dot{\psi} r_x
  =(\Delta +2D\varepsilon^2 \cos^2\psi) r_x 
   -2\mu \varepsilon^2 \cos^2\psi \sin(2\psi) r_z .
\end{equation}
The, performing an averaging of Eqs. (\ref{A4.305})-(\ref{A4.307}) over the 
phase $\psi$  using the relations $\langle \cos^2\psi \cos(2\psi) \rangle=1/4$ 
and $\langle \cos^2\psi \sin(2\psi) \rangle=0$ yields the truncated Bloch 
equations
\begin{equation}
\label{A4.124}
   \dot r_x =-(\delta +2\dot\phi +D\varepsilon^2) r_y 
             -\frac{\mu}{2} \varepsilon^2 r_z , \qquad
   \dot r_y=+(\delta +2\dot\phi +D\varepsilon^2)r_x , \qquad
   \dot r_z= \frac{\mu}{2} \varepsilon^2  r_y ,
\end{equation}
where $\delta =\Delta -2\omega$. Eqs. (\ref{A4.124}) possess a dynamic invariant 
that has a form similar to that of the original Bloch equations (\ref{A4.119}), 
i.e., $r^2_x +r^2_y +r^2_z =1$. The truncated Bloch equations (\ref{A4.124}), 
along with Eqs. (\ref{A4.123}) for $\varepsilon(x,t)$ and $\phi(x,t)$ of the EM 
vector potential pulse, constitute a closed system of equations describing the 
approximate dynamics of the SCQMM. Its solutions are obtained in the next 
Section.

\subsection{Exact Integration of the Truncated Equations}
The combination of Eq. (\ref{A4.123}) and the third of Eqs. (\ref{A4.124}) 
provides a relation between the slow amplitude and the phase of the EM vector 
potential pulse. Multiplication of the first of Eqs. (\ref{A4.123}) by 
$\varepsilon$ gives
\begin{equation}
\label{A4.127}
  {\left[\frac{\partial}{\partial t} +c \frac{\partial}{\partial x} \right]
       \varepsilon^2 (x,t)} =  -\chi \frac{\mu}{\omega} \varepsilon^2 (x,t) r_y .
\end{equation}
Subsequently, the time-derivative of the second of Eqs. (\ref{A4.123}), in which 
$\dot{R}_z =\dot{r}_z$ is replaced from the third of Eqs. (\ref{A4.124}), gives
\begin{equation}
\label{A4.127b}
   {\left[\frac{\partial}{\partial t} +c \frac{\partial}{\partial x} \right]  
      \dot\phi (x,t)} =-\chi \frac{\mu D}{2 \omega} \varepsilon^2 (x,t) r_y .
\end{equation}
From Eqs. (\ref{A4.127}) and (\ref{A4.127b}), and by taking into account the 
independence of the slow temporal and spatial variables, we get  
\begin{equation}
\label{A4.128}
   2\dot\phi(x,t) = D \varepsilon^2 (x,t) +const. ,
\end{equation}
where the constant of integration can be set equal to zero. Using Eq. 
(\ref{A4.128}), the truncated Bloch equations (\ref{A4.124}) can be written as
\begin{equation}
\label{A4.129}
  \dot r_x=-(\delta +2 D \varepsilon^2) r_y , \qquad
  \dot r_y=+(\delta +2 D \varepsilon^2) r_x -\frac{\mu}{2} \varepsilon^2  r_z , \qquad
  \dot r_z= \frac{\mu}{2} \varepsilon^2  r_y .
\end{equation}
The latter can be written in a simpler form using the unitary transformation
\begin{equation}
\label{A4.308}
   r_x =S_x \cos\Phi -S_z \sin\Phi , \qquad  r_y =S_y , \qquad 
   r_z =S_z \cos\Phi +S_x \sin\Phi , 
\end{equation}
where $\Phi$ is a constant transformation angle (to be determined). The 
truncated Bloch equations for the $r_i$, $i=x,y,z$, can be written in terms of 
the new Bloch vector components $S_i$, using a procedure similar to that used in 
the previous Section to obtain Eqs. (\ref{A4.124}). Sustituting Eqs. 
(\ref{A4.308}) into Eqs. (\ref{A4.129}), we get 
\begin{eqnarray}
\label{A4.310}
   \dot S_x \cos\Phi -\dot S_z \sin\phi=-(\delta +2 D \varepsilon^2) S_y , \\
\label{A4.309}
   \dot S_y=\left[ (\delta+2D\varepsilon^2) \cos\Phi -\frac{\mu}{2} \varepsilon^2
    \sin\Phi\right] S_x
   -\left[ (\delta + 2 D \varepsilon^2) \sin\Phi +\frac{\mu}{2} \varepsilon^2
    \cos\Phi \right] S_z , \\
\label{A4.311}
   \dot S_x\sin\Phi +\dot S_z\cos\Phi=-\frac{\mu}{2} \varepsilon^2  S_y .
\end{eqnarray}
Multiplying Eqs. (\ref{A4.310}) and (\ref{A4.311}) by $\cos\Phi$ and $\sin\Phi$, 
respectively, and then adding them together, we get 
\begin{equation}
\label{A4.4000}
   \dot S_x =\left\{ \varepsilon^2 \left[ \frac{1}{2}\mu \sin\Phi -2 D \cos\Phi \right]
                    -\delta \cos\Phi \right\}  S_y .
\end{equation}
Similarly, multiplying Eqs. (\ref{A4.310}) and (\ref{A4.311}) by $\sin\Phi$ and 
$\cos\Phi$, respectively, and then subtracting the one equation from the other, 
we get
\begin{equation}
\label{A4.401}
   \dot S_z =\left\{ \varepsilon^2 \left[ \frac{1}{2}\mu \cos\Phi +2 D \sin\Phi \right]
                    -\delta \sin\Phi \right\}  S_y .
\end{equation}
Let us define the transformation angle through the relation 
$\tan \Phi \equiv \gamma =\frac{4 D}{\mu}$, so that $\cos\Phi=\pm \sigma$ and 
$\sin\Phi=\pm \sigma \gamma$ where $\sigma =1 / \sqrt{1+\gamma^2}$. The choice 
of the sign is irrelevant and here we pick positive sign for both functions. 
Using that $\Phi$ and the definitions $W=\sqrt{(4D)^2 +\mu^2}$ and 
$\eta= -\delta \mu / W$, Eqs. (\ref{A4.4000}), (\ref{A4.309}), and (\ref{A4.401}) 
obtain their final form 
\begin{eqnarray}
\label{A4.197}
   \dot S_x =+\eta S_y , \qquad
   \dot S_y =-\eta S_x +\left[ \eta \gamma -\frac{1}{2} W \varepsilon^2 \right] S_z , \qquad
   \dot S_z =\left[-\eta \gamma +\frac{1}{2} W \varepsilon^2 \right] S_y .
\end{eqnarray}
For the investigation of "coherent propagation" of an EM potential pulse, the 
resonance condition is applied, i.e., $\eta=0$ or $\omega =\Delta / 2$, and then
Eqs. (\ref{A4.197}) become
\begin{eqnarray}
\label{A4.402}
   \dot S_x =0 , \qquad
   \dot S_y =-\frac{1}{2} W \varepsilon^2 S_z , \qquad
   \dot S_z =+\frac{1}{2} W \varepsilon^2 S_y .
\end{eqnarray}
Combining the second and third of Eqs. (\ref{A4.402}) and integrating, we obtain 
the "resonant" conservation law $S_y^2 +S_z^2 =const.$. Assuming that all the 
qubits are in the ground state at $t=-\infty$, we have the initial conditions 
$r_x (t=-\infty) =r_y (t=-\infty) =0$ and $r_z (t=-\infty) =-1$ which are 
transformed into $S_x (t=-\infty) =-\gamma \sigma$, $S_y (t=-\infty) =0$, and 
$S_z (t=-\infty) =-\sigma$ through Eq. (\ref{A4.308}). Applying the initial 
conditions to the resonant conservation law, we get 
\begin{equation}
\label{A4.403}
   S_y^2 +S_z^2 =\sigma^2 .
\end{equation}
In the following, we seek solution of the form 
$\varepsilon=\varepsilon(\tau=t-x/v)$ and $S_i =S_i (\tau=t-x/v)$, with 
$i=x,y,z$. By changing the variables in the first of Eqs. (\ref{A4.123}), with 
$r_y$ being replaced by $S_y$, we get after rearramgement    
\begin{equation}
\label{A4.404}
   \frac{\varepsilon_\tau}{\varepsilon} =\chi \frac{\mu}{2\omega} \frac{v}{c-v} S_y .
\end{equation}
Then, combining Eq. (\ref{A4.304}) with the third of Eqs. (\ref{A4.402}) and 
integrating, we get
\begin{equation}
\label{A4.405}
   \varepsilon^2 (\tau) =\chi \frac{2\mu}{\omega W}  \frac{v}{c-v} [S_z (\tau) +\sigma] , 
\end{equation}
where the conditions $\varepsilon (-\infty)$ and $S_z (-\infty) =-\sigma$ were 
used. The system of Eqs. (\ref{A4.403})-(\ref{A4.405}) for $\varepsilon$, $S_y$, 
and $S_z$ can be integrated exactly; the variables $S_y$ and $S_z$ can be 
eliminated in favour of $\varepsilon$ to give 
$\varepsilon_\tau =\lambda \varepsilon^2 \sqrt{a +b \varepsilon^2}$, in which 
the constants are defined as $a=2\sigma/\kappa$, $b=-1/\kappa^2$, 
$\lambda=\chi \frac{\mu}{2\omega} \frac{v}{c-v}$, 
$\kappa=\frac{2\mu}{\omega W}  \frac{v}{c-v}$ to simplify the notation. The 
equation for $\varepsilon$ can be readily integrated
\begin{equation}
\label{A4.406}
   \int_{\varepsilon_0}^\varepsilon 
       \frac{d\varepsilon}{\varepsilon^2 \sqrt{a+b \varepsilon^2}}
          =\lambda\int_{\tau_0}^\tau d\tau \Rightarrow
   -\frac{\sqrt{a+b \varepsilon^2}}{a \varepsilon} =\lambda (\tau -\tau_0) ,
\end{equation}
where we have set 
$\varepsilon_0 \equiv \varepsilon (\tau=\tau_0) =\sqrt{2\sigma \kappa}$ to 
eliminate the boundary term resulting from the integral over $\varepsilon$. 
Solving Eq. (\ref{A4.406}) for $\varepsilon$, we finally get a 
Lorentzian-like slowly varying amplitude 
\begin{equation}
\label{A4.407}
    \varepsilon (\tau) =\frac{\varepsilon_0}{\sqrt{1 +\tau_p^{-2} (\tau -\tau_0)^2  }} , 
\end{equation}
where $\varepsilon_0 =\sqrt{-{a} / {b}} =\sqrt{ 2\sigma \kappa}$ and 
\begin{equation}
\label{A4.4070}
   \tau_p^{-2} =-\frac{a^2 \lambda^2}{b} 
       =\left( \chi \frac{\sigma \mu}{\omega} \right)^2 \left( \frac{v}{c-v} \right)^2 .
\end{equation}
\newpage

\bibliographystyle{model6-num-names}

\end{document}